# Directed Non-Targeted Mass Spectrometry and Chemical Networking for Discovery of Eicosanoids

Jeramie D. Watrous[1*], Teemu J. Niiranen[2,3*], Kim A. Lagerborg[1], Mir Henglin[4], Yong-Jian Xu[1], Sonia Sharma[5], Ramachandran S. Vasan[6,7,8], Martin G. Larson[6,9], Aaron Armando[10], Oswald Quehenberger[1], Edward A. Dennis[10], Susan Cheng[4,5#], Mohit Jain[1#%]

**Affiliations:**

1: Departments of Medicine and Pharmacology, University of California, San Diego, California, USA

2: Department of Medicine, Turku University Hospital and University of Turku, Turku, Finland

3: Department of Public Health Solutions, National Institute for Health and Welfare, Turku, Finland

4: Cardiovascular Division, Department of Medicine, Brigham and Women's Hospital, Harvard Medical School, Boston, Massachusetts, USA

5: La Jolla Institute of Allergy and Immunology, La Jolla, California, USA

6: Framingham Heart Study, Framingham, Massachusetts, USA

7: Department of Epidemiology, School of Public Health, Boston University, Boston, Massachusetts, USA

8: Sections of Preventive Medicine and Cardiovascular Medicine, School of Medicine, Boston University, Boston, Massachusetts, USA

9: Department of Biostatistics, School of Public Health, Boston, Massachusetts, USA

10: Departments of Chemistry and Biochemistry and Pharmacology, University of California, San Diego, California, USA

*: these authors contributed equally
#: senior authors
%: lead contact

**For Correspondence:**
Mohit Jain MD, PhD; University of California, San Diego; mjain@ucsd.edu
Susan Cheng MD, MPH; Brigham and Women's Hospital; Harvard Medical School; scheng@rics.bwh.harvard.edu

## ABSTRACT

Eicosanoids and related species are critical, small bioactive mediators of human physiology and inflammation. While ~1100 distinct eicosanoids have been predicted to exist, to date, less than 150 of these molecules have been measured in humans, limiting our understanding of eicosanoids and their role in human biology. Using a directed non-targeted mass spectrometry approach in conjunction with computational chemical networking of spectral fragmentation patterns, we find over 500 discrete chemical signals highly consistent with known and putative eicosanoids in human plasma, including 46 putative novel molecules not previously described, thereby greatly expanding the breath of prior analytical strategies. In plasma samples from 1500 individuals, we find members of this expanded eicosanoid library hold close association with markers of inflammation, as well as clinical characteristics linked with inflammation, including advancing age and obesity. These experimental and computational approaches enable discovery of new chemical entities and will shed important insight into the role of bioactive molecules in human disease.

## INTRODUCTION

Eicosanoid and eicosanoid-like molecules (collectively termed 'eicosanoids') represent small polar lipid metabolites produced through extensive and variable oxidation of mainly 18 to 22 carbon polyunsaturated fatty acids (PUFA's), including omega-6 fatty acids, such as arachidonic acid, linoleic acid, and adrenic acid, and omega-3 fatty acids, such as docosahexaenoic acid (DHA), docosapenaenoic acid (DPA), and eiocosapentaenoic acid (EPA).[1-7] Given their chemical diversity, hundreds to thousands of theoretically possible eicosanoid chemical compounds have been catalogued in databases worldwide, subdivided into chemical and functional families including prostaglandins, leukotrienes, and resolvins, among others.[8] Eicosanoids are highly conserved among species, originating in elementary microbes and present in fungi, plants and mammals.[9-12] Eicosanoids signal through cell surface G-protein coupled receptors, tyrosine kinase receptors and intracellular nuclear receptors, mediating a number of diverse homeostatic functions in humans including host immune activation, cellular development, ion transport, muscle contraction, thrombosis, and vasomotor tone, as well as likely many yet undiscovered processes.[13-17] In the setting of acute infection, eicosanoids directly regulate the classic inflammatory triad of fever, edema and pain and provide early benefit by promoting clearing of invasive pathogens and wound healing.[3,18-24] Eicosanoids have also been implicated in the setting of chronic aseptic systemic inflammation, as is associated with a number of human diseases including obesity, diabetes, cardiovascular disease, cancer, and autoimmunity.[25-31] Given their profound biological effects, eicosanoids have been extensively targeted for therapeutic purposes with pharmaceutical agents, including non-sterol anti-inflammatory drugs (NSAIDs) such as acetylsalicylic acid (aspirin), being among the most widely utilized medications in medicine today.[3,5]

Current understanding of eicosanoid biology is limited, however, by technical challenges related to the measure of eicosanoids, owing to their low abundance, dynamic nature, and extensive isometry in chemical structure.[32-34] To date, measurement of eicosanoids has required highly sensitive, liquid chromatography-mass spectrometry (LC-MS) systems using triple-quadrupole (QQQ) based targeted analysis of known eicosanoid entities. Given the limited availability of commercial eicosanoid standards, much of eicosanoid biology has therefore focused on a well-established subset of 50-150 eicosanoid compounds, with typical studies reporting ~50 eicosanoids per sample, capturing only a small fraction of total eicosanoid biology.[35-39] These targeted approaches have limited the study of eicosanoids as well as notably preclude discovery of novel eicosanoids that may be mediating tissue and host biology.

Herein, to greatly expand the repertoire of eicosanoids assayed in humans, we develop a "directed non-targeted mass spectrometry" approach using a high mass accuracy LC-MS for measurement of bioactive lipid species. Using chemical networking of MS/MS spectral fragments and system analysis of chemical patterning we find over 500 distinct eicosanoid entities in human plasma, including 362 eicosanoids not previously documented in humans, and 46 putative novel eicosanoids not previously described. In plasma samples across 1500 individuals, we find a number of specific and novel eicosanoids as highly associated with markers of systemic inflammation and common clinical characteristics linked to inflammatory diseases, including advancing age and increasing obesity. These analytical approaches greatly expand the spectrum of eicosanoids observable in complex organisms, enable discovery of new eicosanoids in humans, and provide a foundation for new insight into the role of bioactive lipids in mediating human health and disease.

## RESULTS

### *Chemical Characterization of Eicosanoids*

Comprehensive detection of eicosanoids in a complex milieu, particularly for those chemical species for which no commercial standards or reference data exist, requires distinguishing chemical features. We therefore first sought to pinpoint common chemical characteristics unique to eicosanoids. Review of the LIPID MAPS chemical database (www.lipidmaps.org, March 2017 version) revealed 1125 theoretical, structurally distinct molecules classified as eicosanoids (**Table S-1**).[8] Virtually all eicosanoids were found to lie within a narrow mass range of 300-400 daltons. Additionally, all eicosanoids contain high degrees of oxidation and unsaturation relative to other biomolecules within this same mass window, resulting in abnormally high mass defect, or the non-whole number portion of a compounds mass. Given the combinatorial nature of their biosynthesis, eicosanoids were also found to exhibit extensive isomerism wherein the 1125 theoretical eicosanoids derived from only 168 unique chemical formulas (**Fig. 1a**), and 786 of the 1125 eicosanoid compounds stem from only 23 unique chemical formulas. To determine if these 168 eicosanoid chemical formulas occupy unique masses relative to other biomolecules, the accurate mass for each formula was searched against the 75,000 compounds present in the Human Metabolome Database (HMDB) using a 2 ppm mass error window.[40] Of the 168 formulas, only 9 formulas (~5%) returned additional matches to non-eicosanoid compounds with the majority of these mapping to sterol, bile acid and free fatty acid metabolites, thereby suggesting that eicosanoids exhibit fairly specific and unique mass characteristics.

To determine whether eicosanoids may be further experimentally distinguished particularly from isobaric, non-eicosanoid species including sterol, bile acid and free fatty acid metabolites, we utilized a ‘directed, non-targeted’ SPE-LC-MS approach with offline solid phase extraction for enrichment of eicosanoid species, reverse phase chromatography, and high mass accuracy mass spectrometry operating in full scan mode (see Methods). Commercial standards for 238 eicosanoids, 18 bile acids, 21 sterols, 15 free fatty acids, and 16 endocannabinoids (since their in-source fragments may be isobaric with eicosanoids) were injected as neat solutions (**Table S-2**). LC-MS analysis revealed that all 238 eicosanoid standards eluted between 1.0 and 5.75 minutes on the LC gradient (**Fig. 1b**), exhibited consistent peak shapes (FWHM < 0.1mins and tailing factor 0.8-1.2), and were chromatographically resolved from sterols (elution between 0.5 and 1.25 minutes), as well as fatty acids and endocannabinoids (elution between 6.0 to 7.0 minutes) (**Fig. 1c**). Bile acids exhibited chromatographic overlap with eicosanoids, though they exhibited noticeably wider peak widths and increased peak tailing (>1.5), particularly tauro-variants (**Fig. 1c**). Collectively this suggests that potential eicosanoids may be distinguished from other known metabolites using chemical characteristics, including accurate mass features, chromatographic retention, and peak shape.

*Monitoring of Eicosanoids in Human Plasma*

In humans, eicosanoids are produced locally in tissues, leak into the vasculature, and enter the systemic circulation.[3,4,21] To determine the full catalog of eicosanoids present in circulation, human plasma was pooled and assayed using our directed, non-targeted LC-MS methods. Human plasma revealed 134 distinct spectral peaks definitively identified as eicosanoids by precise matching of retention time, accurate mass, and tandem MS fragmentation pattern to one of 238 commercially available, known eicosanoid standards (**Fig. 1d**); these include numerous well described members of prostaglandin, HETE, DHET, HODE and additional eicosanoid families. Human plasma also revealed an additional 745 spectral peaks across 56 distinct chemical formulas consistent with eicosanoids based on accurate mass (<2ppm) matching to one of 1125 theoretical eicosanoids in the LIPID MAPS database (**Table S-1**), retention time range (1.0 - 5.75 minutes), and peak shape (FWHM < 0.1min and tailing factor 0.8-1.2) characteristics. These 745 spectral peaks include a number of entities found to be isobaric with the 134 definitively identified eicosanoids (**Fig. 2a**). For example, the extracted ion chromatogram (XIC) for m/z 353.2333 exhibited 18 isobaric chromatographic peaks with commercial standards accounting for 15, XIC for m/z 375.2177 exhibited 32 isobaric chromatographic peaks with commercial standards accounting for only 3, and XIC for m/z 313.2384 exhibited 20 isobaric chromatographic peaks with commercial standards accounting for only 2 (**Fig. 2a**). Due to eicosanoids occupying a unique chemical space, these isobaric species are very likely to represent additional isobaric eicosanoids. For instance, search of the LIPID MAPS database for m/z 313.2384 revealed a possible 29 eicosanoid species which could potentially correspond to these 20 observed features.

*Chemical Networking for Classification of Putative Known Eicosanoids*

The lack of commercial standards or reference tandem mass spectra have to date precluded identification or even chemical classification of unknown metabolites, including eicosanoids. To overcome this limitation, we reasoned that chemical similarity between a known eicosanoid and unknown molecule through comparison of spectral fragmentation patterns may allow for *putative* classification of unknown molecules as eicosanoids. To perform this analysis, we utilized chemical networking of spectral fragmentation for which tandem mass spectra for each unknown signal were compared to reference spectra from known compounds for presence of both identical mass fragments as well as similar mass shifts between fragment peaks.[41,42] This approach allows experimental mass spectra to be matched to identical reference compounds for definitive identification, as well as in the absence of definitive identification, to be matched to related, non-identical compounds for putative chemical classification. To determine the utility of chemical networking for chemical classification, tandem MS fragmentation patterns were obtained for 238 eicosanoid commercial standards and 600 diverse, non-eicosanoid standards. Fragmentation patterns from these standards were chemically networked against over 50,000 metabolites for which spectral fragmentation data were publicly available, including additional tandem MS spectra for the 238 commercial eicosanoid standards collected by third parties. From this analysis, experimentally derived MS/MS spectra for 205 of the 238 eicosanoid standards were found to chemically network to corresponding spectra for the same compound present in the public library suggesting a sensitivity for matching of 86%. For those 33 of the 238 eicosanoids that did not match, manual inspection of the tandem MS patterns revealed poor fragmentation of the parent ion, with often only the parent ion and loss of water present in the fragmentation spectra, precluding chemical matching which required a minimum of 6 matched fragment peaks. Among the 600 non-eicosanoid standards, none were found to chemically network with eicosanoids and were instead found to be chemically similar to one of 50,000 other molecules present in the public libraries, suggesting near perfect specificity for this approach. This specificity for chemical matching is consistent with the fact that eicosanoids, particularly cyclized variants, exhibit unique fragmentation mechanisms compared to other classes of compounds.[35,36,43-46] Importantly, this high specificity is critical in limiting false positive matches when extending the chemical networking to unknown molecules for classification as eicosanoids.

We first applied this chemical networking approach to the 18 unknown isobaric features at m/z 313.2384 (**Fig. 2a**). When tandem mass spectra from these chromatographic features at were subjected to the chemical networking strategy, 16 of the 18 features were specifically matched to one of the 238 eicosanoid standard compounds (**Fig. 2b**). Matched tandem MS spectral pairs were manually inspected and were found to be consistent with sharing a common base structure. **Figure 2c** shows one such comparison between the tandem mass spectra for 9,10-diHOME and an unknown isobaric peak at retention time 3.13 minutes for which nearly identical fragmentation patterns are observed, indicating a very similar chemical structure. As such, matching of unknown compounds based on accurate mass and analogous fragmentation patterns to a known eicosanoid may allow such molecules to be classified as *putative known* eicosanoids. This approach suggests that chemical networking of unknown molecules to known chemical entities may be utilized to identify and study additional, high confidence, putative eicosanoids in human circulation.

We next extended this chemical networking to all 745 spectral features identified as potential eicosanoids, based on accurate mass match to LIPID MAPS database entries, but for which no chemical standard or reference spectra was available. The results from the search were visualized as a mass spectral network (**Fig. 3a**), where top scoring, high confidence, chemical matches between experimental and reference fragmentation data were clustered together and displayed in order to represent chemical relationships among compound families. Of the 745 unknown experimental tandem MS spectra, 362 spectra were chemically networked with at least one eicosanoid standard and did not form secondary connections with any one of over 50,000 non-eicosanoid compounds present in the library repository, thereby classifying these 362 metabolites as *putative known* eicosanoids (**Table S-3**). The remaining 383 tandem MS spectra that did not chemically network with a known eicosanoid were mostly cluster clustered with either non-eicosanoid compounds (such as glucuronides, sterols, bile acids, and fatty acid ethyl esters) or clustered within themselves. The lack of similarity in tandem MS patterns with any reference spectra thereby suggest that these 383 chromatographic peaks cannot be considered as eicosanoid in nature without further evidence.

Tandem MS spectral pairs from the chemical networking of 362 putative knowns with the 238 reference eicosanoid standards were manually inspected and found to be consistent with both compounds sharing a common key substructure. For instance, in the molecular tandem MS network, the tandem MS spectra for 12-oxo LTB4 was clustered with four other tandem MS spectra from unknown signals (**Fig. 3b**). When the tandem MS spectra for 12-oxo LTB4 and one of these unknowns was compared, it was observed that their patterns were highly similar thus indicating a common chemical structure (**Fig. 3c-d**). Importantly, algorithms for automated *in silico* fragmentation, such as CFM-ID [47] and MetFrag [48], were not able to accurately predict fragments for eicosanoid compounds, especially cyclized variants such as prostaglandins (**Fig. S-2**), given the unpredictable nature of eicosanoid fragmentation.[35,43-46,49-52] However, these irregular fragmentation mechanisms allow for far fewer false positive matches to non-eicosanoid compounds. All 362 spectral matched were manually inspected to confirm similarity with 100 of these matches plotted in **Figure S-3**. Collectively, these analyses suggest that directed non-targeted LC-MS/MS in conjunction with chemical networking may be utilized to identify an additional 362 putative known eicosanoids in human circulation, thereby greatly extending the analytical breath of current approaches.

*Systems Chemistry for Identification of Novel Eicosanoids*

In addition to the ~500 known or putative known eicosanoids found in human plasma, we next sought to discover novel eicosanoids yet undescribed in any chemical database including LIPID MAPS, that may also be present in humans. Given the unique chemical properties of eicosanoids, we reasoned additional systems chemistry approaches may prove useful in identification of novel eicosanoids. Eicosanoids typically originate from one of a handful possible PUFAs, which are extensively modified through both enzymatic and non-enzymatic biosynthetic mechanisms. Across the 1125 theoretical eicosanoids, the extent of these modifications is nearly combinatorial where compounds exist at almost every $H_{n+2}$ and/or $O_{n+1}$ formula increment. These chemical patterns within eicosanoid families may be visualized by plotting the nominal mass versus the mass defect, here defined as the non-whole number value of the monoisotopic mass, for each compound within the family.[53] For instance, among the prostaglandin family of eicosanoids, this type of plot reveals four distinct chemical series (**Fig. 4a**), where each series is incremented by addition of two hydrogens and the distance between each series represents the addition of a single oxygen. When extending this analysis beyond prostaglandins, distinct gaps in these formulaic patterns were readily observed. Among docosanoids starting at formula $C_{22}H_{34}O_2$, database entries for eicosanoids were present for every addition of oxygen up to $C_{22}H_{34}O_6$ with the lone exception of $C_{22}H_{34}O_3$ (**Fig. 4b**). When this formula's calculated nominal mass of m/z 345.2435 was searched within the non-targeted LC-MS data collected from human plasma, a spectral peak at 4.72 minutes was matched and produced an tandem MS pattern that was nearly identical with that of the known docosanoid 10-HDoHE (**Fig. 4c**). Manual interpretation of the fragments showed that while 10-HDoHE has four double bonds between carbons 11 and 22, this novel eicosanoid only has three. This result indicated that the gaps in chemical series within the eicosanoid compounds may be utilized to identify *putative novel* eicosanoids.

To determine if additional novel eicosanoids were present in human plasma, all 1125 theoretical eicosanoids listed in the LIPID MAPS database were plotted as nominal mass versus their mass defect and computationally analyzed for missing values along $H_{n+2}$ and $O_{n+1}$ series (**Fig. S-4**). This analysis was extended to include one entry preceding the lowest formula value in the series as well as one entry past the final value as long as the number of oxygens was constrained between 3 and 8 and the number of hydrogens was between 24 and 40, which represents the maximum and minimum values observed for known eicosanoids of 18 to 22 carbons. This analysis revealed 214 missing unique formulas spread across almost all subfamilies of eicosanoids with each one representing a potential novel eicosanoid. We searched in human plasma for compounds with accurate masses corresponding to these 214 chemical formulas using our directed, non-targeted LC-MS approaches, findings 184 distinct chromatographic peaks present at 45 of the 214 chemical formulas, with each of these peaks having accurate mass (<2ppm) matching to a theoretical formula, a retention time range (1.0 - 5.75 minutes), and a peak shape (FWHM < 0.1min and tailing factor 0.8-1.2) consistent with eicosanoids. To determine if these peaks were structurally related to known eicosanoids, targeted tandem MS spectra was collected for each of the 184 chromatographic peaks and chemically networked. The resulting analysis revealed that 46 of the 184 putative novel eicosanoids clustered with at least one known eicosanoid and none of the 50,000 non-eicosanoid metabolites, indicating structural similarity is present between these putative novel eicosanoids with known members of the compound families (**Table S-4**). Each of the 46 spectral pairs resulting from the chemical networking search were manually inspected and found to be consistent with those of the eicosanoid reference standards with 5 examples shown in **Figure S-5**. One such example is an unknown compound m/z 347.2600, which corresponds to the chemical formula $C_{22}H_{36}O_3$, and was spectrally matched to the octadecanoid 13-HoTrE (**Fig. S-5a**). Comparison of the tandem MS fragmentation patterns between the unknown and known compounds revealed that all carboxyl-containing fragments from carbons C7 to C1 were identical between the two compounds while all carboxyl-containing fragments from carbons C12 or C13 were shifted by 54.0472 daltons. This shift indicates that between carbons C12 and C7 there is an additional $C_4H_6$ present in the novel eicosanoid, and its structure is consistent with a novel set of HDoHE family of compounds in which two double bonds are converted to single bonds. This systems chemistry approach thereby enables discovery of novel eicosanoids, identifying an additional 46 putative novel eicosanoids in human circulation.

*Relation of eicosanoids to human inflammation*

Eicosanoid lipid mediators play critical roles in regulating systemic pro- and anti-inflammatory responses in the setting of both acute infection and chronic inflammation.[3,5,22] To date, however, biological functions have only been described for a fraction of the ~150 commonly studied eicosanoids found in human plasma. To determine if any of the hundreds proposed putative known and putative novel eicosanoids have roles in human biology similar to those of known eicosanoids, plasma samples from 1500 fasting, community-dwelling individuals were assayed for eicosanoids using directed, non-targeted LC-MS. Definitive known, putative known, and putative novel eicosanoids were mapped to plasma samples by matching of retention time, accurate mass, and tandem MS fragmentation patterns, with focus on those eicosanoids previously observed in pooled plasma and present with signal-to-noise ratios above LOQ level in >95% of individuals. Eicosanoids were statistically associated with common clinical features linked to systemic inflammation, including advancing age and obesity (measured by body mass index [BMI]), as well as high sensitivity C-reactive protein (CRP), a well-established marker of systemic inflammation (**Fig. 5**).[54,55] For both clinical features and CRP measures, dozens of both known and putative eicosanoids were found to be highly related at 'metabolome-wide' adjusted statistical thresholds of $p<1x10^{-5}$, with a number of molecules even reaching 'high confidence levels' of $p<1x10^{-20}$. For each clinical phenotype, both positive and negative associations were observed (**Table S-5**), suggesting both pro- and anti-inflammatory activities for eicosanoids, consistent with the described function of eicosanoids. For identifiable eicosanoids, production of arachidonic acid metabolites (e.g. HETE's and HpETE's) showed positive correlation with BMI and CRP levels, suggesting a pro-inflammatory role for these molecules, while DiHDPA and DiHETrE metabolites showed negative correlations, suggesting an anti-inflammatory role; similar patterns have been observed in cellular systems upon acute activation of inflammatory pathways.[56] Moreover, particular putative eicosanoids, such as *m/z* 327.2176 were found to be commonly associated with multiple correlates of inflammation (p-value $8e^{-35}$ for BMI, $7e^{-27}$ for CRP), whereas other putative eicosanoids, such as *m/z* 349.2020 (p-value $1e^{-35}$), m/z 327.2162 (p-value $2e^{-20}$), and *m/z* 347.2590 (p-value $2e^{-12}$) were specifically associated with age, BMI or CRP levels, respectively, suggesting both universal and specific mediators of inflammation. Interestingly, for all three clinical phenotypes examined a number of putative known and putative novel eicosanoids were found to be highly associated, highlighting the discovery potential.

## DISCUSSION

In this report, we describe new approaches for the assay and identification of eicosanoids in human biosamples using directed, non-targeted LC-MS coupled to computational chemical networking. While traditional targeted methods for studying eicosanoids can produce very sensitive and quantitative measurements, directed, non-targeted analysis greatly expands upon the spectrum of eicosanoids measurable in complex organisms, enable discovery of hundreds of new eicosanoids in humans, and provide a foundation for new insight into the role of bioactive lipids in mediating human health and disease. This approach makes use of directed extraction whereby samples were prepared using methods designed specifically for isolation of eicosanoids (versus a general Bligh-Dyer lipid extraction), chromatographic separation designed to separate the extensive array of eicosanoid isomers, non-targeted mass spectrometric measured allowing capture of all possible eicosanoid species, and MS/MS chemical networking allowing for identification of previously unreported eicosanoids. Application of these approaches to human plasma revealed hundreds of putative eicosanoids, advancing upon the current state of the art analytical techniques and greatly expanding the current chemical repertoire, revealing 134 definitively known, 362 putative known, and 46 putative novel eicosanoids observable in human plasma. These molecules were validated by a number of methodologies, including extensive chemical networking of mass spectral fragmentation patterns and manual annotation of a subset of eicosanoids. These methods were also found to have high specificity for eicosanoid compounds given the unique and irregular mechanisms by which they fragment. Many of the putative eicosanoids described herein were found to be highly associated with both pro- and anti-inflammatory clinical characteristics and measures in a cohort of 1500 individuals. These eicosanoids and tandem MS spectral patterns have been provided in a database as a rich resource to the scientific community for further study and discovery (**Supplemental Information**). Additionally, as many compound classes exhibit similar chemical patterning as observed in eicosanoids, such as PUFAs, endocannabinoids, FAHFAs and sterols, this approach could be adapted to perform similar discovery based work in these other chemical families potentially expanding their number of measureable entities and revealing additional interesting biology.

While greatly expanding upon the number of eicosanoids reported in human plasma, the current experimental and computational approaches will continue to shed even more insight into eicosanoid biology as they are applied in complementary systems. While the current studies focus on human plasma, undoubtedly additional eicosanoids will be found localized to particular cells or tissues, or induced under particular biological conditions. Moreover, it is important to note that our described approaches exclude eicosanoids with novel fragmentation mechanisms different from those exhibited by chemical standards and therefore it is possible that a number of additional eicosanoids may exist; assay of these molecules will however require orthogonal systems. Finally, while our human studies across 1500 individuals focus upon those eicosanoids that were initially discovered in pooled human plasma and universally present, a number of additional known and putative novel eicosanoids were observed in specific individuals; lack of additional patient plasma however for follow-up validation preclude study of these signals.

As commercial standards only cover about ~20% of total unique theoretical eicosanoids, targeted studies on inflammatory pathways have been confined to studying the same subset of compounds despite potentially hundreds more eicosanoids being present in human circulation. As many of these compounds are likely modulated under particular conditions such as acute systemic inflammation or during development, or may occur in particular tissues and cells in a localized manner, their study is critical for advancing our understanding of these complex signaling processes. Furthermore, as a number of these putative eicosanoids were found to be highly associated with clinical characteristics and markers of inflammation, future studies will aim to further characterize these molecules including identifying the biochemical pathways responsible for their production and degradation, localizing individual eicosanoids to specific tissues of origin, establishing causal pro- and anti-inflammatory molecules, and elucidating their downstream signaling mechanisms. Importantly, the studies described herein highlight the discovery potential of directed, non-targeted LC-MS and corresponding computational chemical networking approaches, underscore the breath, complexity and diversity of eicosanoids present in human circulation, and provide a foundation for new insight into the role of bioactive lipids in mediating human health and disease.

**METHODS**

Chemicals and consumables

Non-deuterated and deuterated chemical standards (See **Table S-2**) were purchased from Cayman Chemical and IROA Technologies. LCMS grade solvents used for sample preparation and metabolomic analysis including methanol, acetonitrile, water, and isopropanol were purchased from Honeywell International Inc while LCMS grade ethanol and acetic acid were purchased from Sigma-Aldrich. All pipetting instruments and consumables were purchased from Eppendorf. For sample preparation, 250μL V-bottom, 500μL V-bottom, and 1.2mL deep well 96-well plates were purchased from Thermo Scientific, Axygen and Greiner, respectively. LCMS amber autosampler vials and tri-layer vial caps were purchased from Agilent Technologies and 300μL glass inserts were purchased from Wheaton. Strata-X polymeric 96-well (10mg/well) solid phase extraction plates as well as Kinetex C18 1.8μm (100 x 2.1 mm) UPLC columns were purchased from Phenomenex Inc. UPLC BEH RP-18 guard columns were purchased from Waters Inc. EZ-Pierce 20μm aluminum foil was purchased from Thermo Scientific. Pooled plasma was purchased from BioreclamationIVT.

LC-MS and spectral data handling

LC-MS was performed using a Thermo Vanquish UPLC coupled to a Thermo QExactive Orbitrap mass spectrometer. For data processing, in-house R-scripts were used to perform initial bulk feature alignment, MS1-MS2 data parsing, pseudo DIA-to-DDA MS2 deconvolution, and CSV-to-MGF file generation. RAW to mzXML file conversion was performed using MSconvert version 3.0.9393 (part of the ProteoWizard Software Suite). Feature extraction, secondary alignment, and compound identification were performed using both mzMine 2.21 as well as Progenesis QI software suites. Statistical analysis was performed using R (3.3.3).

Plasma eicosanoid extraction

For preparation of plasma, samples were thawed at 4ºC over 8 hours in light free conditions. Once thawed, plasma was mixed thoroughly by orbital shaking at 500 rpm at 4ºC for 15 minutes. Cold (-20ºC) ethanol containing 20 deuterated eicosanoid internal standards was used to precipitate proteins and non-polar lipids as well as extract eicosanoids from the plasma through the addition of 1:4 plasma:ethanol-standard mixture to each well, shaking at 500rpm at 4ºC for 15 minutes, and centrifugation at 4200 rpm at 4ºC for 10 minutes. To allow for more efficient SPE loading, 65μL of supernatant was transferred to an Axygen 500μL V-bottom 96-well plate where each well contained 350μL of water. A second extraction of the protein pellet was performed by gently adding 65μL of -20ºC ethanol (containing no standards), gently hand vortex mixing for 30 seconds, and transferring another 65μL aliquot of supernatant to the same Axygen V-bottom 96-well plate, at which point samples are ready for SPE plate loading. The Phenomenex Strata-X polymeric 10mg/mL 96-well SPE plate was washed by sequentially adding and vacuum assisted eluting 600 μL of methanol and 600μL of ethanol and equilibrated by addition of 1.2mL of water per well. All solvents were added while taking care not to let wells run dry between pull through. The entire volume of the water-diluted samples were added to the SPE wells and allowed to gravity elute. Once the sample volume reached the top of the SPE bed, 600μL of 90:10 water:methanol was added to each well and slowly pulled through (5 mmHg vacuum) until no liquid was visualizen in the wells. Wells were further dried by increasing vacuum to 20 mmHg for 60 seconds. Bound metabolites were then eluted into an Axygen 500μL V-bottom 96-well plate by addition of 450μL of ethanol to each SPE plate well, with slow pull through (5 mmHg vacuum) after 120 second equilibration. Eluent was immediately dried in vacuo using a Thermo Savant vacuum concentrator operated at 35ºC. Once dry, samples were resuspended in a solution containing 40:60:0.1 methanol:water:acetic acid as well as 10μM CUDA internal standard by addition of 50μL of solution to each well, immediately sealing the plate, vortex mixing at 500rpm at 4ºC for 10 minutes. Samples were immediately transferred to a Greiner deep-well 96-well plate with each well containing a Wheaton 300μL glass insert (to prevent the rapid degradation of eicosanoid signals typically observed when eicosanoids are prepared in polypropylene plates). The sample plate was then immediately sealed and centrifuged at 500 rpm at 4ºC for 5 minutes to removed any air bubbles that might have formed within the glass inserts. Samples were stored at less than 2 hours at 4ºC until LC-MS analysis.

LC-MS/MS data acquisition

Once prepared, 20μL of sample was injected onto a Phenomenex Kinetex C18 (1.7μm particle size, 100 x 2.1 mm) column and separated using mobile phases A (70% water, 30% acetonitrile and 0.1% acetic acid) and B (50% acetonitrile, 50% isopropanol, 0.02% acetic acid) running the following gradient: 1% B from -1.00 to 0.25 minutes, 1% to 55% B from 0.25 to 5.00 minutes, 55% to 99% B from 5.00 to 5.50 minutes, and 99% B from 5.50 to 7.00 minutes. Flow rate was set at 0.375 mL/min, column temperature and mobile phase pre-heater was set at 50ºC, needle wash was set for 5 seconds post-draw using 50:25:25:0.1 water:acetonitrile:isopropanol:acetic acid. Mass detection was performed using a Thermo QExactive orbitrap mass spectrometer equipped with a heated electrospray ionization (HESI) source and collision-induced dissociation (CID) fragmentation. HESI probe geometry was manually optimized about the x, y and z axis by infusing CUDA and PGA2 standards into the mobile phase stream at 1%, 50% and 99% B. Final geometry was probe height of 0.75 B-C (between B and C markers about 3/4 the way to C), x-offset of 1 mm left of center, and a Y-offset of 1.70 according to the micrometer. Source settings used for all samples were: negative ion mode profile data, sheath gas flow of 40 units, aux gas flow of 15 units, sweep gas flow of 2 units, spray voltage of -3.5kV, capillary temperature of 265ºC, aux gas temp of 350ºC, S-lens RF at 45. Data was collected using an MS1 scan event followed by 4 DDA scan events using an isolation window of 1.0 m/z and a normalized collision energy of 35 arbitrary units. For MS1 scan events, scan range of m/z 225-650, mass resolution of 17.5k, AGC of 1e6 and inject time of 50ms was used. For tandem MS acquisition, mass resolution of 17.5k, AGC 3e5 and inject time of 40ms was used. Once target eicosanoids were discovered, additional tandem MS spectra was collected in a targeted fashion using Inclusion Lists to obtain high quality spectra for each compound.

Chemical networking of tandem MS spectra

Tandem MS spectra was collected in a targeted fashion for all observed chromatographic peaks in pooled plasma matching all known, putative known and putative novel eicosanoids based on accurate mass (m/z for [M-H]- within 5ppm mass error). Tandem MS spectra for each chromatographic peak were averaged, extracted and saved as individual MGF files. These MGF files (1063 in total) were uploaded to GNPS (http://gnps.ucsd.edu) for chemical networking analysis. Analysis was performed using both Networking and Dereplication workflows. For Networking the following thresholds: Precursor Ion Mass Tolerance of +/-0.05Da, Fragment Ion Mass Tolerance of +/-0.05Da, Minimum Cosine Score of 0.60, Minimum Matched peaks of 6, TopK of 10, Maximum Connected Component Size of 500, Minimum Cluster Size of 1, Library Score Threshold of 0.85, Minimum Matched Library Peaks of 6, Analog Search turned On, Maximum Analog Mass Difference of 85Da, Filter Precursor Window turned On, all other filters turned off. For Dereplication the following thresholds were used: Precursor Ion Mass Tolerance of +/- 0.05Da, Fragment Ion Mass Tolerance of +/-0.05Da, Minimum Cosine Score of 0.60, Minimum Matched peaks of 6, Analog Search turned On, Maximum Analog Mass Difference of 85Da, Filter Precursor Window turned On, all other filters turned off. All primary and third party databases, including our in-house database from the 238 eicosanoid commercial standards were used for matching. For all matches to library reference spectra, only the top scoring hit was considered. All resulting matches were manually checked for consistency in tandem MS fragmentation patterns between the library reference spectra and the experimental spectra.

Eicosanoid analysis of human samples

For evaluation of eicosanoids across a human cohort, we used fasting plasma specimens collected in 2005–2008 from N=1500 participants (mean age 66.2±9.0 years, 55% women) of the community-based Framingham Heart Study Offspring Cohort (**Table 1**).[57] All participants provided written informed consent. All study protocols were approved by the Brigham and Women's Hospital, Boston University School of Medicine, National Institute for Health and Welfare (Finland) and UC San Diego Institutional Review Boards. At the time of specimen collection, the participants underwent a medical history, physical, and laboratory assessment for high sensitivity C-reactive protein (CRP). CRP was measured using an immunoturbidometric method on a Roche Cobas 501. The intra- and inter-assay CVs for CRP were 2.5% and 4.5%, respectively.

Plasma samples were prepared and LC-MS analysis performed as described above. All data was converted to 32-bit mzXML data format using MSconvert version 3.0.9393 (part of the Proteowizard software suite). Initial bulk retention time correction was performed as described previously[58] whereby all data files for a given sample set were loaded into Mzmine 2.21 and the sample specific retention times for all deuterated internal standards as well as approximately 50 endogenous landmark peaks (defined as chromatographic features with %CV < 50% across all samples, minimum peak height of 500,000 counts and a no isobaric peaks within +/- 30 seconds of target) were exported and used to model non-linear retention time drift within each sample using cubic smoothing splines with 8 to 16 degrees of freedom within the model. Using these sample specific models, the retention times for all MS1 and MS2 scans with each mzXML file were adjusted so that the maximum drift observed was reduced from (on average) 0.15 minutes to 0.025 minutes across as many as 3000 sample injections. Parameters used for Mzmine peak extraction are as follows: Mass Detection threshold of 250k counts, Chromatogram Builder 0.025 minute min time span, min height of 500k counts and m/z tolerance of 5ppm, Chromatogram Deconvolution (Local Minimum Search) chromatographic threshold of 80%, search minimum RT range of 0.03 min, minimum relative height of 0.4%, min height of 500k counts, top-edge ratio of 1.2, peak duration of 0.025 to 0.5 minutes, Join Aligner mz tolerance of 5ppm, RT tolerance of 0.4 minutes, weight for retention and mass both at 90, Peak List Rows Filter retaining only peaks present in at least 50% of data files. Custom Library Search for all known, putative known and putative novel Eicosanoids was performed using an RT tolerance of 0.1 minutes and a mass tolerance of 5 ppm. Resulting features were manually denoised by visual inspection using the Mzmine peak list viewer where features exhibiting abnormal/poor peak shapes, inconsistent peak shapes and/or drastic shifts in retention time were deleted.

Exported peak lists for each file group were normalized by centering the batch mean for each m/z-feature to the global mean for all batches within the run set. Once normalized, features present in the MS1 data were filtered based on signal reliability by the used of the 'dtwclust' R-module as described previously where essentially all features were hierarchical clustered and automatically parsed based on patterns exhibited within their intensity profile across all samples when plotted with respect to injection order with features exhibiting stable patterns being retained while those exhibiting patterns indicative of chemical instability, misalignment or highly random distributions were removed.

For statistical analyses of human data, we constructed linear regression models relating each of the known, putative known, and putative novel eicosanoids with age, body mass index, and C-reactive protein. Prior to entry into regression analyses, eicosanoid variables were logarithmically transformed and standardized. Statistical analyses were performed using R v3.2.2 (R Foundation for Statistical Computing, Vienna, Austria).

## ACKNOWLEDGEMENTS

This work was supported by grants from the National Institutes of Health (K01DK116917 to J.D.W.; S10OD020025, R03AG053287, R01ES027595 to M.J.; R01HL134168 to S.C.; R01GM020501, R01DK105961, U19AI135972, P30DK063491, and U54GM069338 to

E.A.D. and O.Q.), support from the NHLBI to the Framingham Heart Study (contracts N01HC25195; HHSN268201500001I) as well as generous awards from Foundations, including the American Heart Association (CVGPS Pathway Award to S.C., M.J.), the Doris Duke Charitable Foundation #2015092 to M.J., Tobacco-Related Disease Research Program #24RT-0032 to M.J. and 24FT-0010 to J.D.W., UC San Diego Frontiers of Innovation Scholars Program to K.A.L., Emil Aaltonen Foundation to T.N., the Finnish Medical Foundation to T.N., and the Paavo Nurmi Foundation to T.N.. The funders play no role in the design of the study; the collection, analysis, and interpretation of the data; and the decision to approve publication of the finished manuscript. All authors had full access to all the data (including statistical reports and tables) in the study and can take responsibility for the integrity of the data and the accuracy of the data analysis.

## AUTHOR CONTRIBUTIONS

M.J., S.C., J.D.W., and T.N. designed the experiments. J.D.W., A.A., Y.X., and K.A.L. performed the experiments. J.D.W., T.N., M.J., S.C., V.S.R., M.G.L. and S.S. analyzed and interpreted data. J.D.W., T.N., M.J., S.C., E.A.D., and O.Q. wrote and edited the manuscript.

## DECLARATION OF INTERESTS

The authors declare no competing interests.

**FIGURE LEGENDS**

**Fig. 1: Directed non-targeted LC-MS based measure of eicosanoids in humans. a,** Histogram illustrating the high degree of isomerism amongst eicosanoids with number of database entries for each mass to charge (m/z). **b,** Extracted ion chromatograms for all 238 commercial eicosanoid standards. **c,** A mass defect versus retention time plot for eicosanoids and related compounds. **d,** Extracted ion chromatograms for all 134 spectral peaks observed in pooled plasma that were matched to commercial eicosanoid standards.

**Fig. 2: Identification of eicosanoids in human plasma. a,** Extracted ion chromatograms for m/z's 353.2333, 375.2177 and 313.2384 with the red asterisks indicating peaks for which commercial standards are available. **b,** Tandem MS spectra for all peaks present at m/z 313.2384, excluding the two peaks for which standards were available. Tandem MS spectra were clustered against known eicosanoids, with top 7 correlations shown. **c,** Comparison of tandem MS spectra from m/z 313.2384 chromatographic peak at 3.27 minutes with that of 9,10-DiHOME, revealing similar chemical structures.

**Fig. 3: Chemical networking for discovery of eicosanoids**. **a,** Tandem MS spectra for the 745 chromatographic features matching to chemical formulas of known eicosanoids was clustered against tandem MS for the 238 commercial standards with the resulting correlations displayed as a simplified molecular network. Here, red nodes indicate tandem MS spectra for commercial standards with blue nodes indicating tandem MS spectra from the 745 unknown compounds, with the size of the nodes reflecting number of unique spectra contained within each node. **b,** Expanded view of the 12-oxo-LTB4 node (red box from 3a) showing the observed m/z and retention time in minutes within parenthesis. **c,** Overlay of tandem MS spectra from 12-oxo-LTB4 and m/z 367 at 1.62 minutes showing high degree of similarity. **d,** Manual annotation of tandem MS fragments of m/z 367 at 1.62 minutes and how they relate to the structure of 12-oxo-LTB with blue arrows indicating identical fragments and orange arrows indicating a shift of $+CH_4O$ between carbons 11 and 13. See also Figure S3.

**Fig. 4: Systems chemistry for identification of novel eicosanoids. a,** Mass defect versus nominal mass plot of 15 chemical formulas, which compose the prostaglandin series of eicosanoids, illustrating common chemical patterns within eicosanoid families. **b,** Example of a series of chemical formulas that exhibit a gap, or missing formula entry, in the overall pattern. Untargeted data was searched again at these formula masses to determine if chromatographic features were present. **c,** Tandem MS spectra from one of these missing entries ($C_{22}H_{34}O_3$) was found to be highly similar to the tandem MS spectra for 10-HDoHE with both sharing common fragments (blue arrows) as well as analogous fragments that were shifted by the mass of two hydrogens (orange arrows). Manual annotation indicates that the novel eicosanoid molecule is identical to 10-HDoHE with the exception that it contains 3 instead of 4 double bonds between carbons 11 and 22. See also Figure S4 and S5.

**Fig. 5: Eicosanoid correlates of human inflammation.** Associations among known (blue), putative known (red), and putative novel (green) plasma eicosanoids with age, body mass index, and C-reactive protein across 1500 individuals. Data points represent the negative Log P values derived from linear regression models relating each eicosanoid with a given clinical variable.

## TABLES LEGEND

**Table 1: Clinical characteristics of human cohort.** Values are displayed as means ± SD, frequencies, or median and 25th, 75th percentiles.

## SUPPLEMENTAL FIGURE LEGENDS

**Fig. S1. Flow chart for eicosanoid discovery.** Flow chart showing experimental design of how directed non-targeted LC-MS/MS in conjunction with chemical networking was used to identify definitive known, putative known and putative novel eicosanoids.

**Fig. S2. Performance of automated tandem MS fragment assignment tools.** (A)Tandem MS spectra from commercial standards were inputed into the MetFrag webtool with LipidMaps serving as the library. On average, Metfrag was only able to assign ~5% of fragment peaks to corresponding chemical structures. Here are two examples using a linear and a cyclic eicosanoid where both results showed poor ability to assign spectral fragments to chemical structures with almost all of the dominant fragments not assigned. Note that matched peaks were re-colored red for visibility. (B) *In sili*co prediction of tandem MS fragments of commercial standards were subjected to fragmentation using CFM-id. Here is a typical example using a cyclic eicosanoid where many of the observed fragments fail to be predicted. CFM-id performed notably better at predicting fragments for linear eicosanoids, however, many fragments were still left unpredicted.

**Fig. S3. Examples of experimental tandem MS spectral matches to tandem MS spectra from eicosanoid commercial standards.** Each example shows the tandem MS spectra from the unknown compound (black) above the tandem MS spectra of the standard (green). Comparison of the spectra reveal presence of identical MS fragment peaks as well as, in some cases, peaks consistently shifted by a given mass indicating that the unkown compounds share a similar structure to that of the eicosanoid commercial standard.

**Fig. S4. Mass defect plots used for identification of chemical formulas potentially belonging to novel eicosanoid compounds.** (A) For each eicosanoid entry in the LipidMaps library, the nominal mass was plotted against the mass defect to reveal patterns in chemical formulas which compose the 1149 unique compounds. (B) By searching for gaps in the chemical iterations within each chemical family, 'missing' entries were identified and plotted in red. (C) Once data for each of the missing formulas was extracted and searched via chemical spectral networking, all formulas resulting in positive analog matches were left plotted in red relative to all entries from eicosanoids in Lipid MAPS.

**Fig. S5. Annotation of tandem MS spectra for potentially novel eicosanoids.** (A) m/z 347.2591 at 5.58 minutes. (B) m/z 345.2434 at 5.37 minutes. (C) m/z 345.2434 at 5.47 minutes. (D) m/z 345.2434 at 4.94 minutes. (E) m/z 357.2646 at 2.32 minutes.

## SUPPLEMENTAL INFORMATION TABLE LEGENDS

**Table S1.** List of 1125 LIPID MAPS database entries used as source database for mining non-targeted data for putative known and putative novel eicosanoids.

**Table S2.** List of all commercial standards used for assignment of definite known eicosanoids as well as identification of putative known and putative novel eicosanoids through chemical networking.

**Table S3.** Results for chemical networking for putative known eicosanoids.

**Table S4.** Results for chemical networking for putative novel eicosanoids.

**Table S5.** Results from statistical analysis for Figure 5.

## SUPPLEMENTAL INFORMATION DATA PACKAGE

**Data_Package.zip -** All MS/MS spectra (1 spectra per file) of eicosanoid commercial standards, the 362 putative known eicosanoids and the 46 putative novel eicosanoids. The file names here correspond with the input file names in Tables S3 and S4.

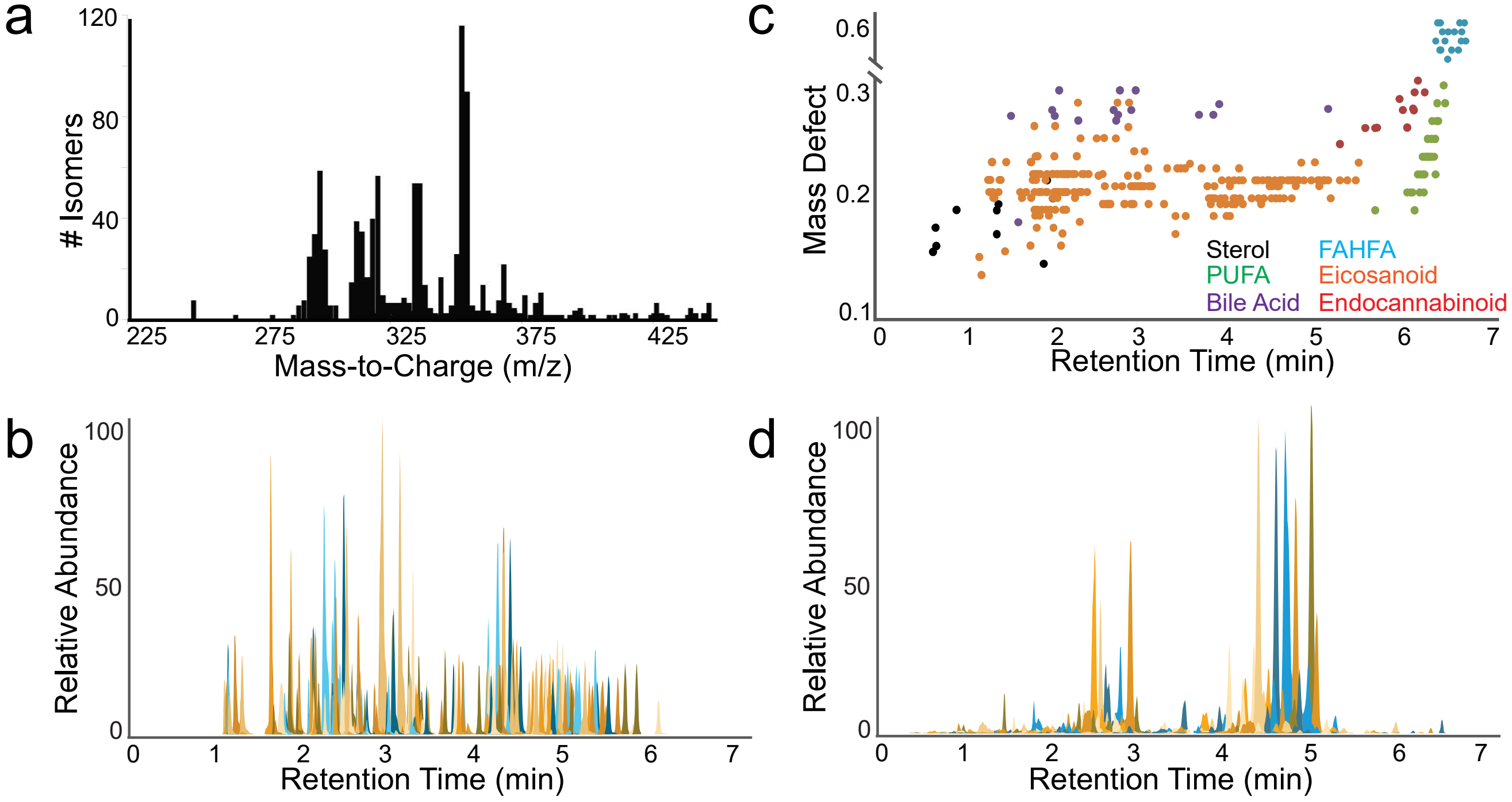

a

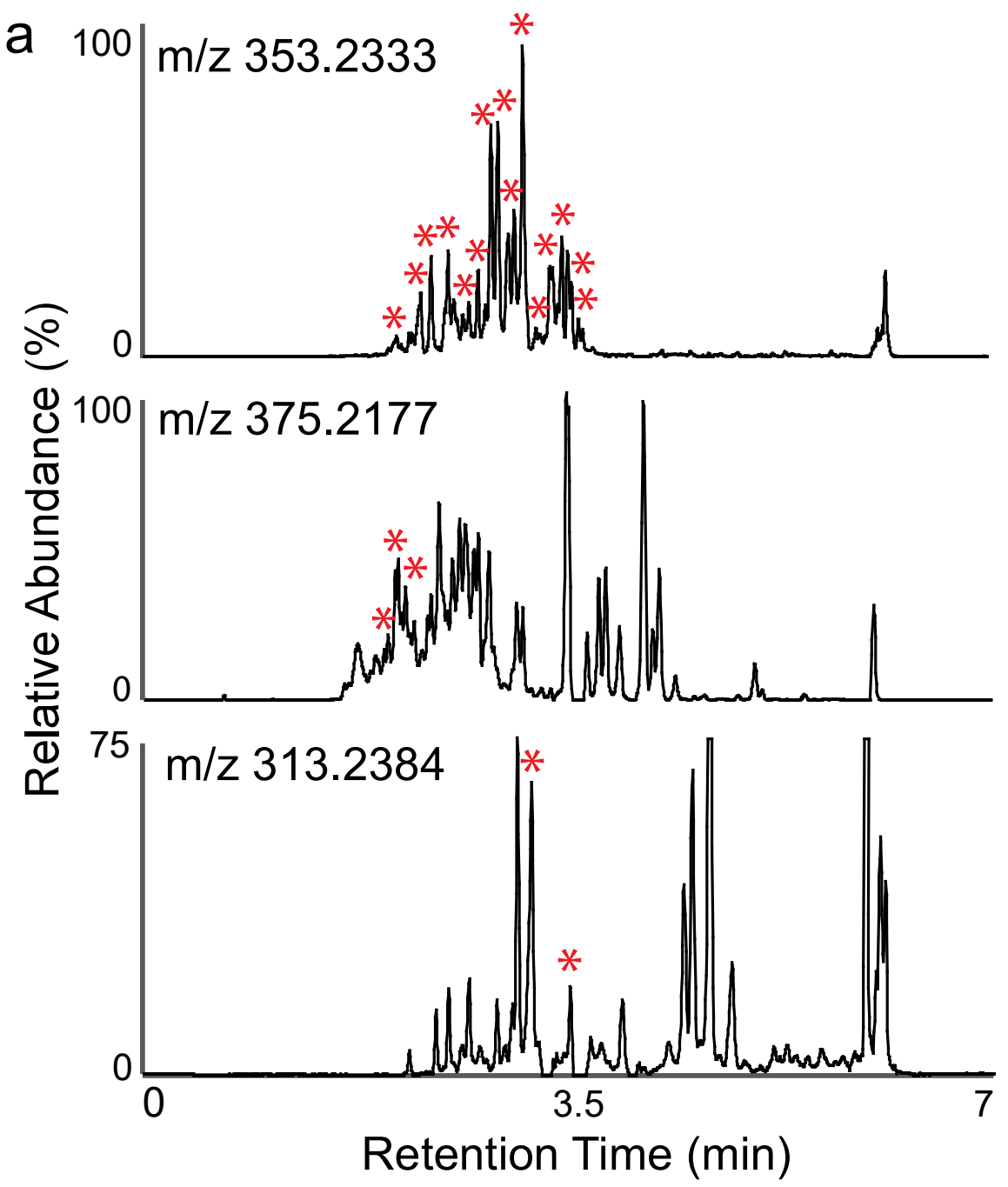

b

| Compound Name | RT | Cosine |
|---|---|---|
| 11,12-DiHETrE | 2.88 | 0.95 |
| 12,13-DiHOME | 3.31 | 0.93 |
| 9,10-DiHOME | 3.42 | 0.93 |
| 8-iso-Prostaglandin A2 | 5.20 | 0.93 |
| 12,13-DiHOME | 3.13 | 0.92 |
| Leukotriene B3 | 4.22 | 0.92 |
| 9,10-DiHOME | 3.27 | 0.91 |

c

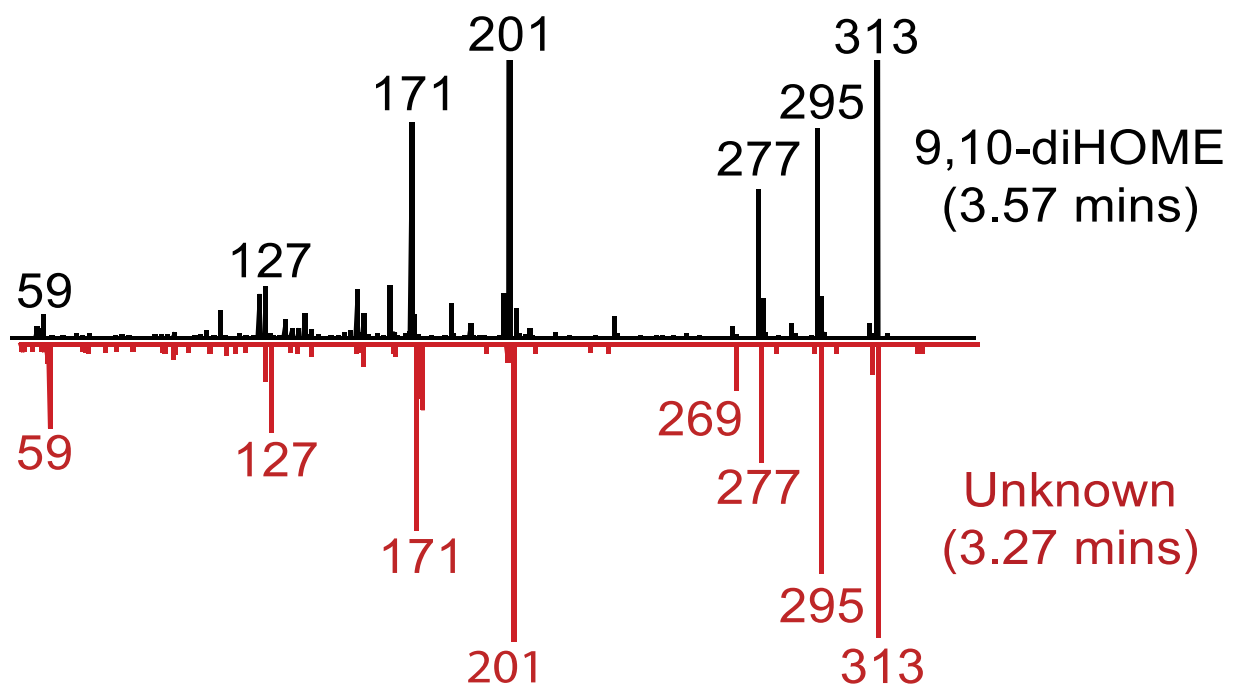

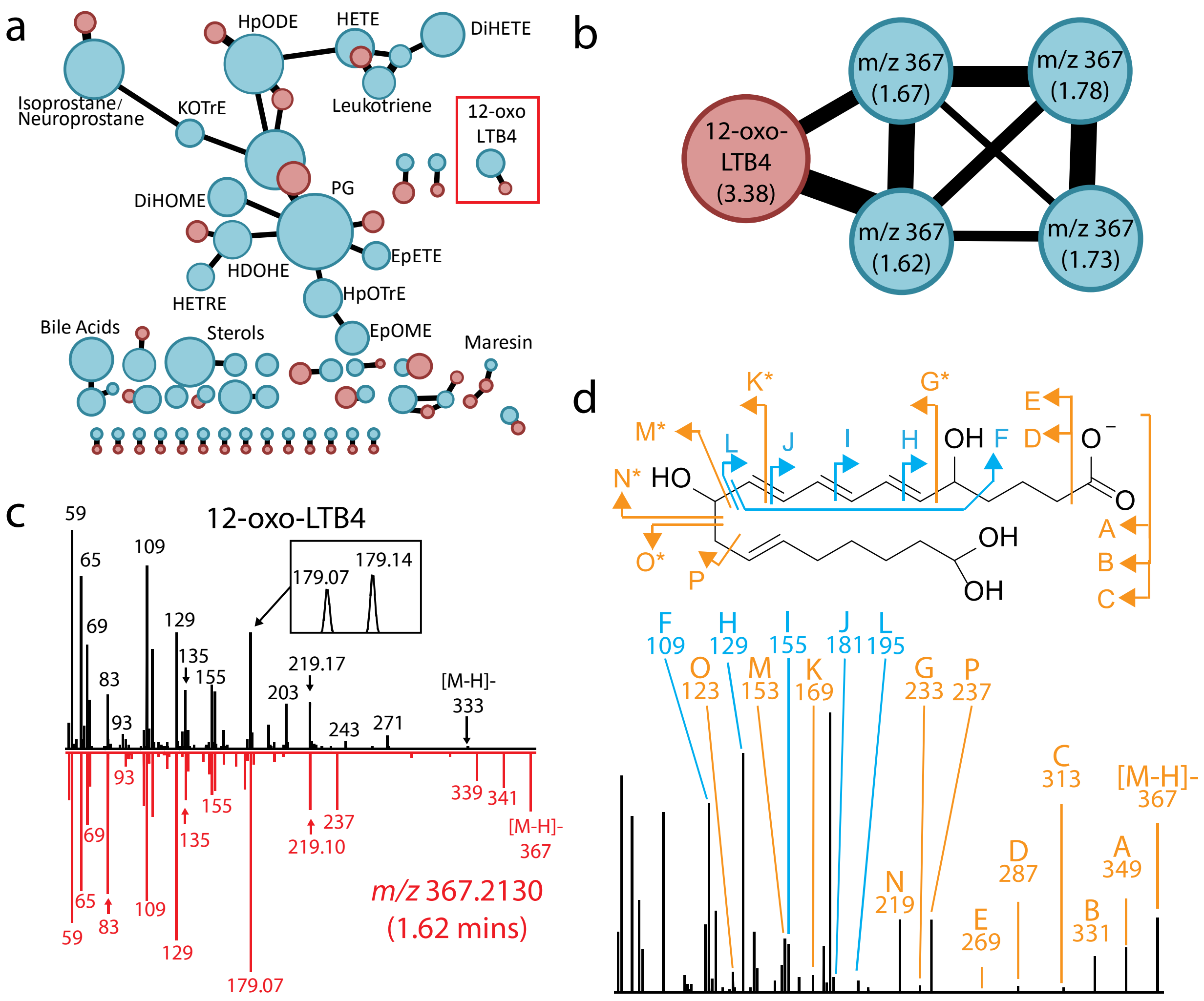

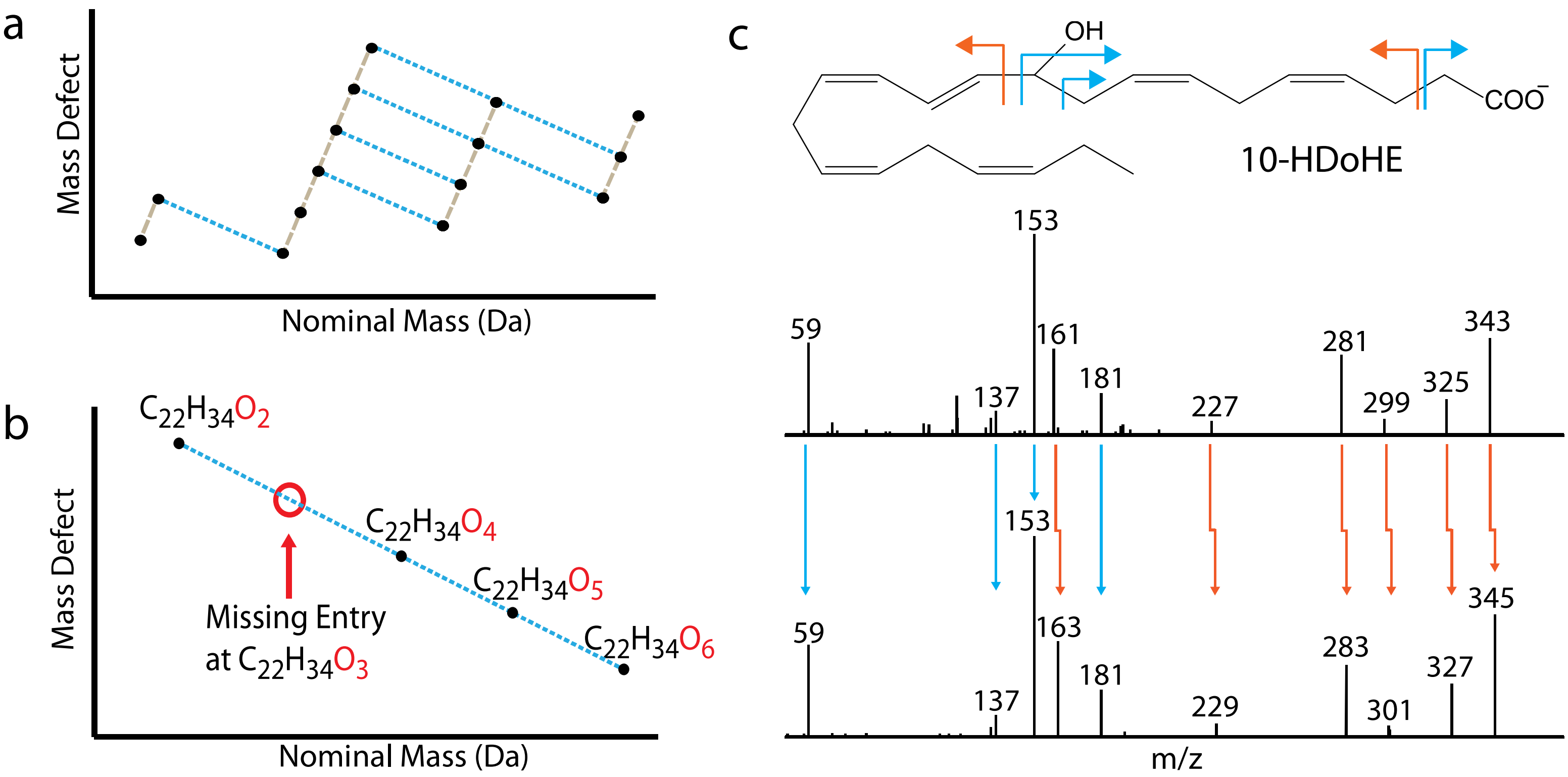

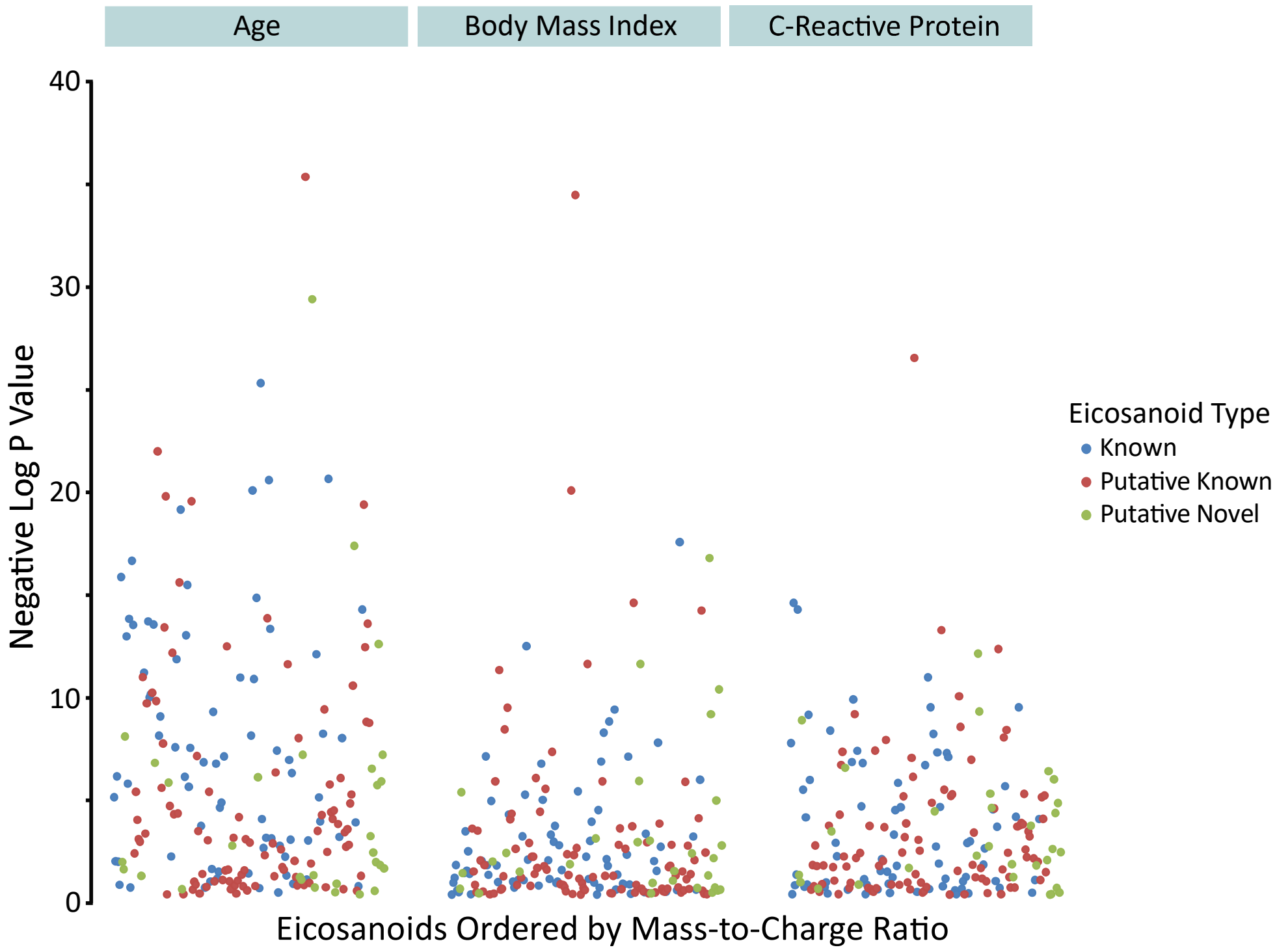

| Characteristic | Total Sample (N=1500) |
|---|---|
| Age, years | 66.1 ± 8.8 |
| Women, % | 52.7 |
| Body mass index, $kg/m^2$ | 28.5 ± 5.6 |
| C-reactive protein | 1.57 (0.76, 3.29) |

## Figure S1

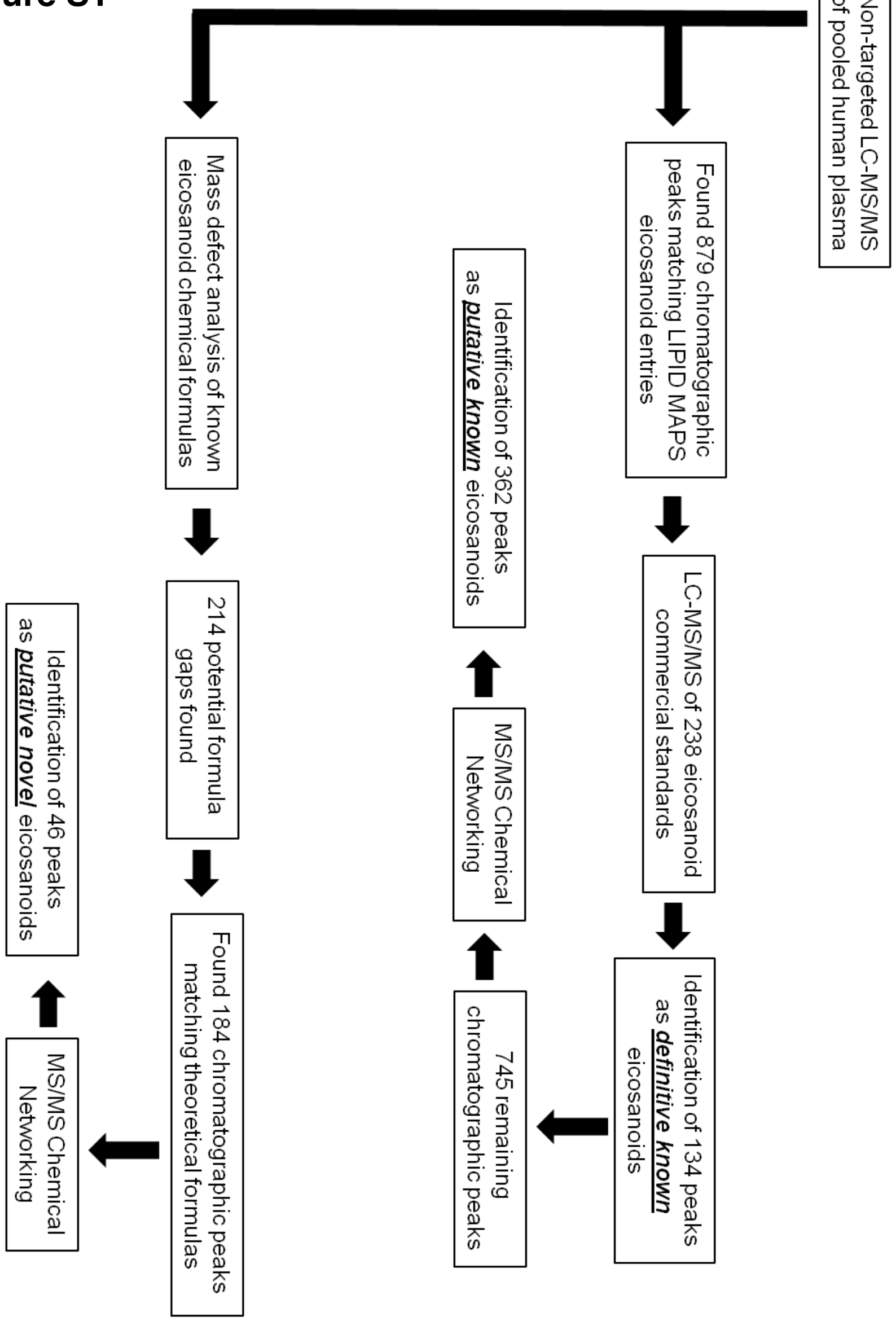

**Figure S1.** Flow chart showing experimental design of how directed non-targeted LC-MS/MS in conjunction with chemical networking was used to identify definitive known, putative known and putative novel eicosanoids.

**Figure S2.** <u>Performance of automated tandem MS fragment assignment tools.</u> (A)Tandem MS spectra from commercial standards were inputed into the MetFrag webtool with LipidMaps serving as the library. On average, Metfrag was only able to assign ~5% of fragment peaks to corresponding chemical structures. Here are two examples using a linear and a cyclic eicosanoid where both results showed poor ability to assign spectral fragments to chemical structures with almost all of the dominant fragments not assigned. Note that matched peaks were re-colored red for visibility. (B) *In silic*o prediction of tandem MS fragments of commercial standards were subjected to fragmentation using CFM-id. Here is a typical example using a cyclic eicosanoid where many of the observed fragments fail to be predicted. CFM-id performed notably better at predicting fragments for linear eicosanoids, however, many fragments were still left unpredicted.

## Figure S2-A

**Prostaglandin A2**
MS/MS peaks matched: 5 of 115

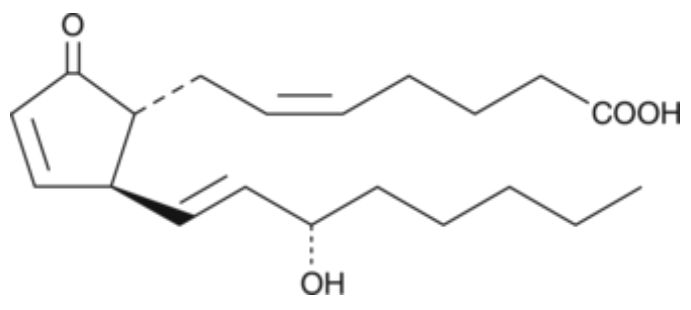

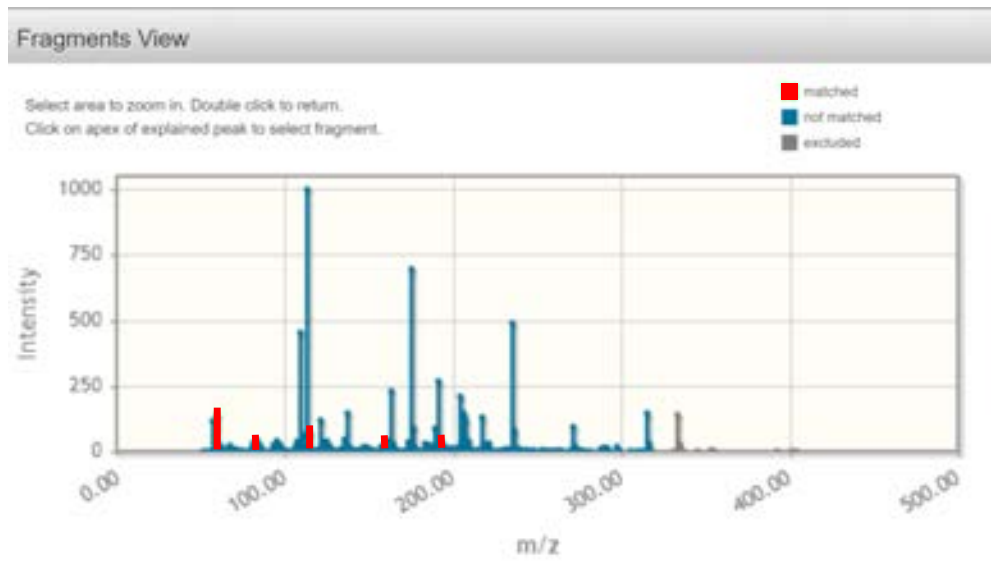

**5-HETE**
MS/MS peaks matched: 6 of 84

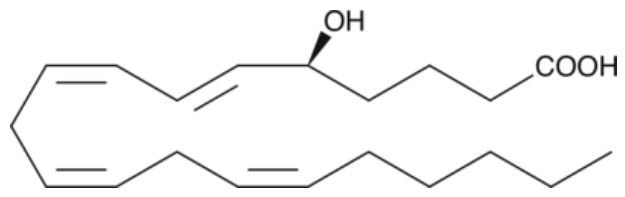

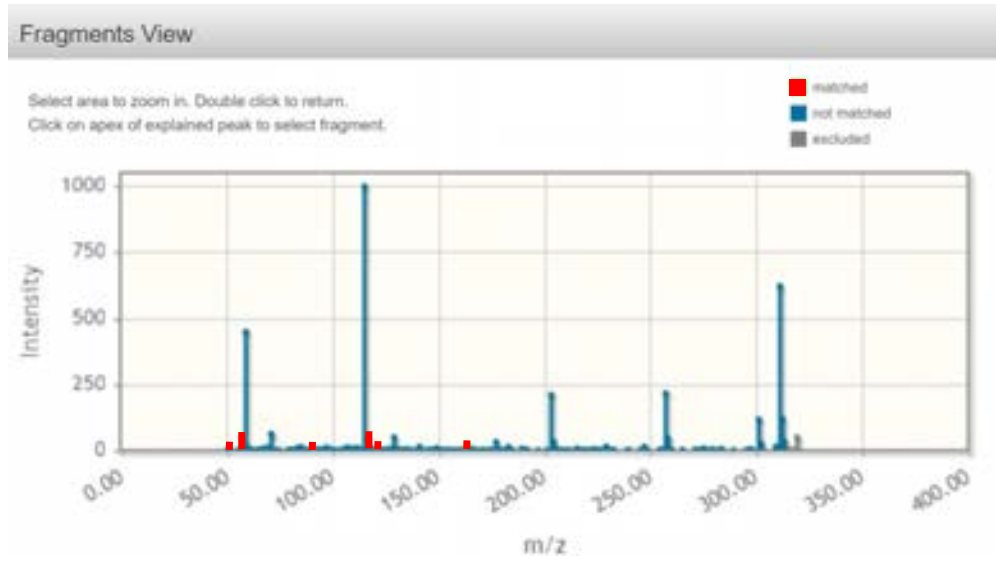

## Figure S2-B

**Prostaglandin F2α**

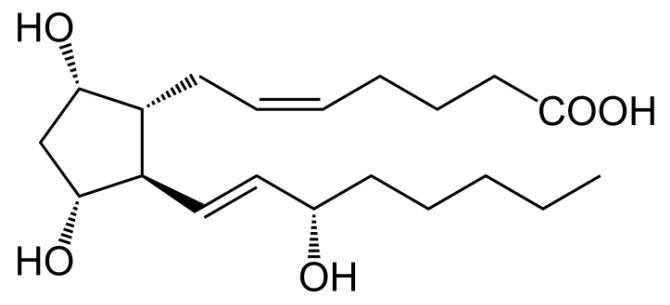

Predicted Tandem MS Spectra

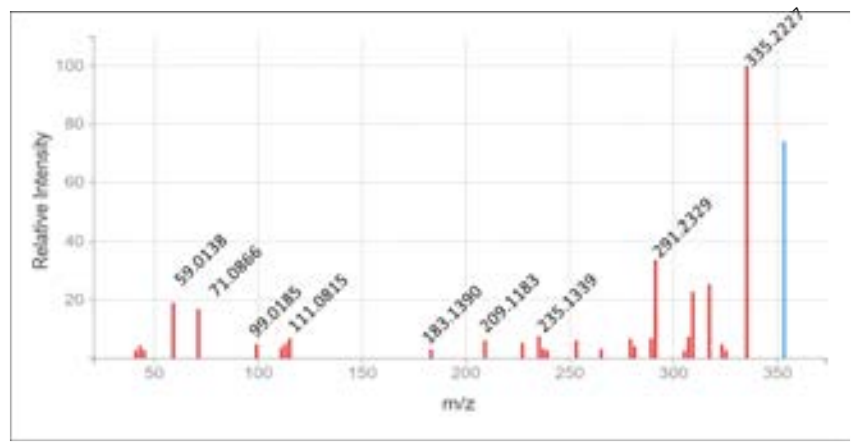

Experimental Tandem MS Spectra

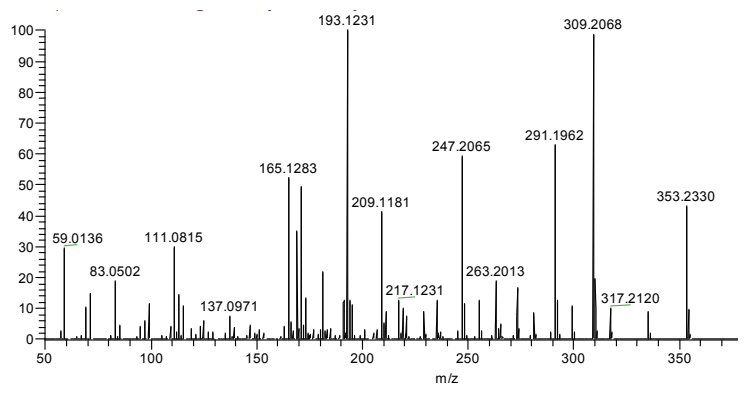

# Figure S3

**Figure S3. Related to Figure 3.** Examples of experimental tandem MS spectral matches to tandem MS spectra from eicosanoid commercial standards. Each example shows the tandem MS spectra from the unknown compound (black) above the tandem MS spectra of the standard (green). Comparison of the spectra reveal presence of identical MS fragment peaks as well as, in some cases, peaks consistently shifted by a given mass indicating that the unkown compounds share a similar structure to that of the eicosanoid commercial standard.

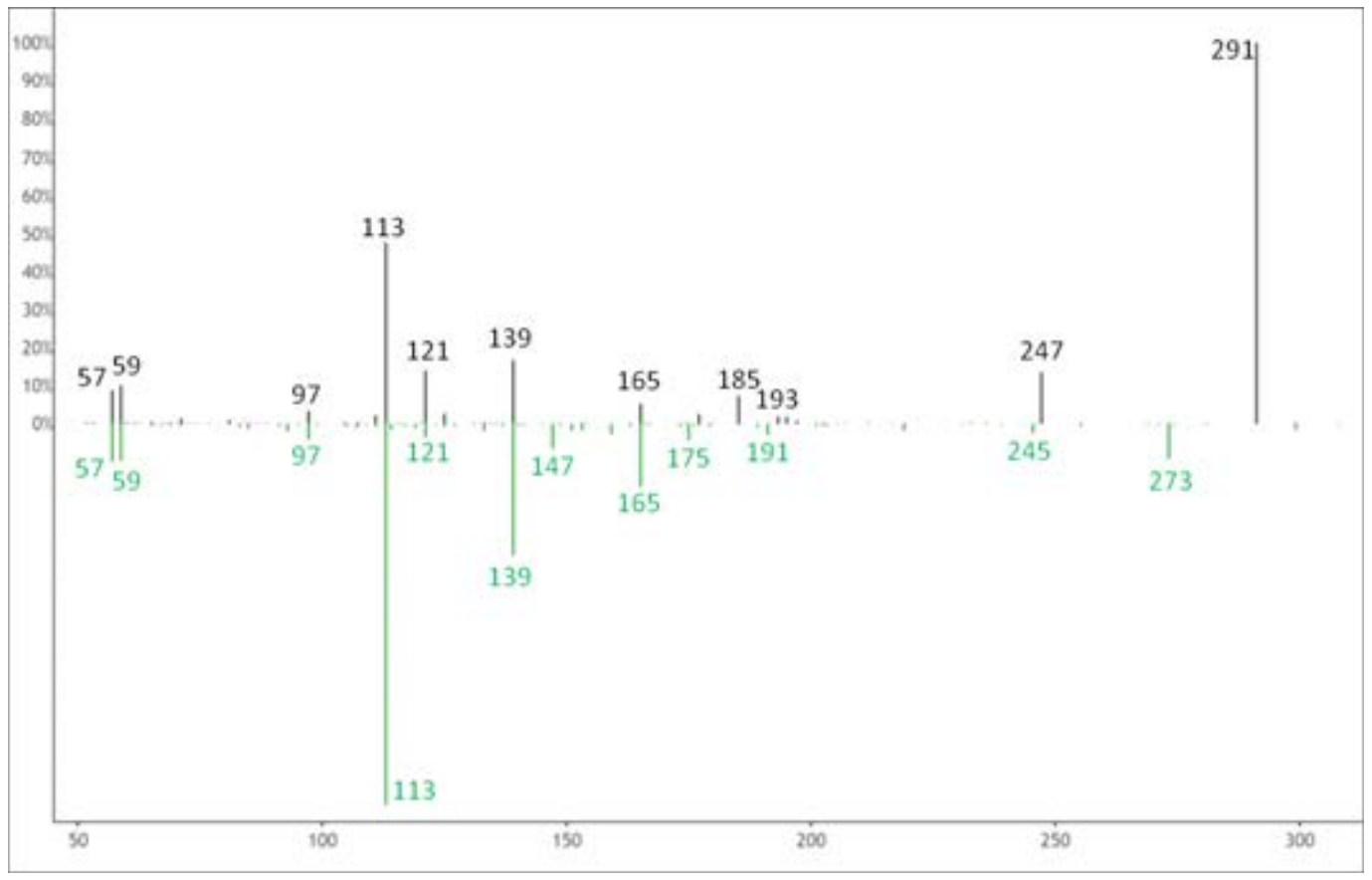

m/z 291.1965
at 4.57 minutes

15-oxoETE
reference spectra

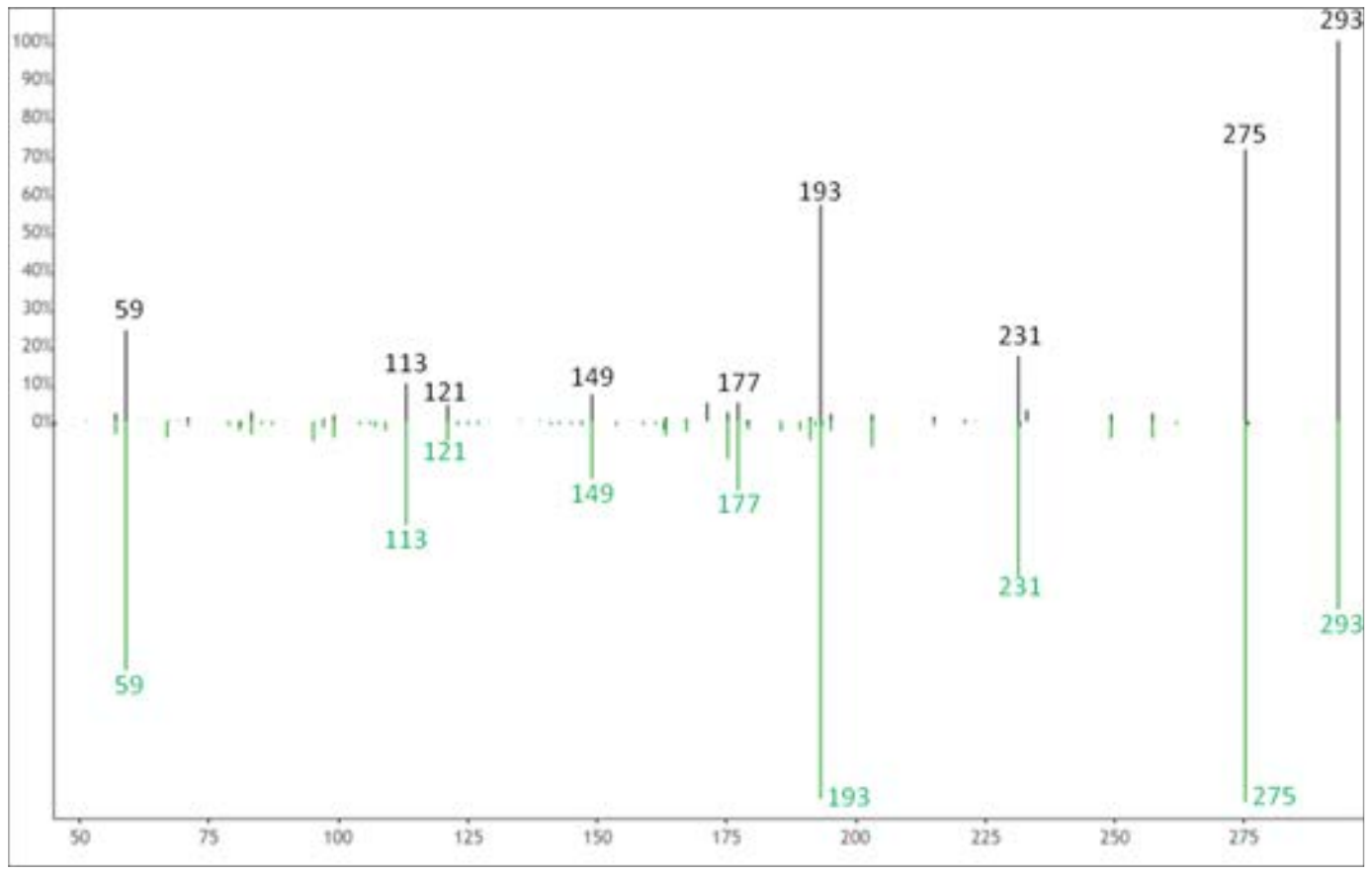

m/z 293.2122
at 4.54 minutes

13-HOTrE
reference spectra

## Figure S3 (Continued)

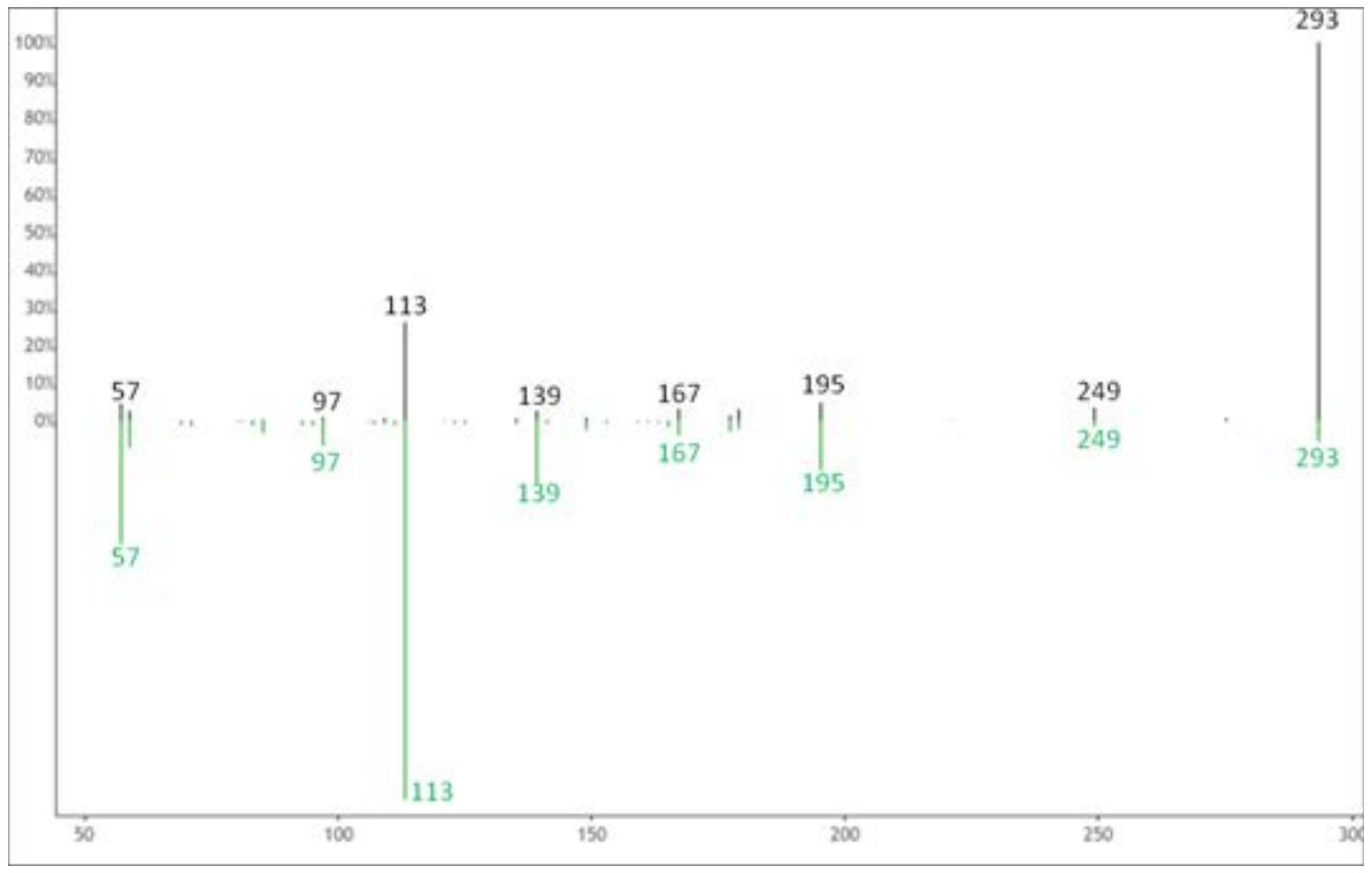

m/z 293.2122
at 5.00 minutes

13-oxo-ODE
reference spectra

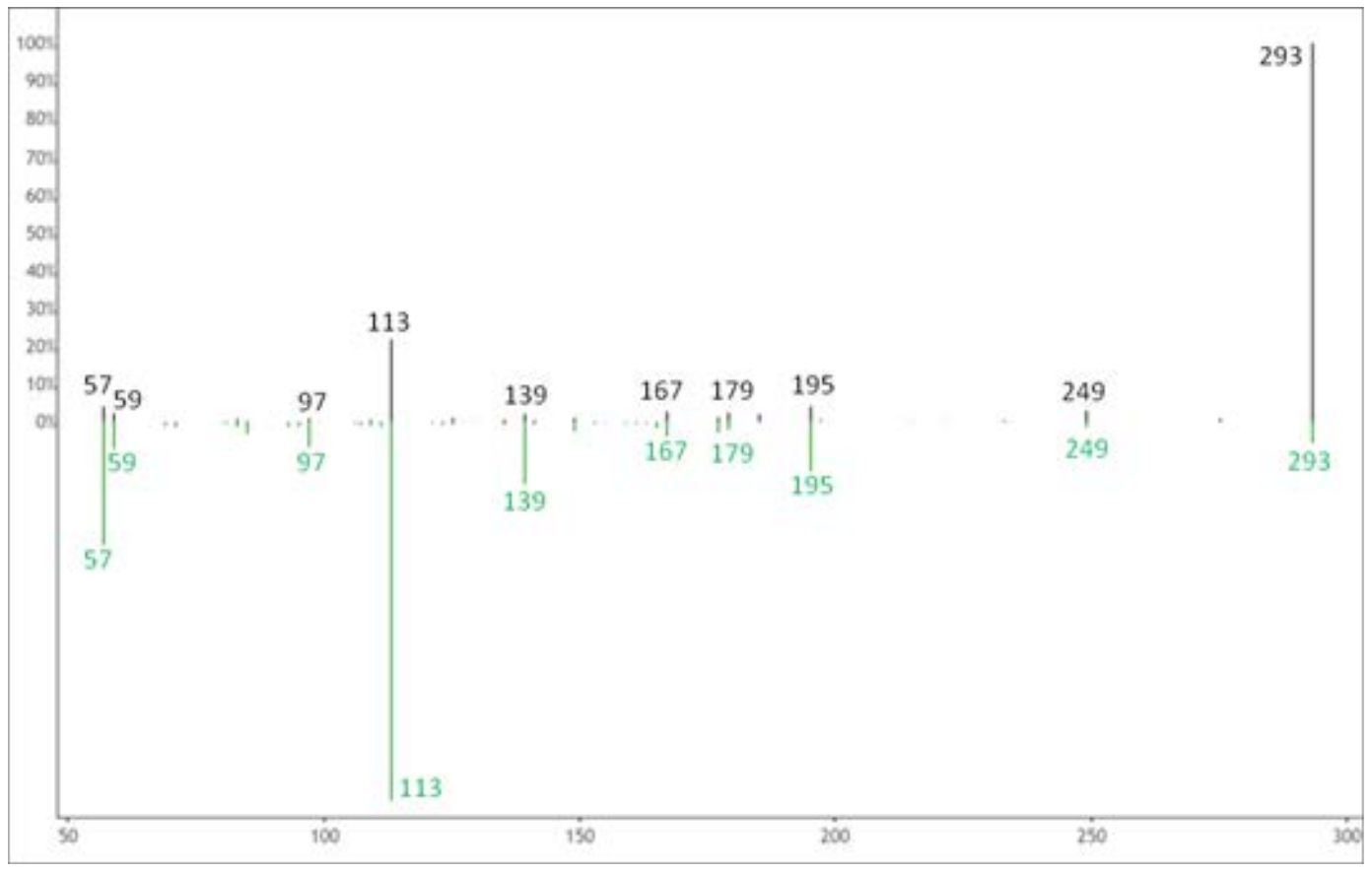

m/z 293.2122
at 5.09 minutes

13-OxoODE
reference spectra

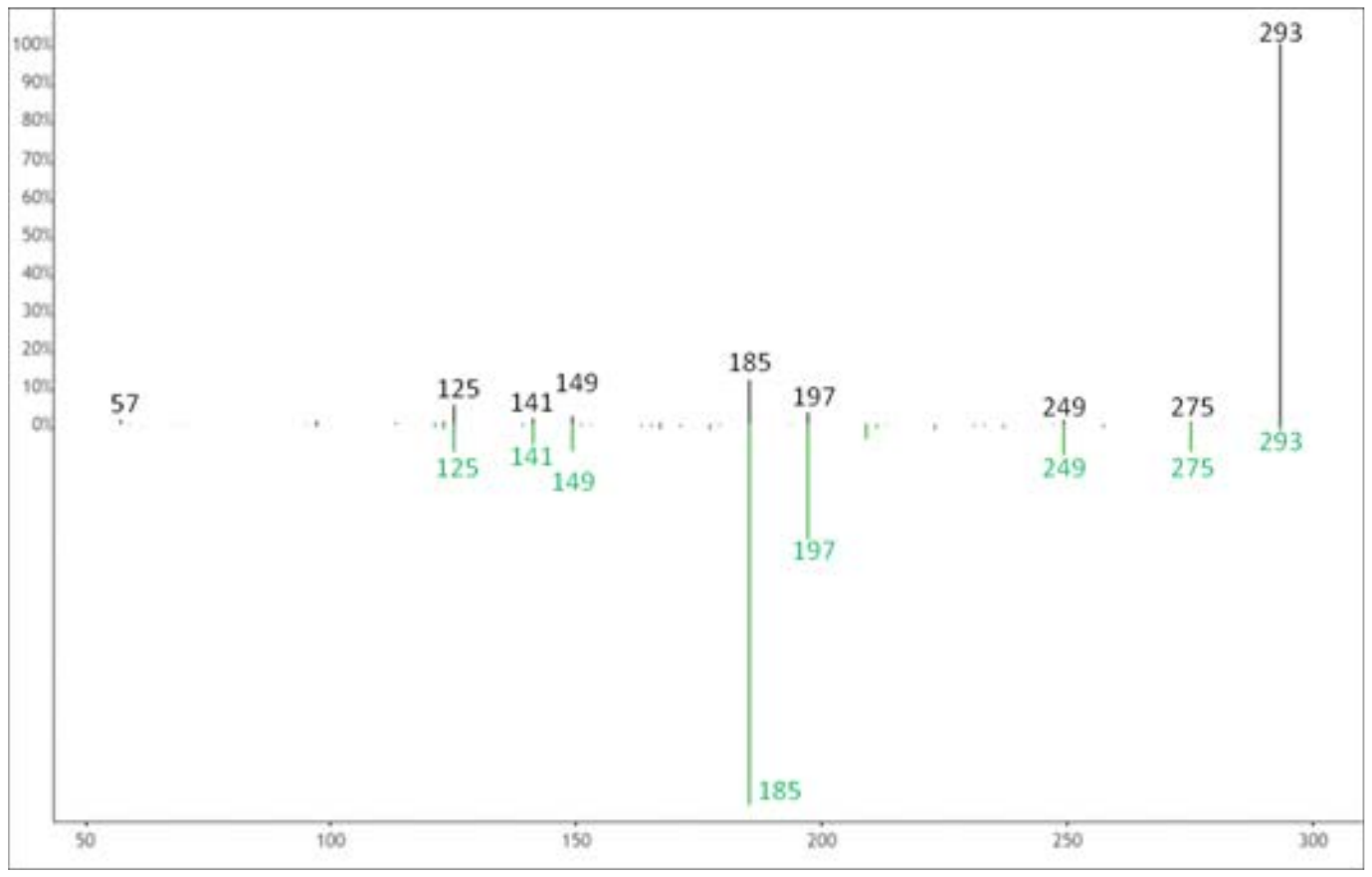

m/z 293.2122
at 5.14 minutes

9-oxo-ODE
reference spectra

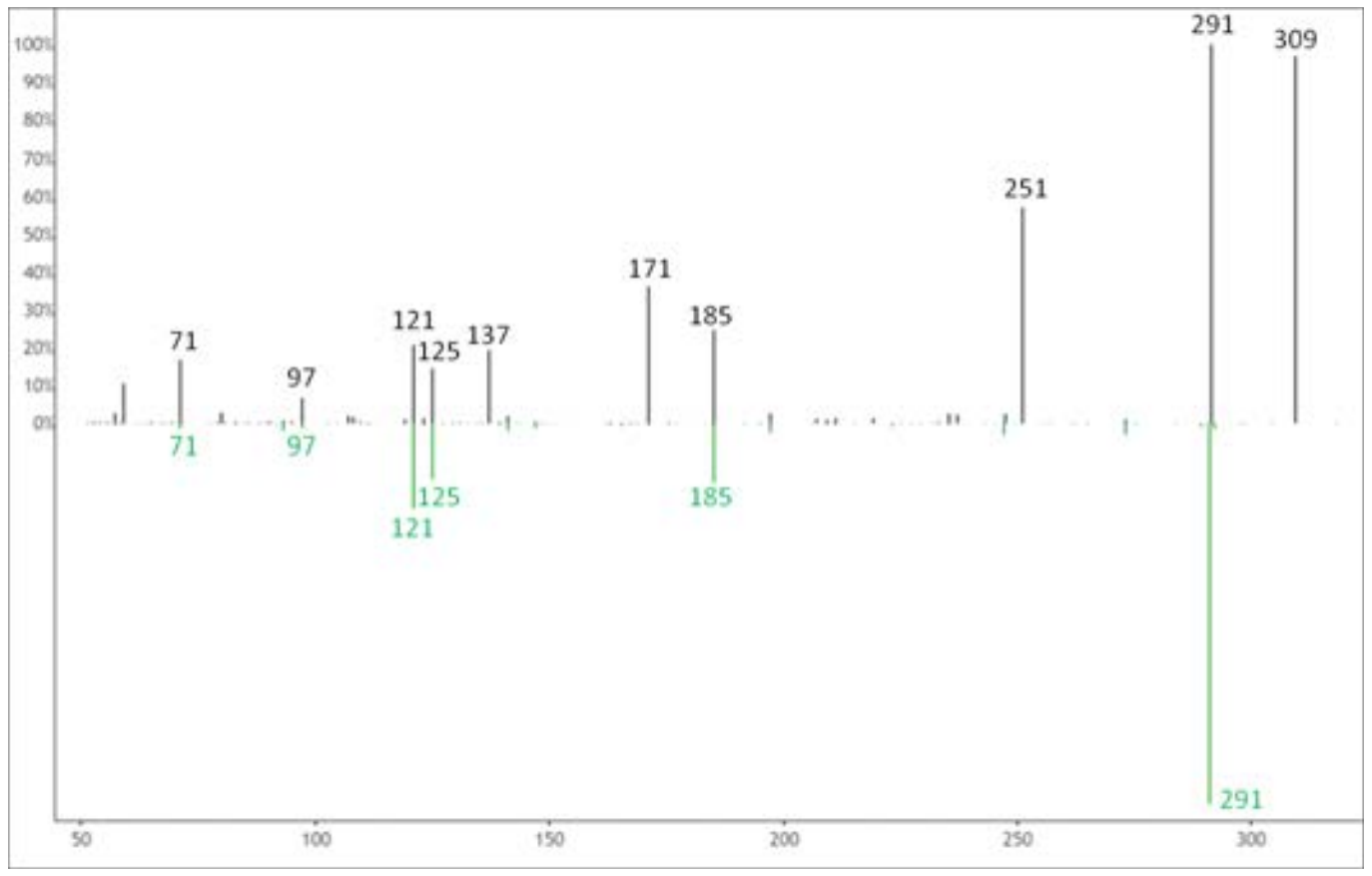

m/z 309.2071
at 2.16 minutes

9-OxoOTrE
reference spectra

## Figure S3 (Continued)

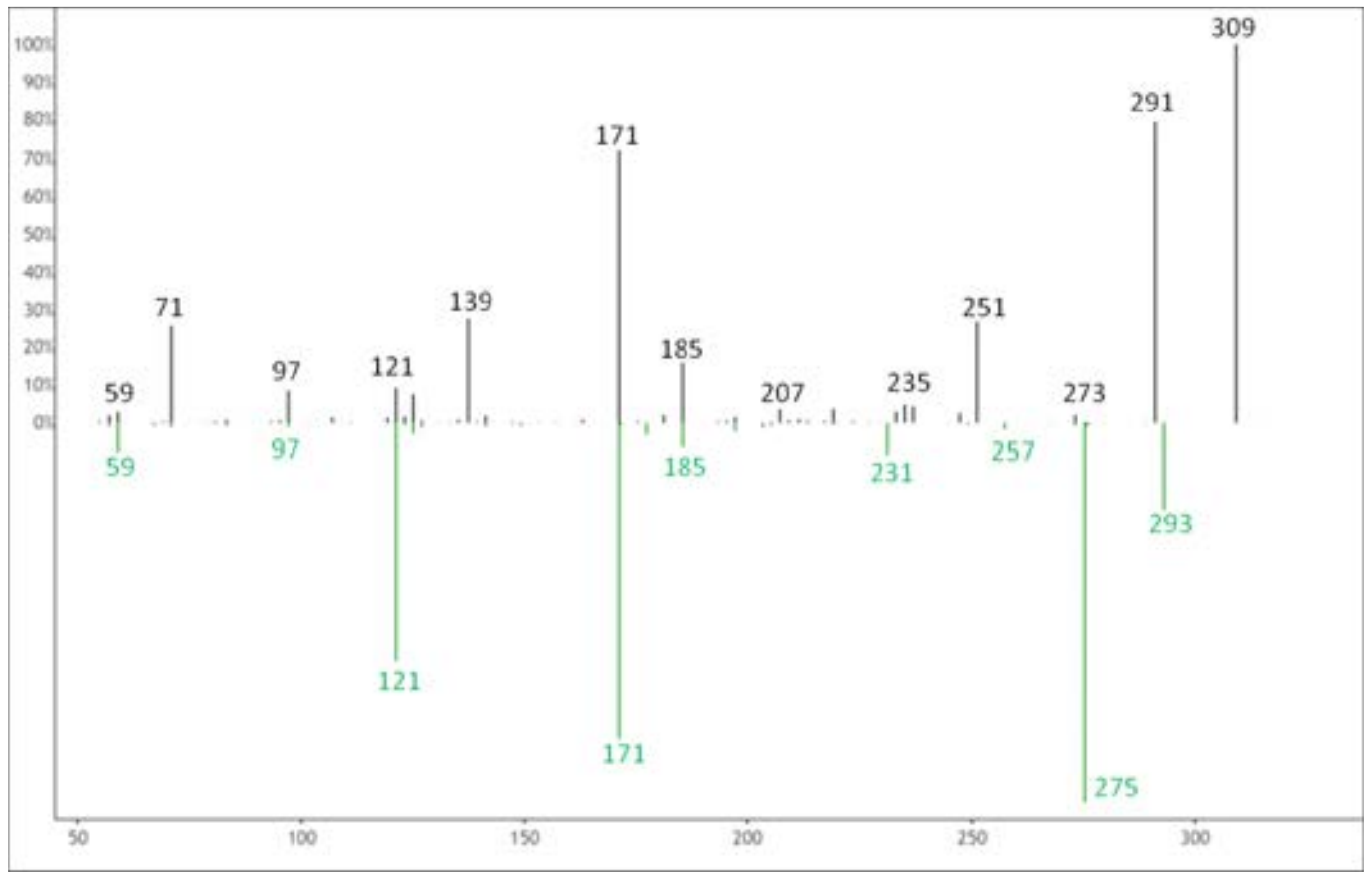

m/z 309.2071
at 2.32 minutes

9-HOTrE
reference spectra

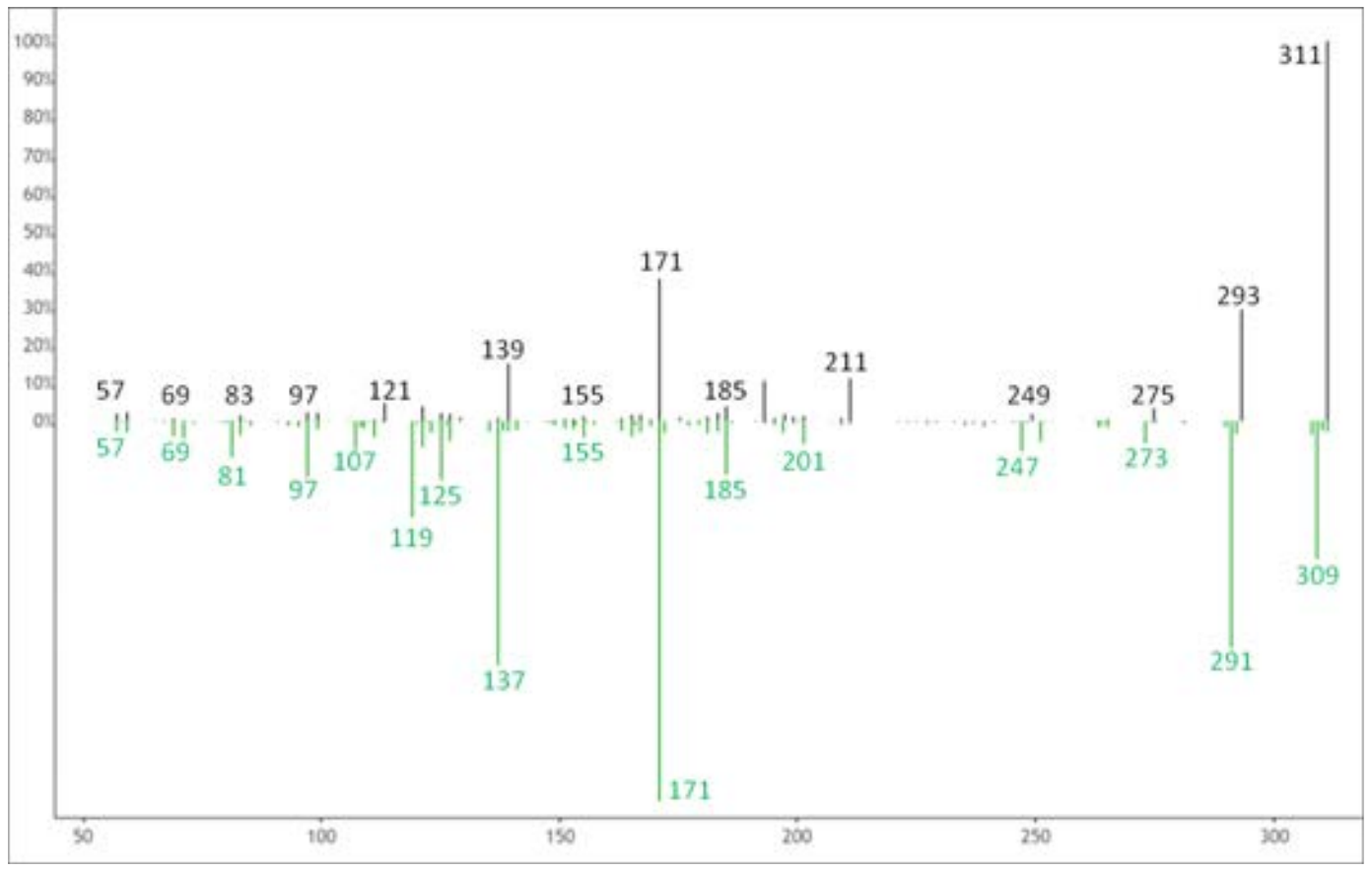

m/z 311.2228
at 3.02 minutes

9-HpOTrE
reference spectra

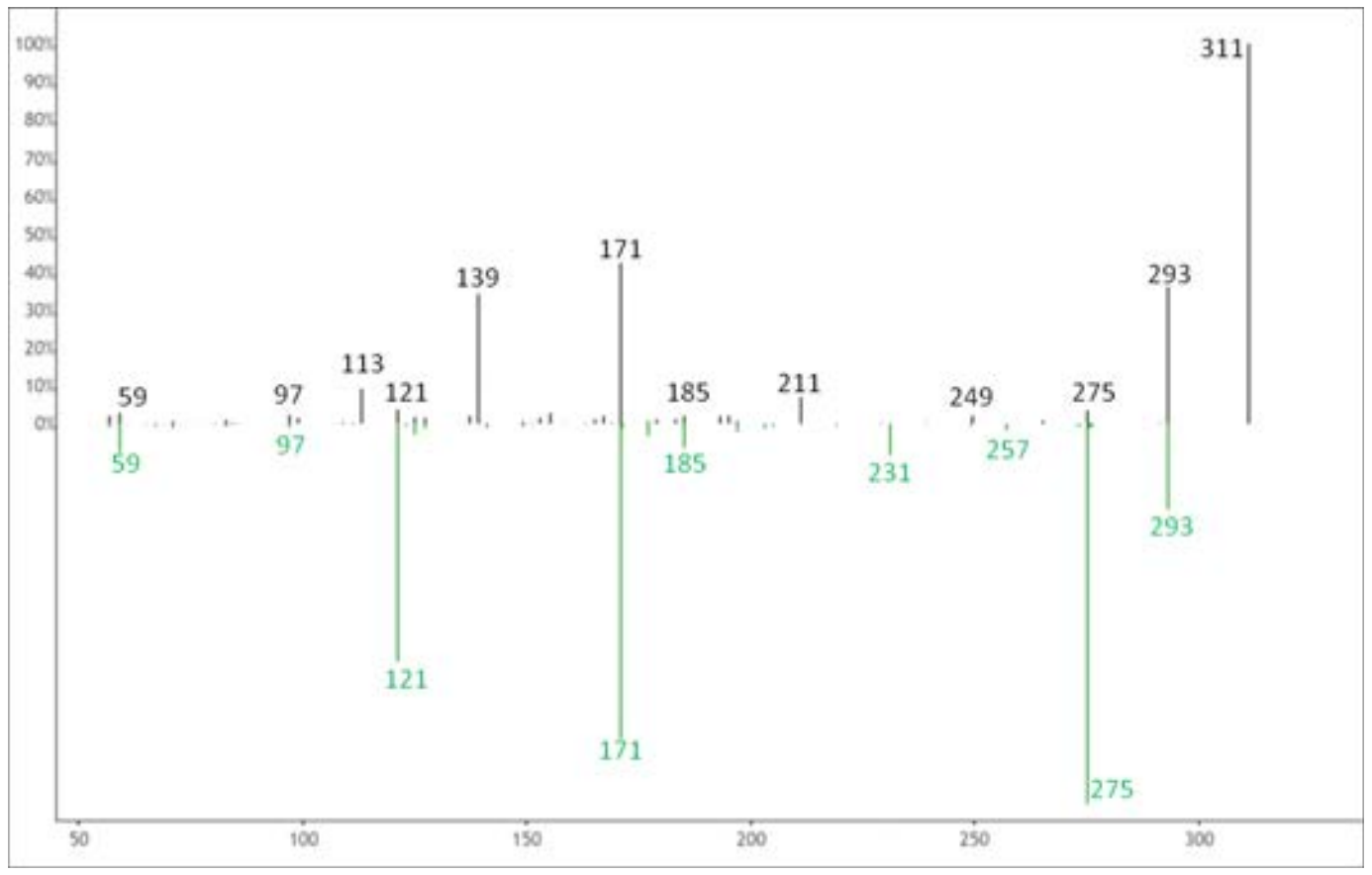

m/z 311.2228
at 3.22 minutes

9-HOTrE
reference spectra

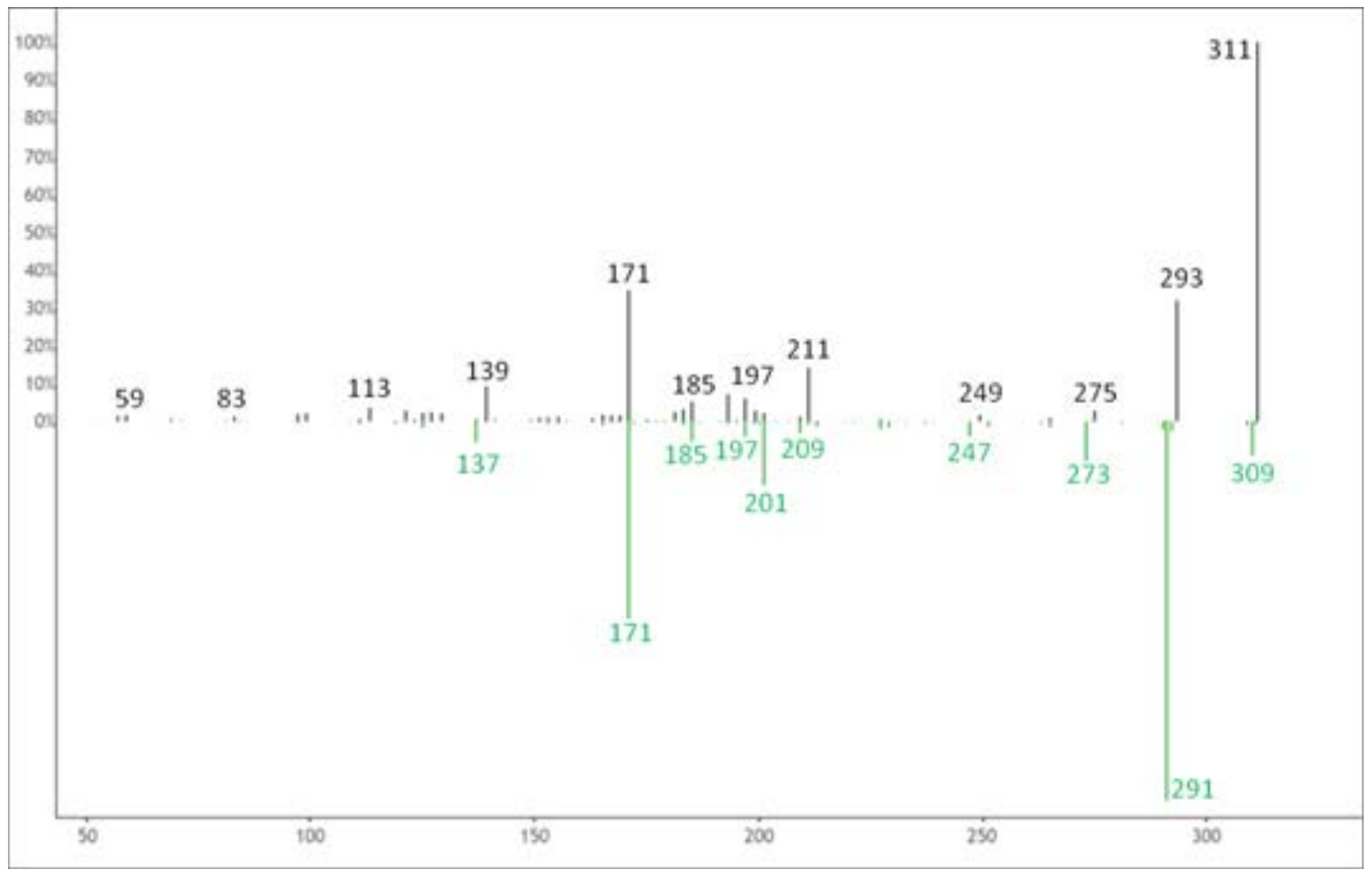

m/z 311.2228
at 3.26 minutes

9-HpOTrE
reference spectra

## Figure S3 (Continued)

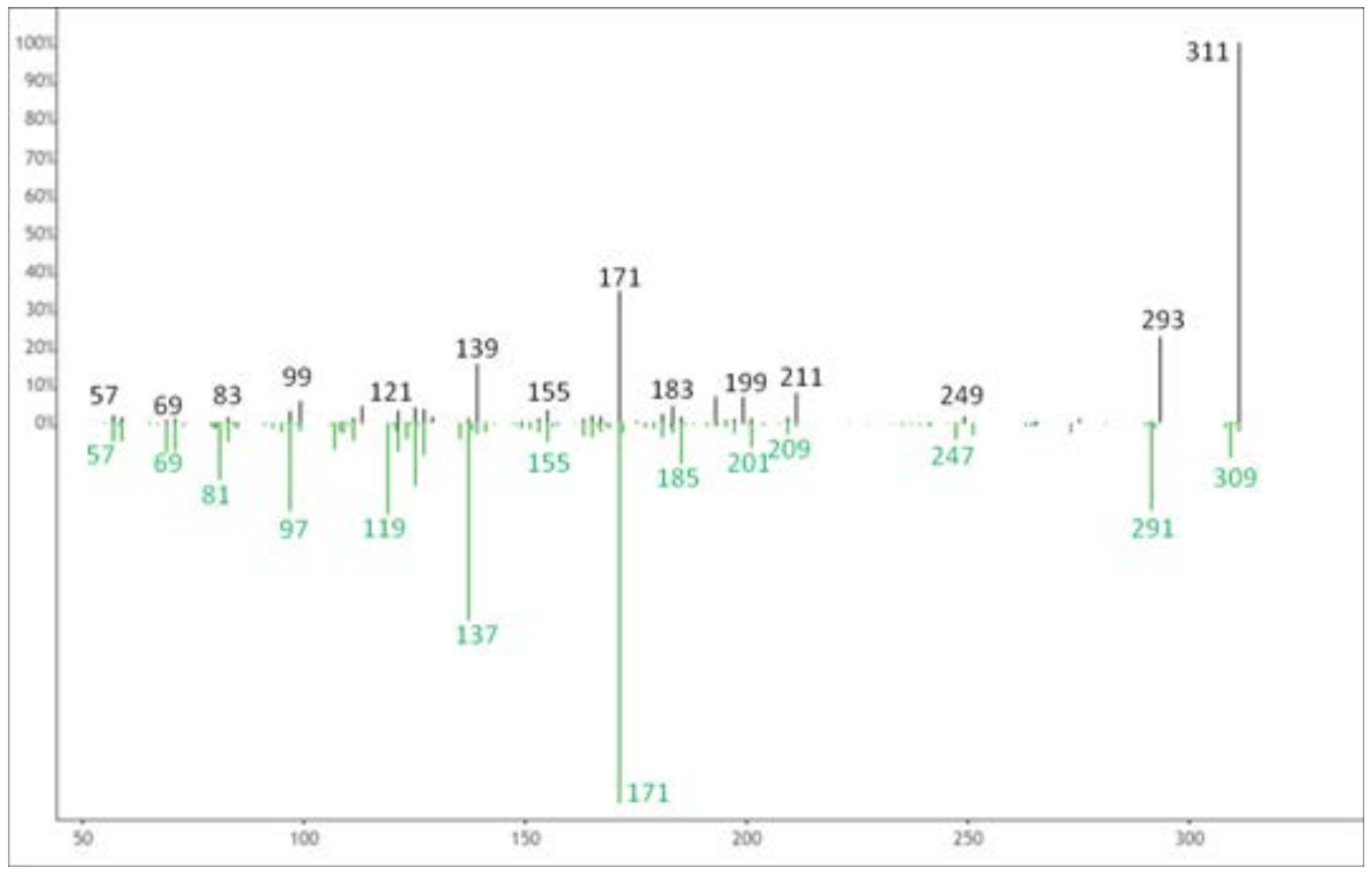

m/z 311.2228
at 3.55 minutes

9-HpOTrE
reference spectra

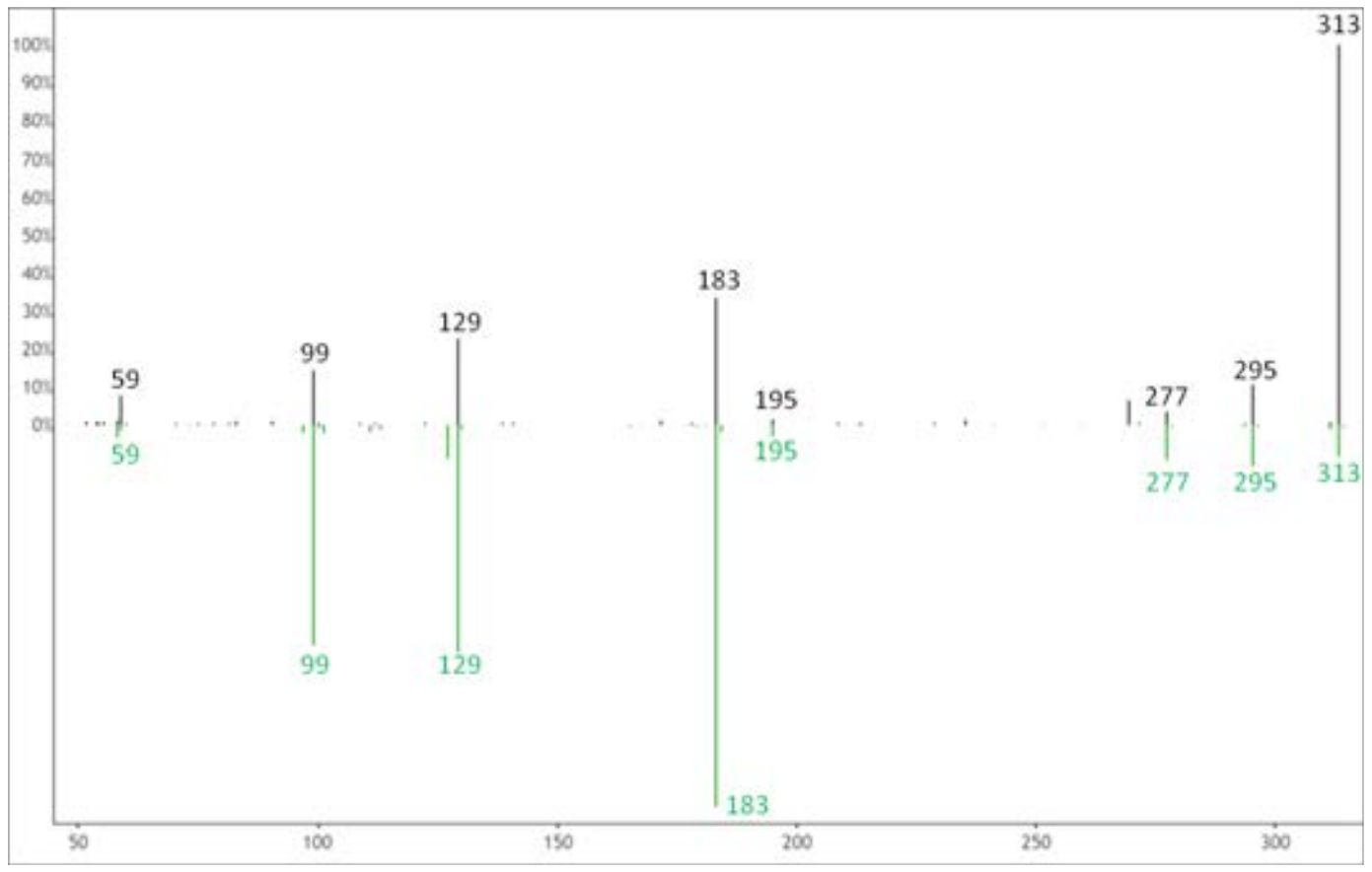

m/z 313.2384
at 3.13 minutes

12,13-DiHOME
reference spectra

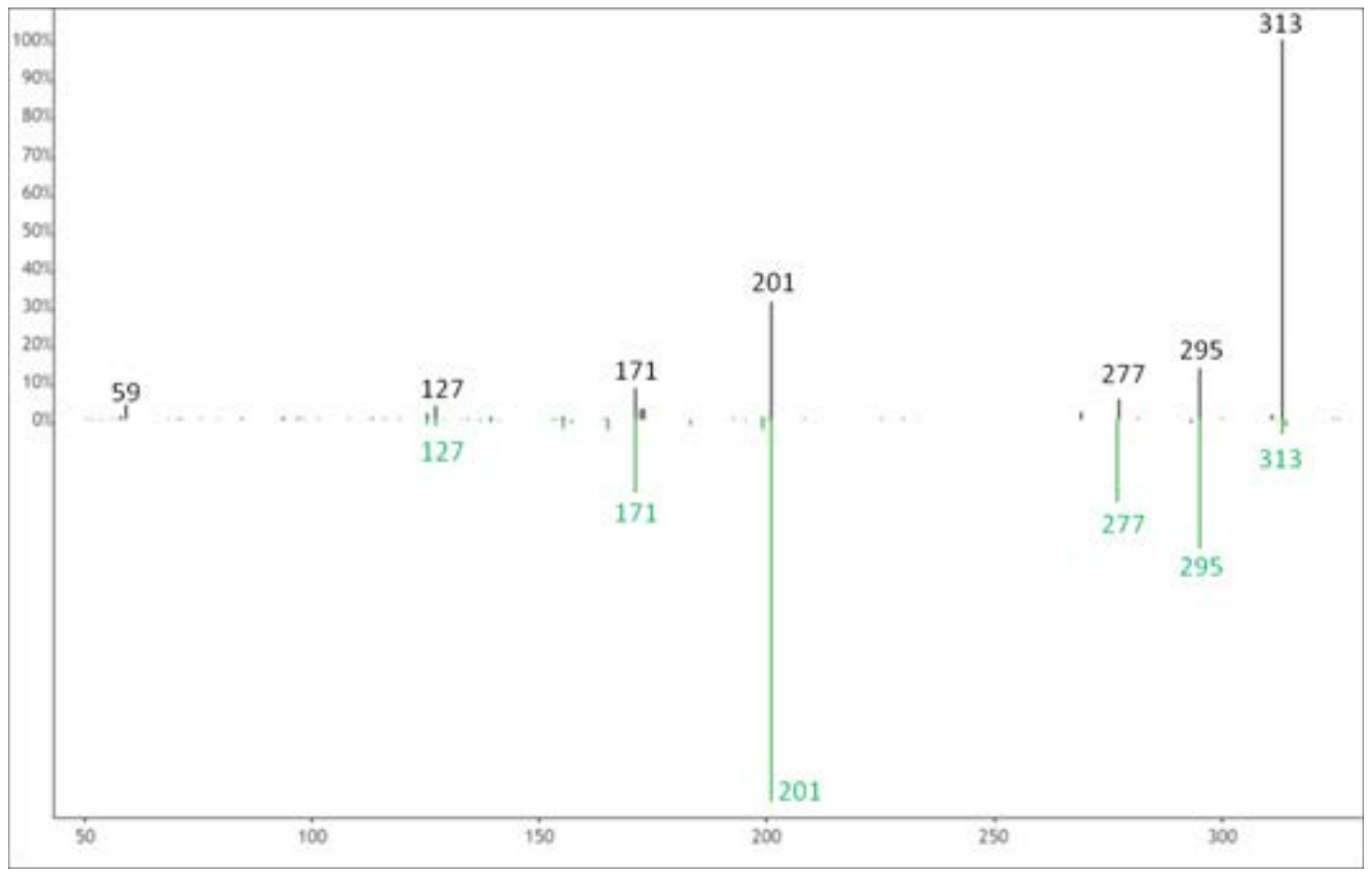

m/z 313.2384
at 3.27 minutes

9,10-DiHOME
reference spectra

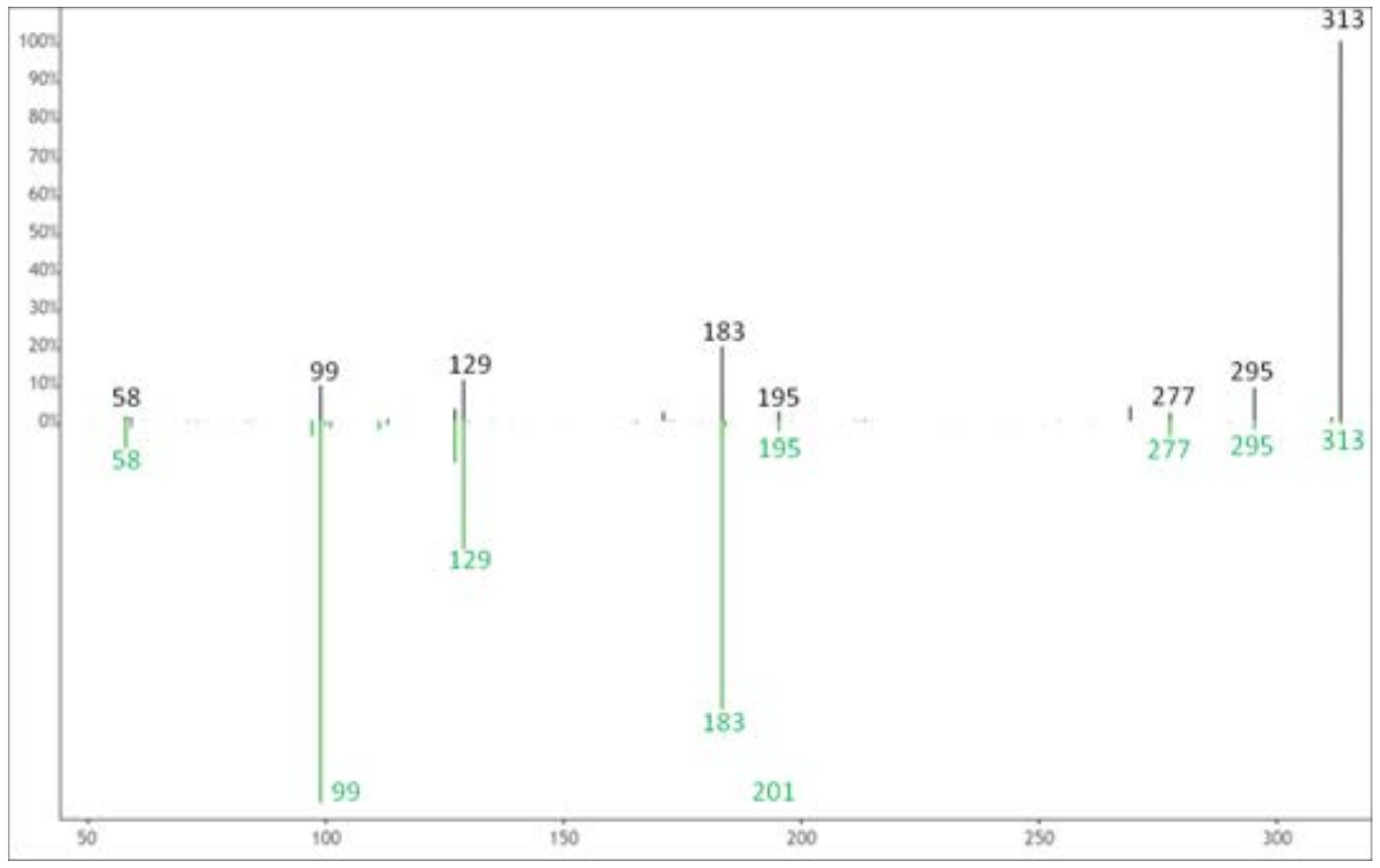

m/z 313.2384
at 3.31 minutes

12,13-DiHOME
reference spectra

## Figure S3 (Continued)

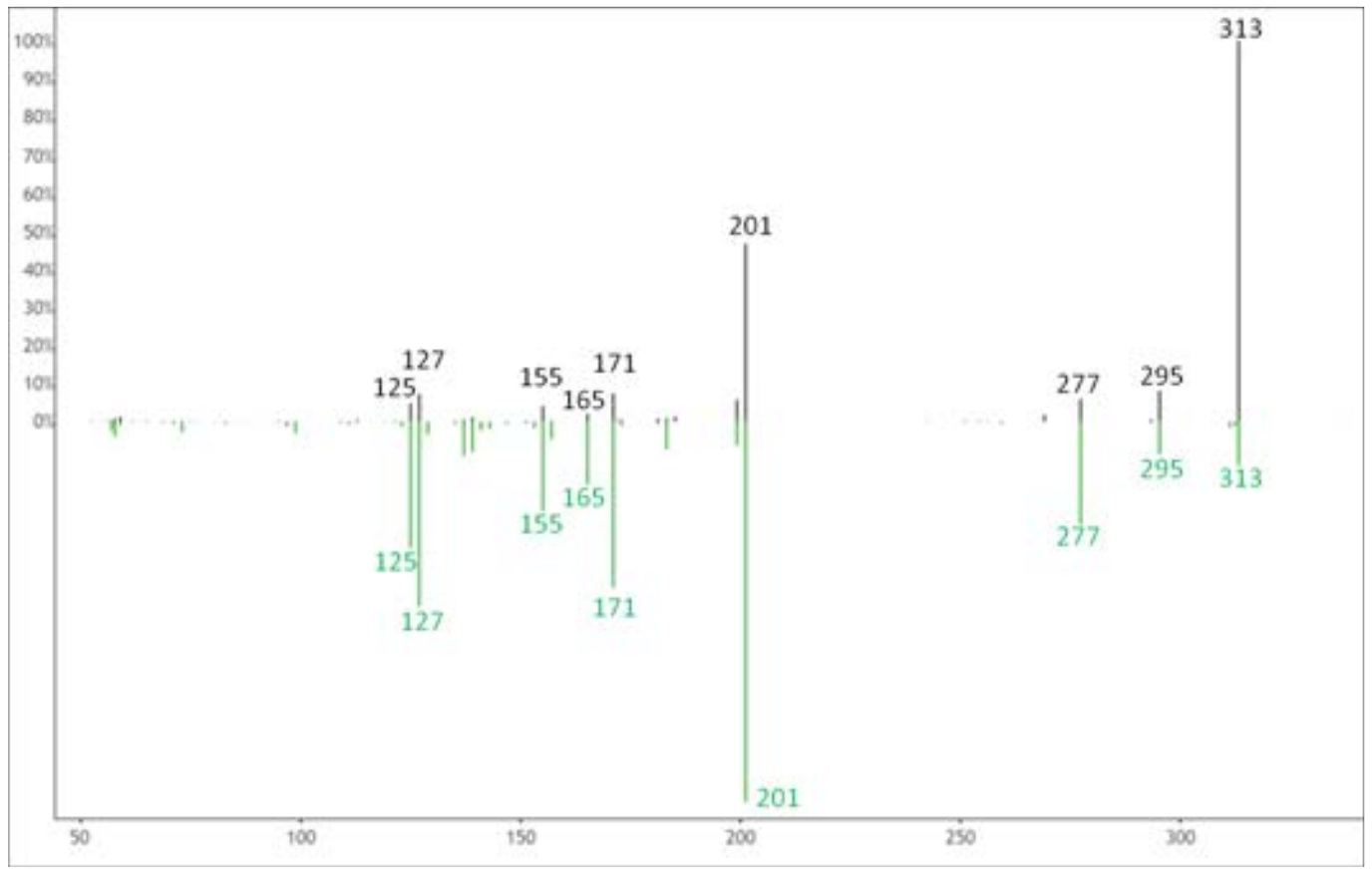

m/z 313.2384
at 3.42 minutes

9,10-DiHOME
reference spectra

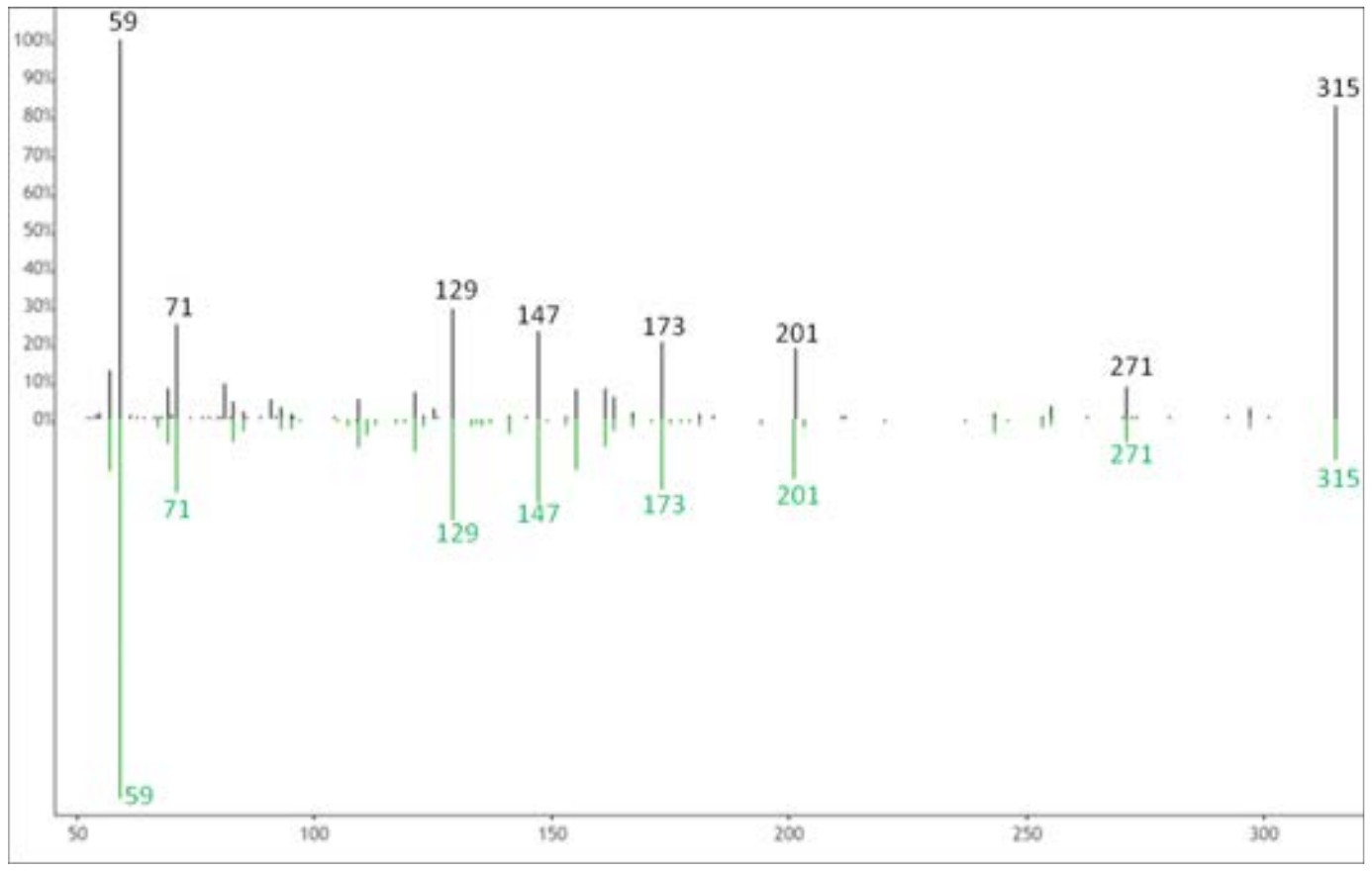

m/z 315.1965
at 4.51 minutes

5-HpEPE
reference spectra

## Figure S3 (Continued)

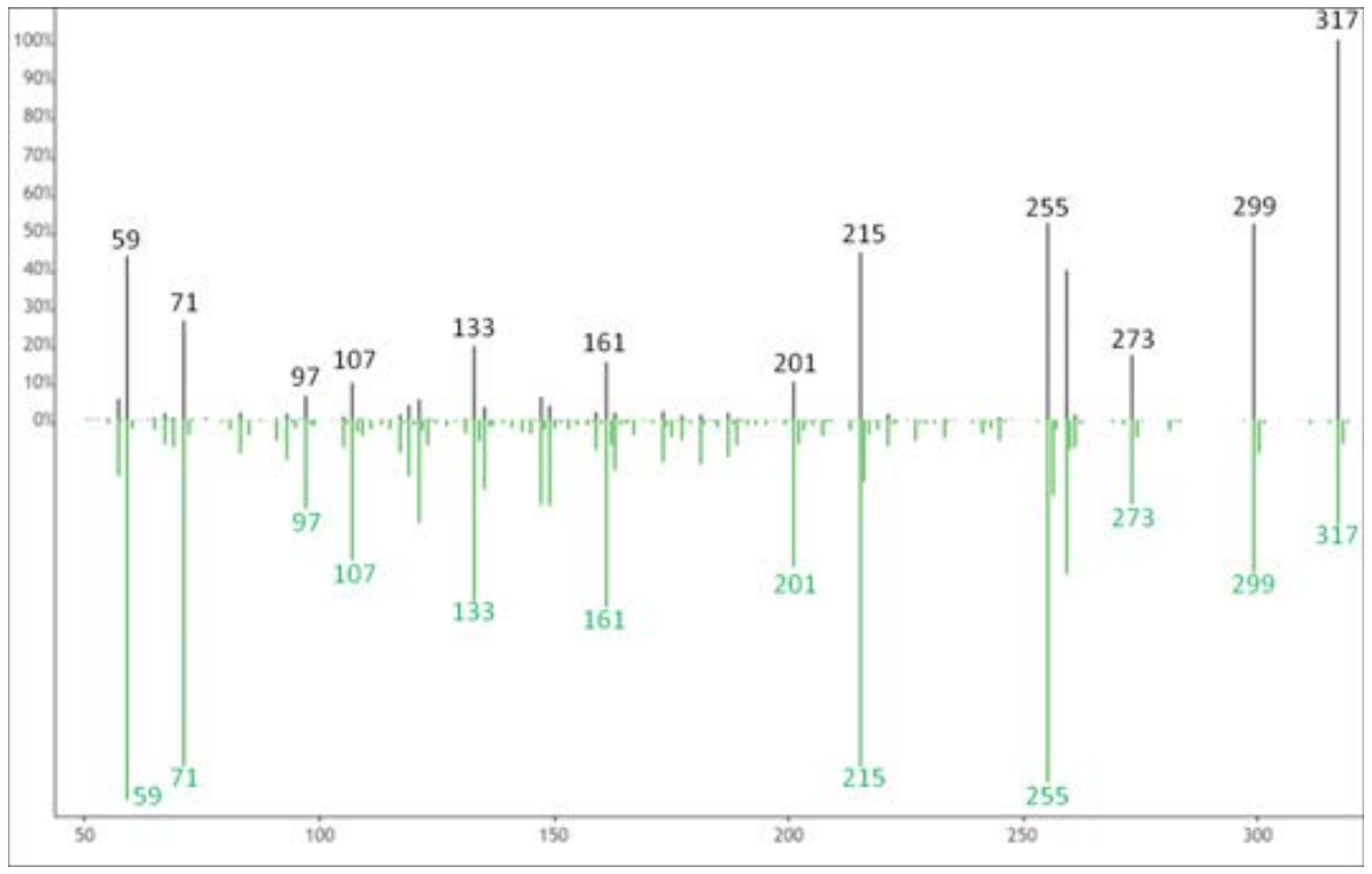

m/z 317.2122
at 3.80 minutes

18-HEPE
reference spectra

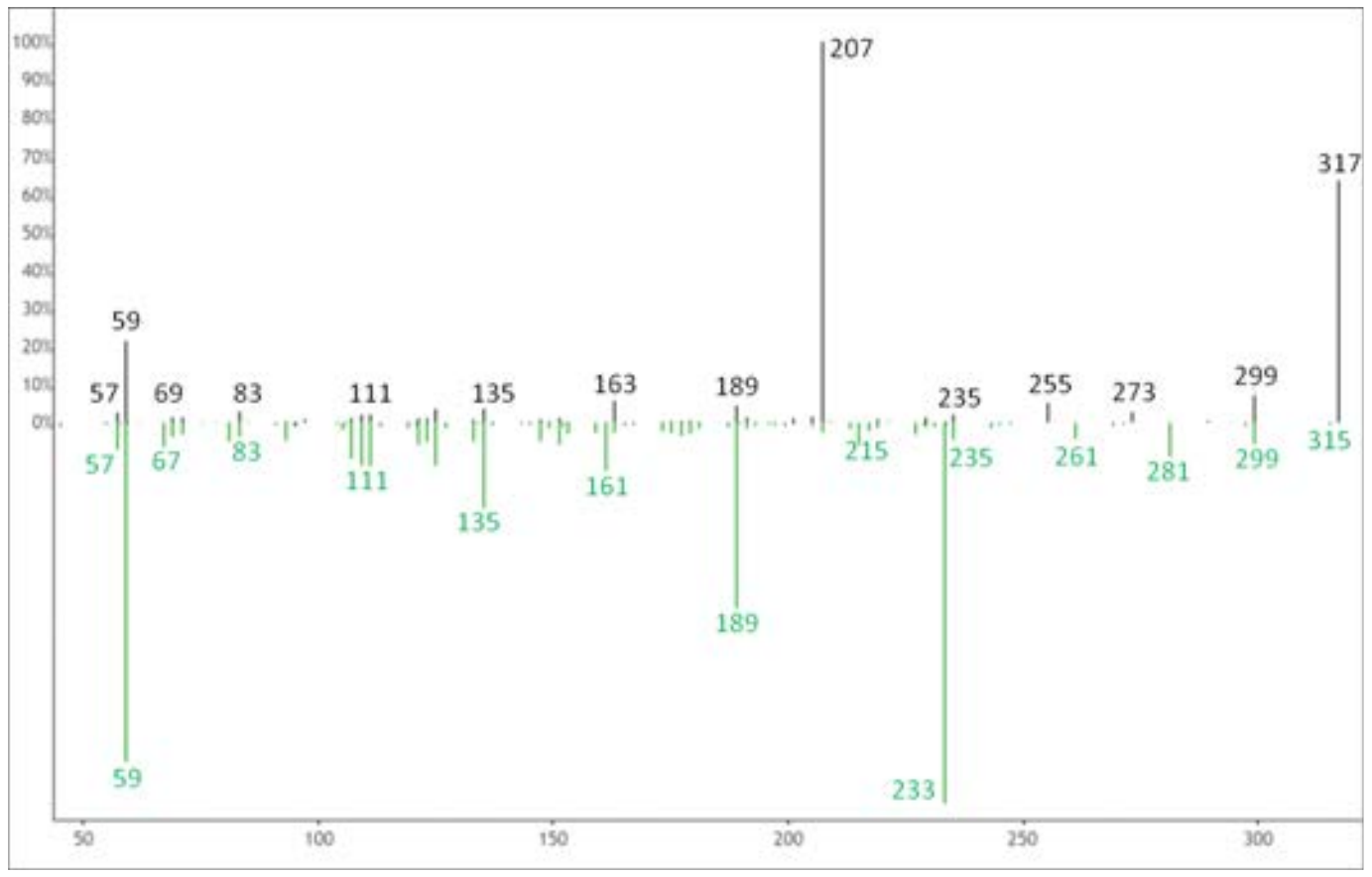

m/z 317.2122
at 3.94 minutes

16-HDoHE
reference spectra

## Figure S3 (Continued)

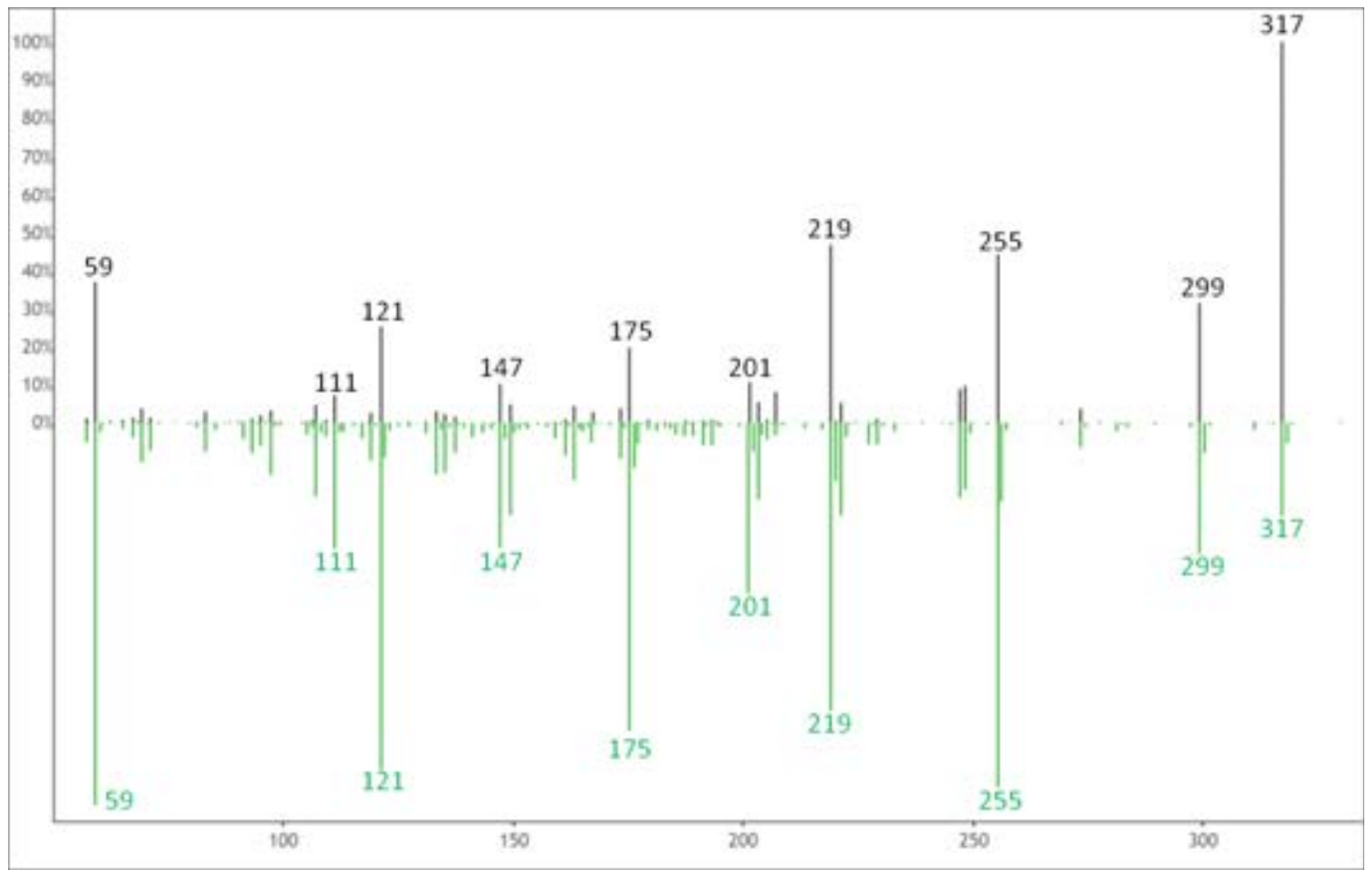

m/z 317.2122
at 3.97 minutes

15-HEPE
reference spectra

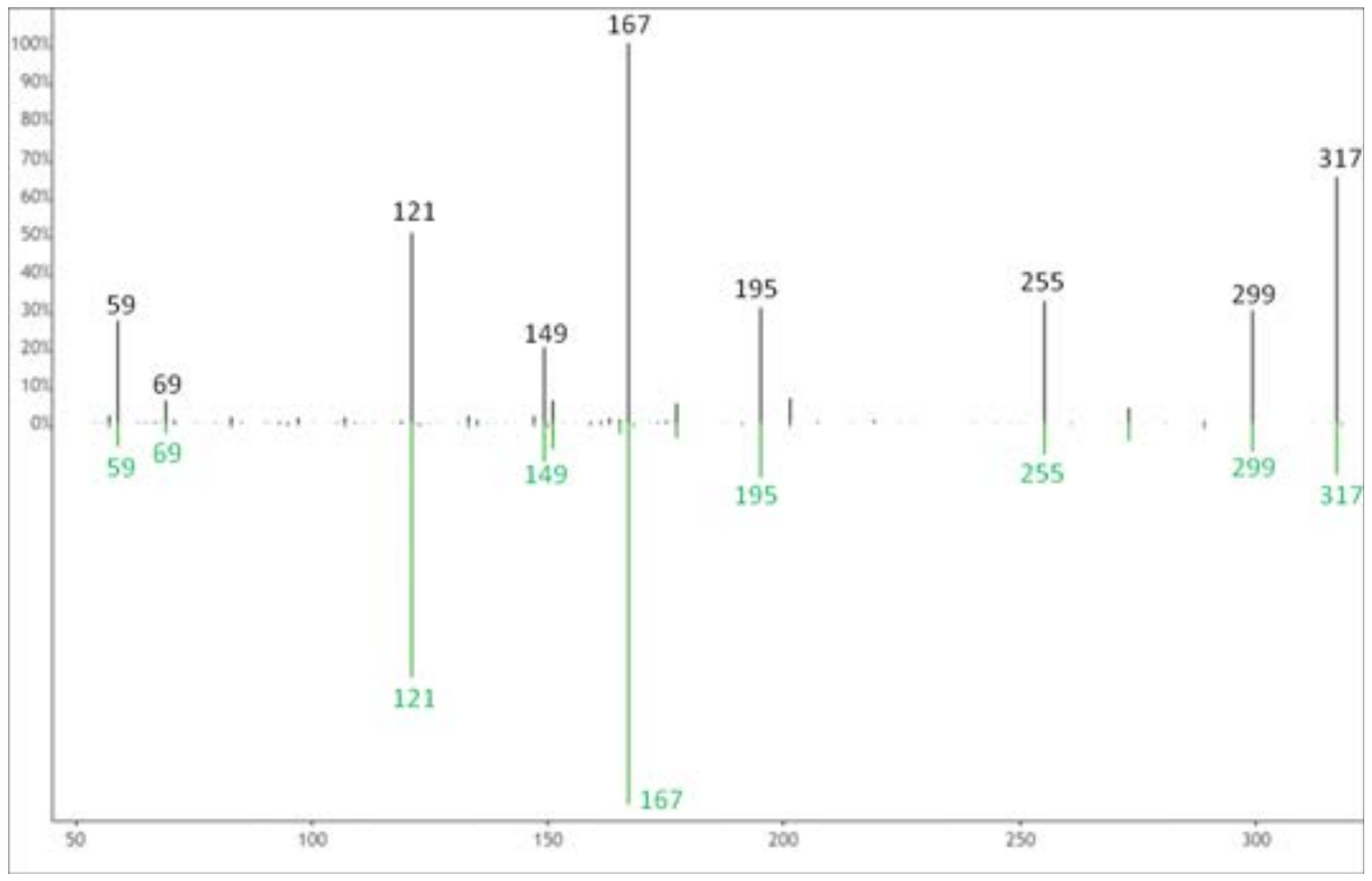

m/z 317.2122
at 4.02 minutes

11-HEPE
reference spectra

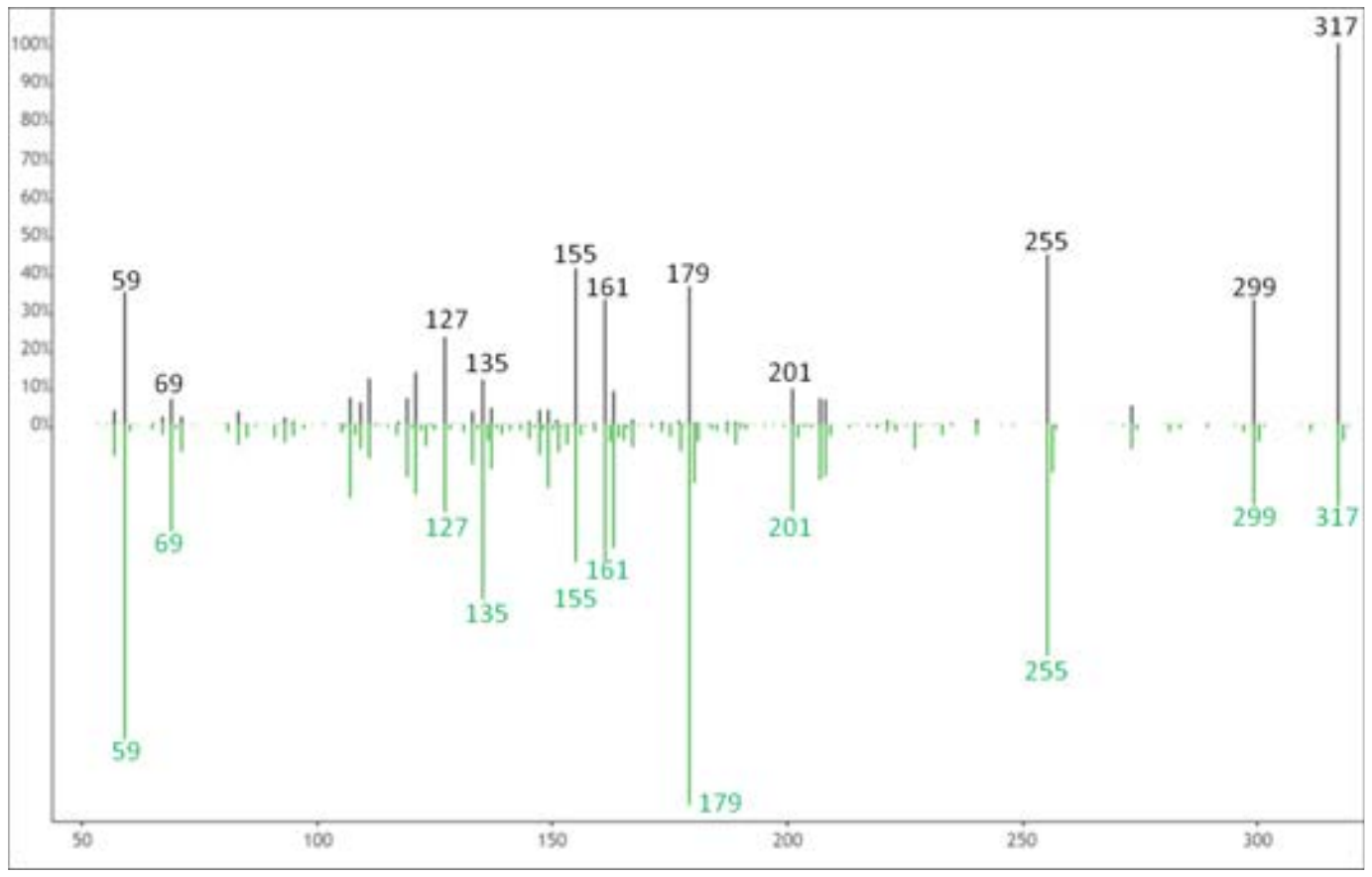

m/z 317.2122
at 4.10 minutes

12-HEPE
reference spectra

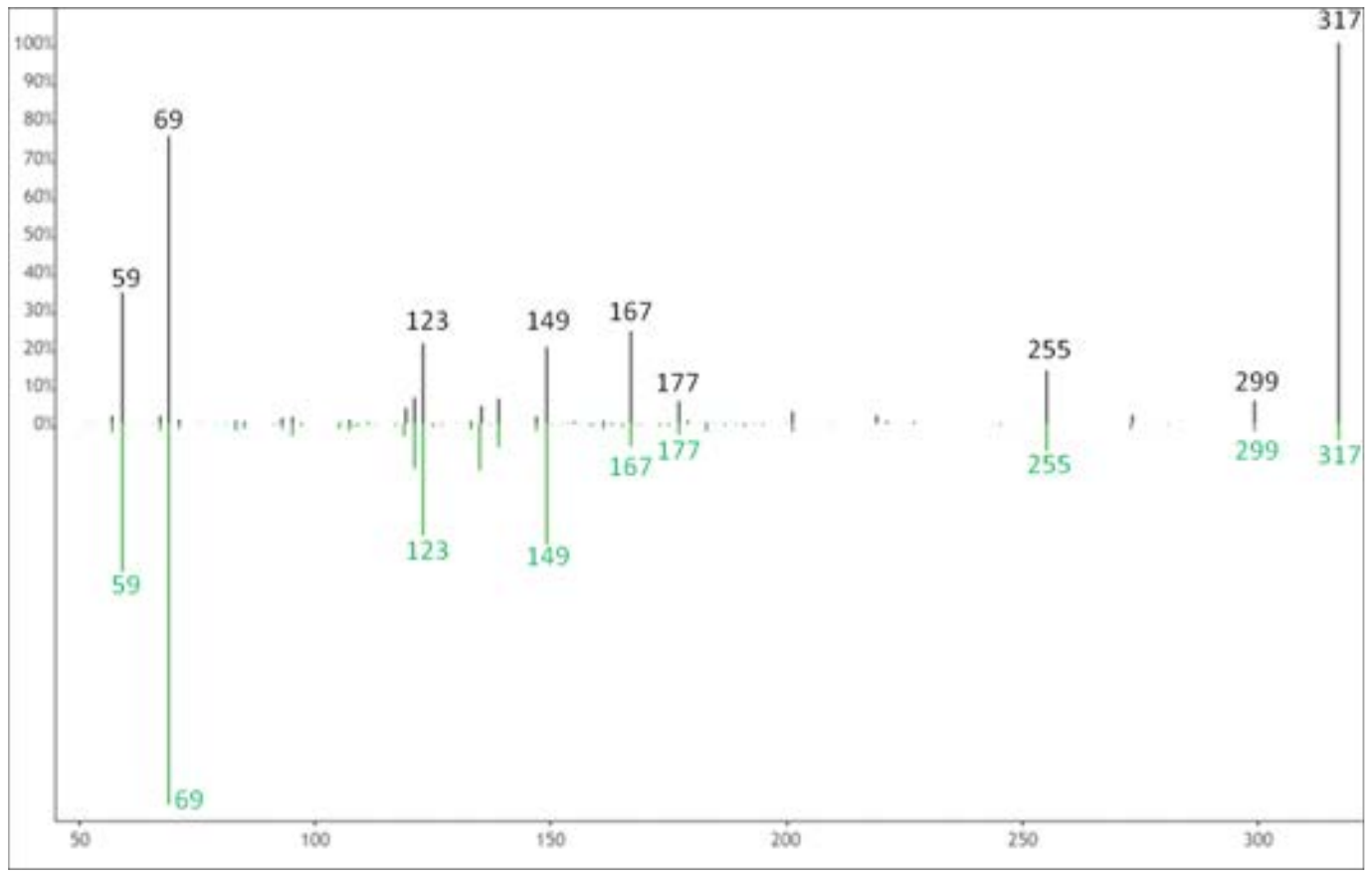

m/z 317.2122
at 4.15 minutes

9-HEPE
reference spectra

## Figure S3 (Continued)

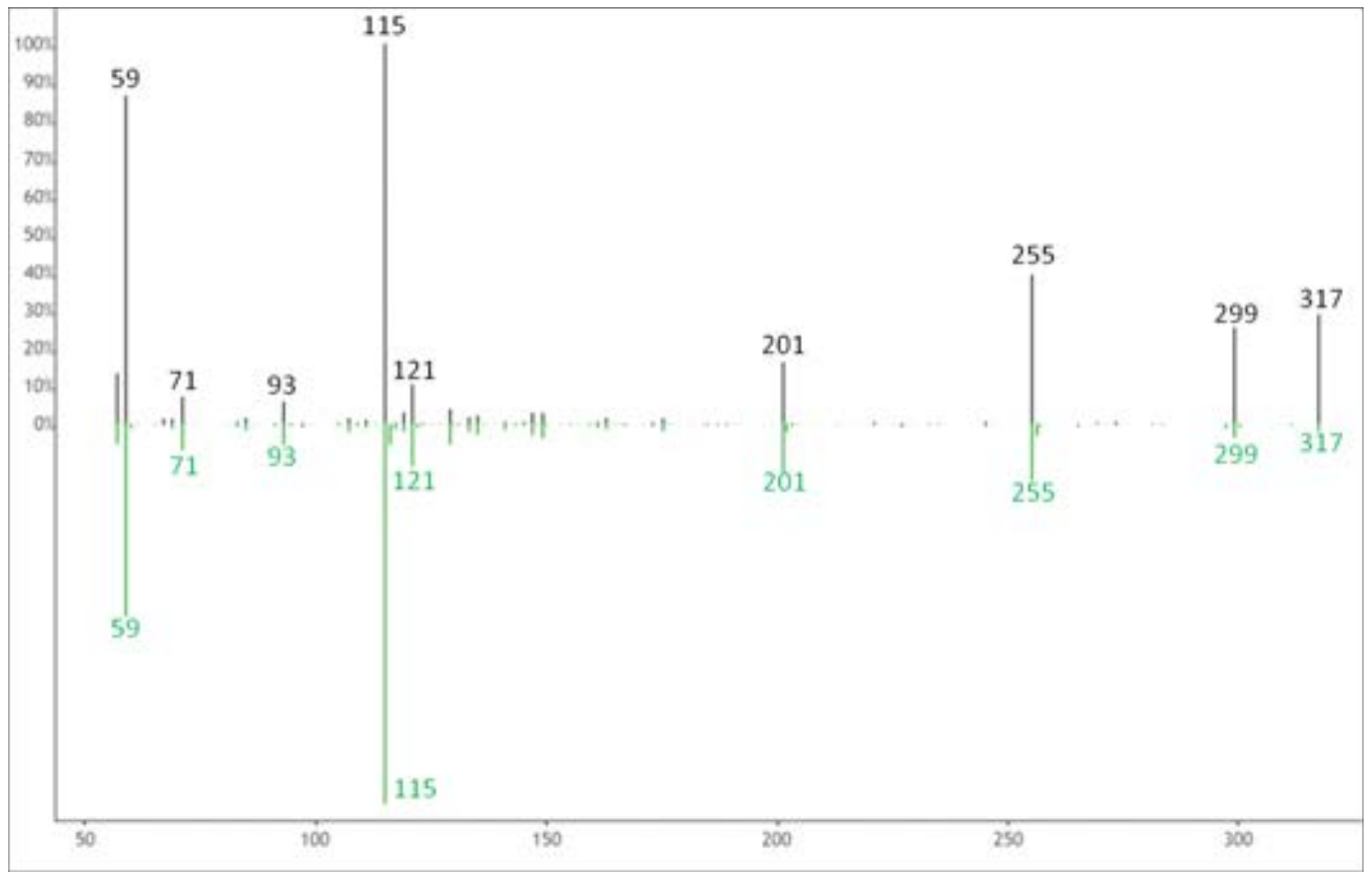

m/z 317.2122
at 4.29 minutes

5-HEPE
reference spectra

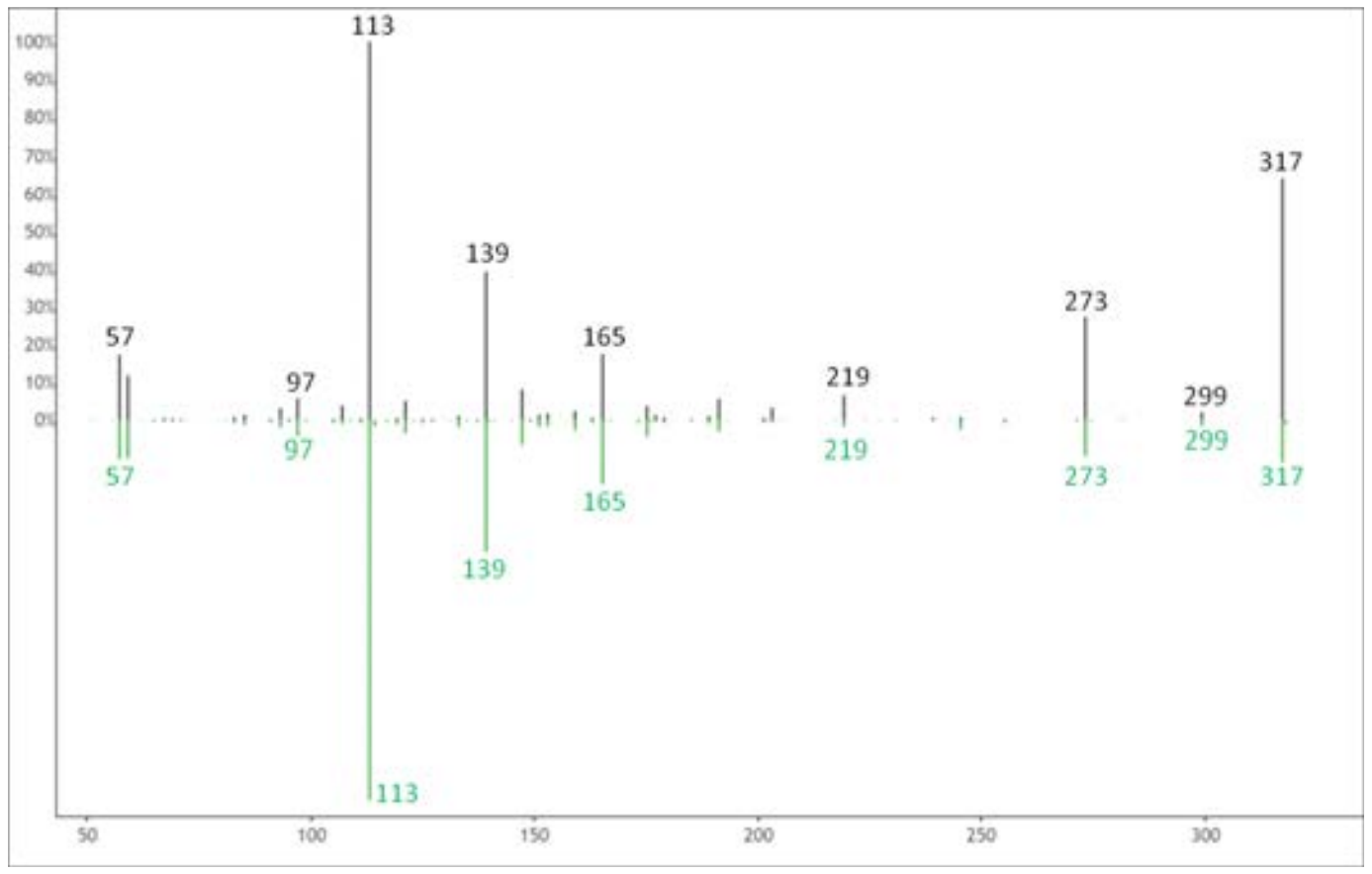

m/z 317.2122
at 4.52 minutes

15-oxo-EET
reference spectra

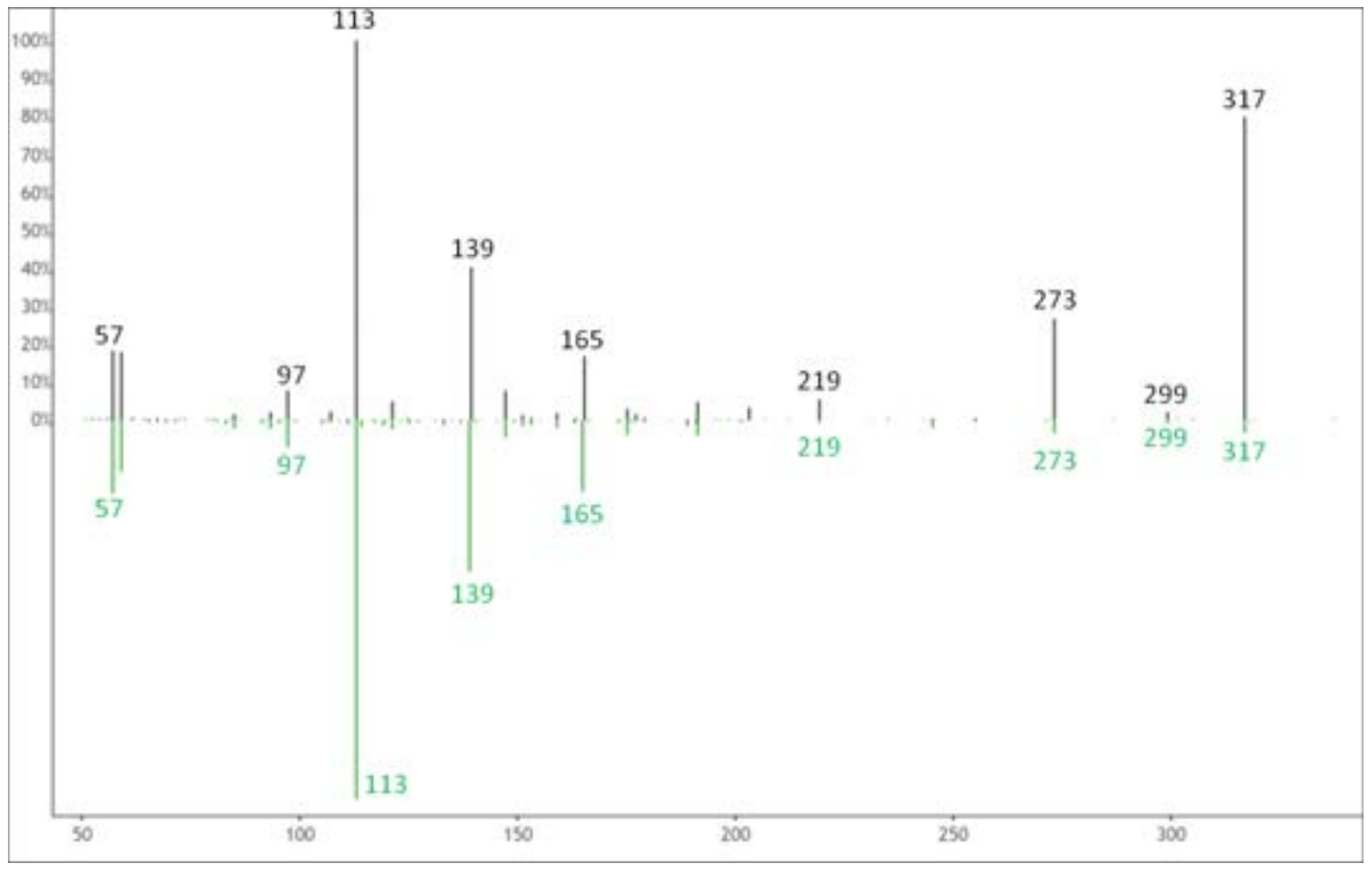

m/z 317.2122
at 4.59 minutes

15-oxo-ETE
reference spectra

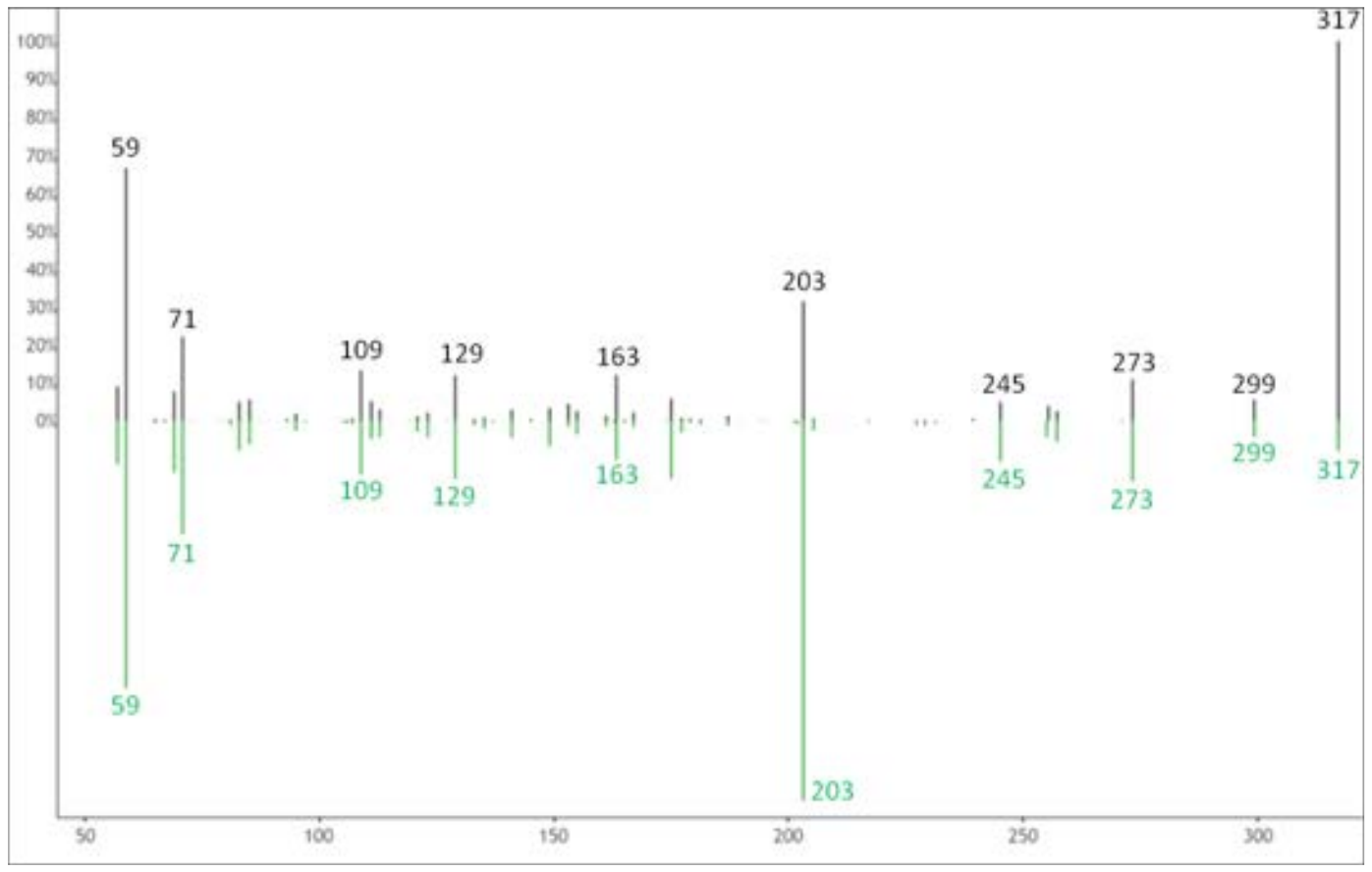

m/z 317.2122
at 5.22 minutes

5-oxo-EET
reference spectra

## Figure S3 (Continued)

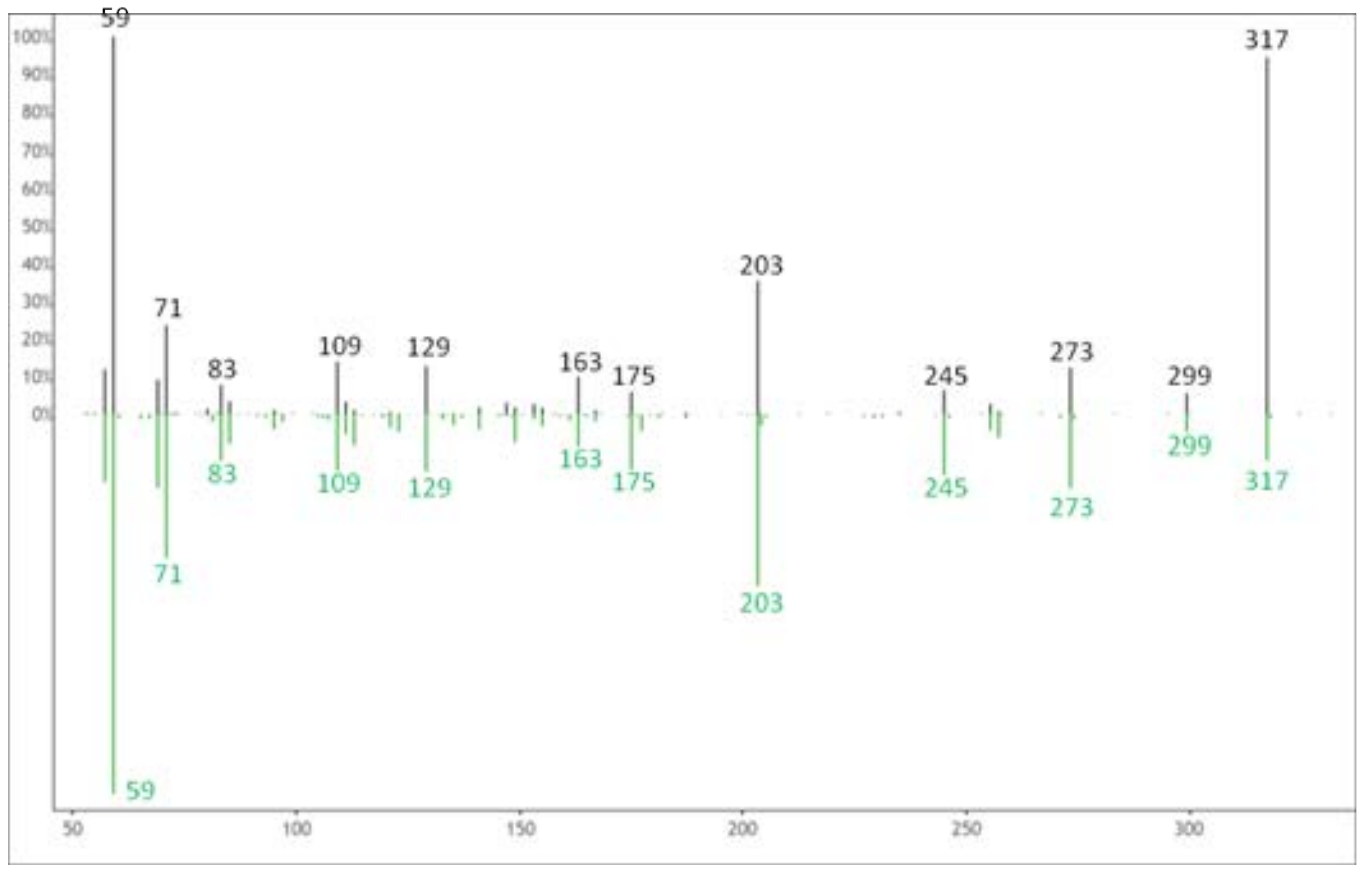

m/z 317.2122
at 5.30 minutes

5-HpETE
reference spectra

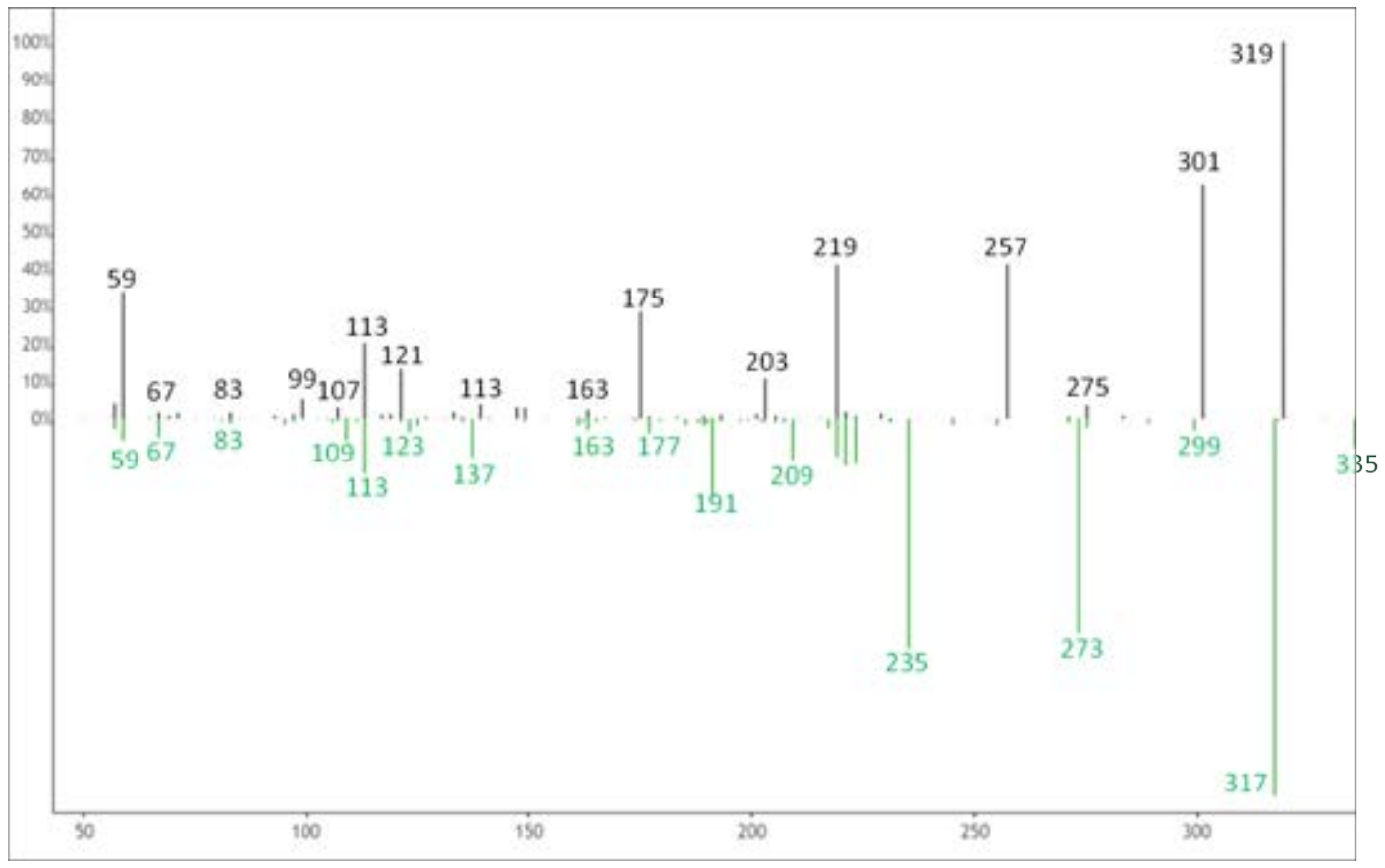

m/z 319.2278
at 4.47 minutes

15-*epi*-PGA1
reference spectra

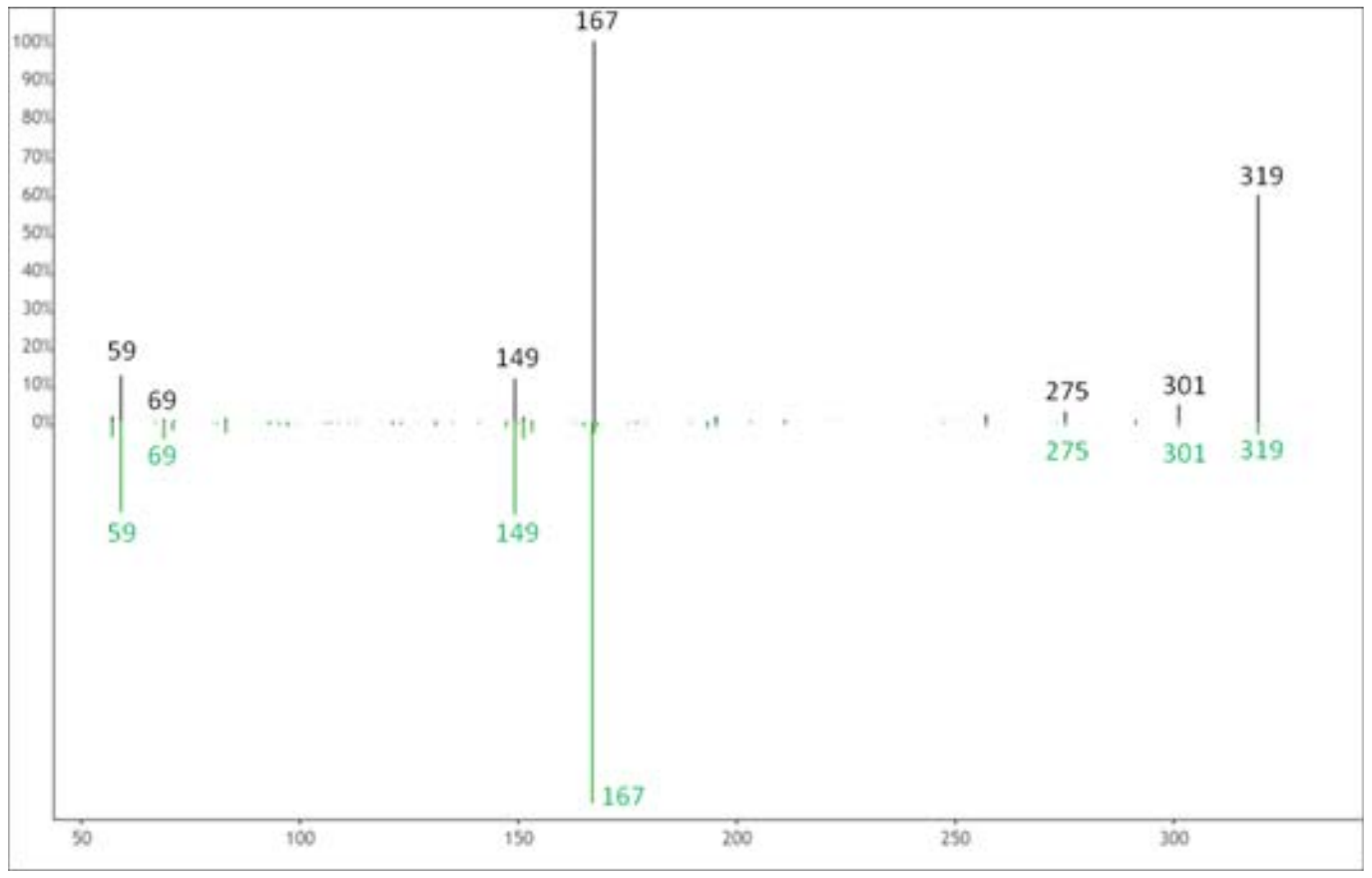

m/z 319.2278
at 4.62 minutes

11-HETE
reference spectra

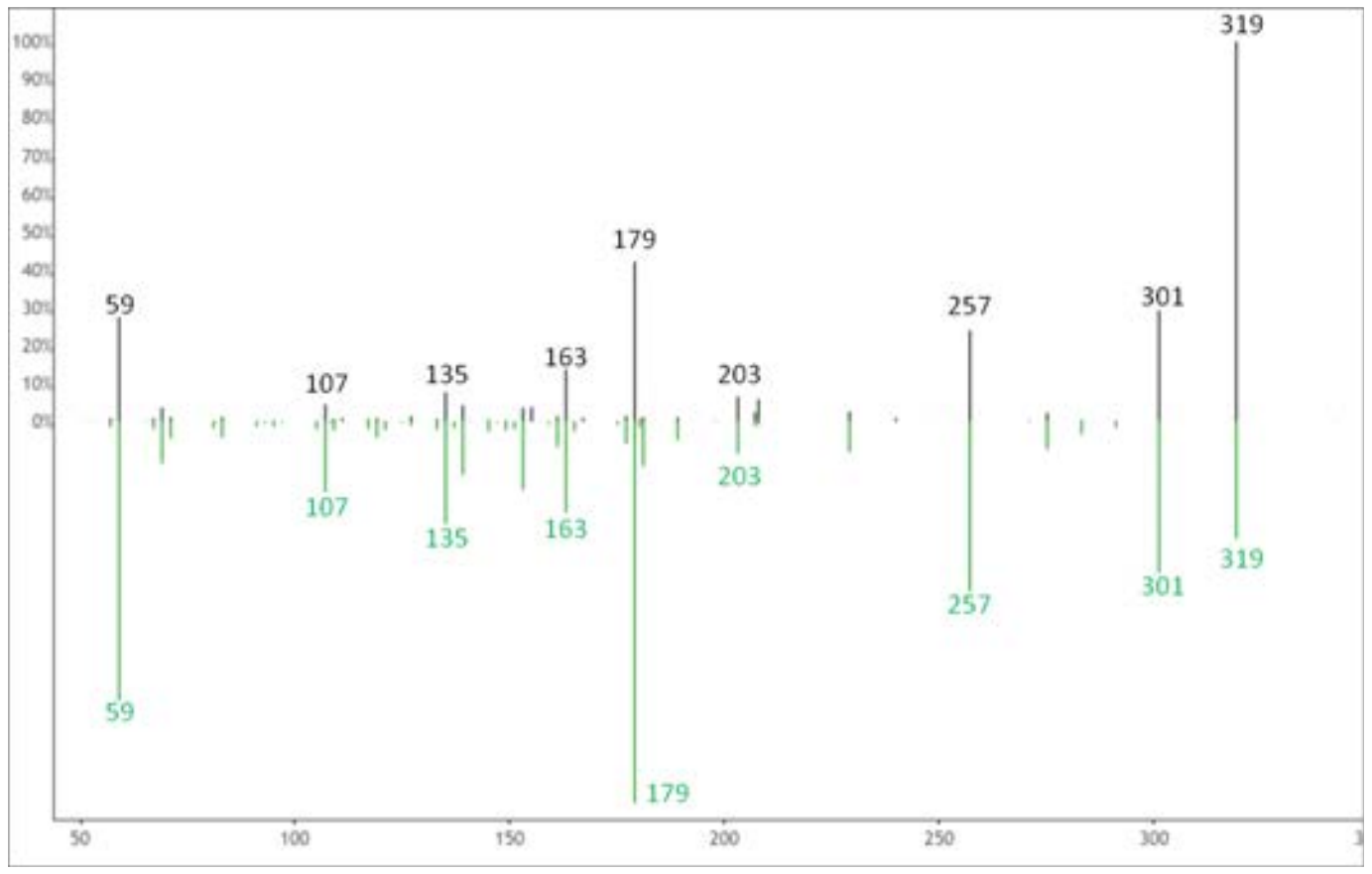

m/z 319.2278
at 4.72 minutes

12-HETE
reference spectra

## Figure S3 (Continued)

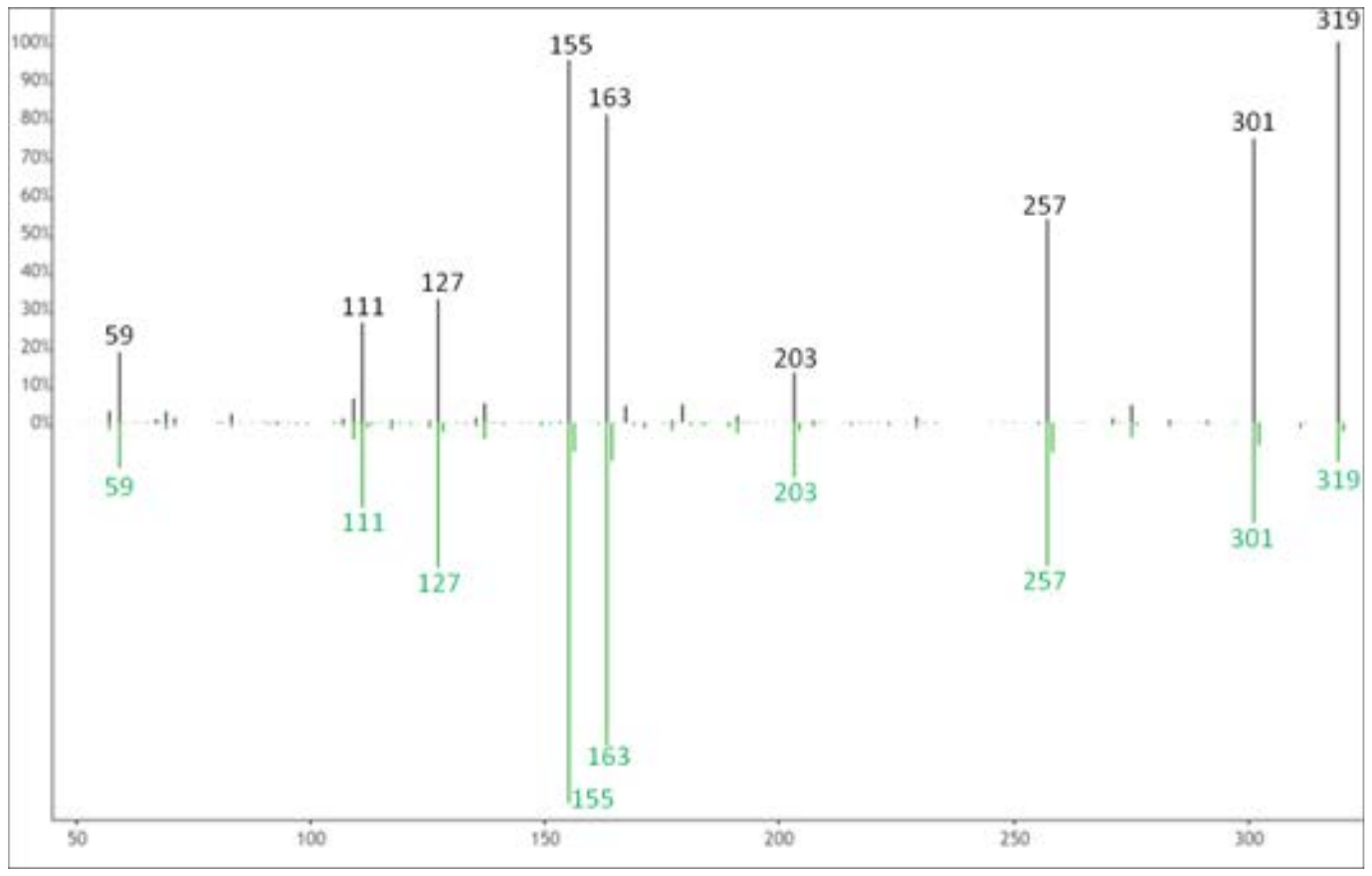

m/z 319.2278
at 4.76 minutes

8-HETE
reference spectra

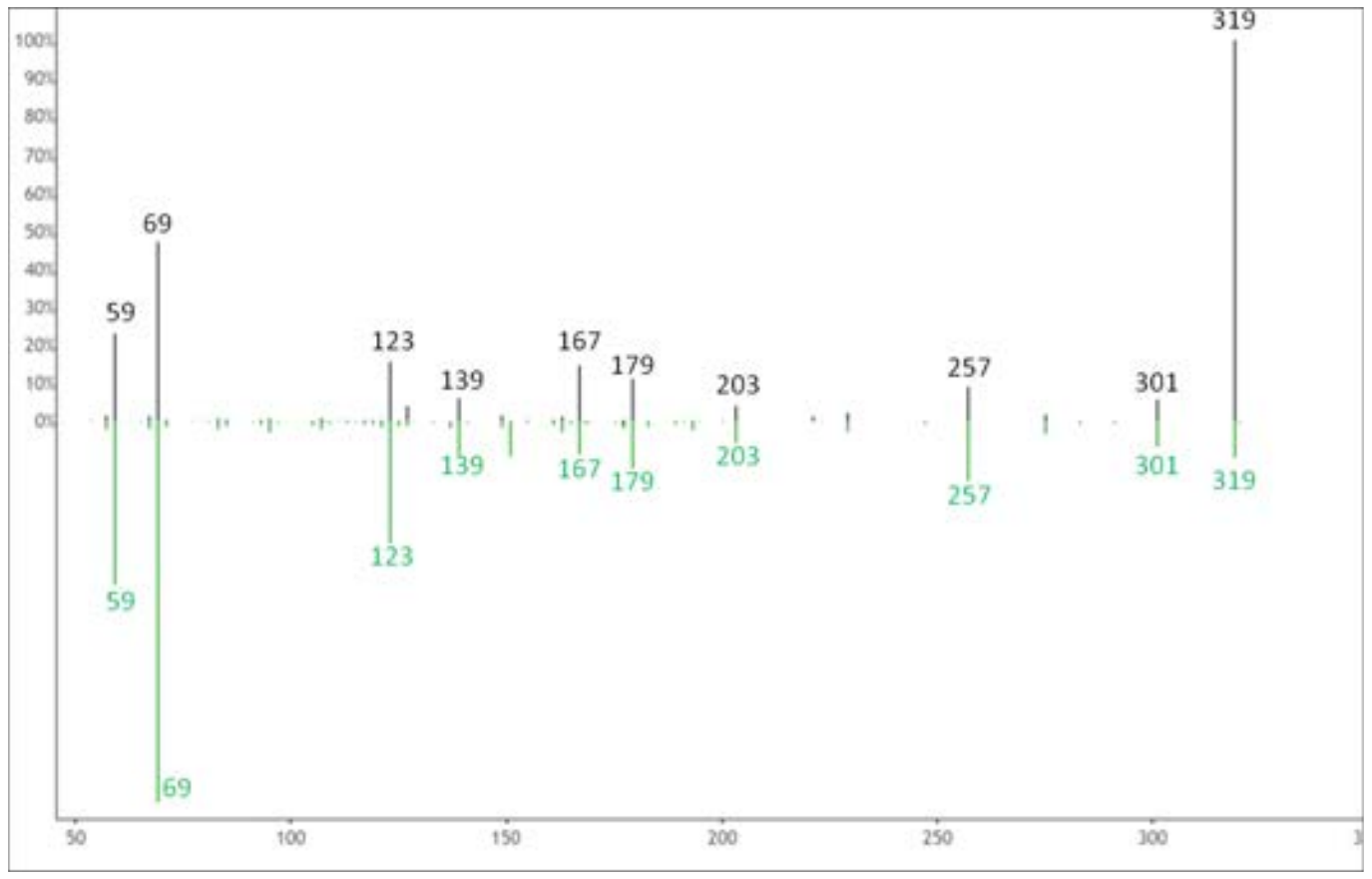

m/z 319.2278
at 4.82 minutes

9-HETE
reference spectra

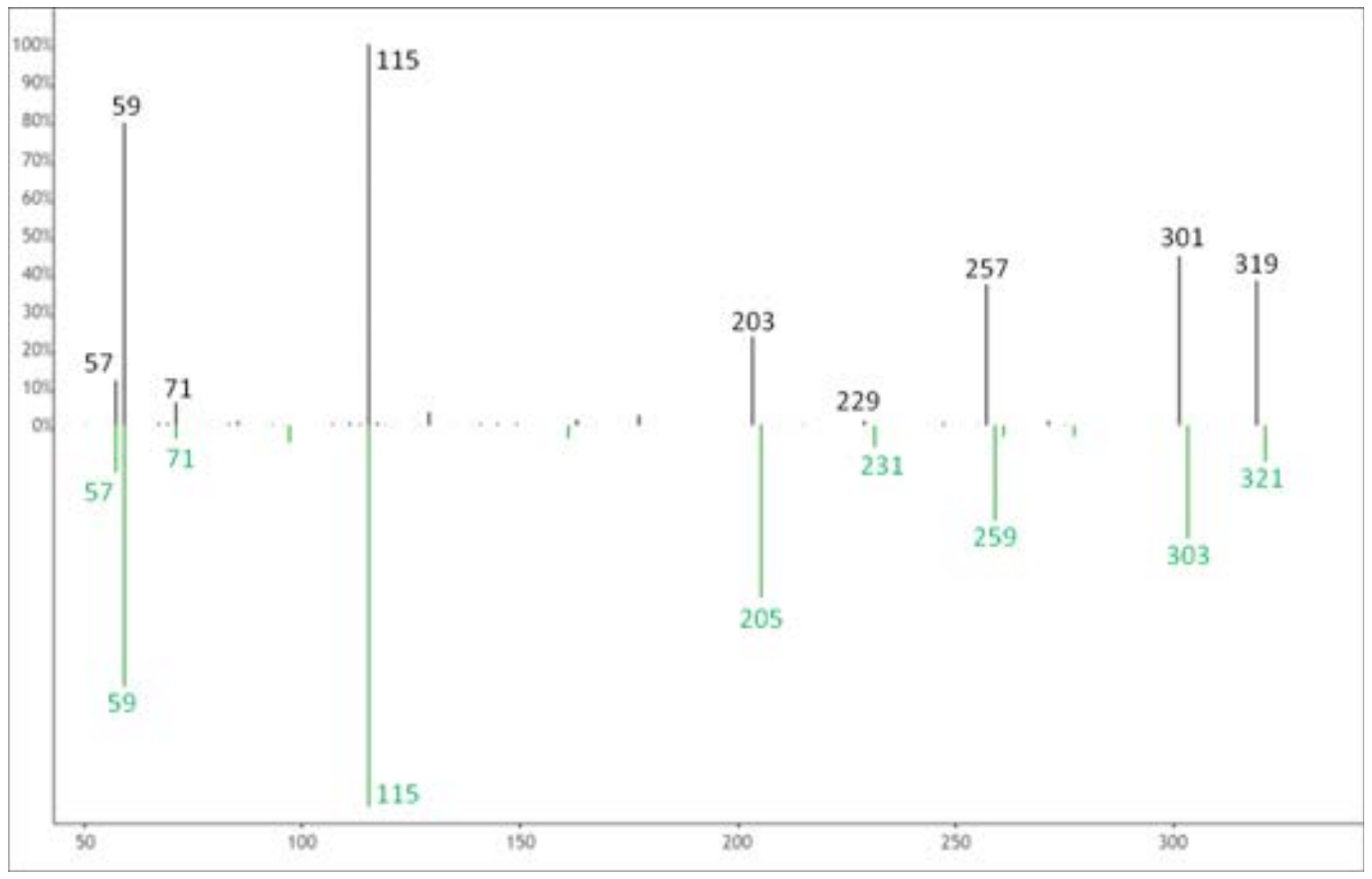

m/z 319.2278
at 4.98 minutes

5-HETrE
reference spectra

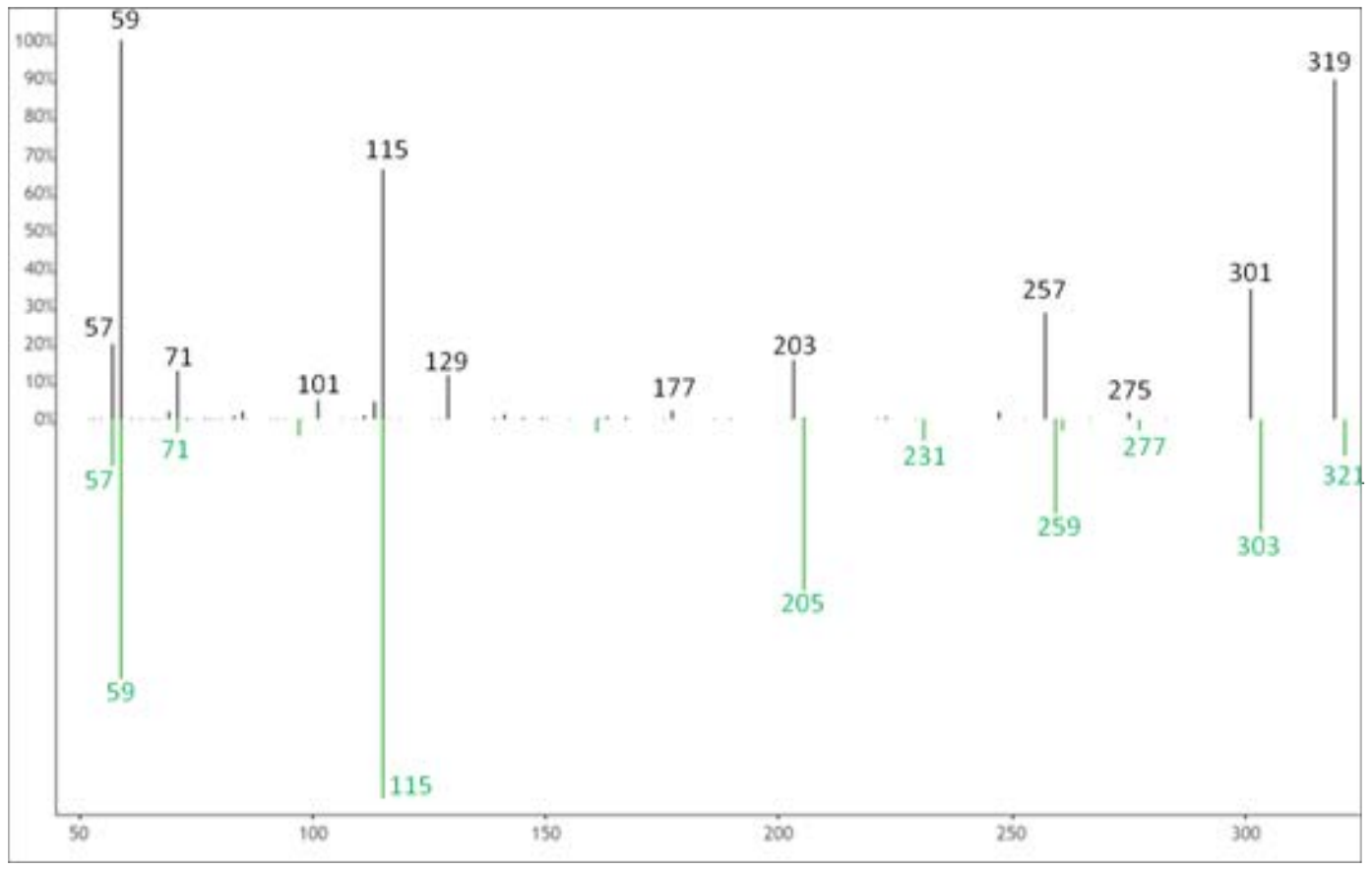

m/z 319.2278
at 5.05 minutes

5-HETrE
reference spectra

## Figure S3 (Continued)

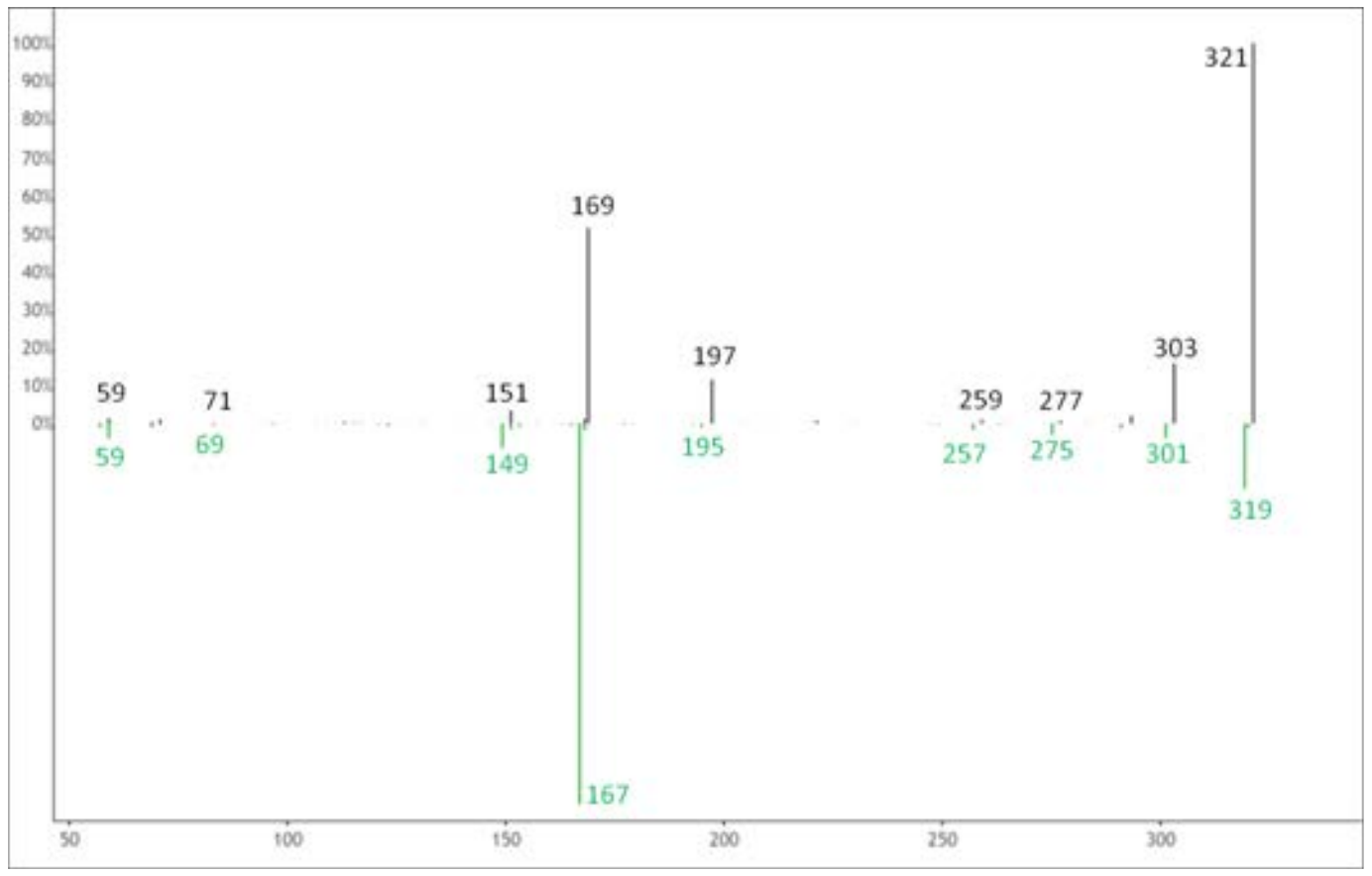

m/z 321.2435
at 4.94 minutes

11-HETE
reference spectra

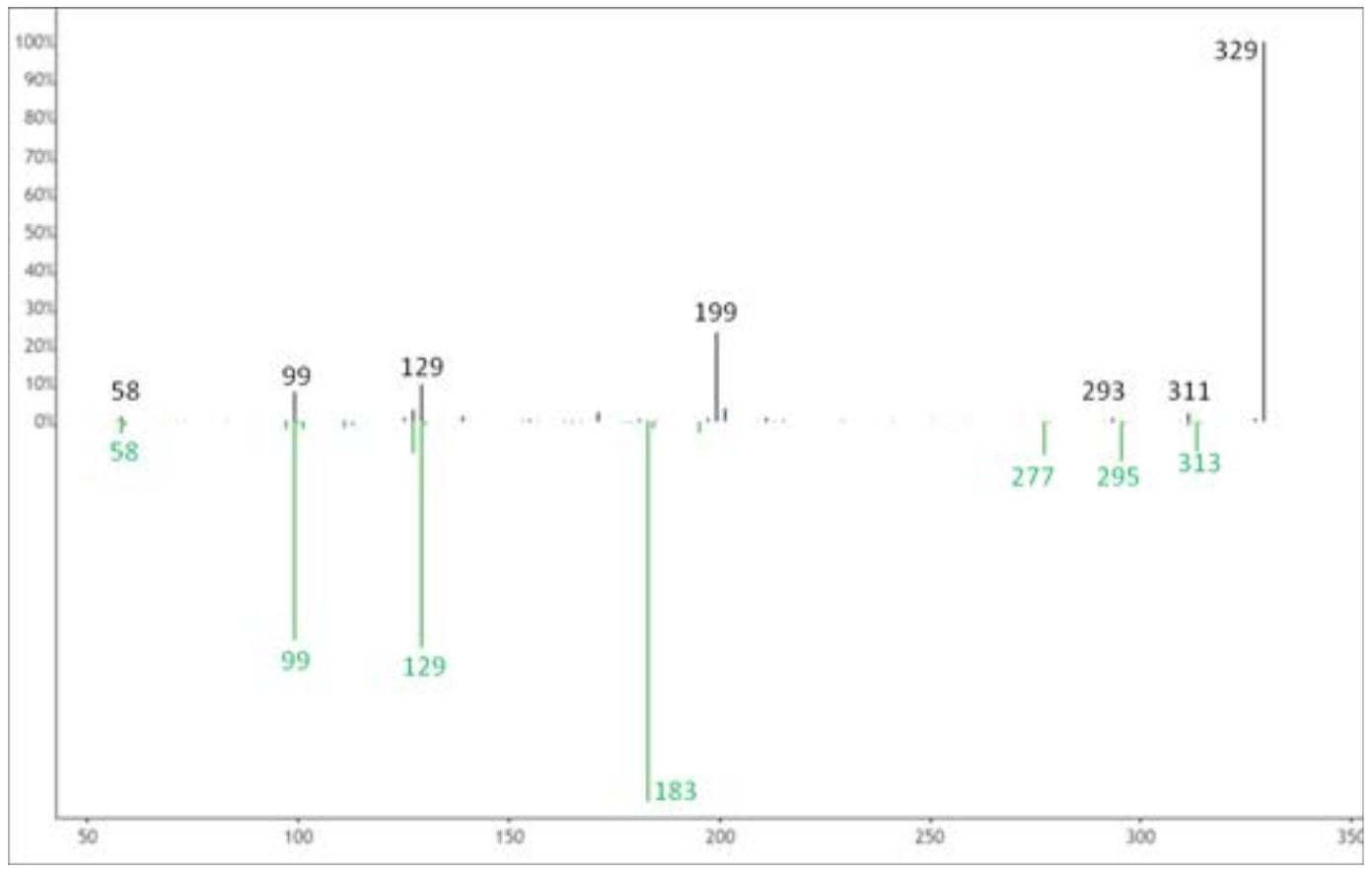

m/z 329.2333
at 2.30 minutes

12,13-DiHOME
reference spectra

## Figure S3 (Continued)

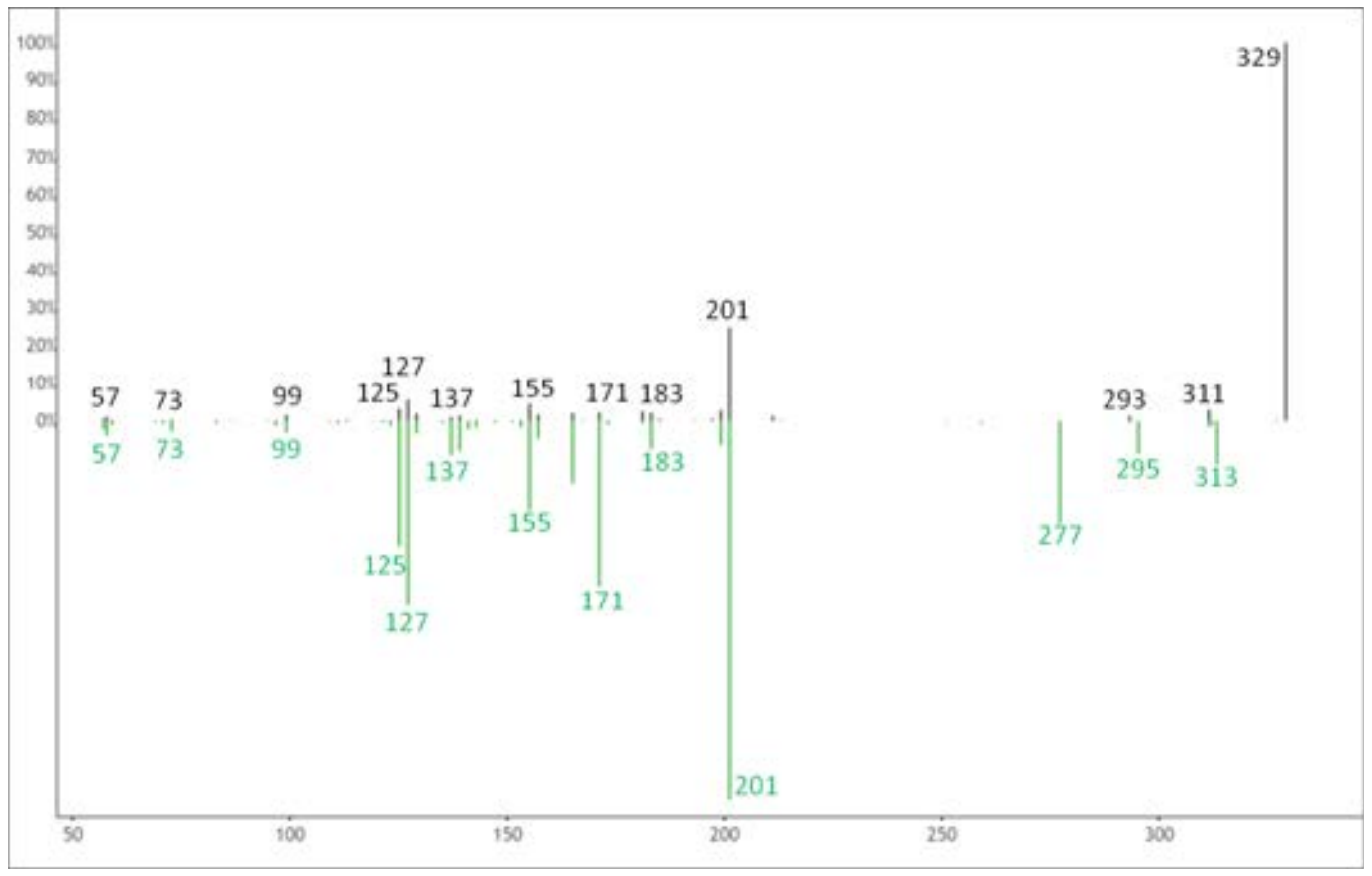

m/z 329.2333
at 2.37 minutes

9,10-DiHOME
reference spectra

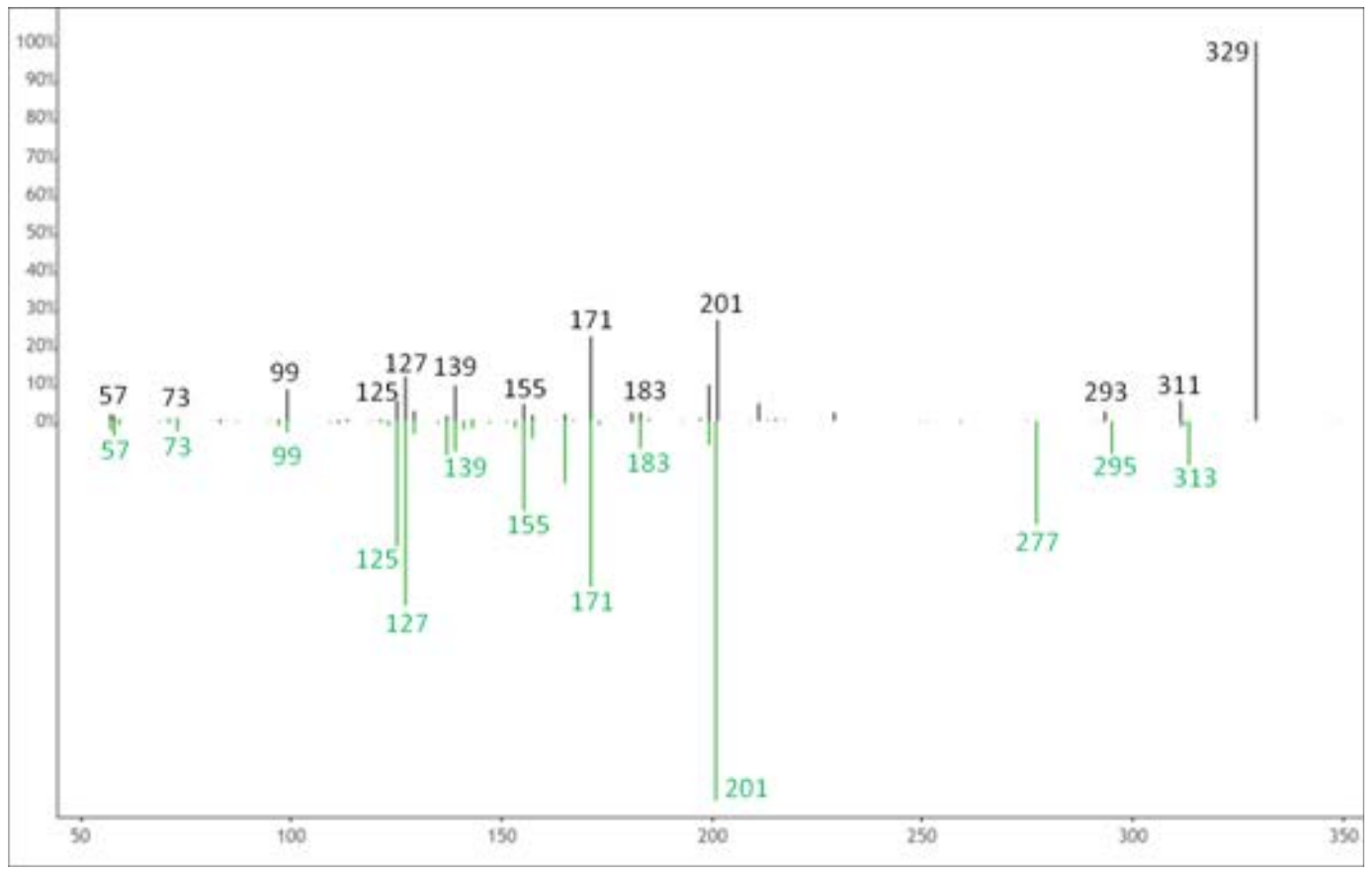

m/z 329.2333
at 2.47 minutes

9.10-DiHOME
reference spectra

## Figure S3 (Continued)

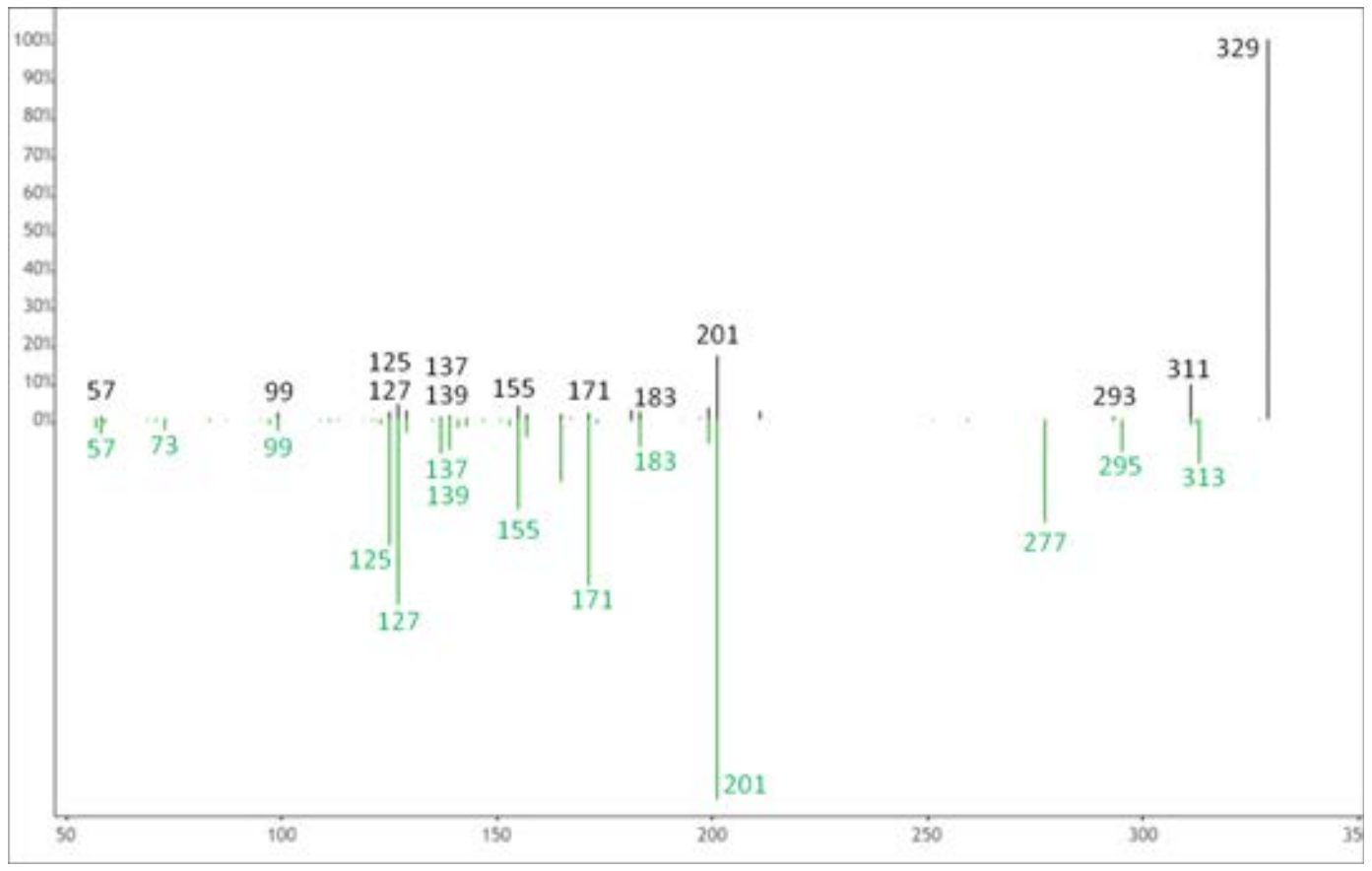

m/z 329.2333
at 2.79 minutes

9-HODE
reference spectra

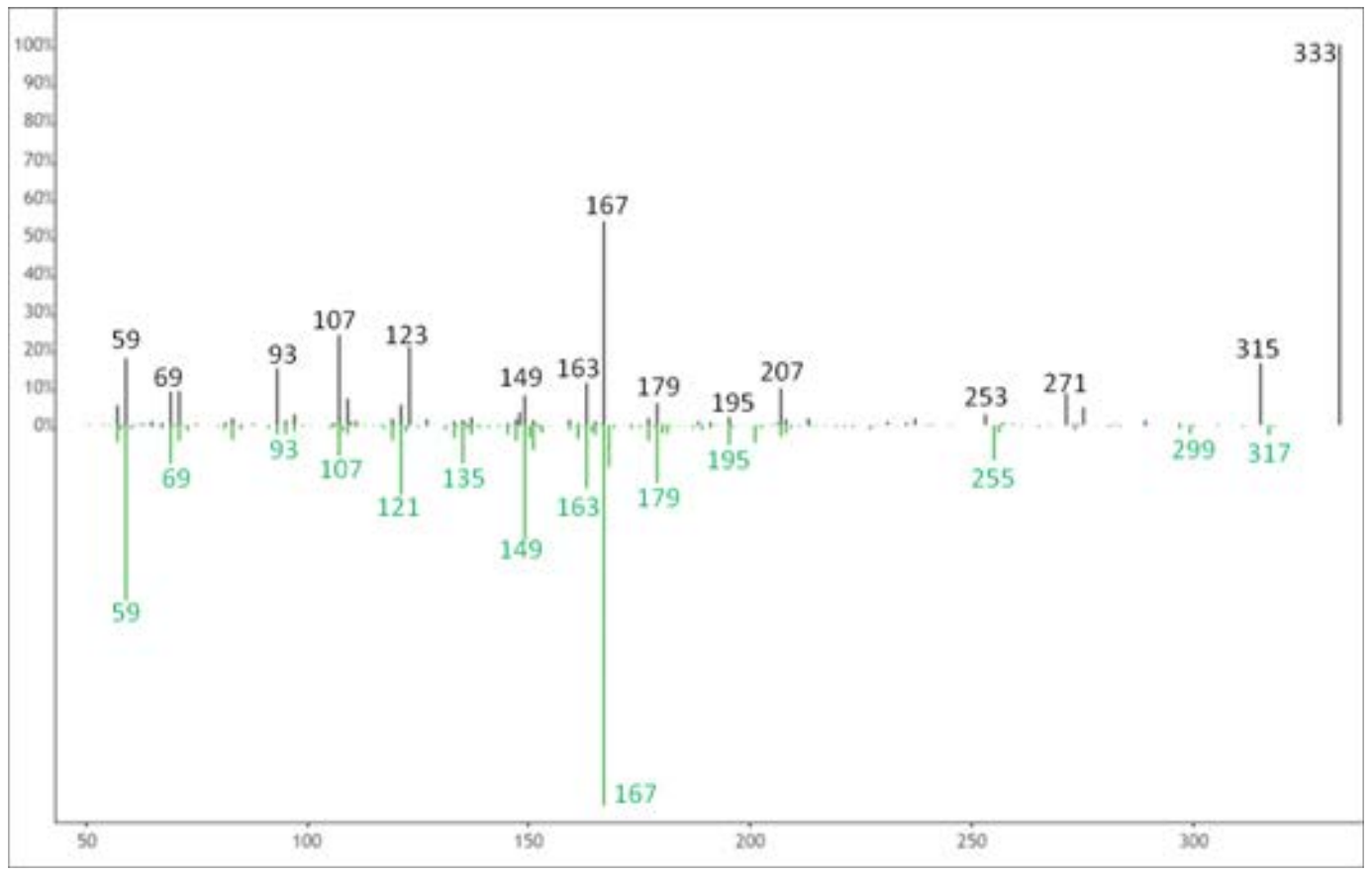

m/z 333.2071
at 2.45 minutes

11,12-EpETE
reference spectra

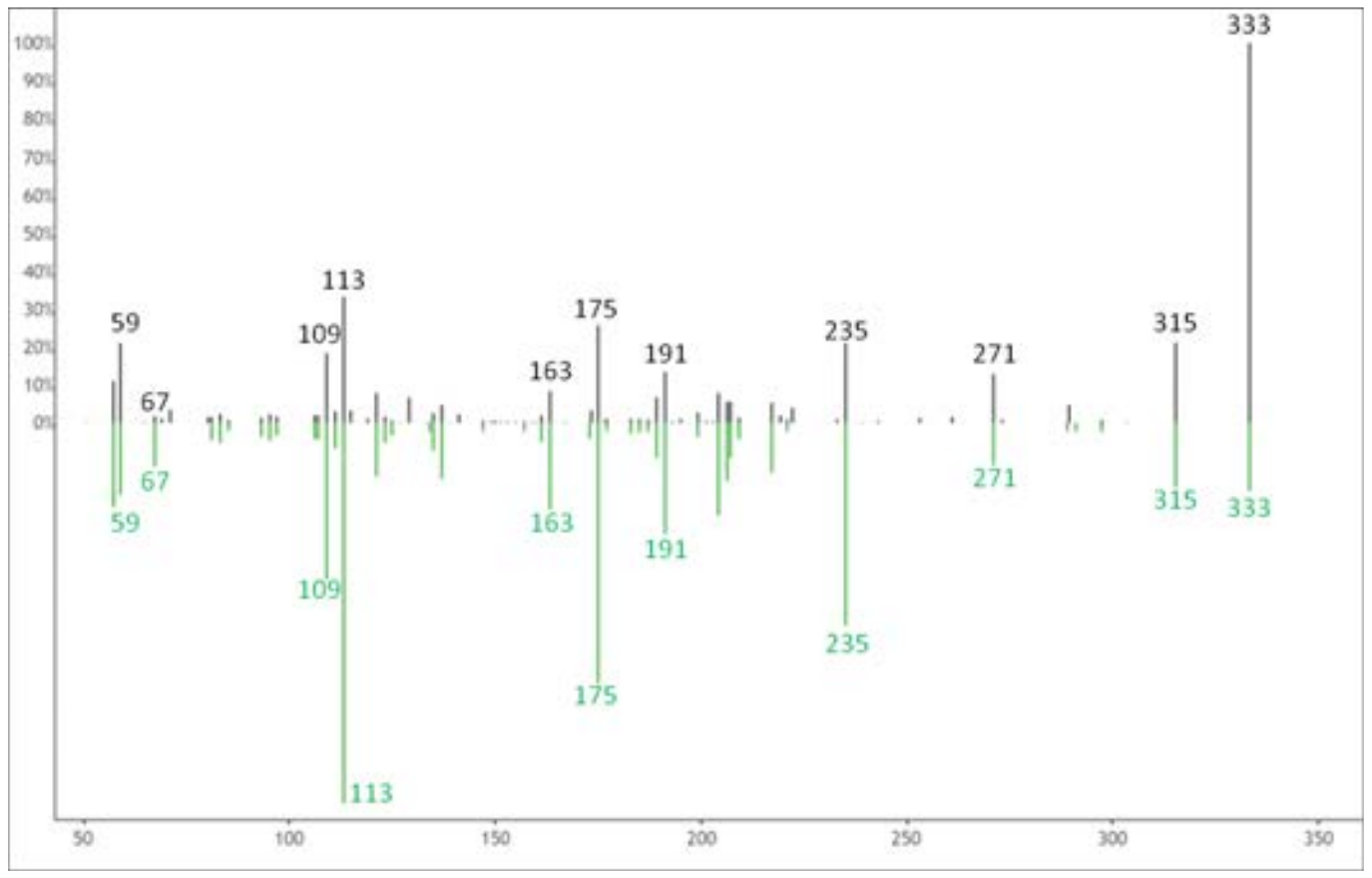

m/z 333.2071
at 2.66 minutes

PGB2
reference spectra

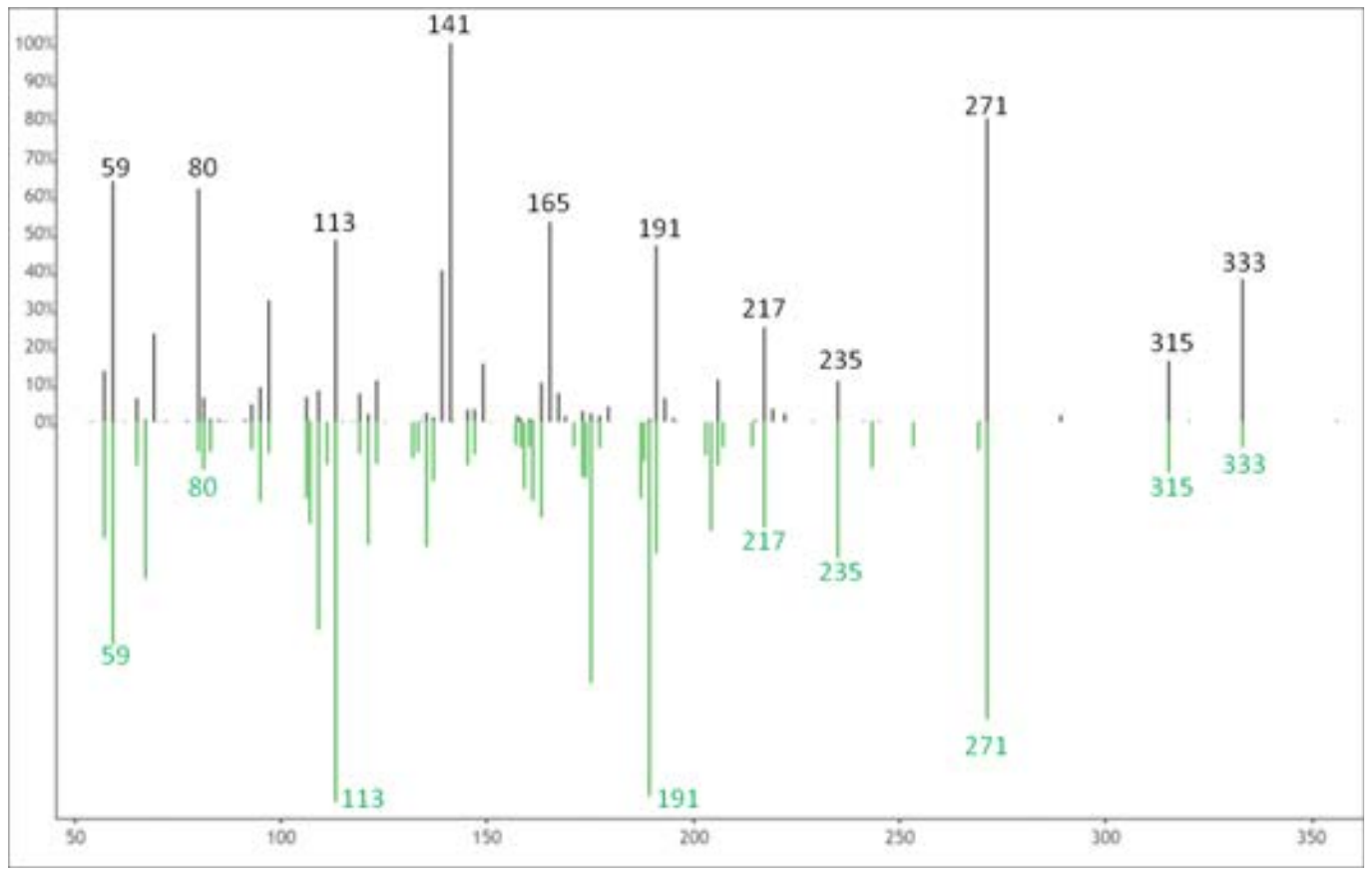

m/z 333.2071
at 2.99 minutes

8-*epi* $PGA_2$
reference spectra

## Figure S3 (Continued)

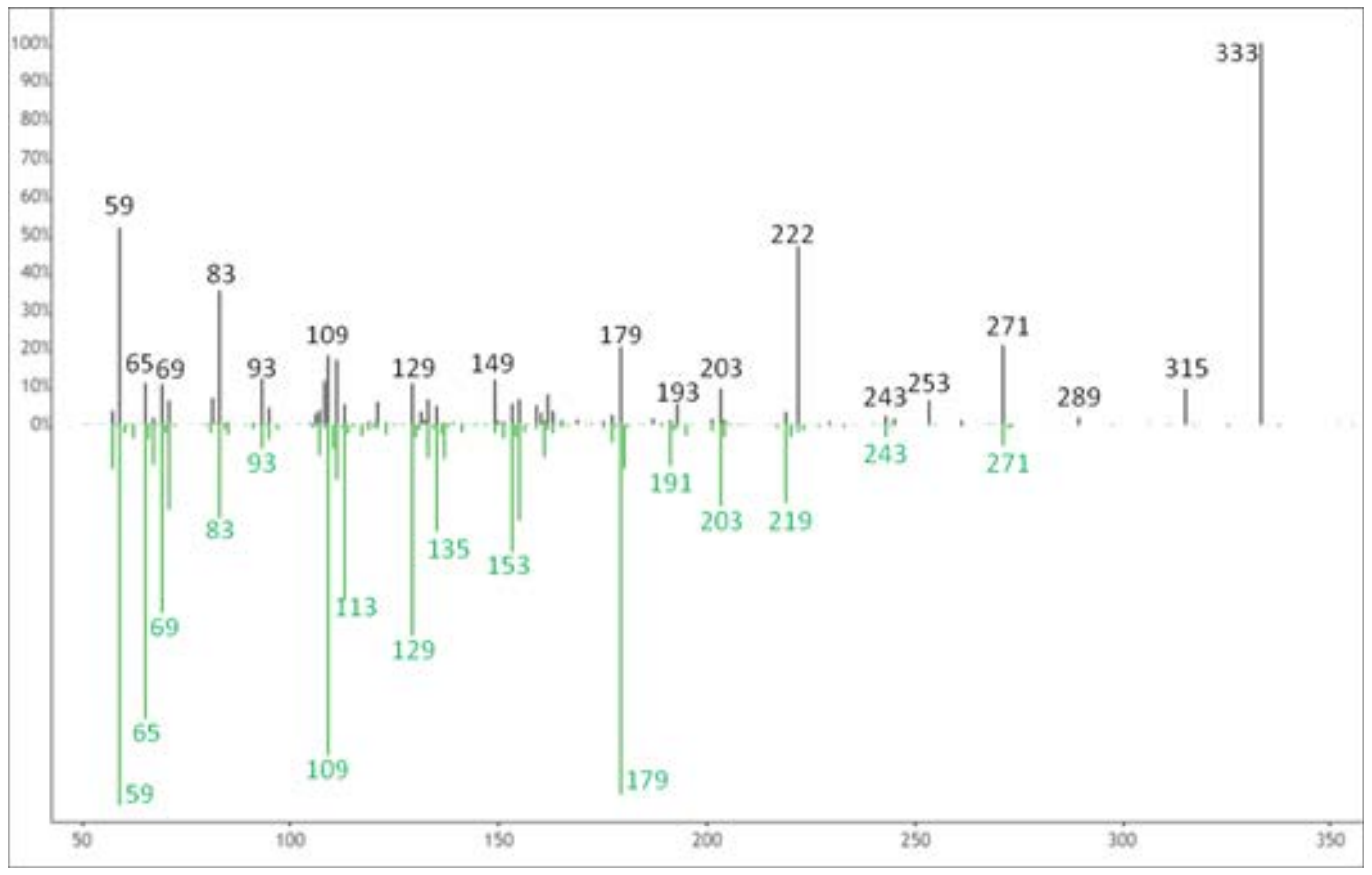

m/z 333.2071
at 3.25 minutes

12-oxo-LTB4
reference spectra

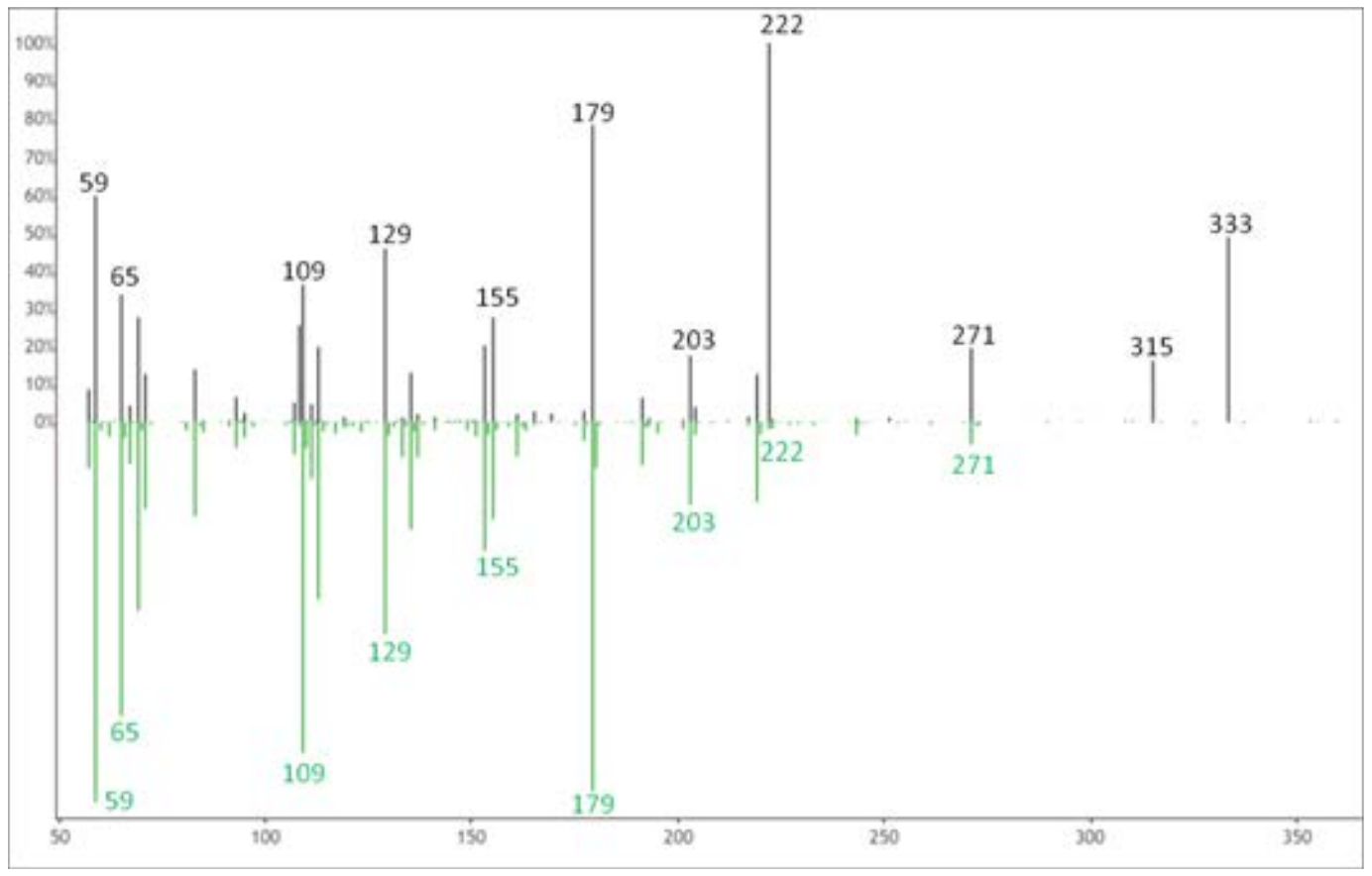

m/z 333.2071
at 3.49 minutes

12-oxo-LTB4
reference spectra

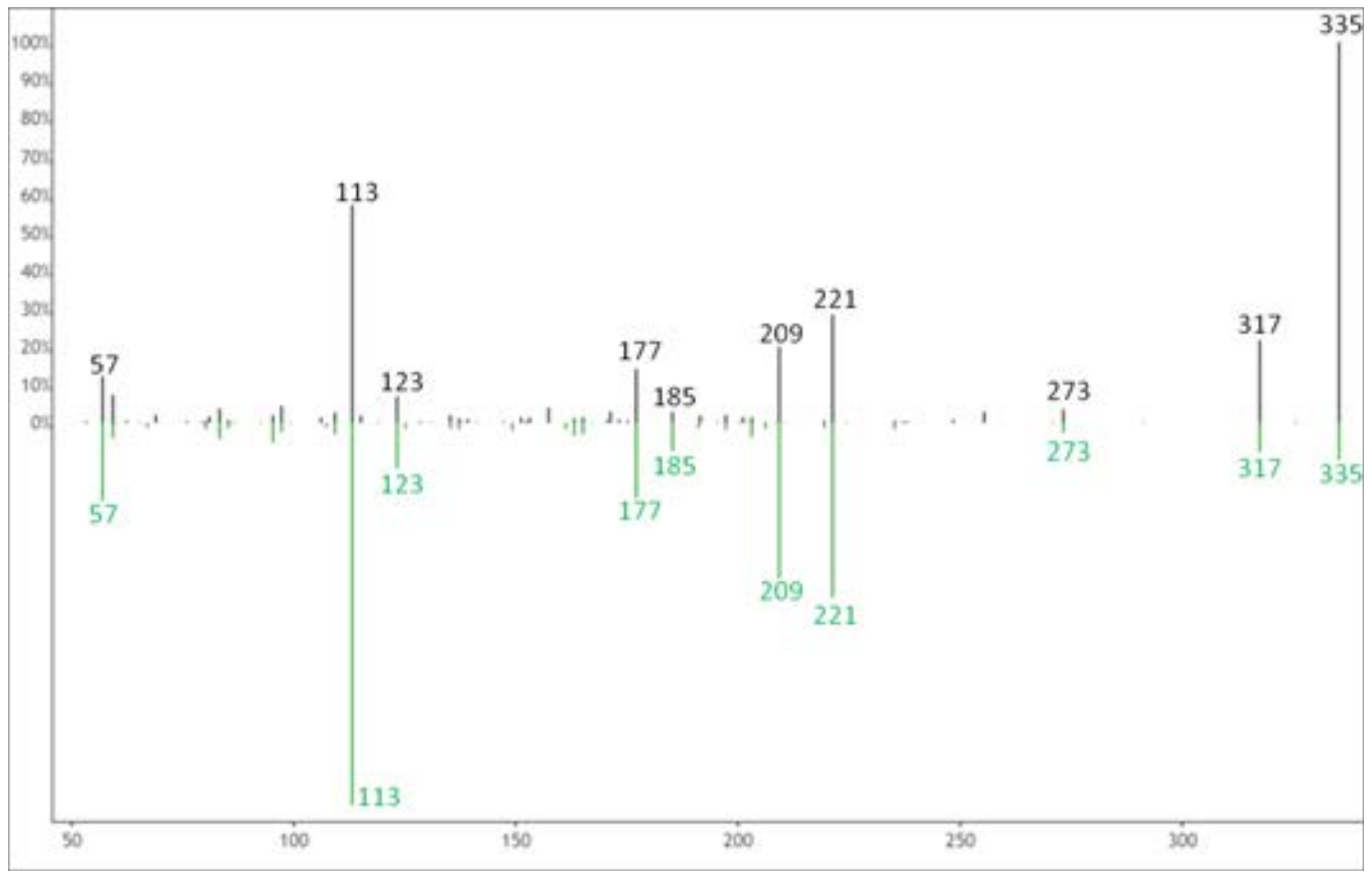

m/z 335.2228
at 2.74 minutes

PGB1
reference spectra

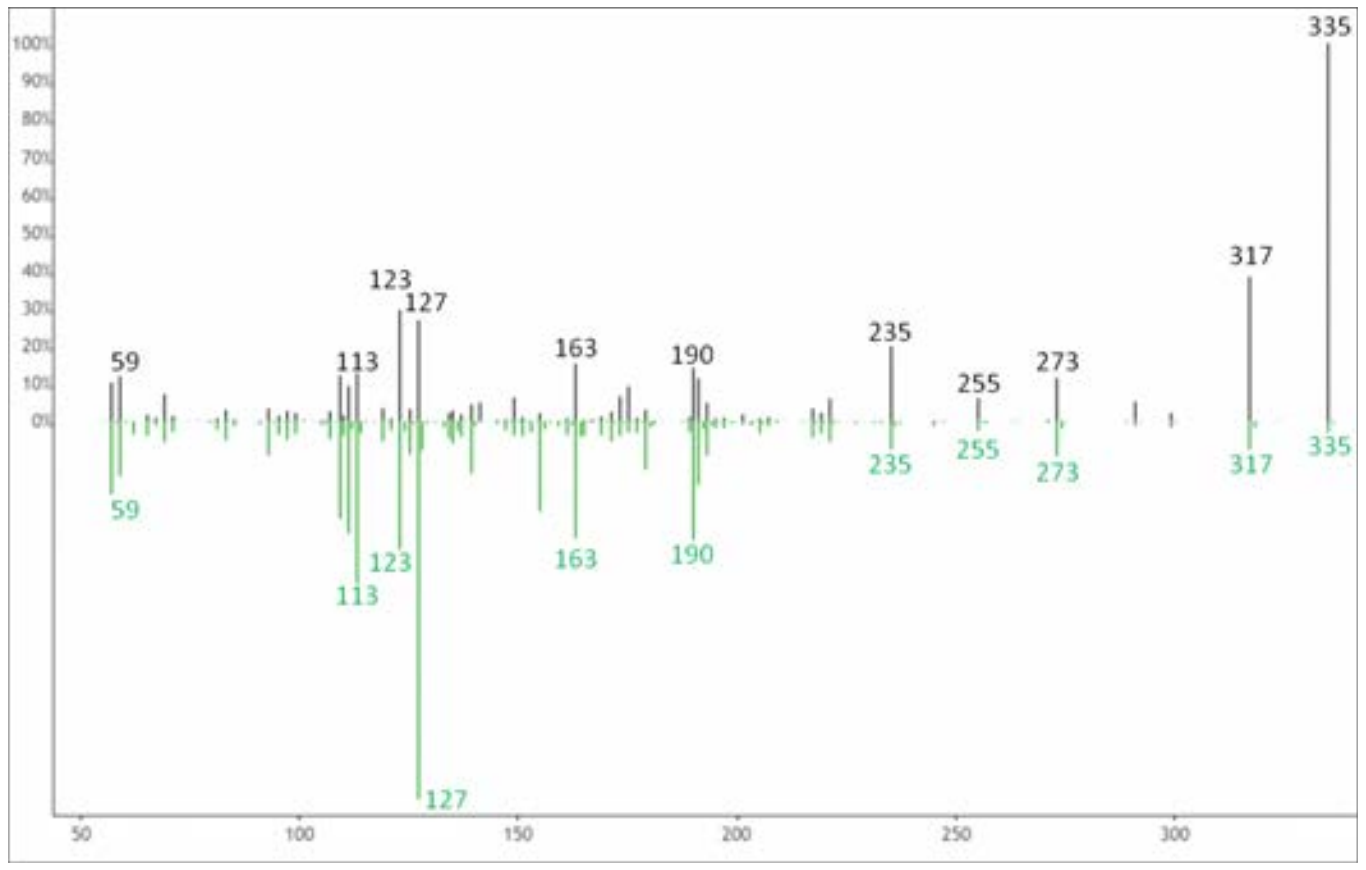

m/z 335.2228
at 2.80 minutes

8,15-DiHETE
reference spectra

## Figure S3 (Continued)

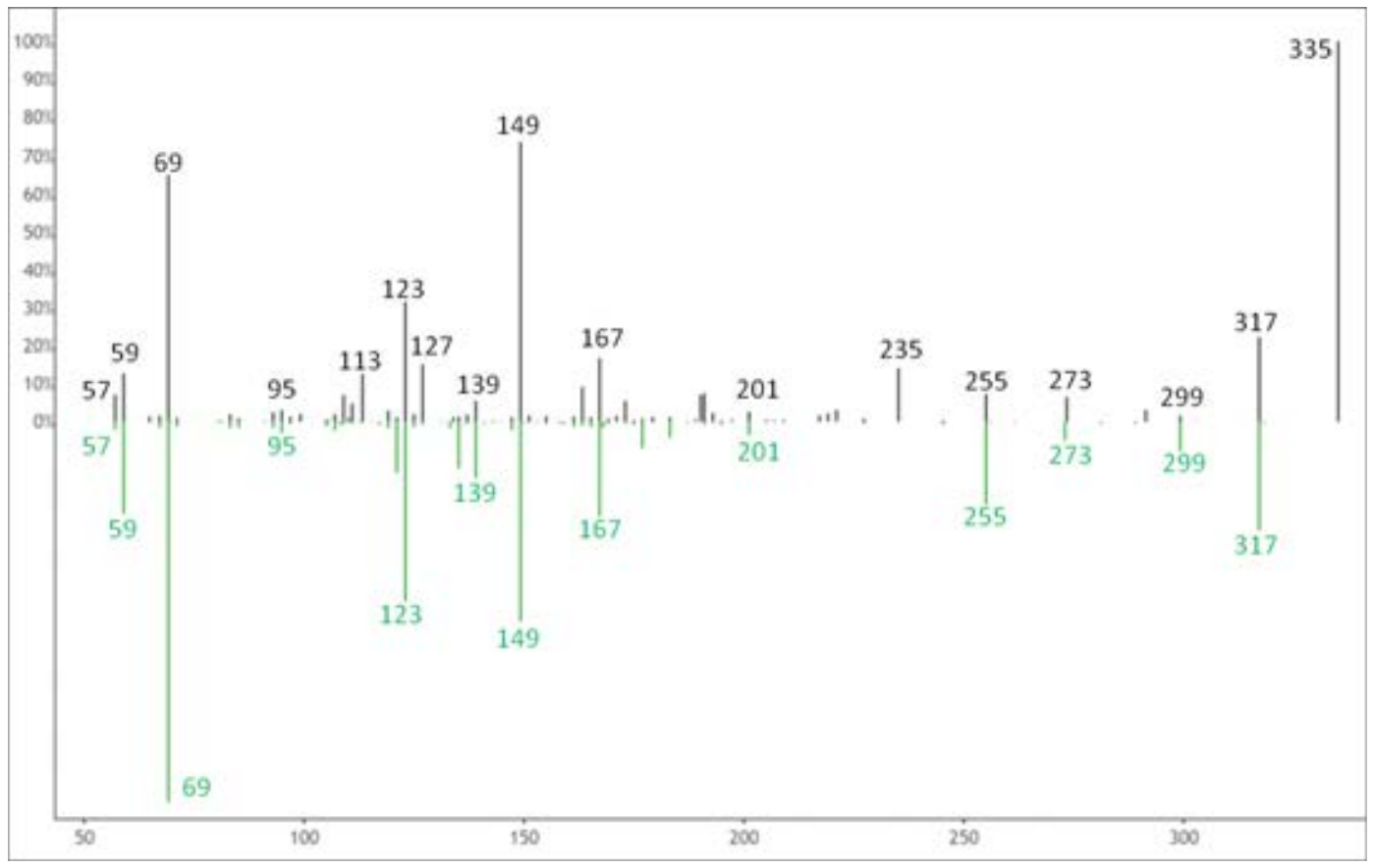

m/z 335.2228
at 2.88 minutes

9-HEPE
reference spectra

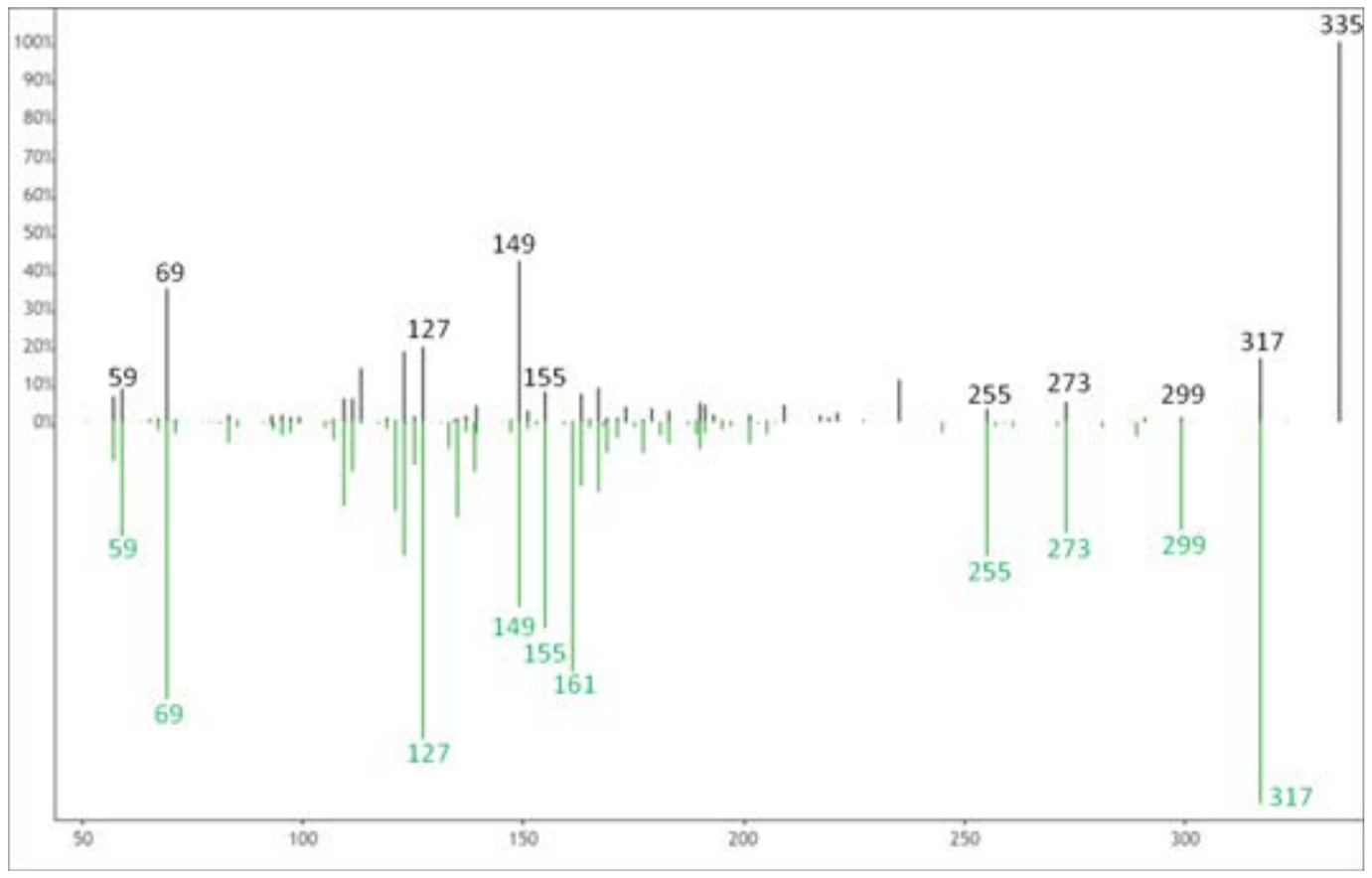

m/z 335.2228
at 2.92 minutes

8,9-EpETE
reference spectra

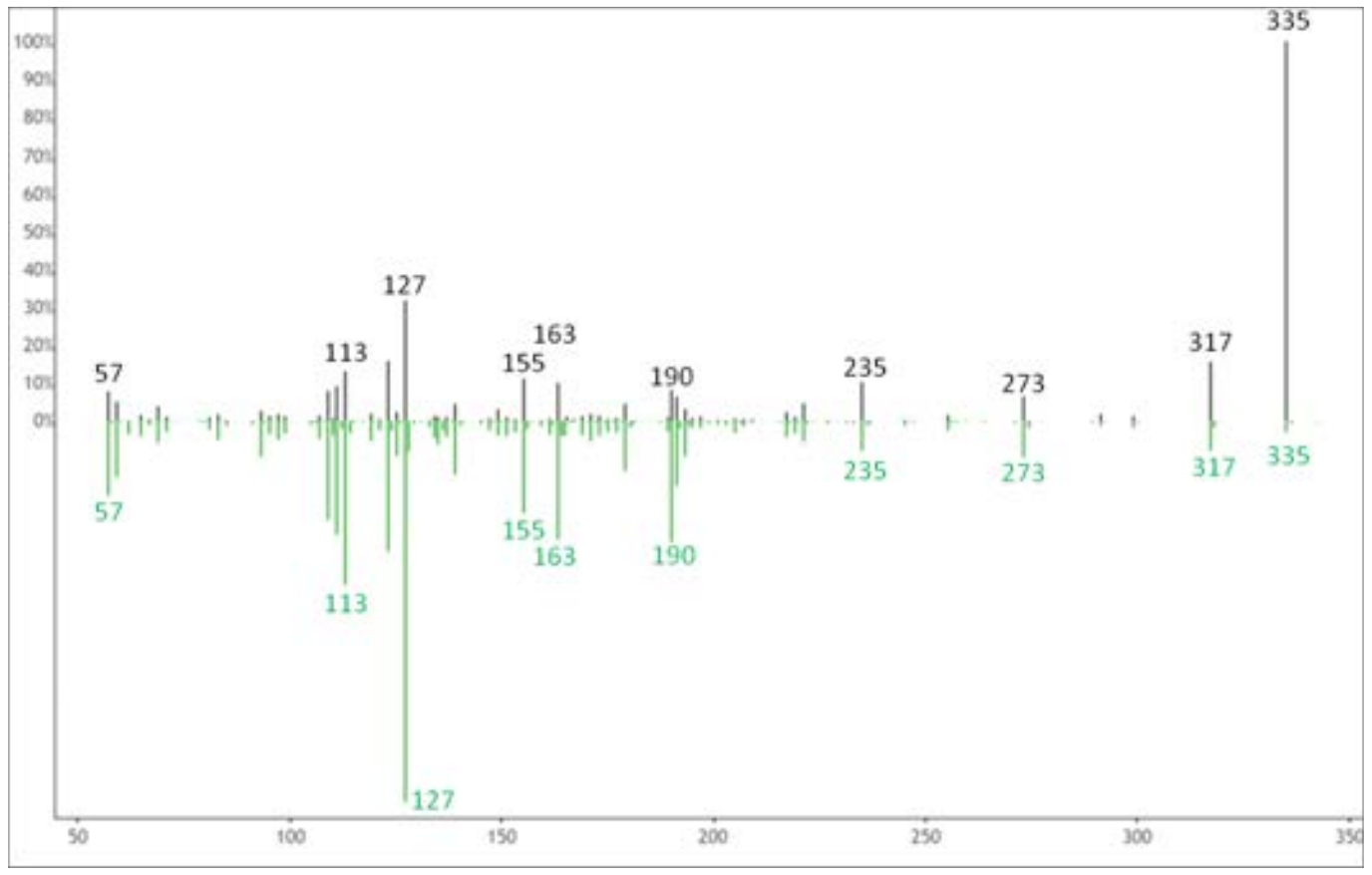

m/z 335.2228
at 2.96 minutes

8,15-DiHETE
reference spectra

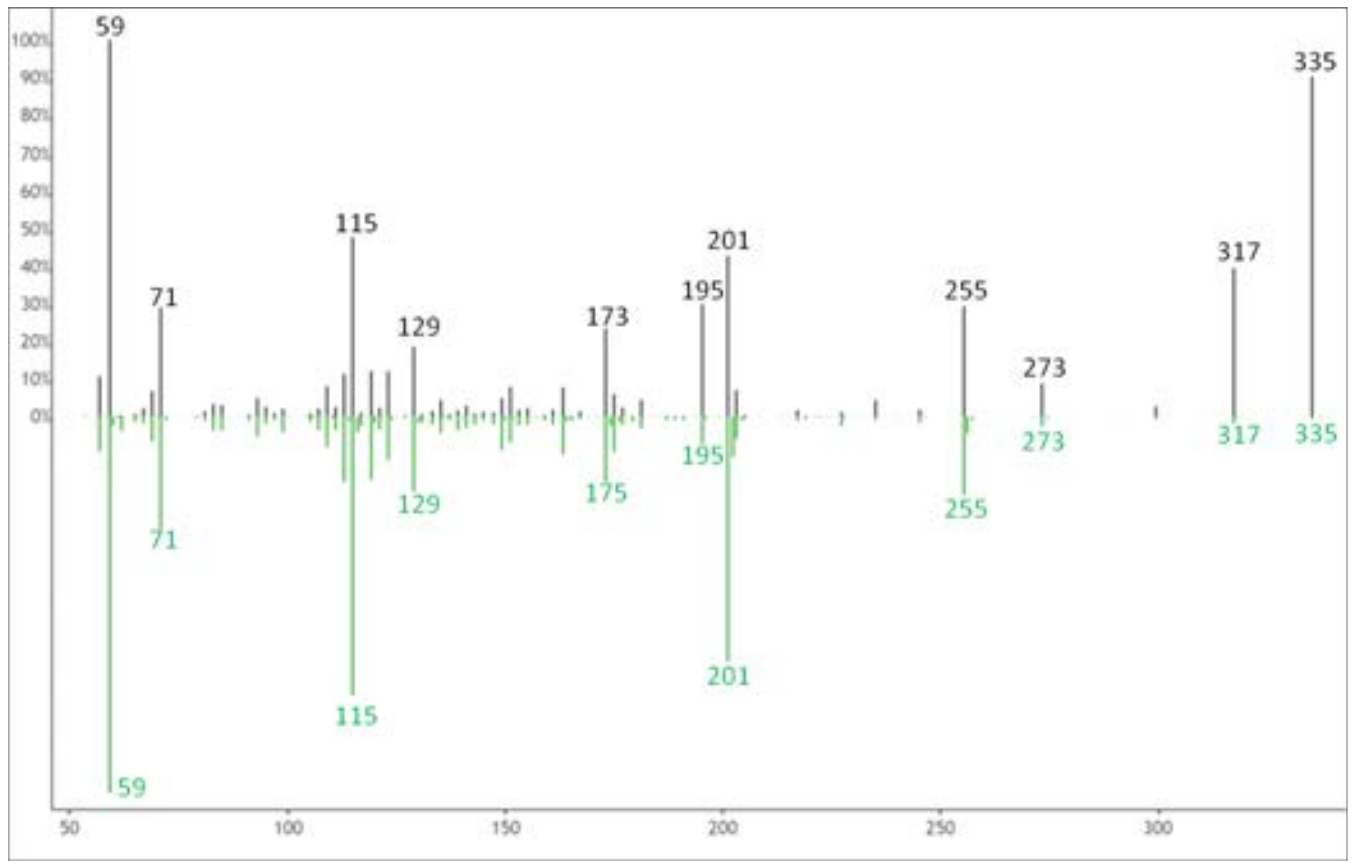

m/z 335.2228
at 3.01 minutes

5,15-DiHETE
reference spectra

## Figure S3 (Continued)

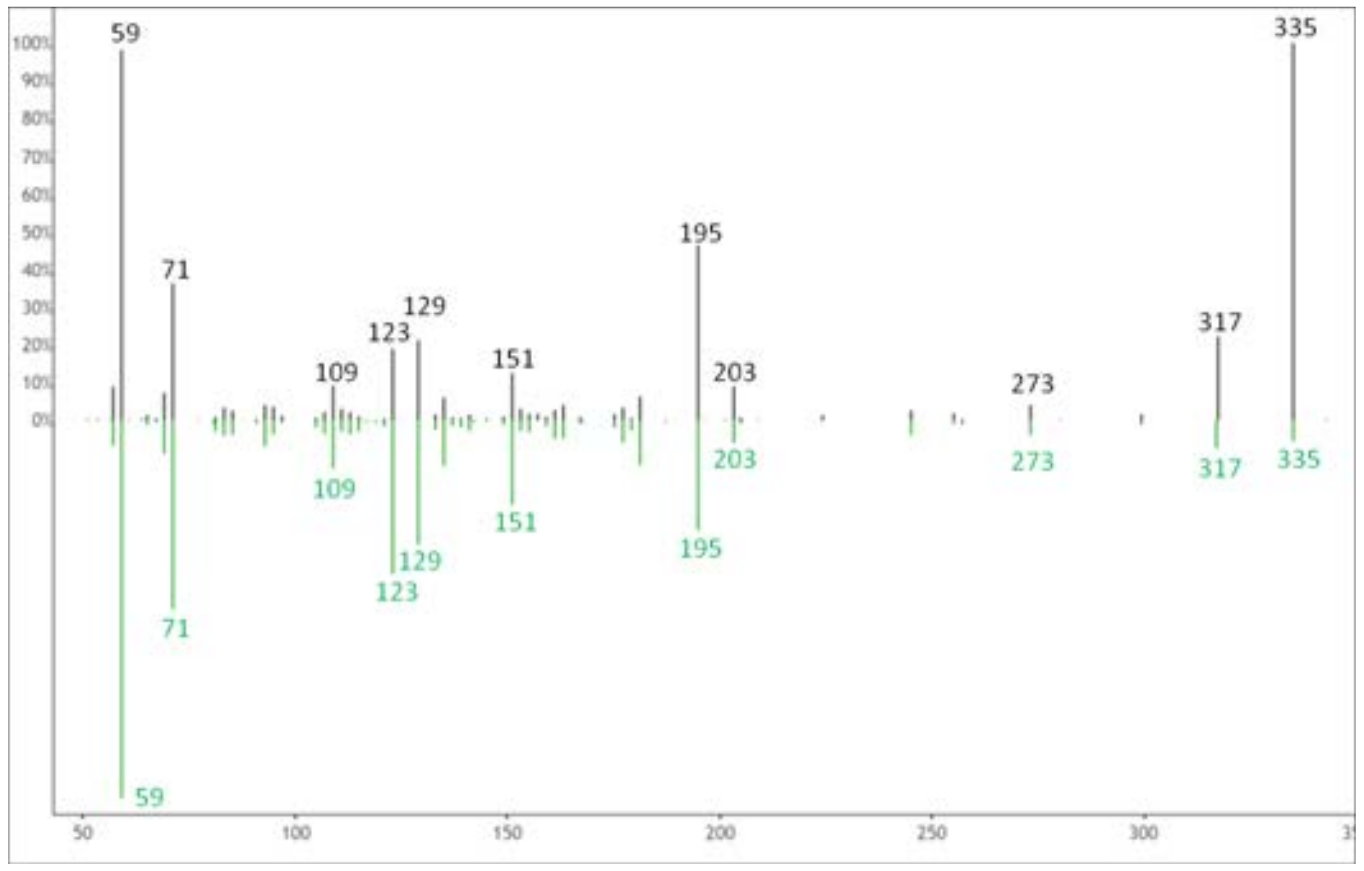

m/z 335.2228
at 3.04 minutes

6-trans-12-epi-LTB4
reference spectra

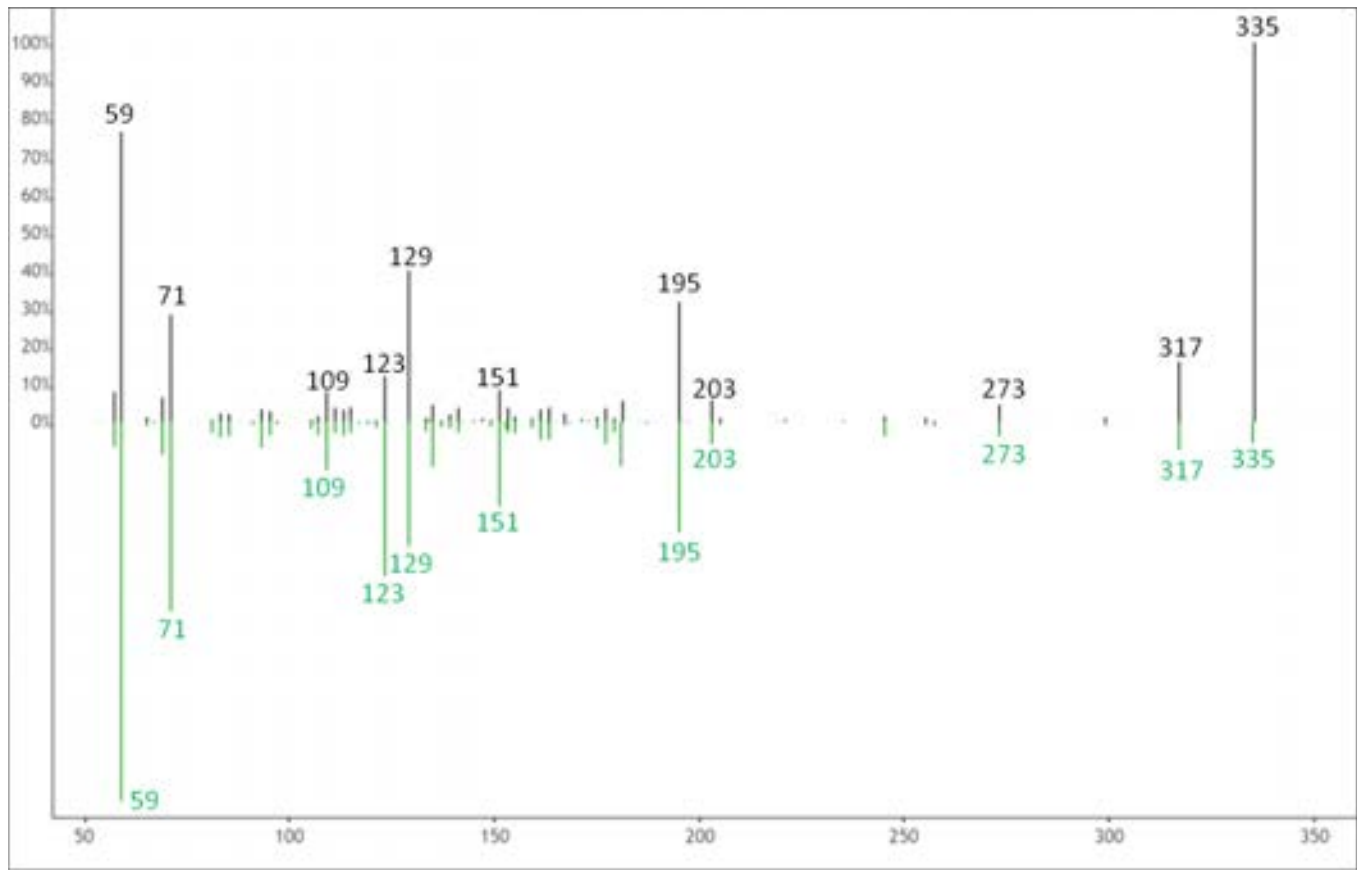

m/z 335.2228
at 3.22 minutes

6-trans-12-epi-LTB4
reference spectra

## Figure S3 (Continued)

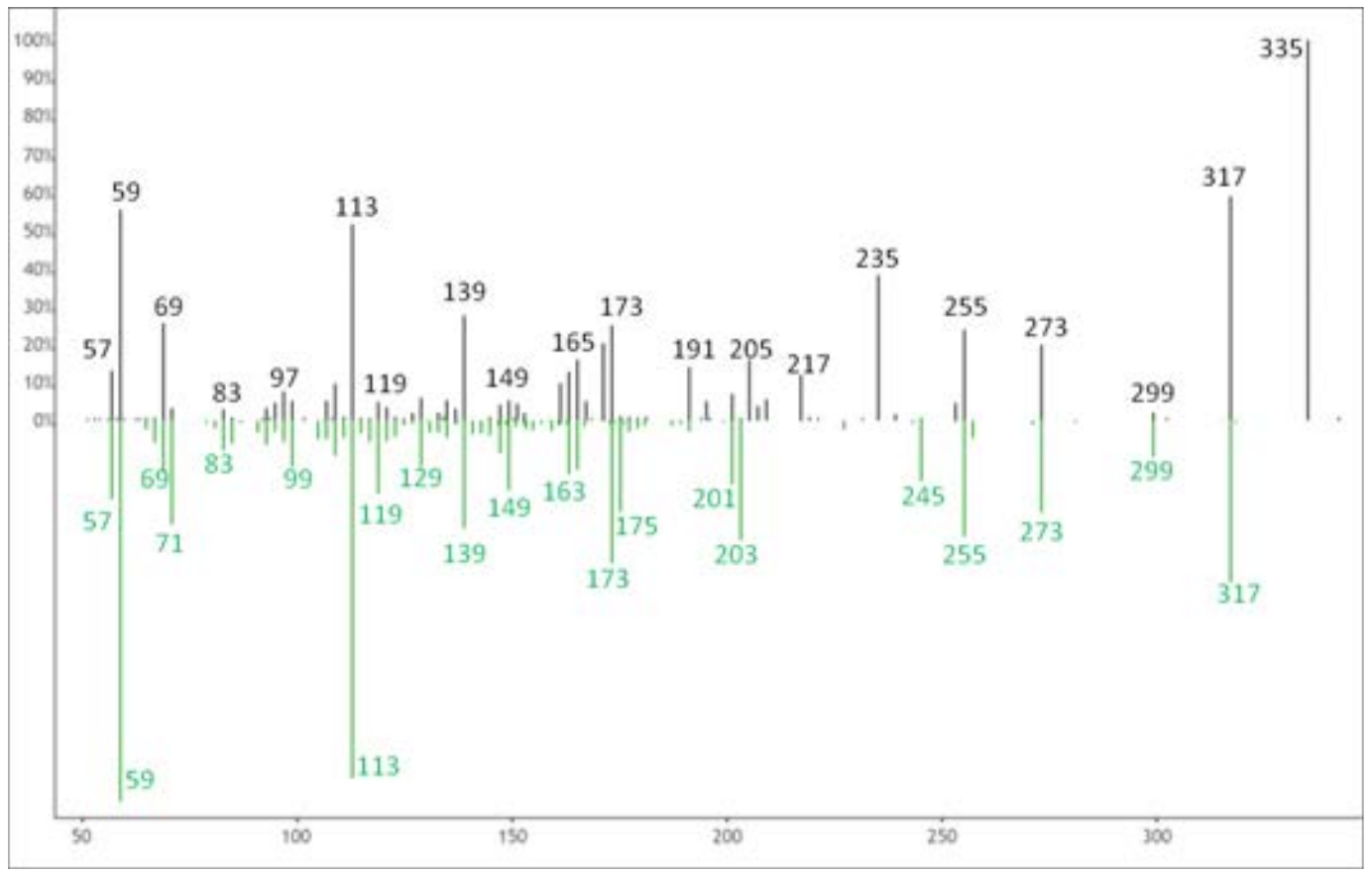

m/z 335.2228
at 3.43 minutes

5,15-DiHETE
reference spectra

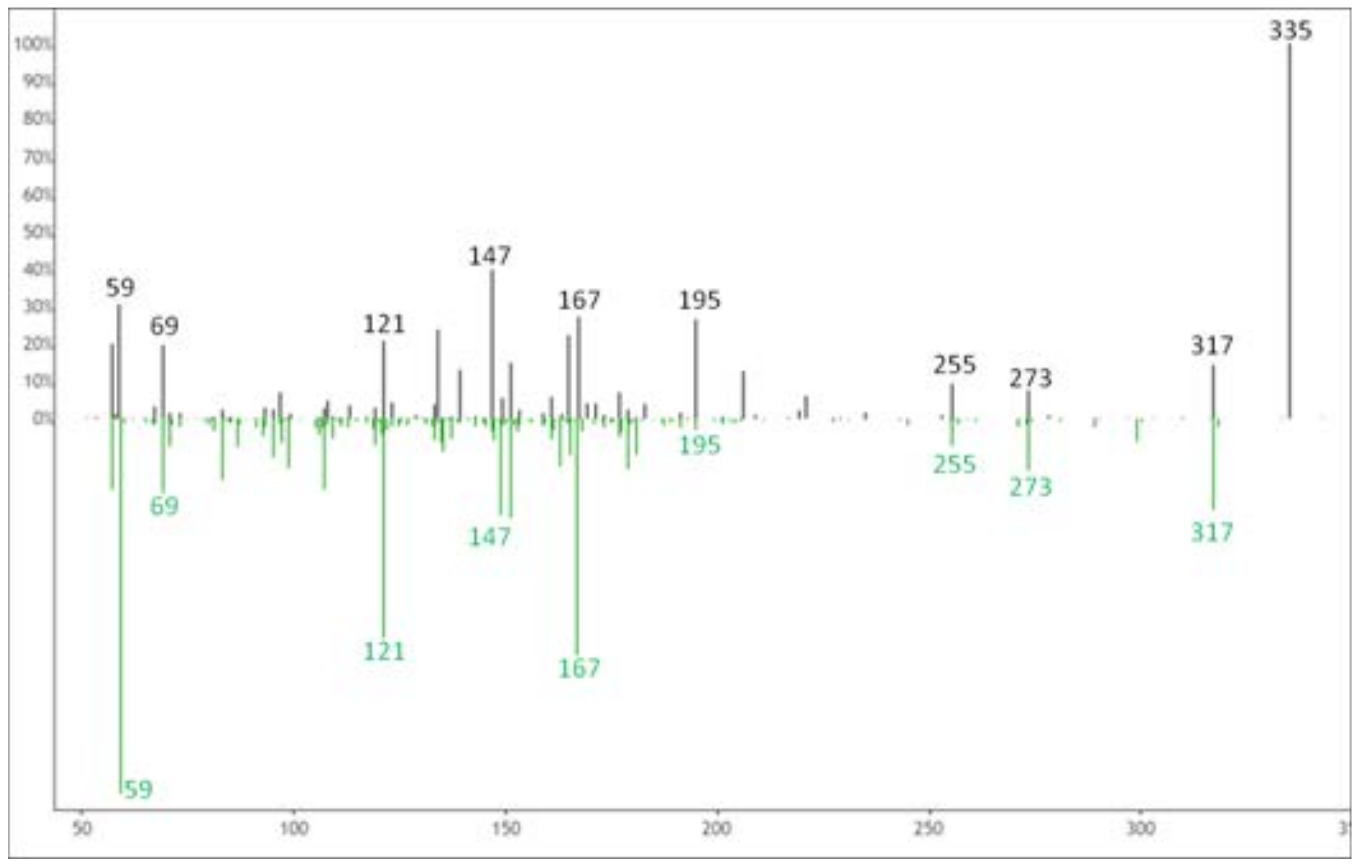

m/z 335.2228
at 3.47 minutes

11,12-EpETE
reference spectra

## Figure S3 (Continued)

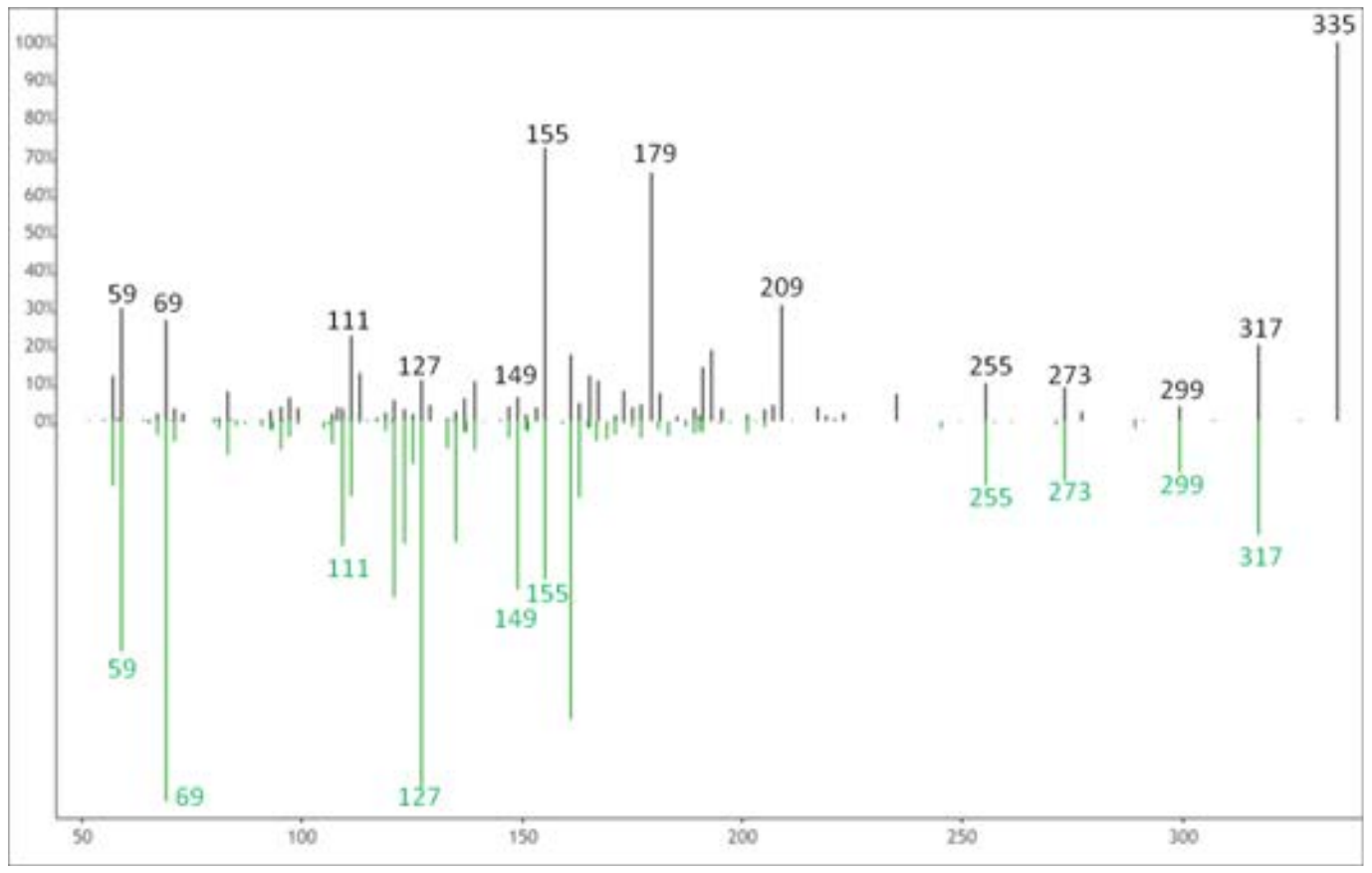

m/z 335.2228
at 3.60 minutes

8,9-EpETE
reference spectra

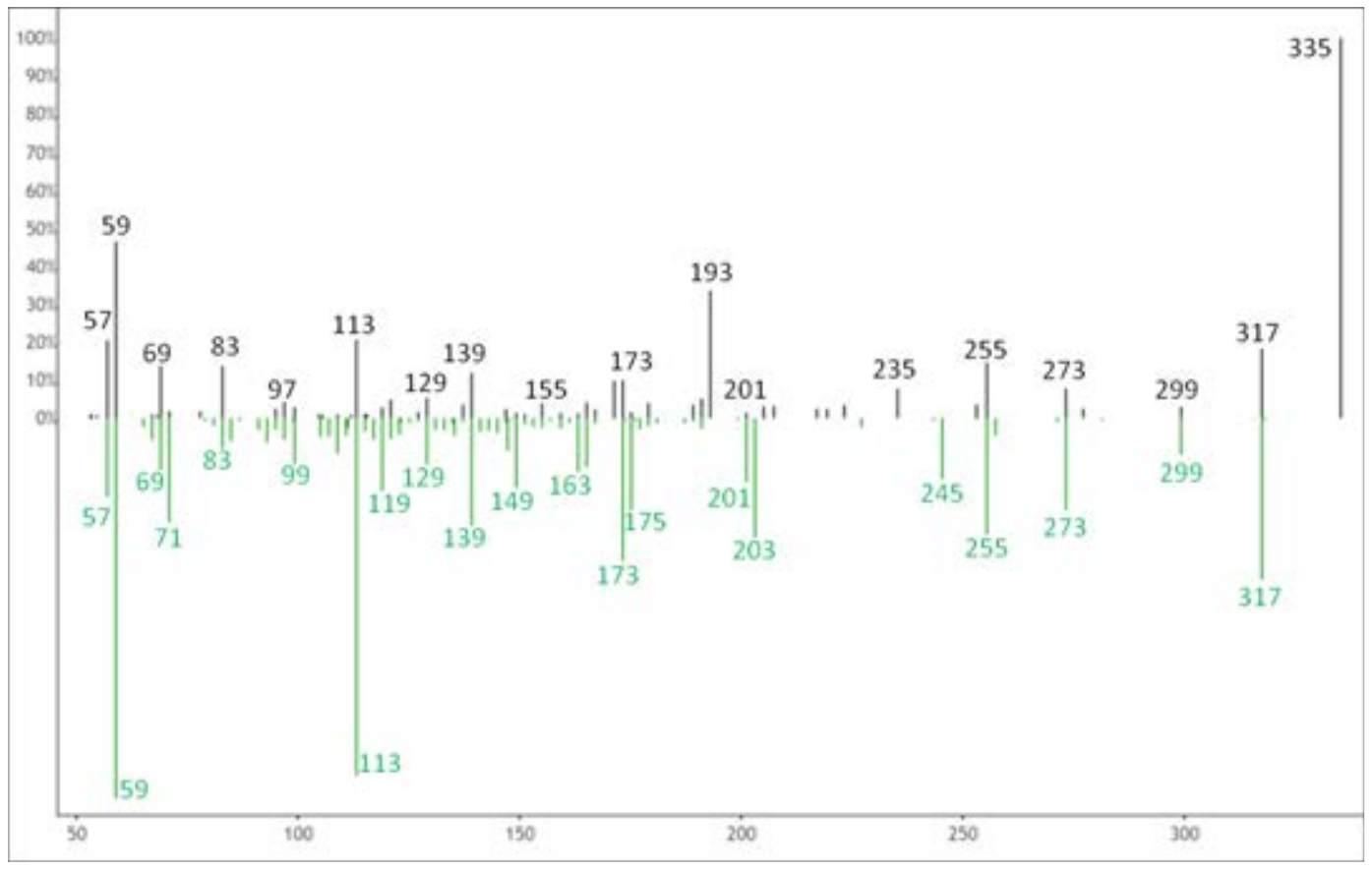

m/z 335.2228
at 3.65 minutes

5,15-DiHETE
reference spectra

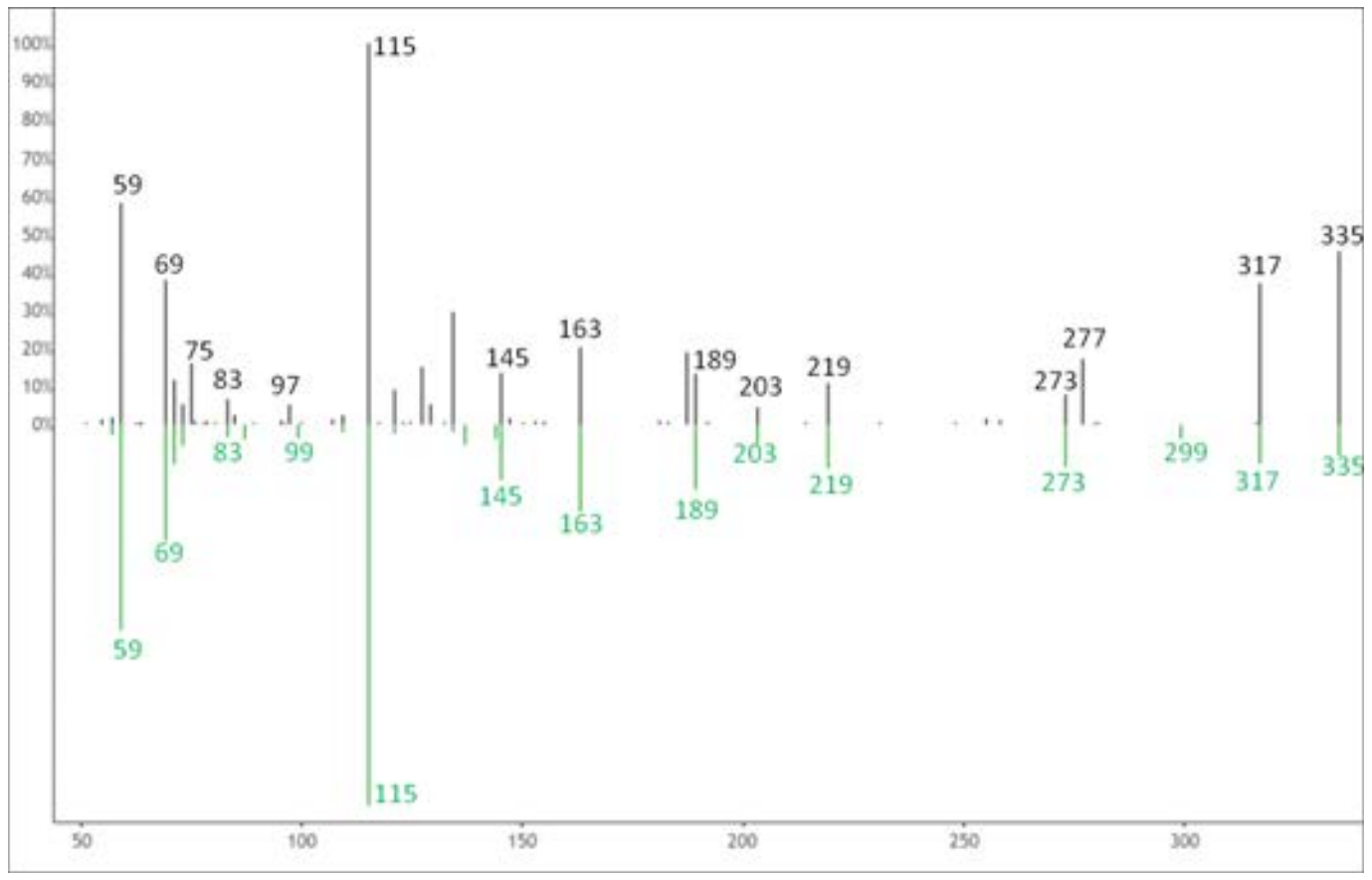

m/z 335.2228
at 3.93 minutes

5,6-DiHETE
reference spectra

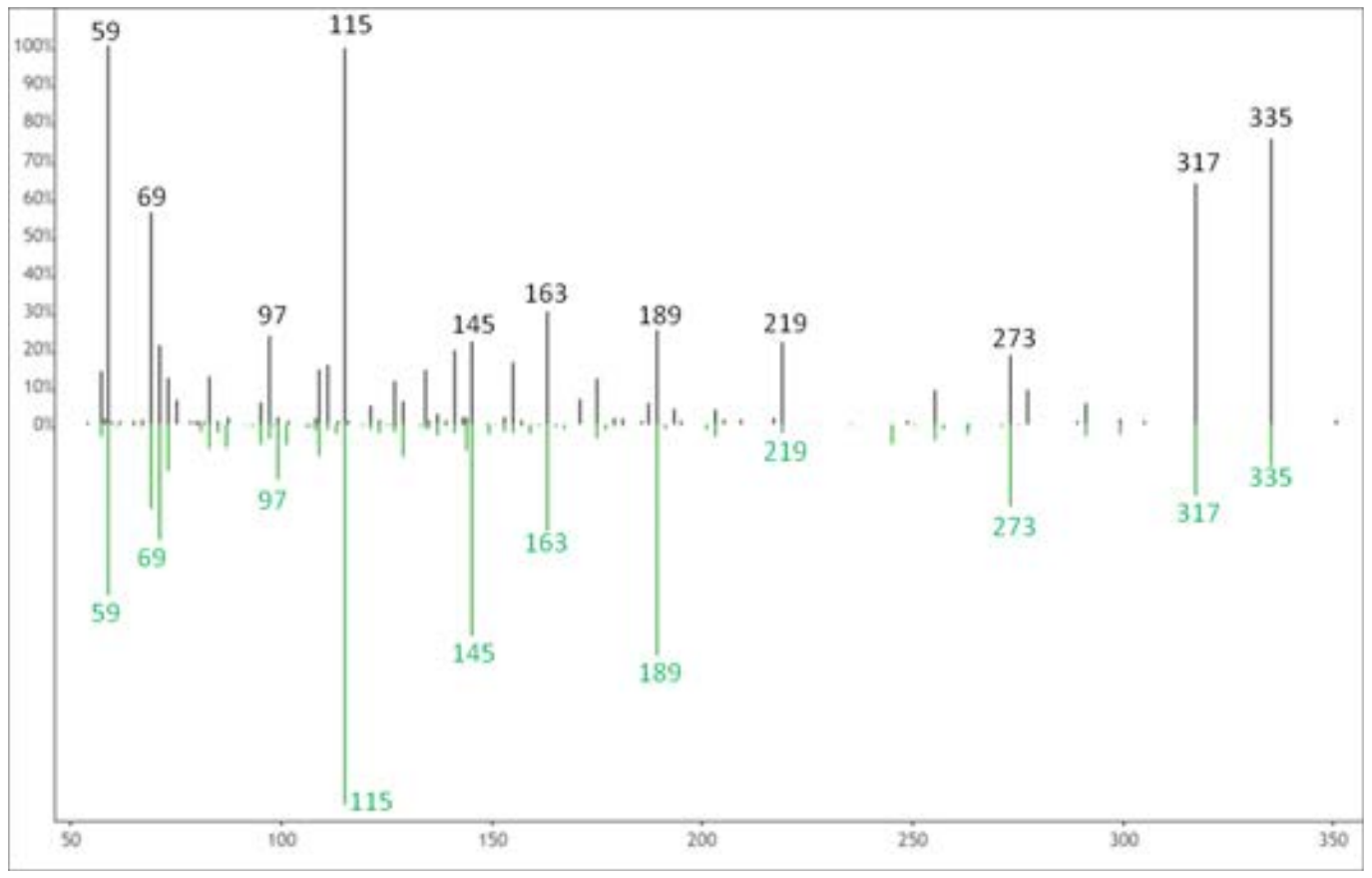

m/z 335.2228
at 4.02 minutes

5,6-DiHETE
reference spectra

## Figure S3 (Continued)

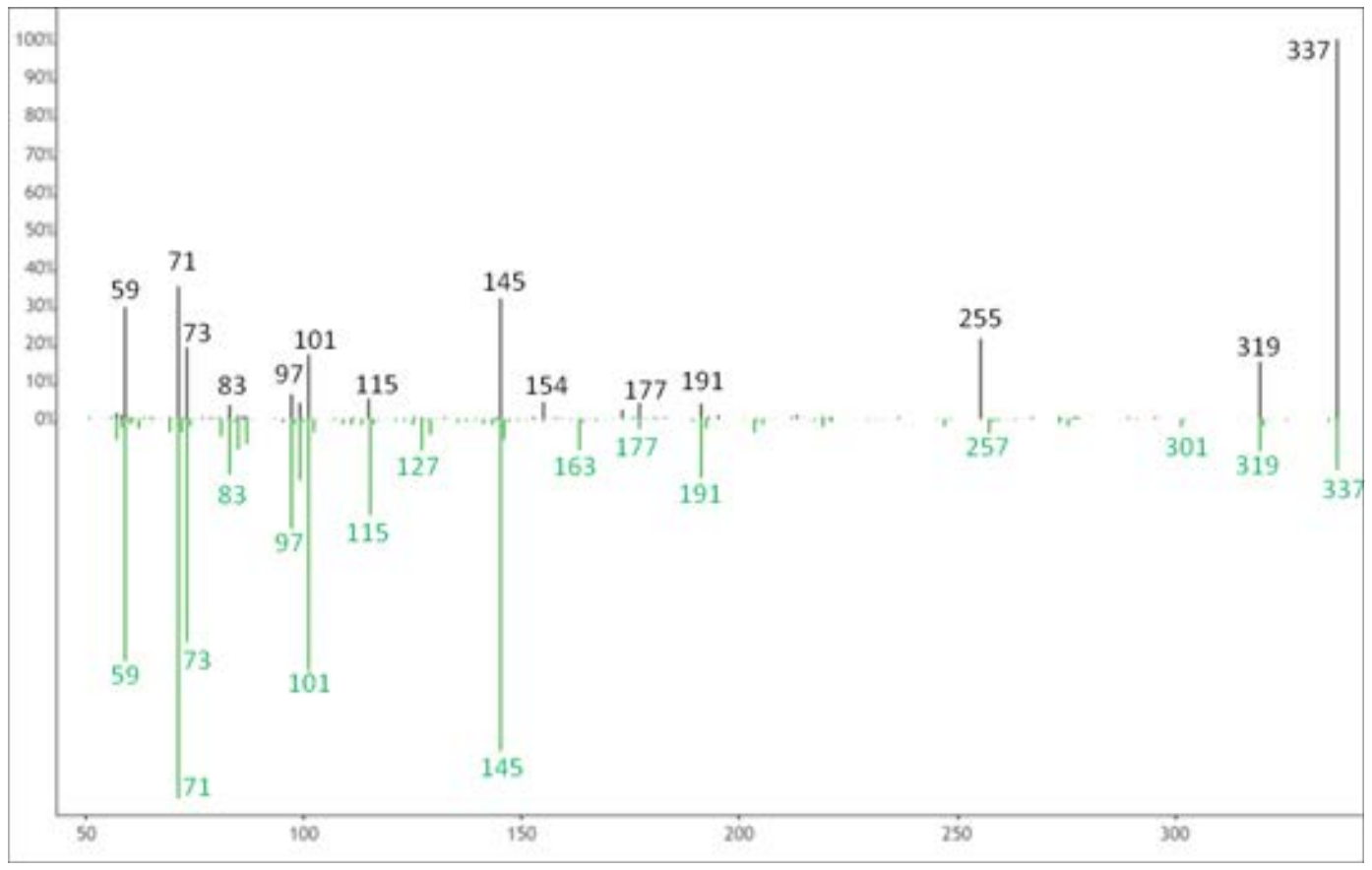

m/z 337.2384
at 4.14 minutes

5,6-DiHETrE
reference spectra

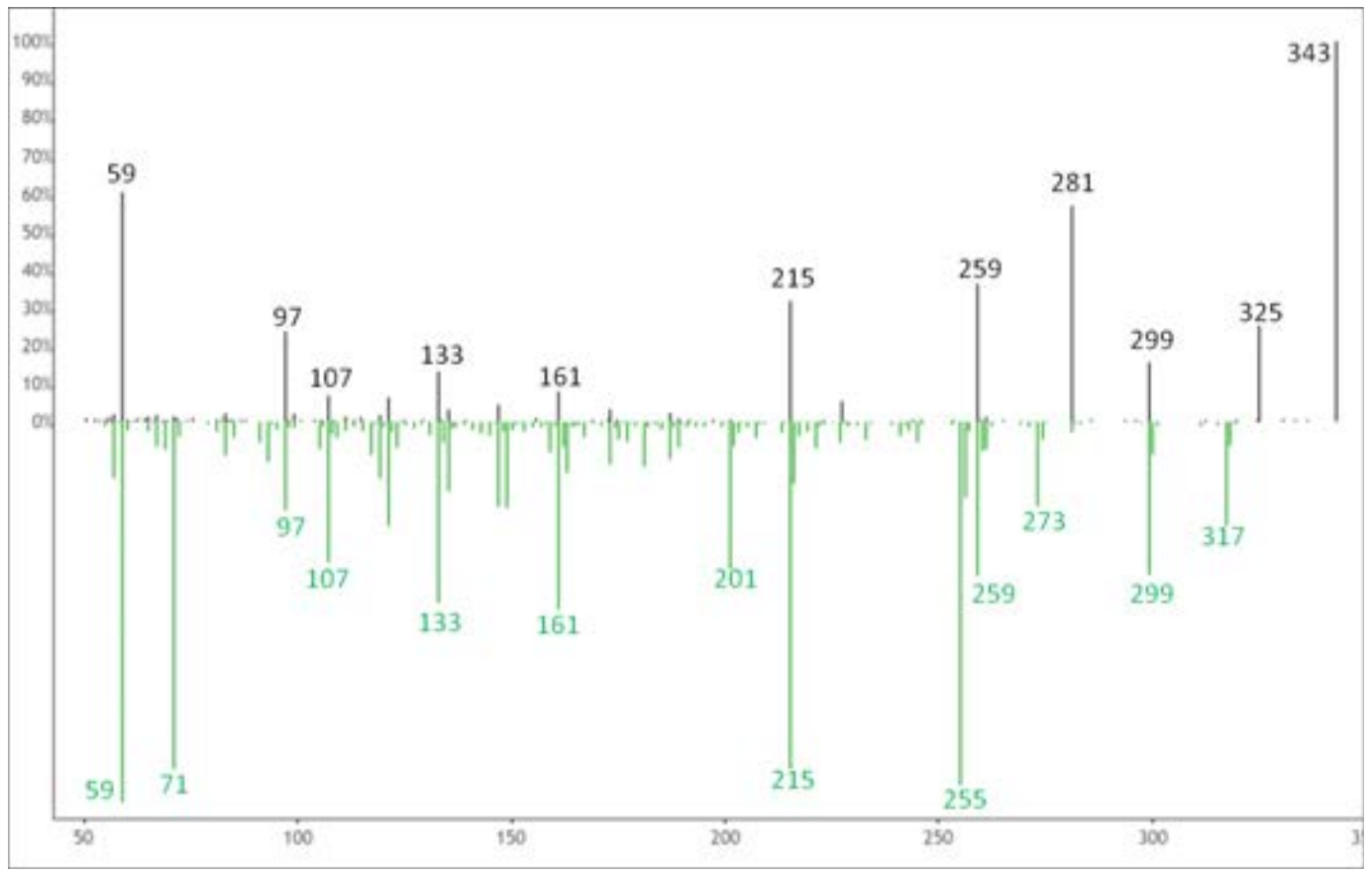

m/z 343.2278
at 4.16 minutes

18-HEPE
reference spectra

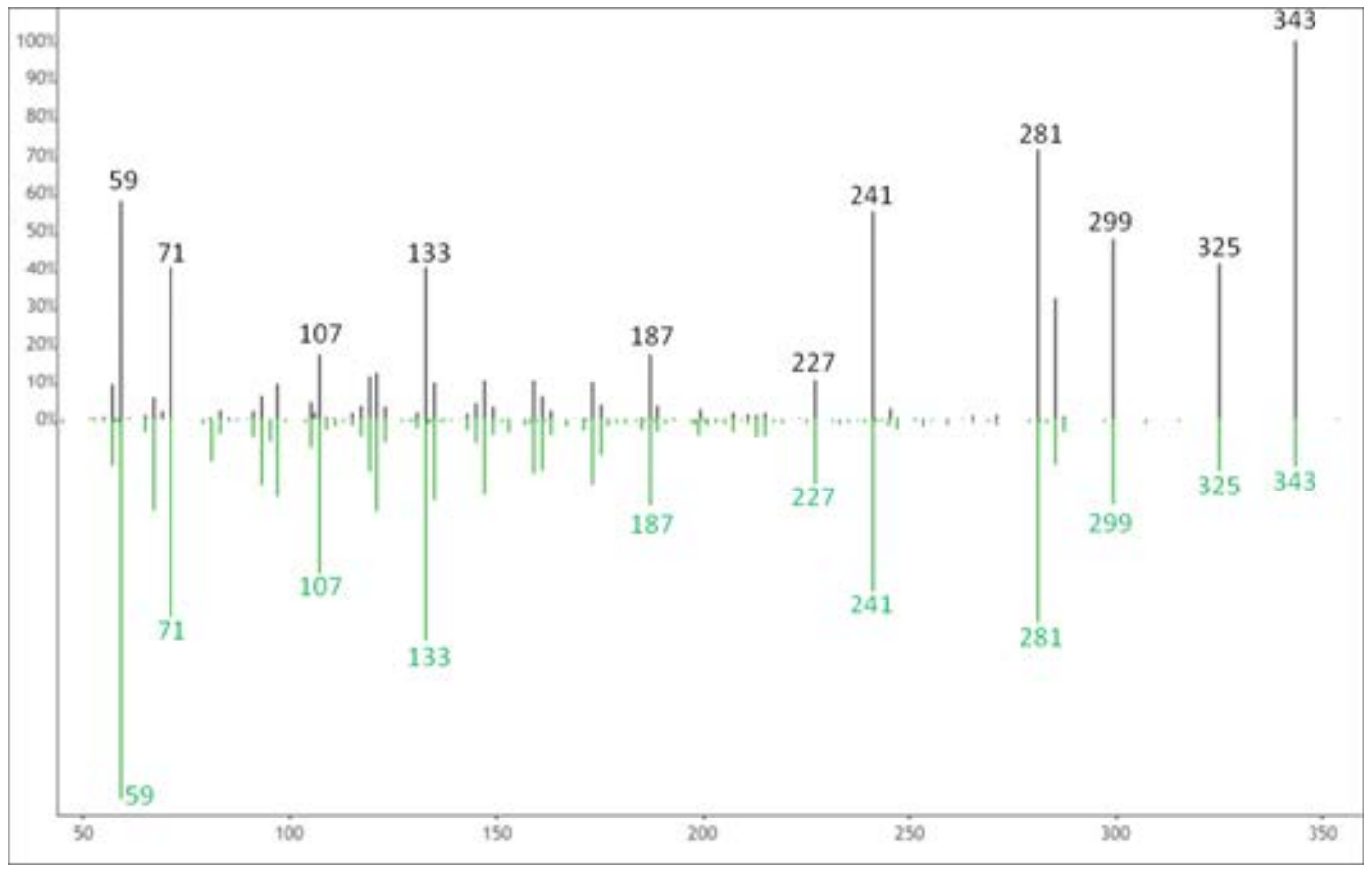

m/z 343.2278
at 4.31 minutes

20-HDoHE
reference spectra

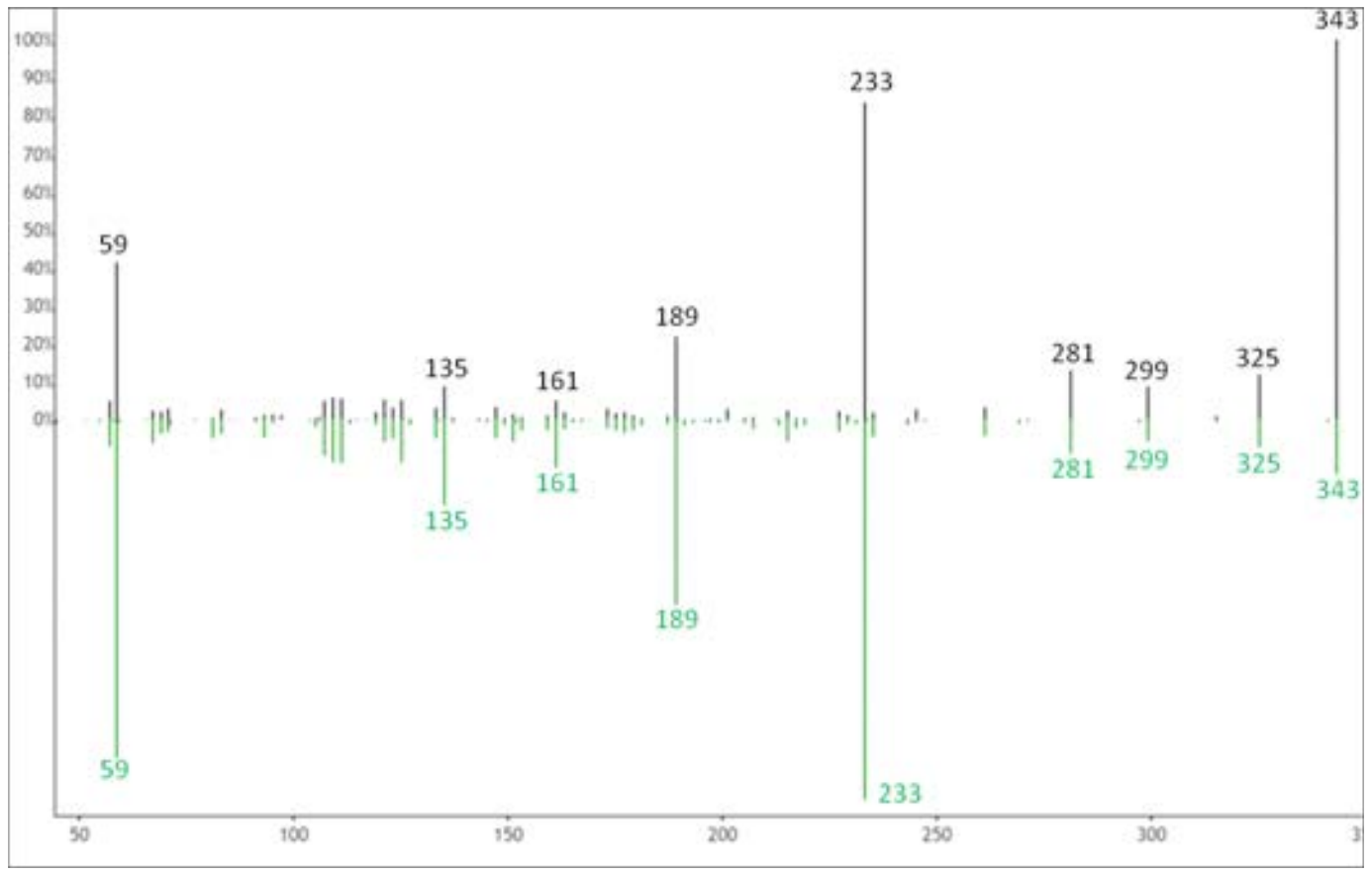

m/z 343.2278
at 4.45 minutes

16-HDoHE
reference spectra

## Figure S3 (Continued)

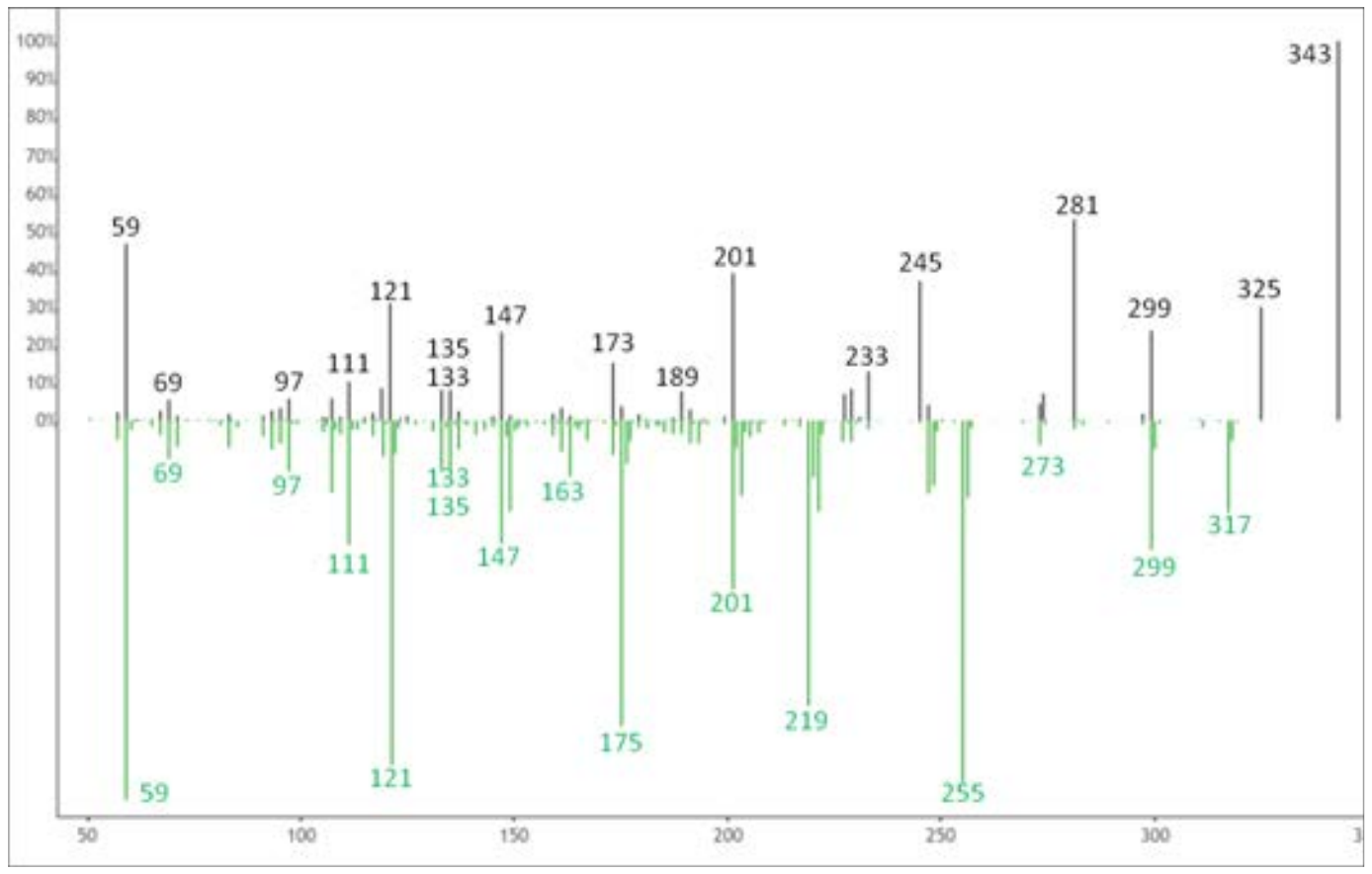

m/z 343.2278
at 4.50 minutes

15-HEPE
reference spectra

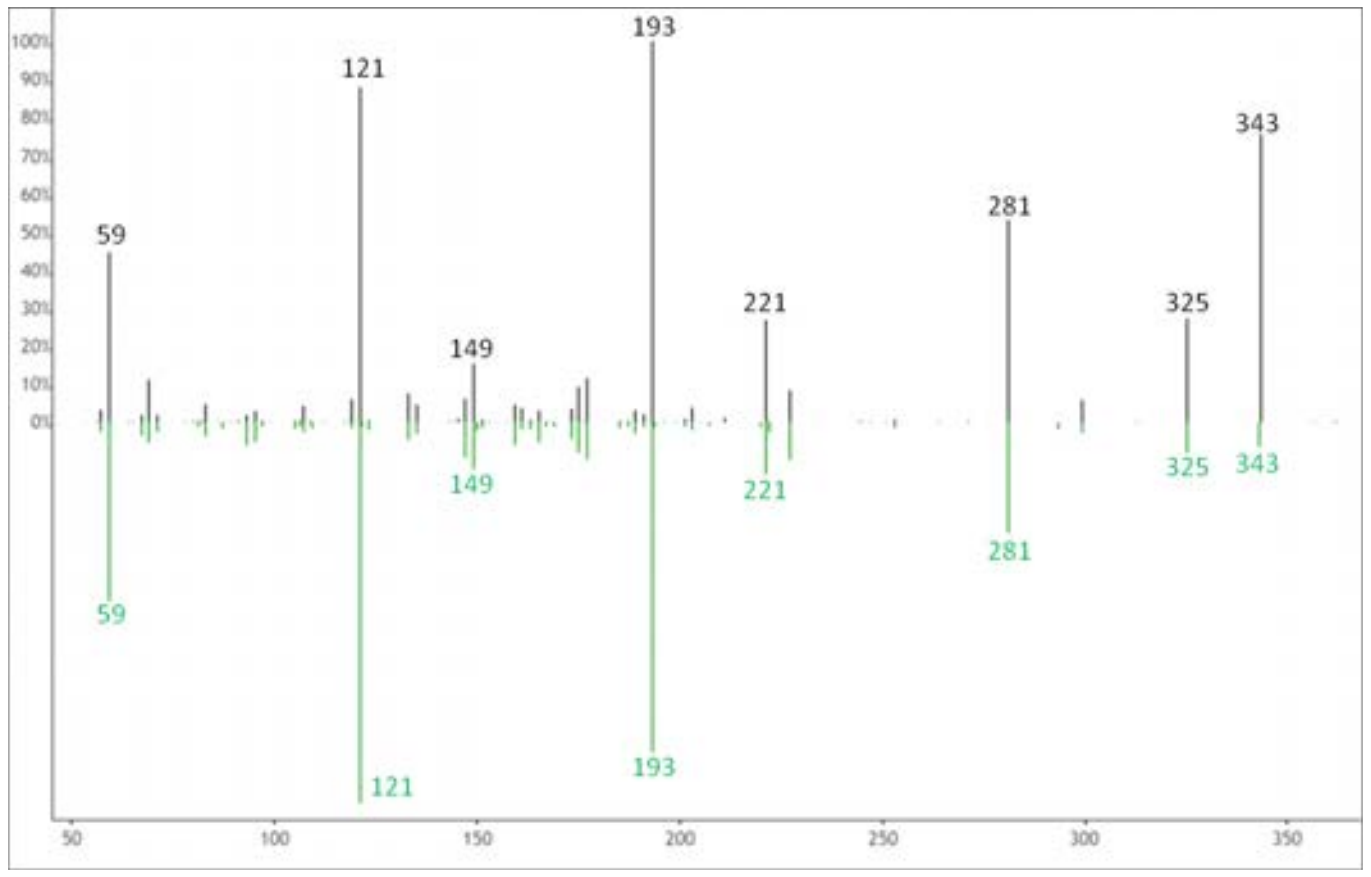

m/z 343.2278
at 4.55 minutes

13-HDoHE
reference spectra

## Figure S3 (Continued)

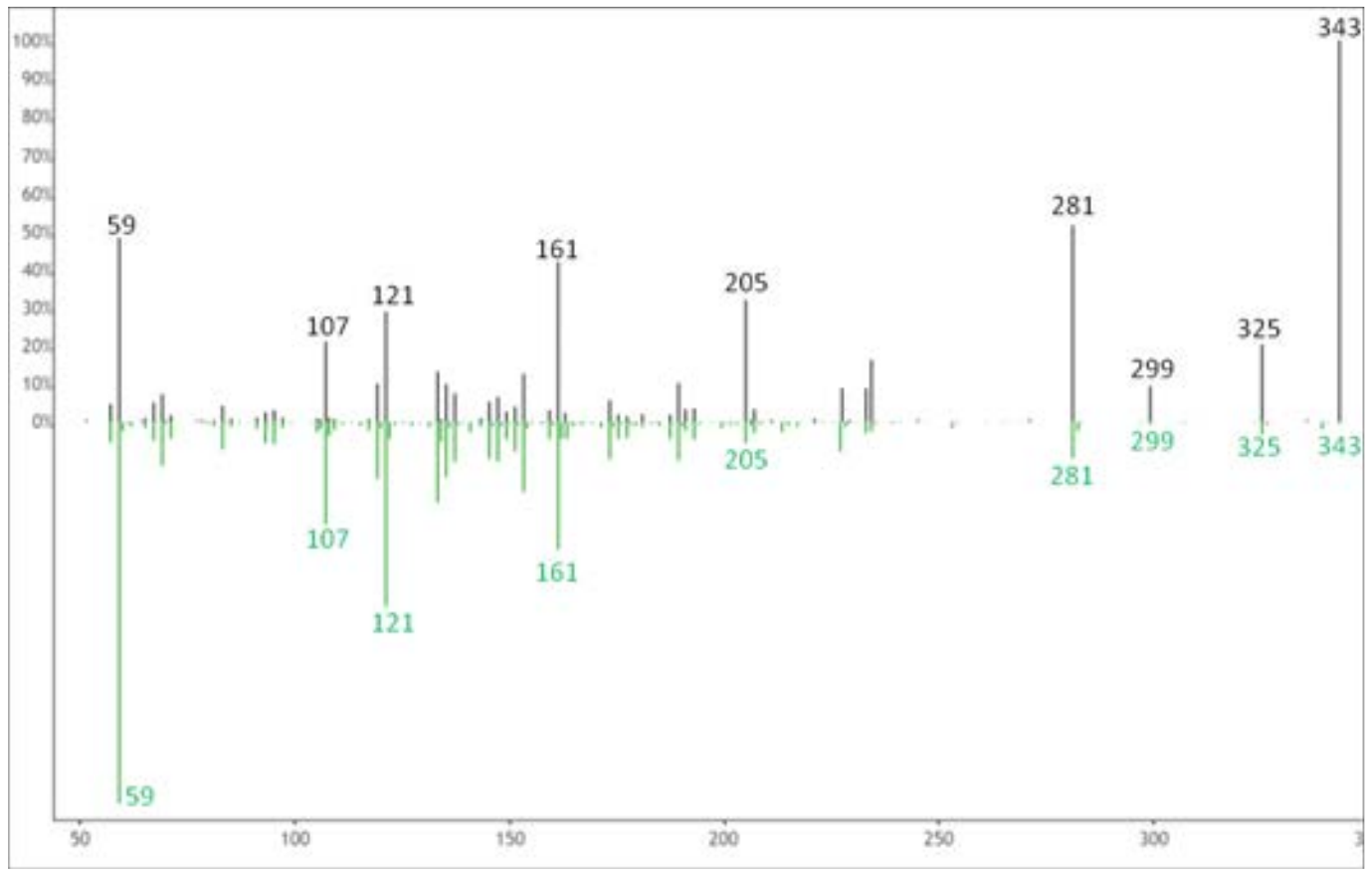

m/z 343.2278
at 4.59 minutes

14-HDoHE
reference spectra

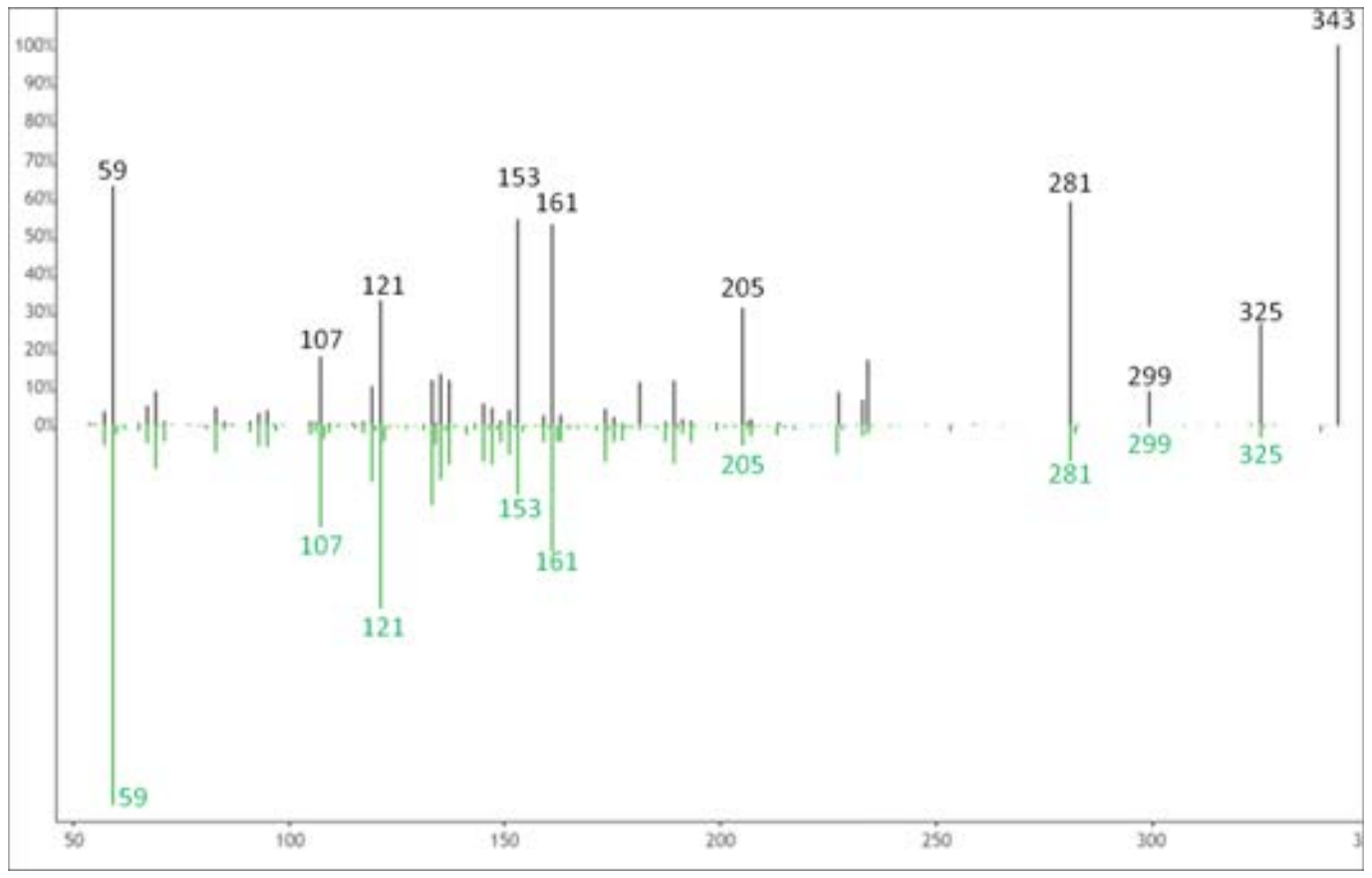

m/z 343.2278
at 4.61 minutes

14-HDoHE
reference spectra

**Figure S3 (Continued)**

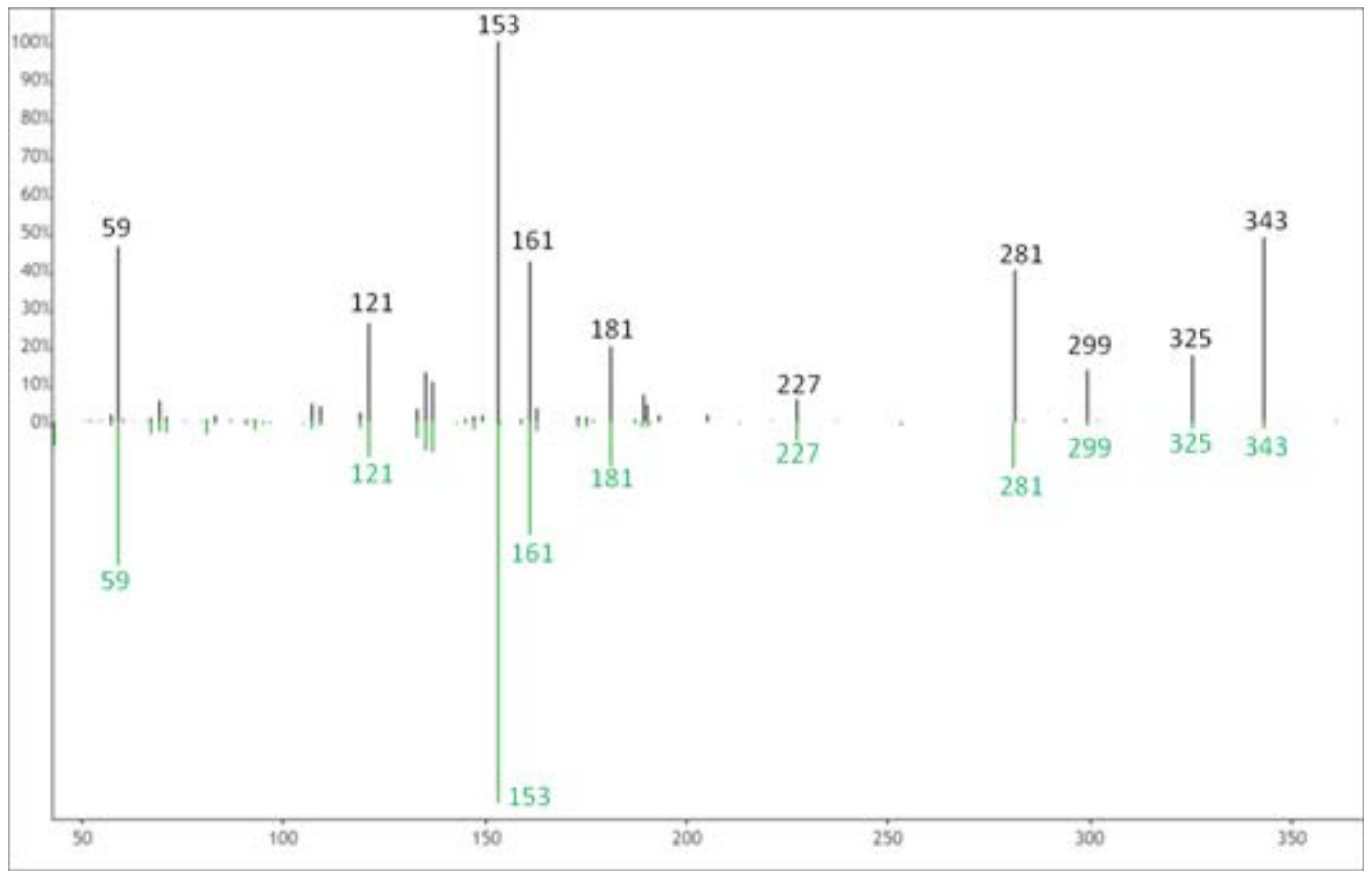

m/z 343.2278
at 4.65 minutes

10-HDoHE
reference spectra

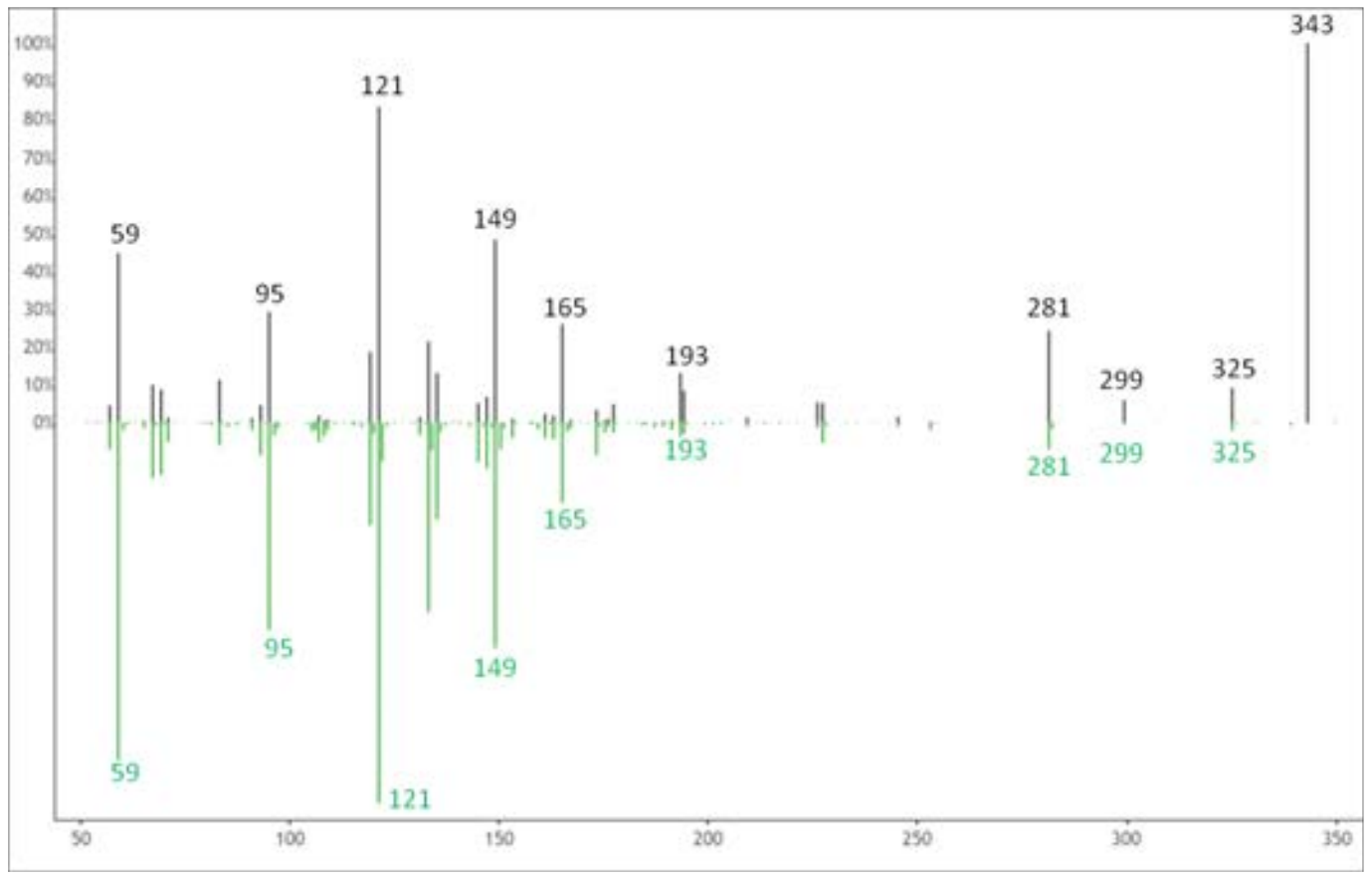

m/z 343.2278
at 4.70 minutes

11-HDoHE
reference spectra

**Figure S3 (Continued)**

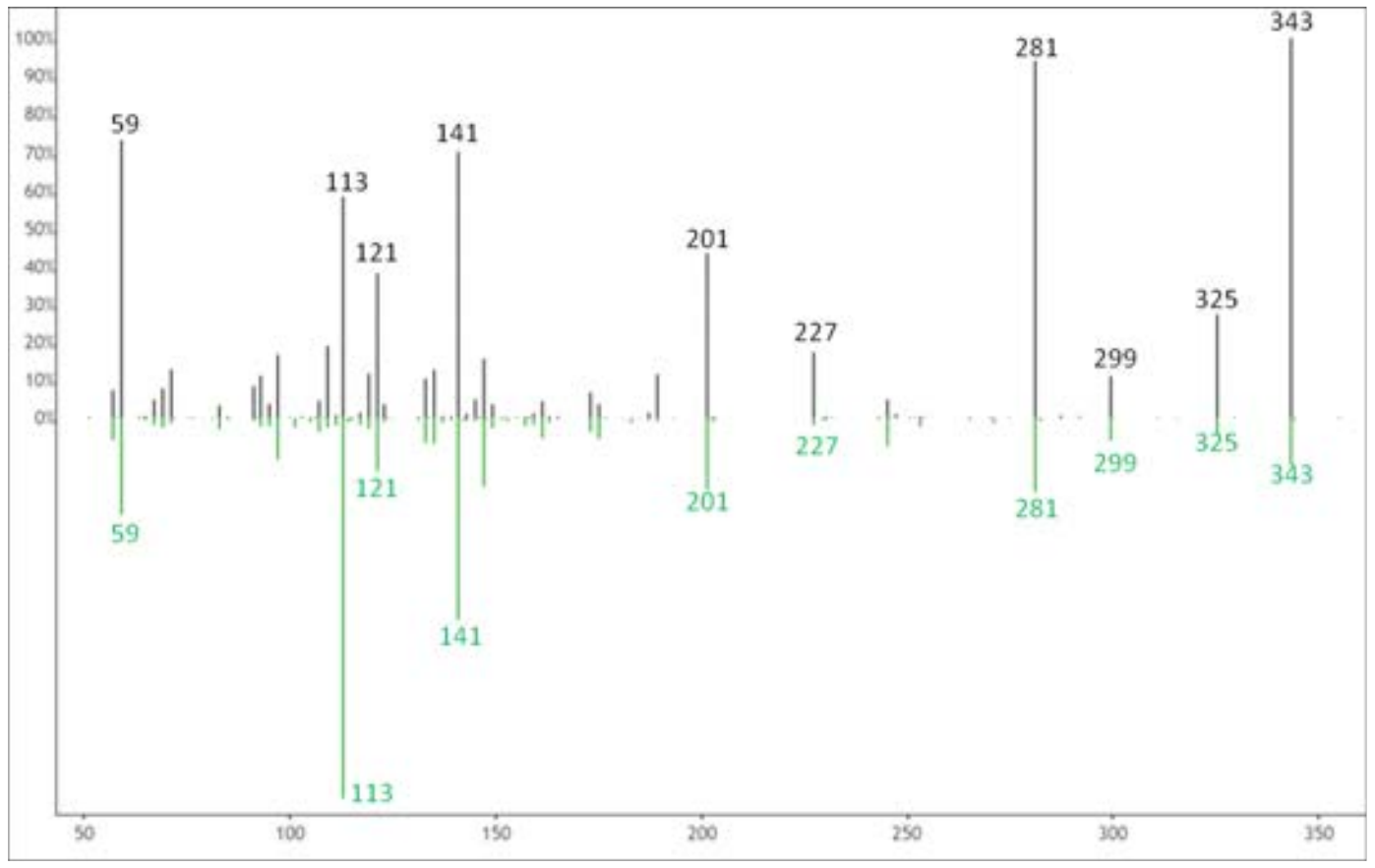

m/z 343.2278
at 4.78 minutes

7-HDoHE
reference spectra

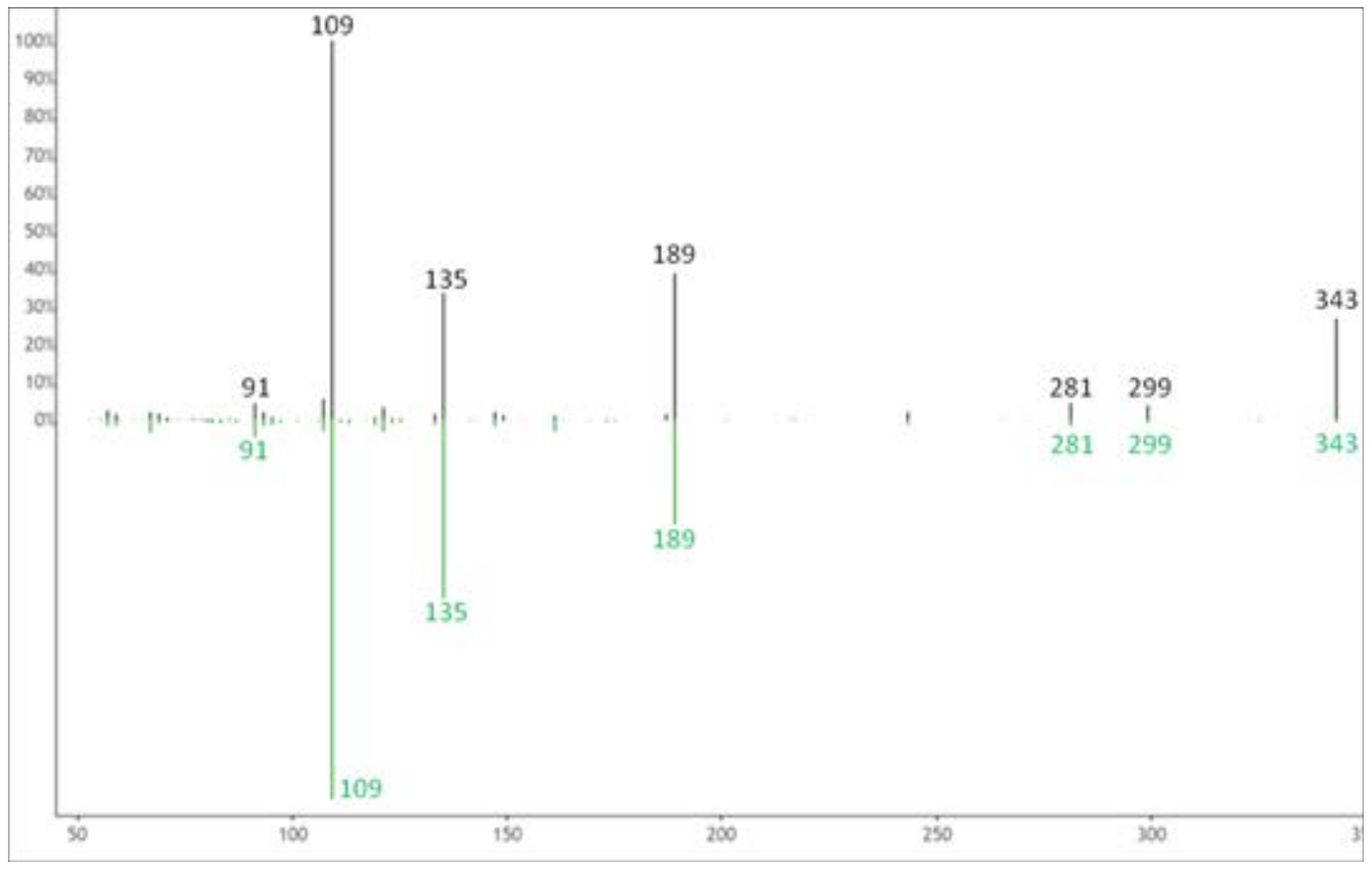

m/z 343.2278
at 4.82 minutes

8-HDoHE
reference spectra

## Figure S3 (Continued)

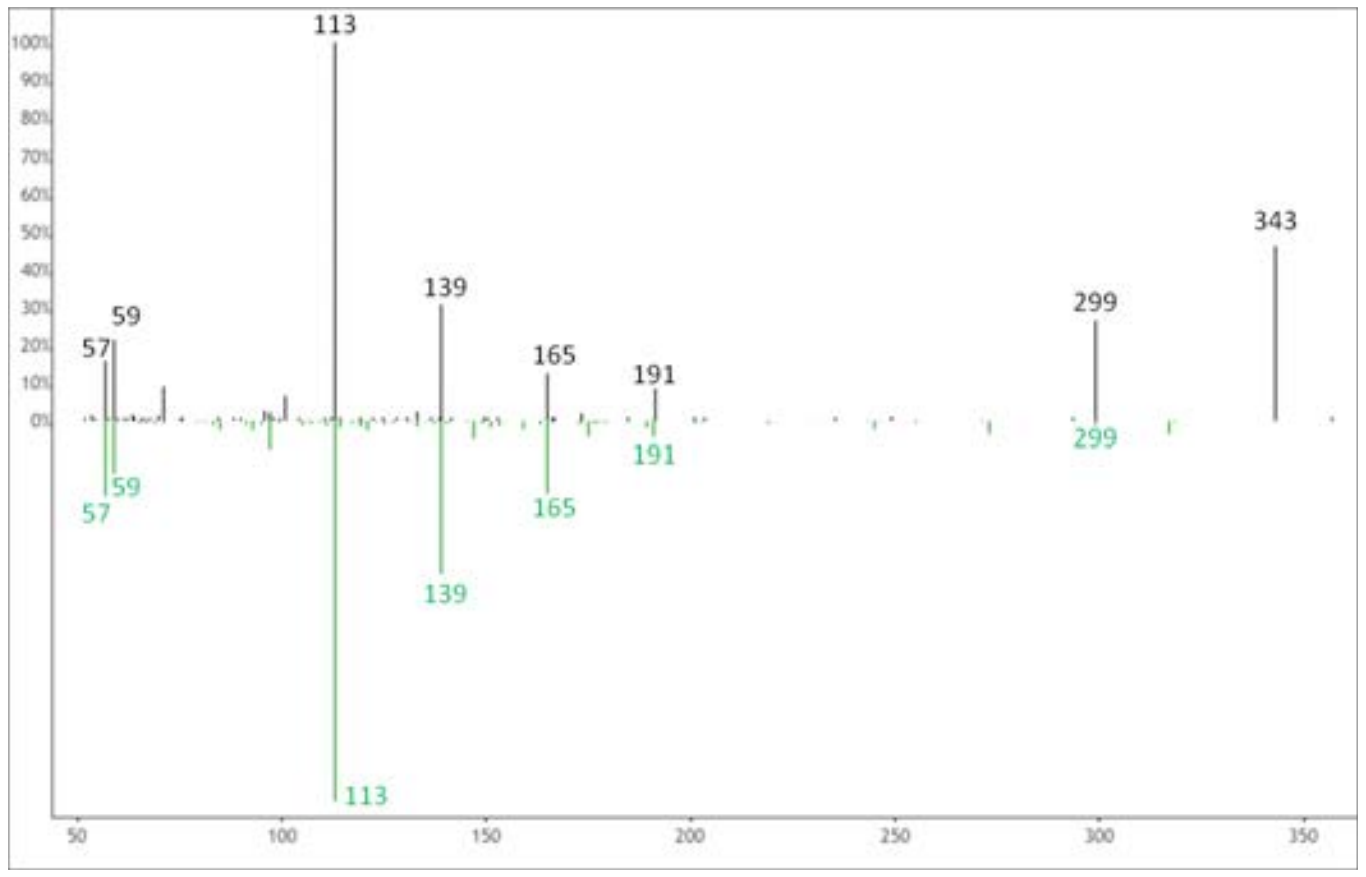

m/z 343.2278
at 5.06 minutes

15-OxoETE
reference spectra

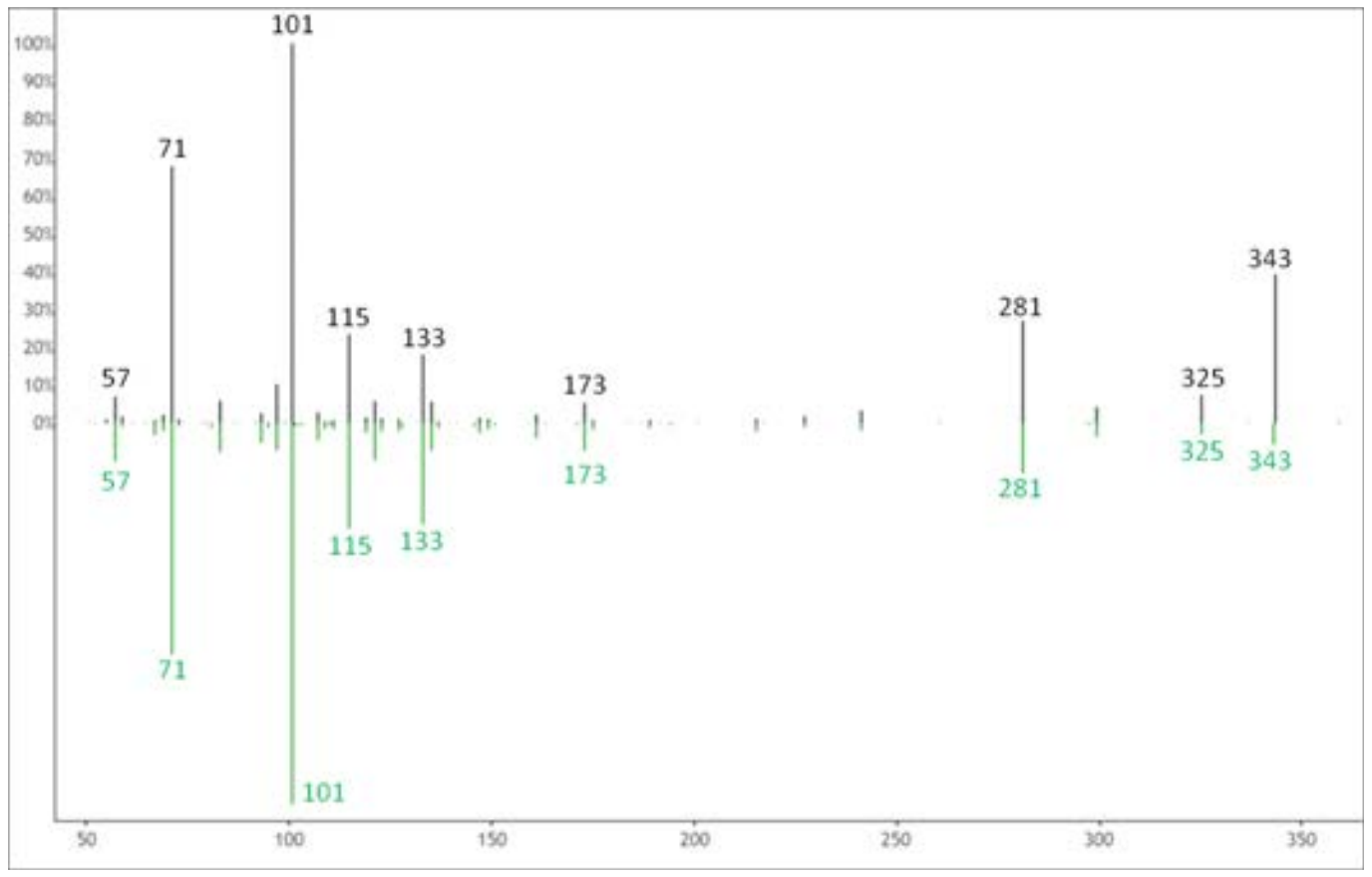

m/z 343.2278
at 5.13 minutes

4-HDoHE
reference spectra

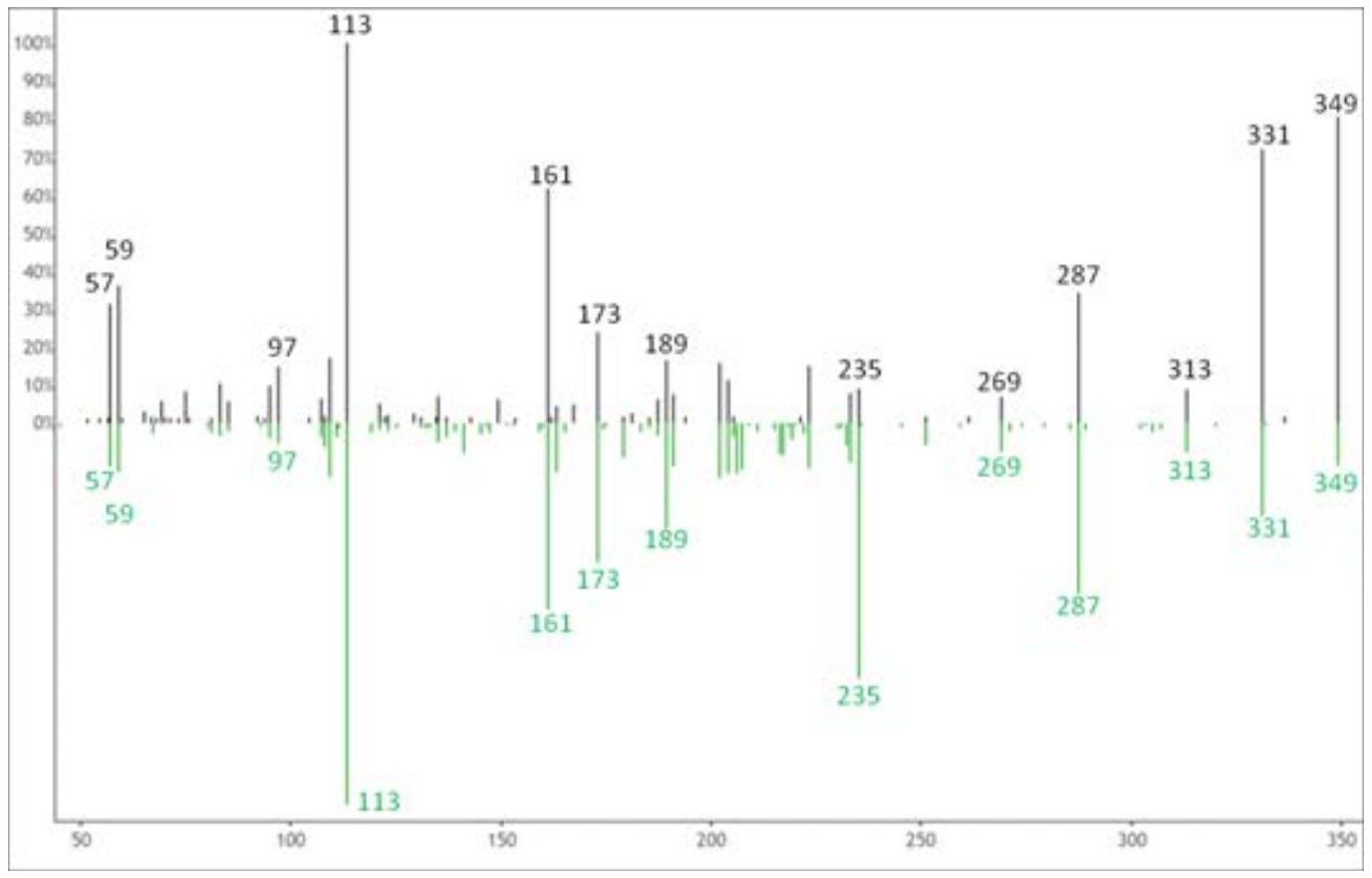

m/z 349.2020
at 2.20 minutes

8-iso-15-keto-PGE2
reference spectra

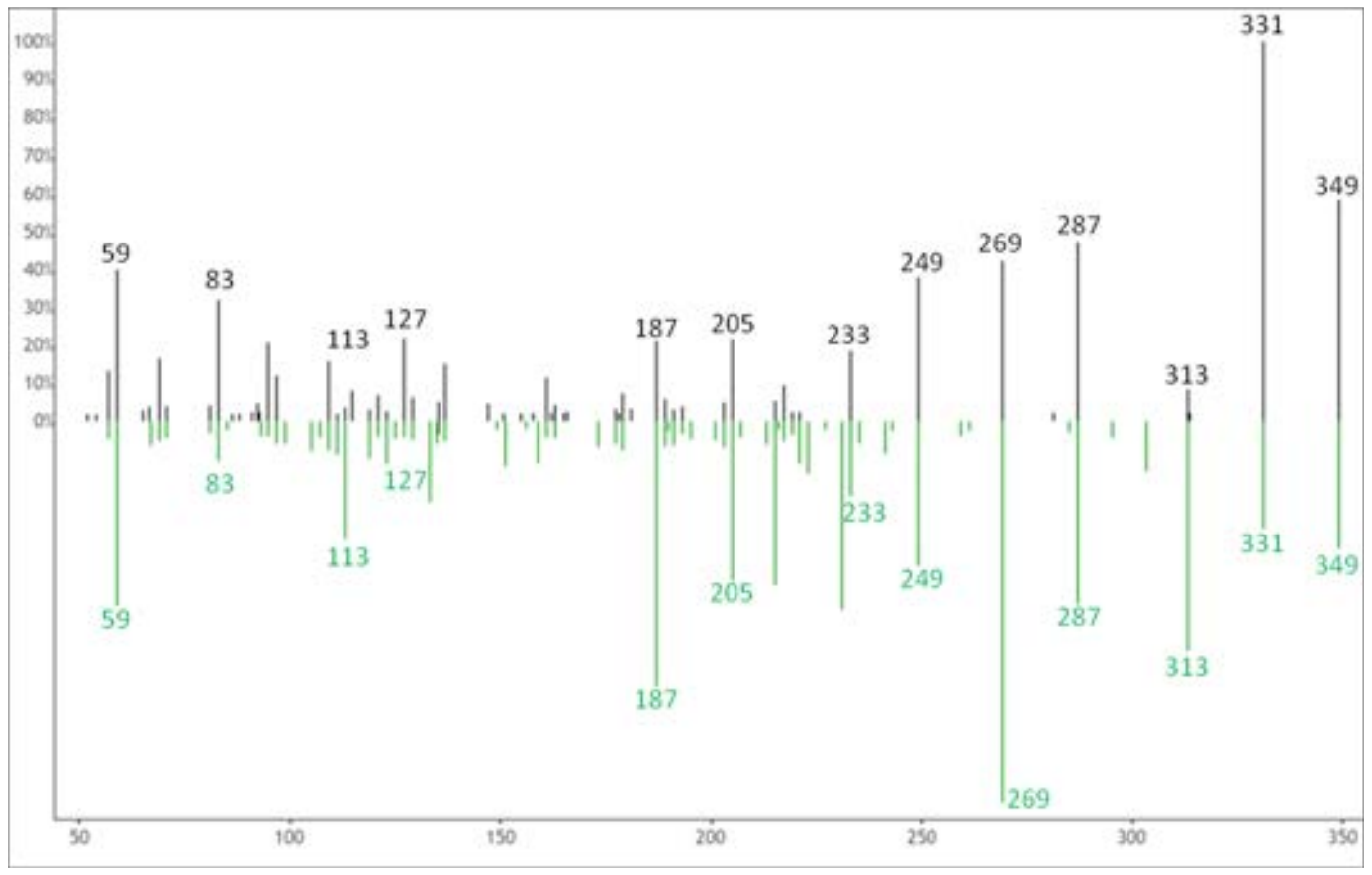

m/z 349.2020
at 2.43 minutes

11-deoxy-11-
methylene $PGD_2$
reference spectra

## Figure S3 (Continued)

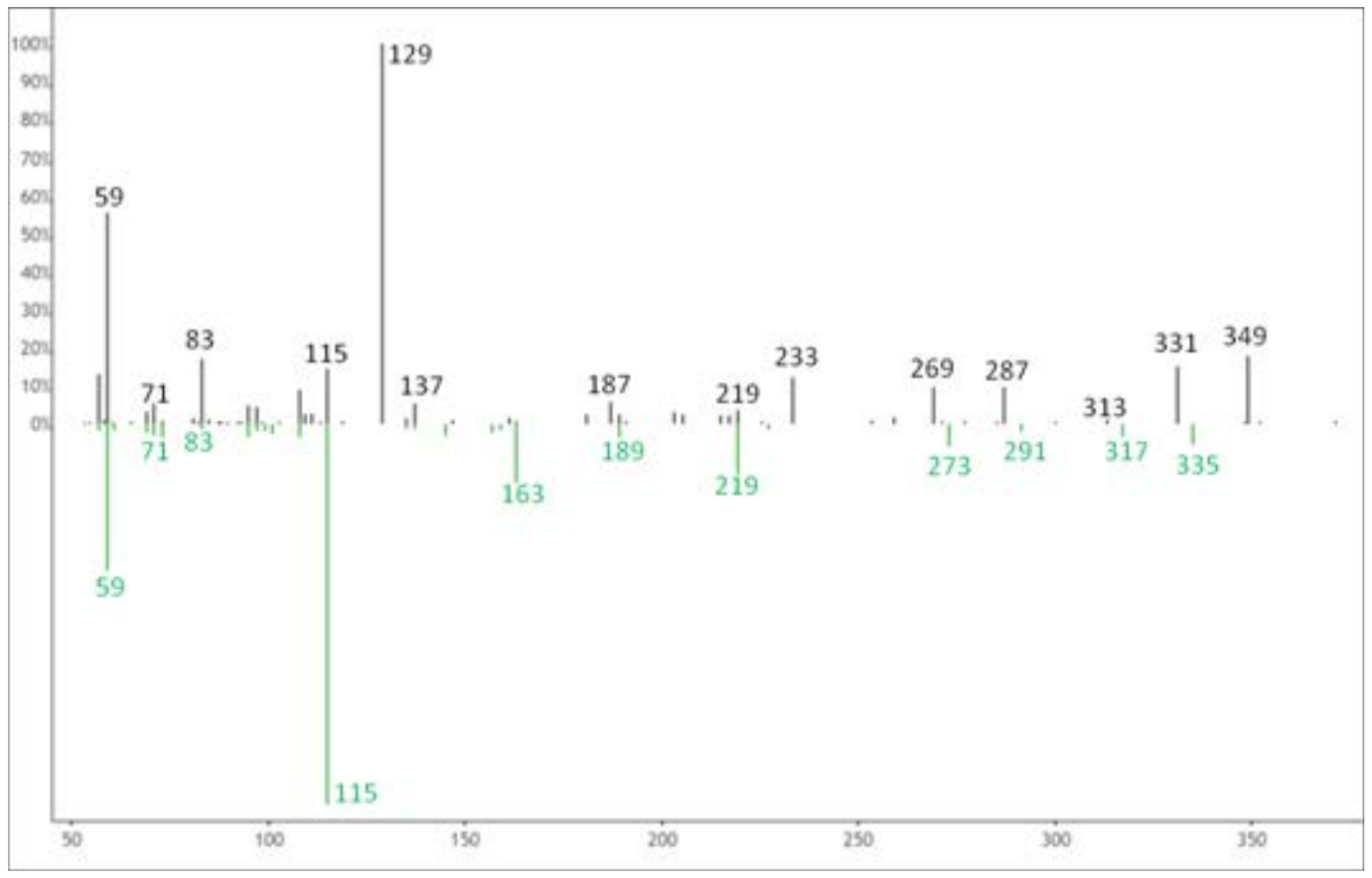

m/z 349.2020
at 2.48 minutes

5,6-DiHETE
reference spectra

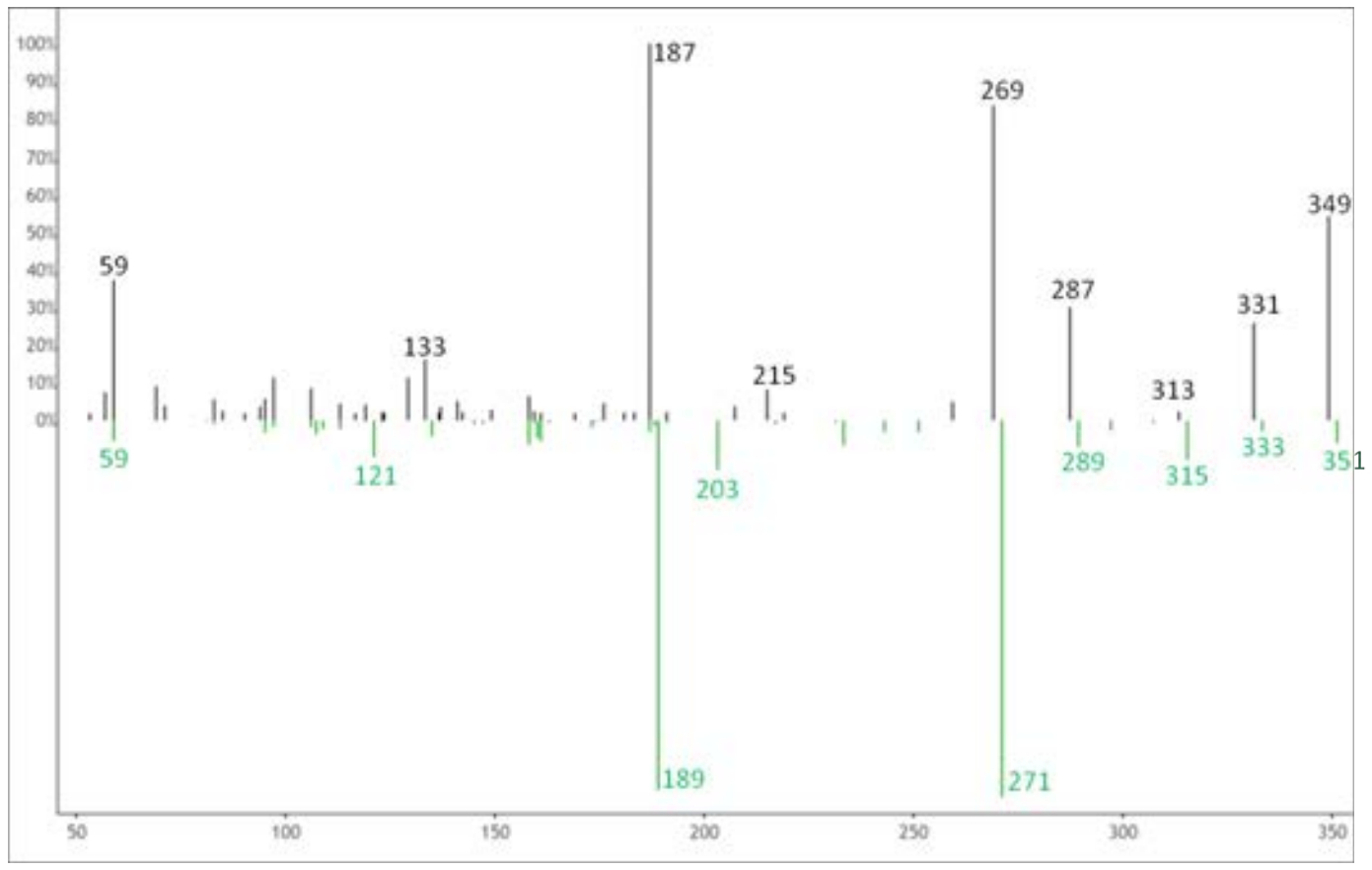

m/z 349.2020
at 2.52 minutes

5-trans-
Prostaglandin D2
reference spectra

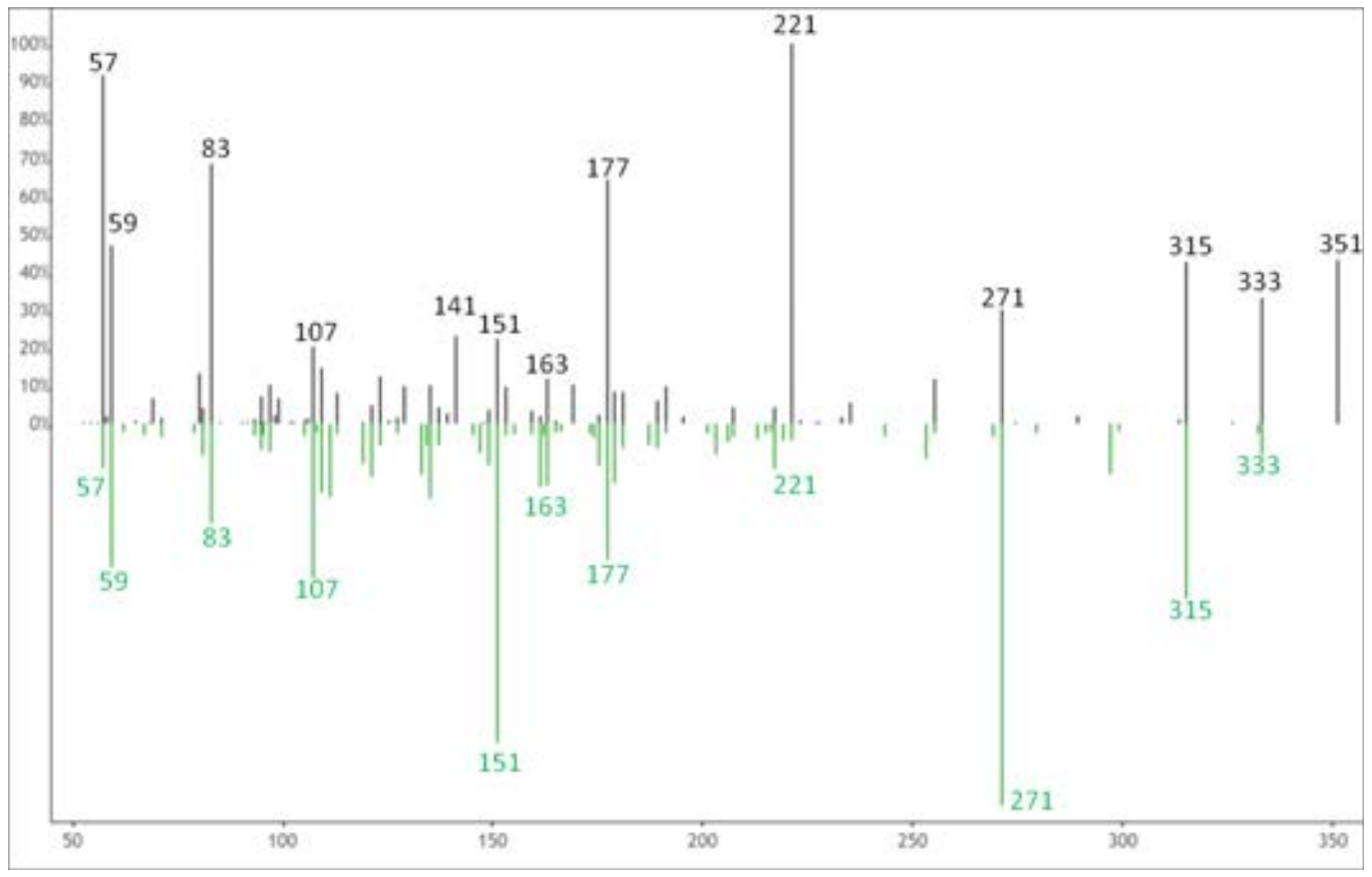

m/z 351.2177
at 2.52 minutes

12-HpEPE
reference spectra

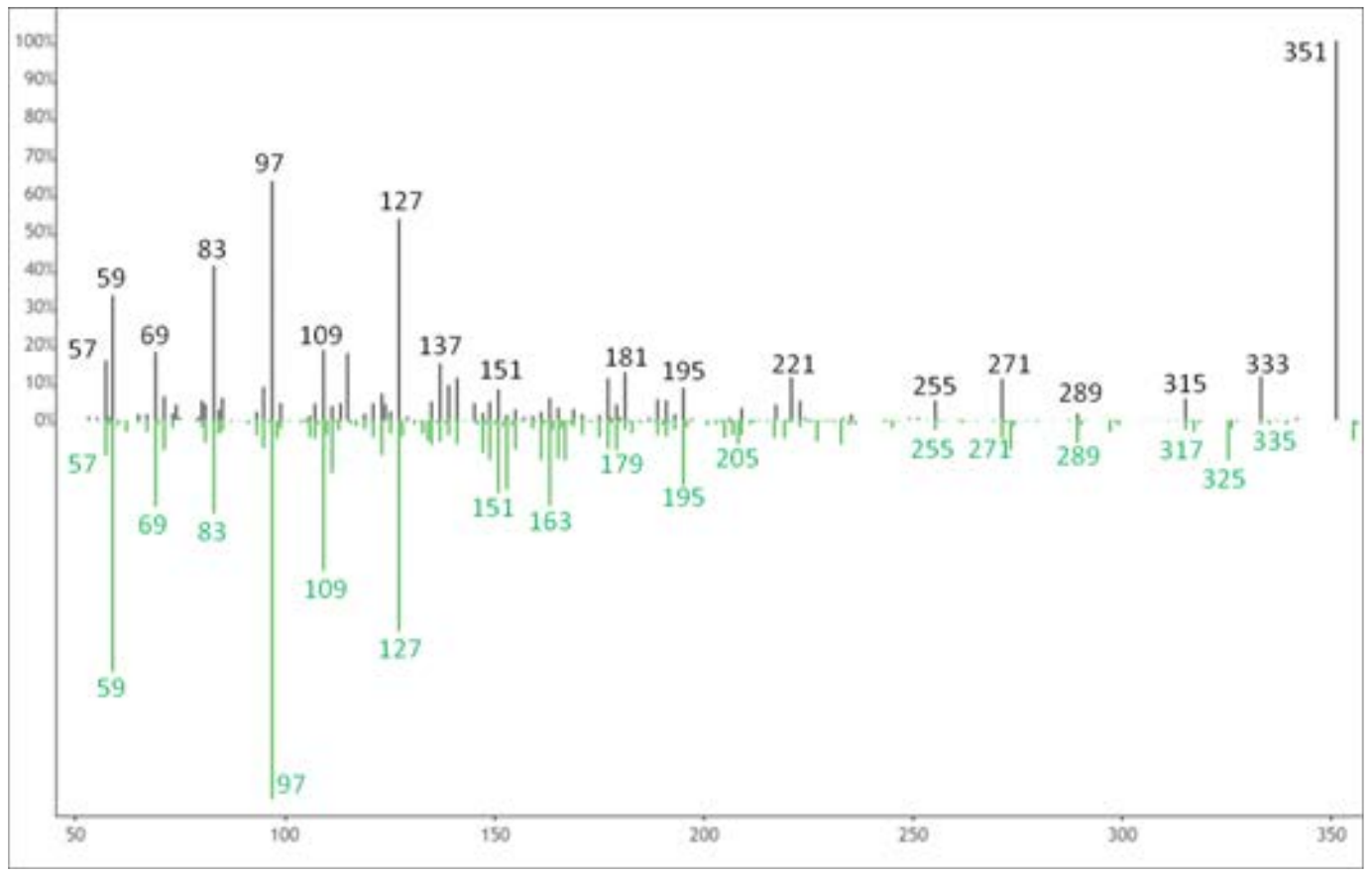

m/z 351.2177
at 2.71 minutes

Hepoxilin A3
reference spectra

## Figure S3 (Continued)

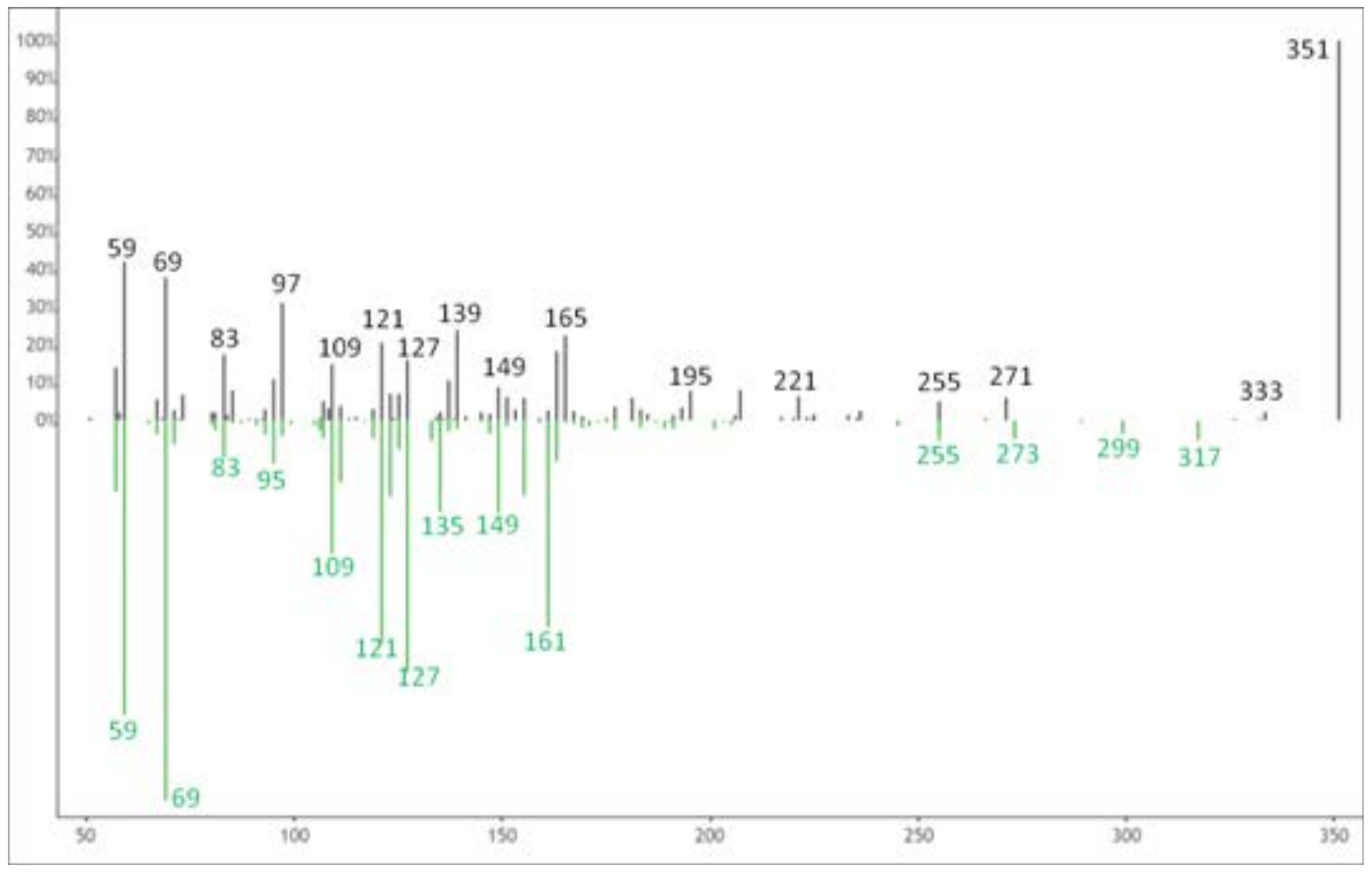

m/z 351.2177
at 2.77 minutes

8,9-EpETE
reference spectra

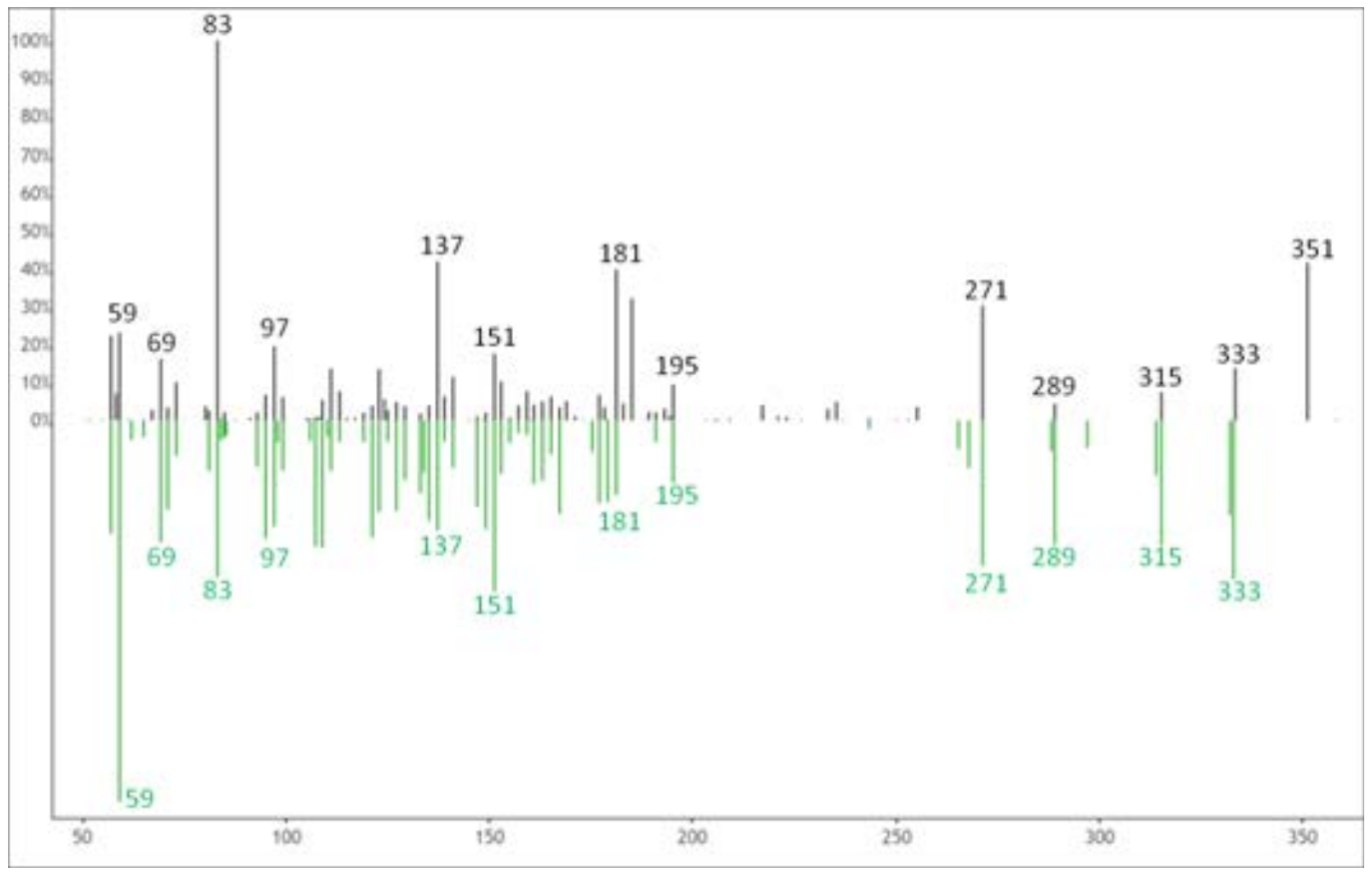

m/z 351.2177
at 2.81 minutes

12-HpEPE
reference spectra

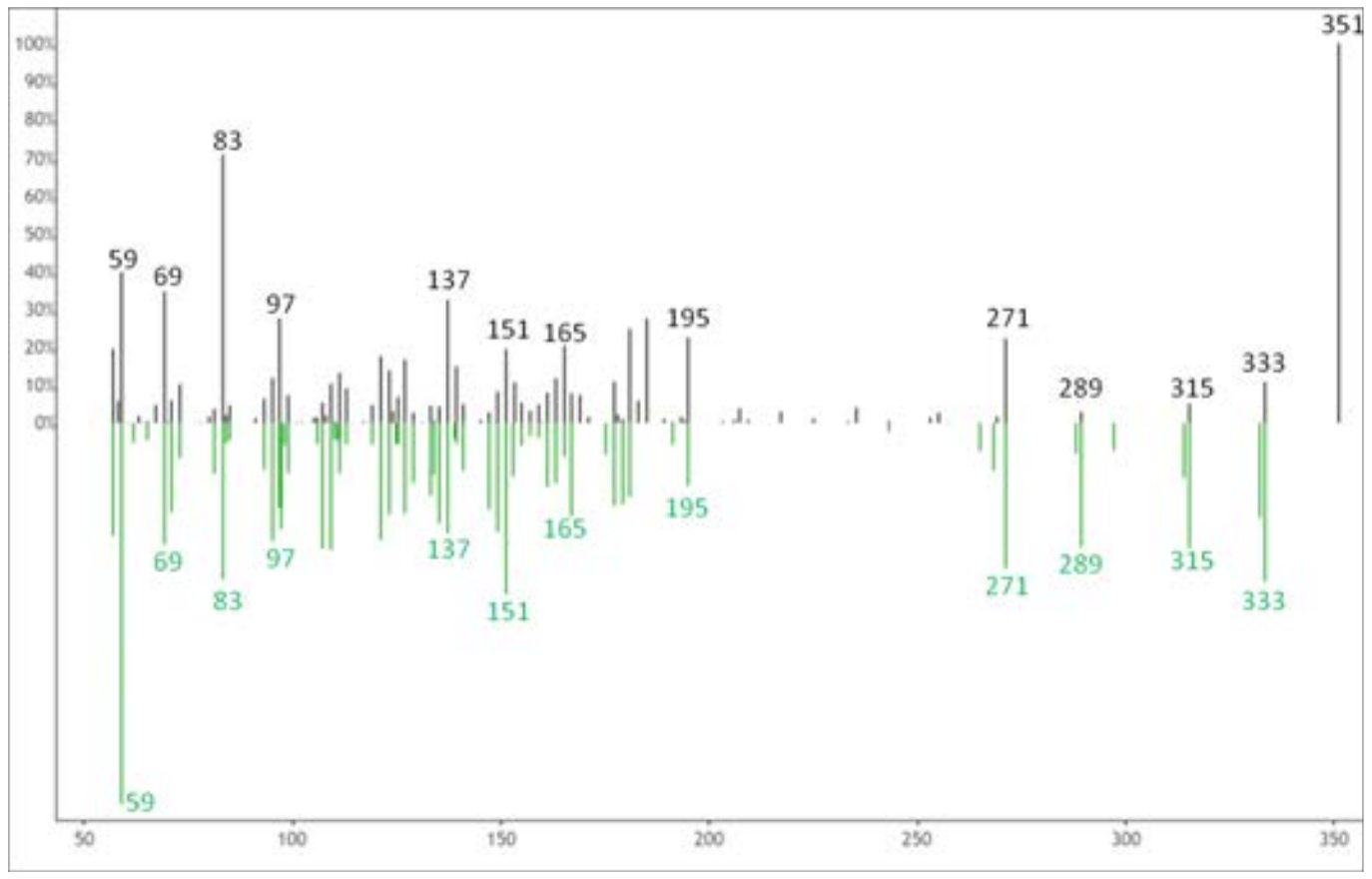

m/z 351.2177
at 2.86 minutes

12-HpEPE
reference spectra

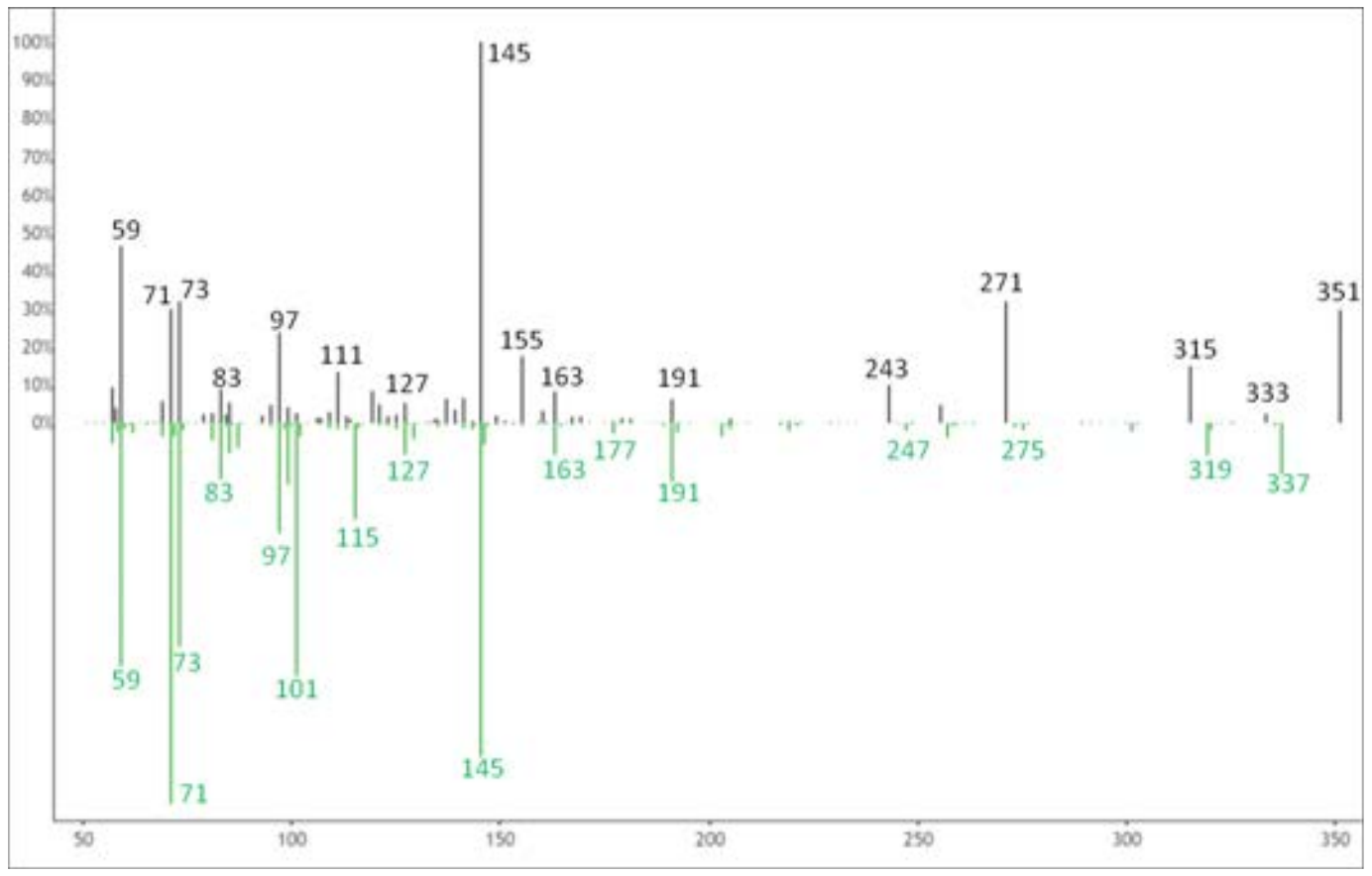

m/z 351.2177
at 3.17 minutes

5,6-DiHETrE
reference spectra

## Figure S3 (Continued)

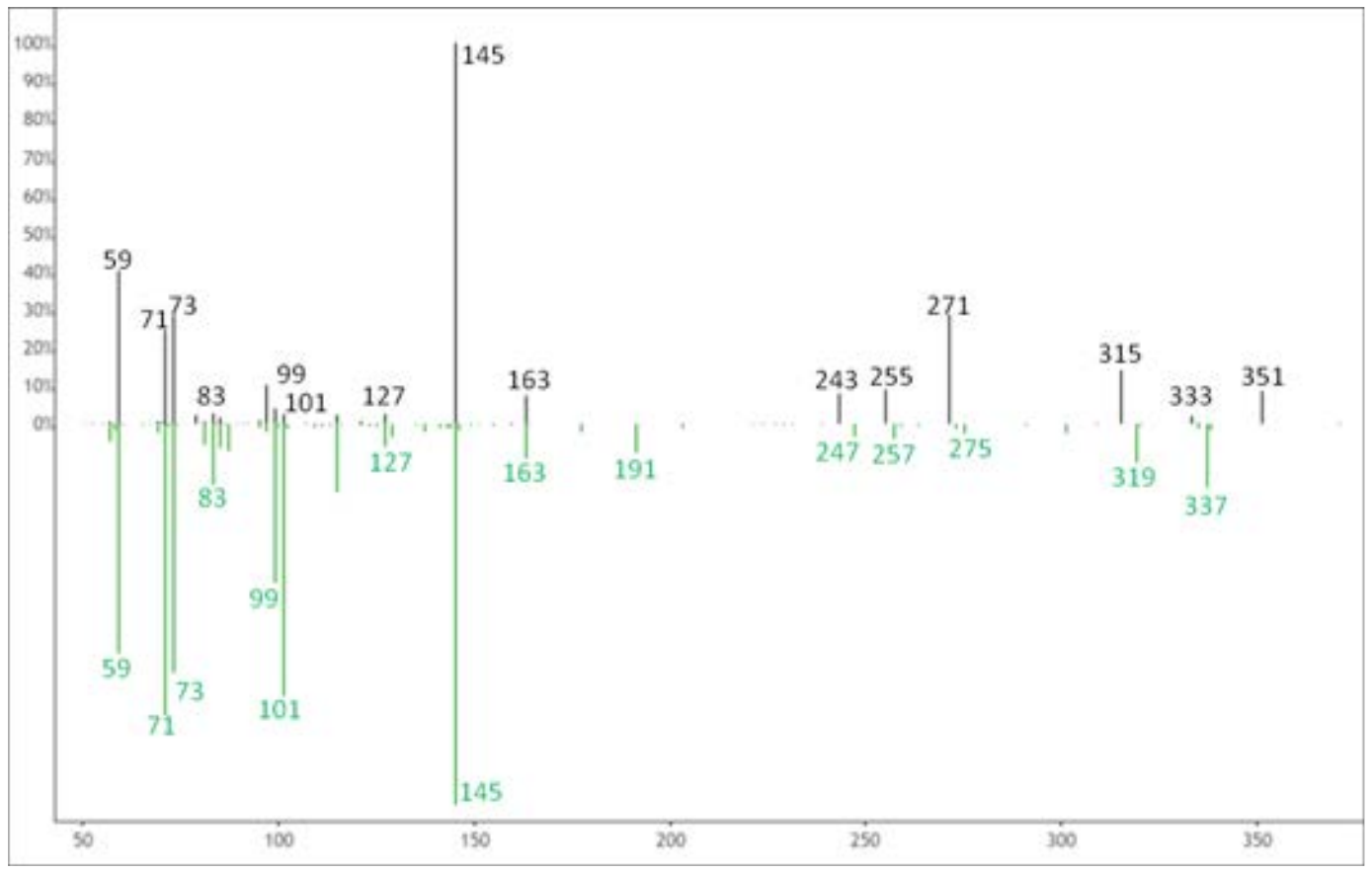

m/z 351.2177
at 3.27 minutes

5,6-DiHETrE
reference spectra

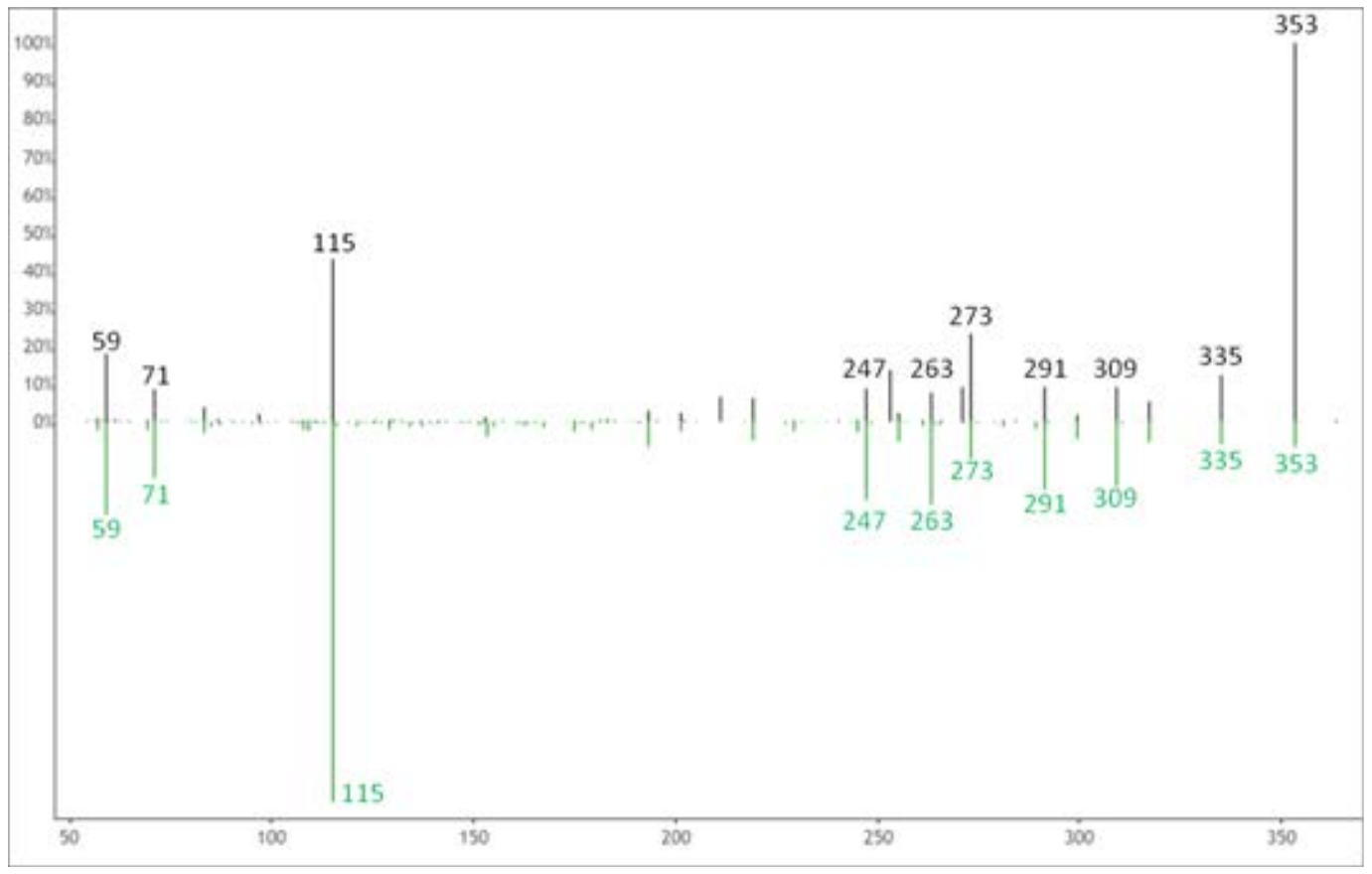

m/z 353.2333
at 1.93 minutes

5-iPF$_{2\alpha}$-VI
reference spectra

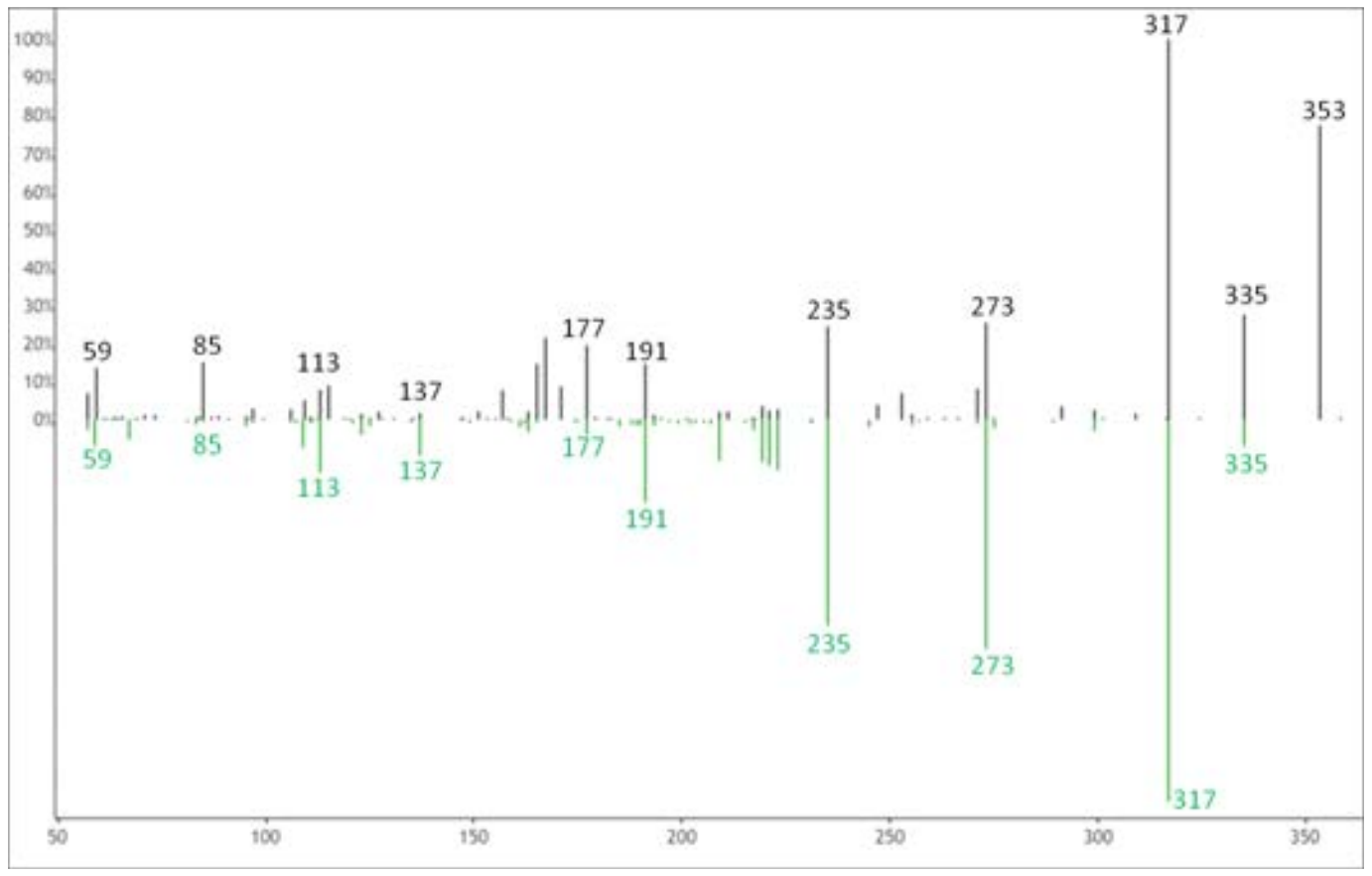

m/z 353.2333
at 2.27 minutes

PGA1 reference
spectra

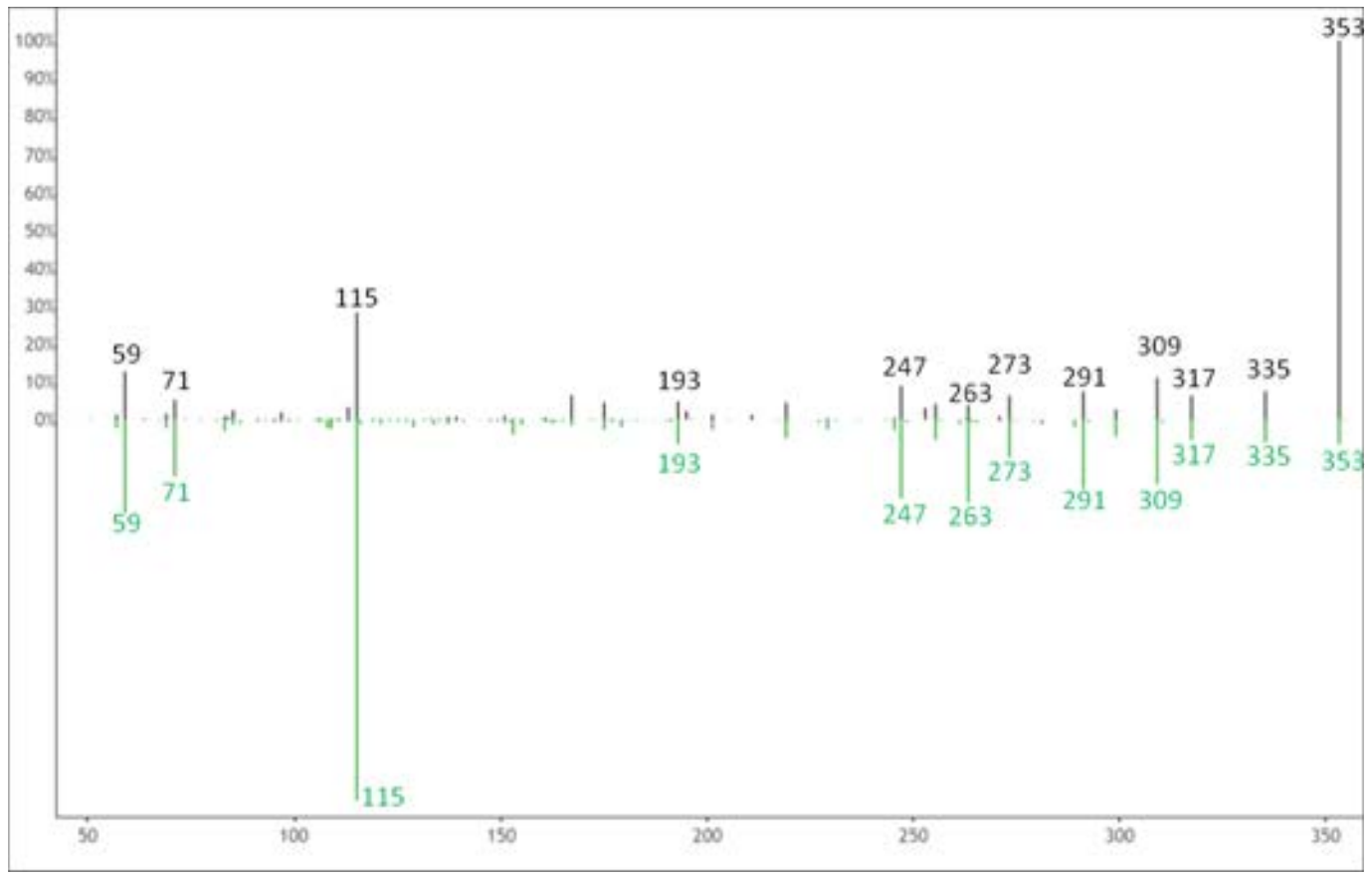

m/z 353.2333
at 2.35 minutes

5-iPF$_{2\alpha}$-VI
reference spectra

**Figure S3 (Continued)**

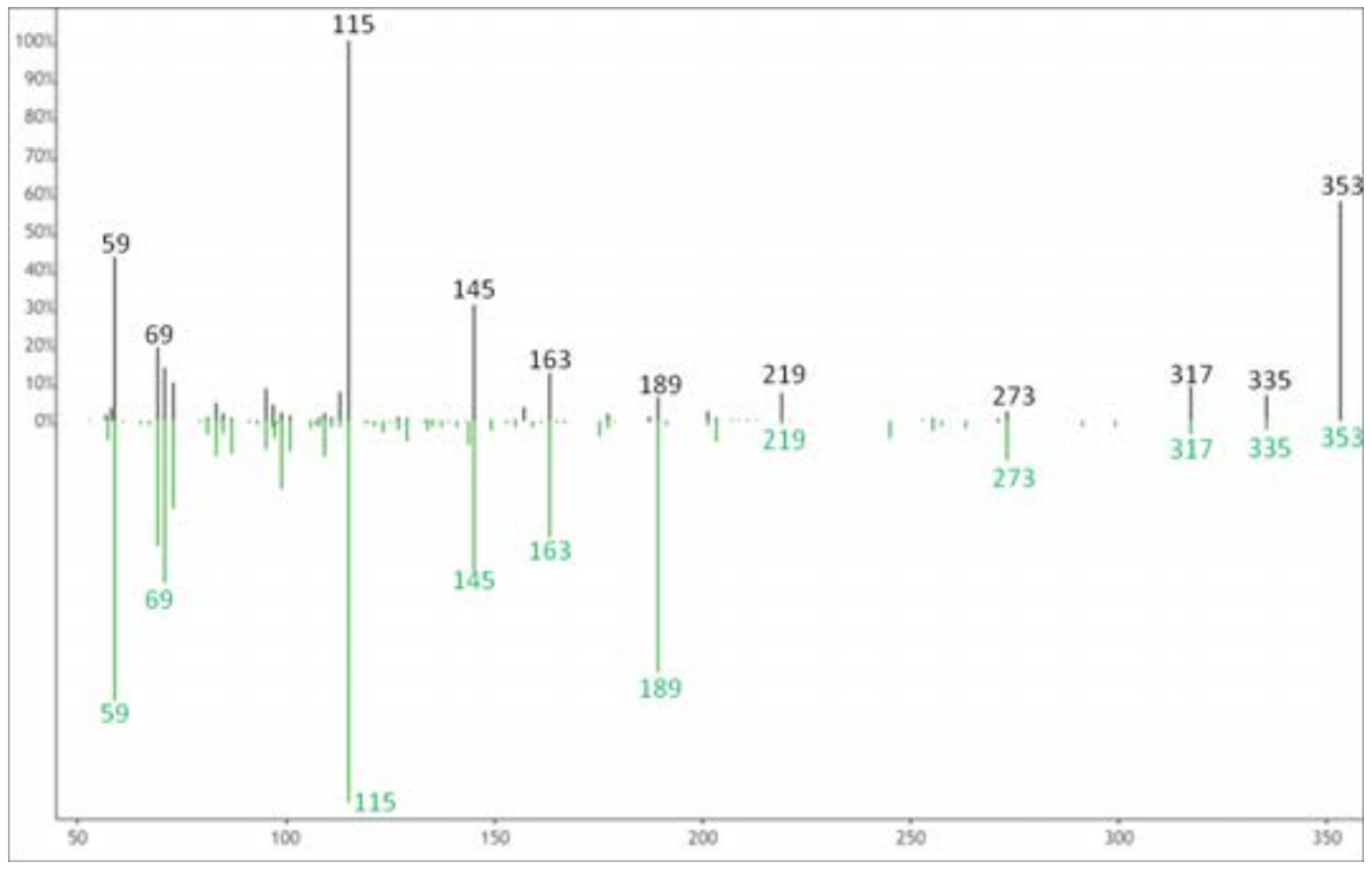

m/z 353.2333
at 3.15 minutes

5,6-DiHETE
reference spectra

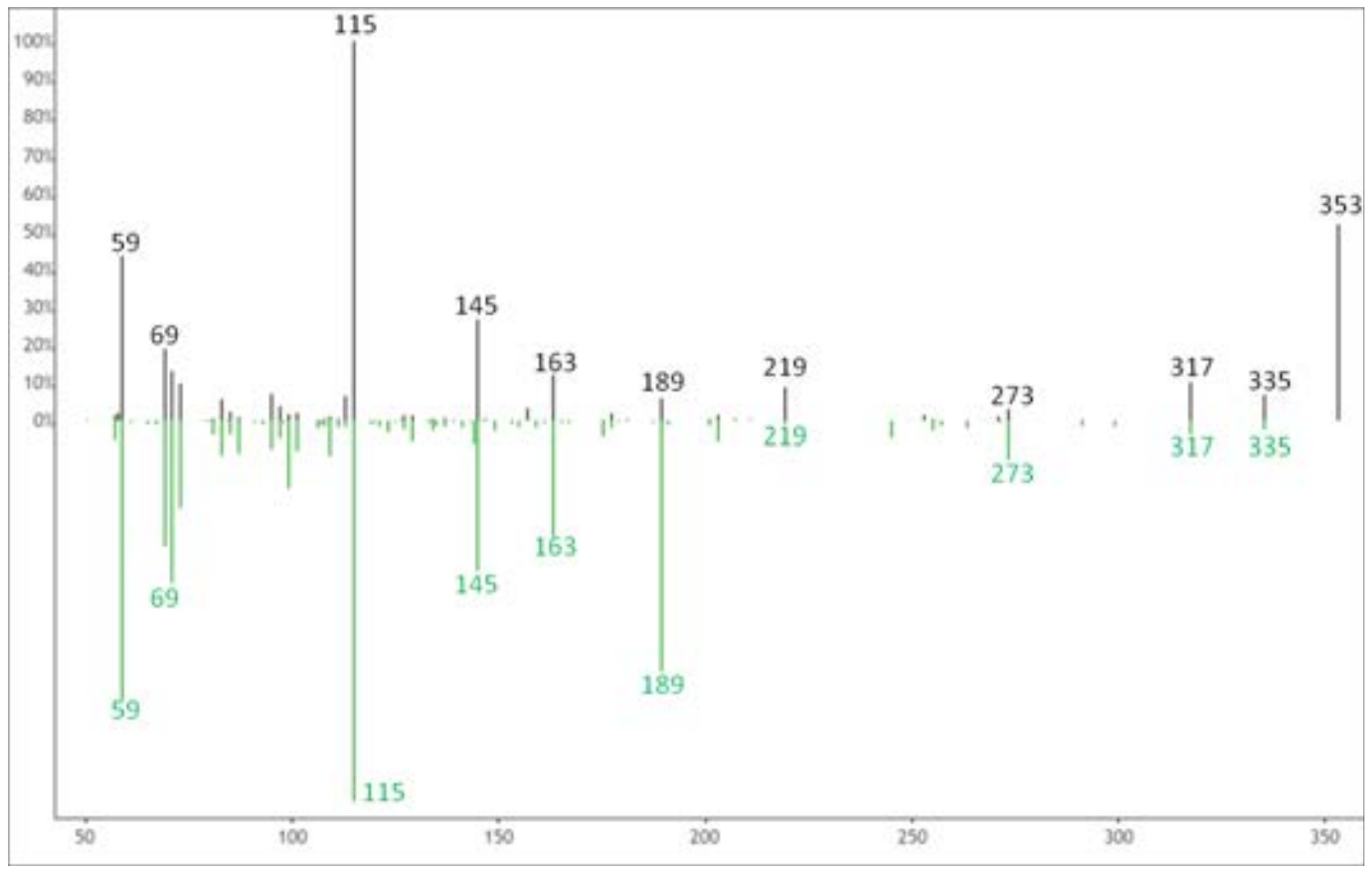

m/z 353.2333
at 3.19 minutes

5,6-DiHETE
reference spectra

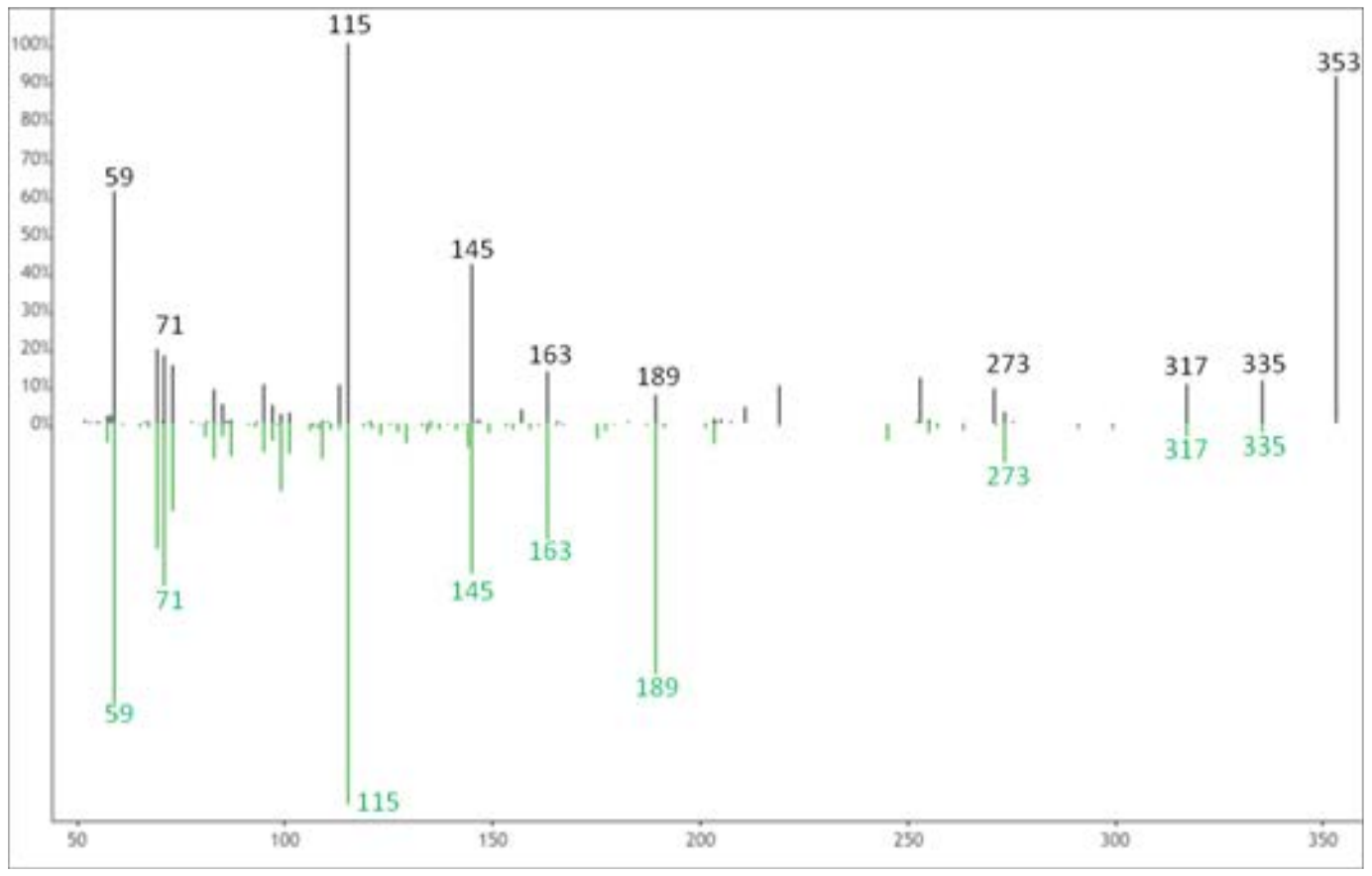

m/z 353.2333
at 3.23 minutes

5,6-DiHETE
reference spectra

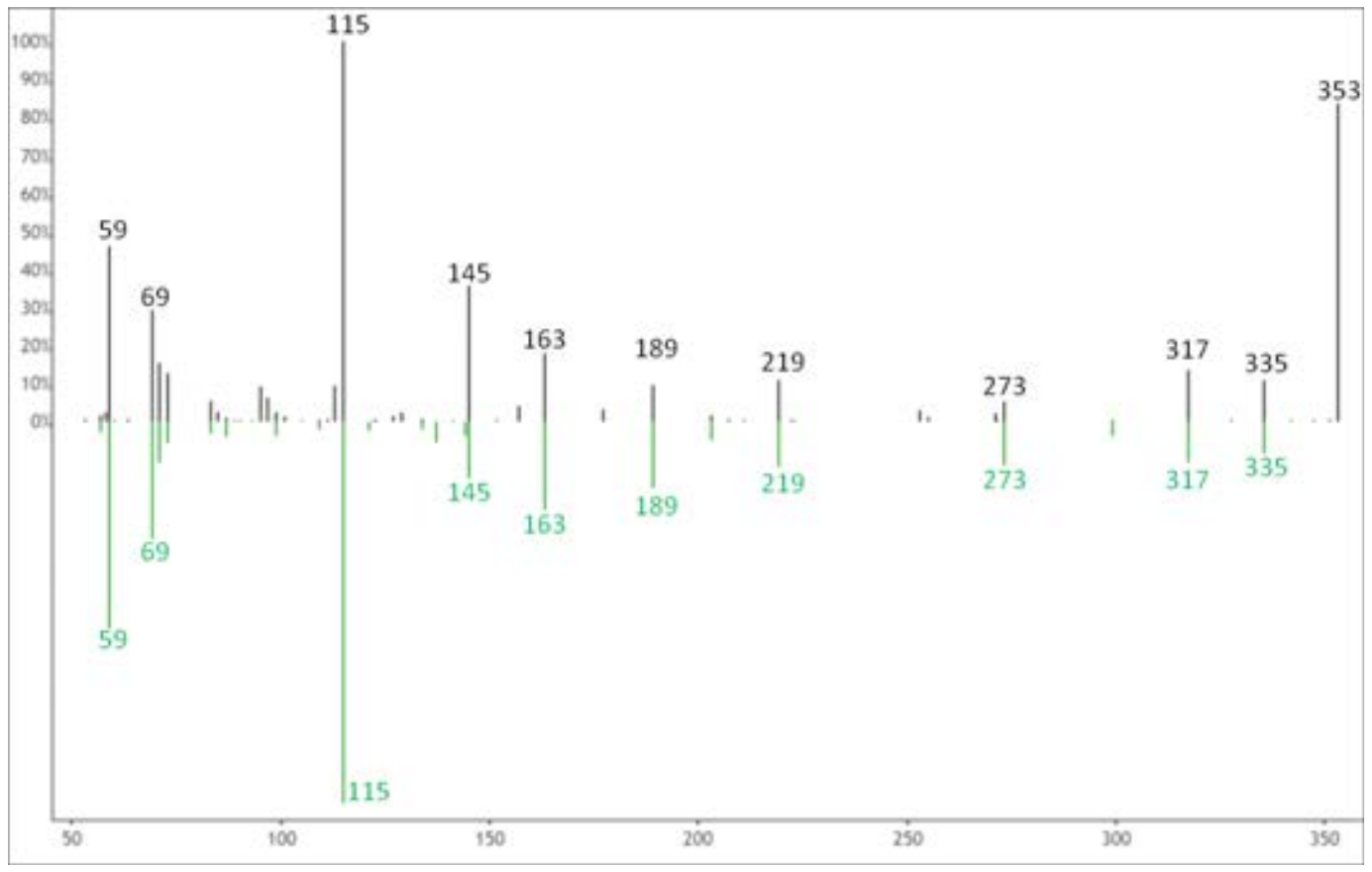

m/z 353.2333
at 3.29 minutes

5,6-DiHETE
reference spectra

## Figure S3 (Continued)

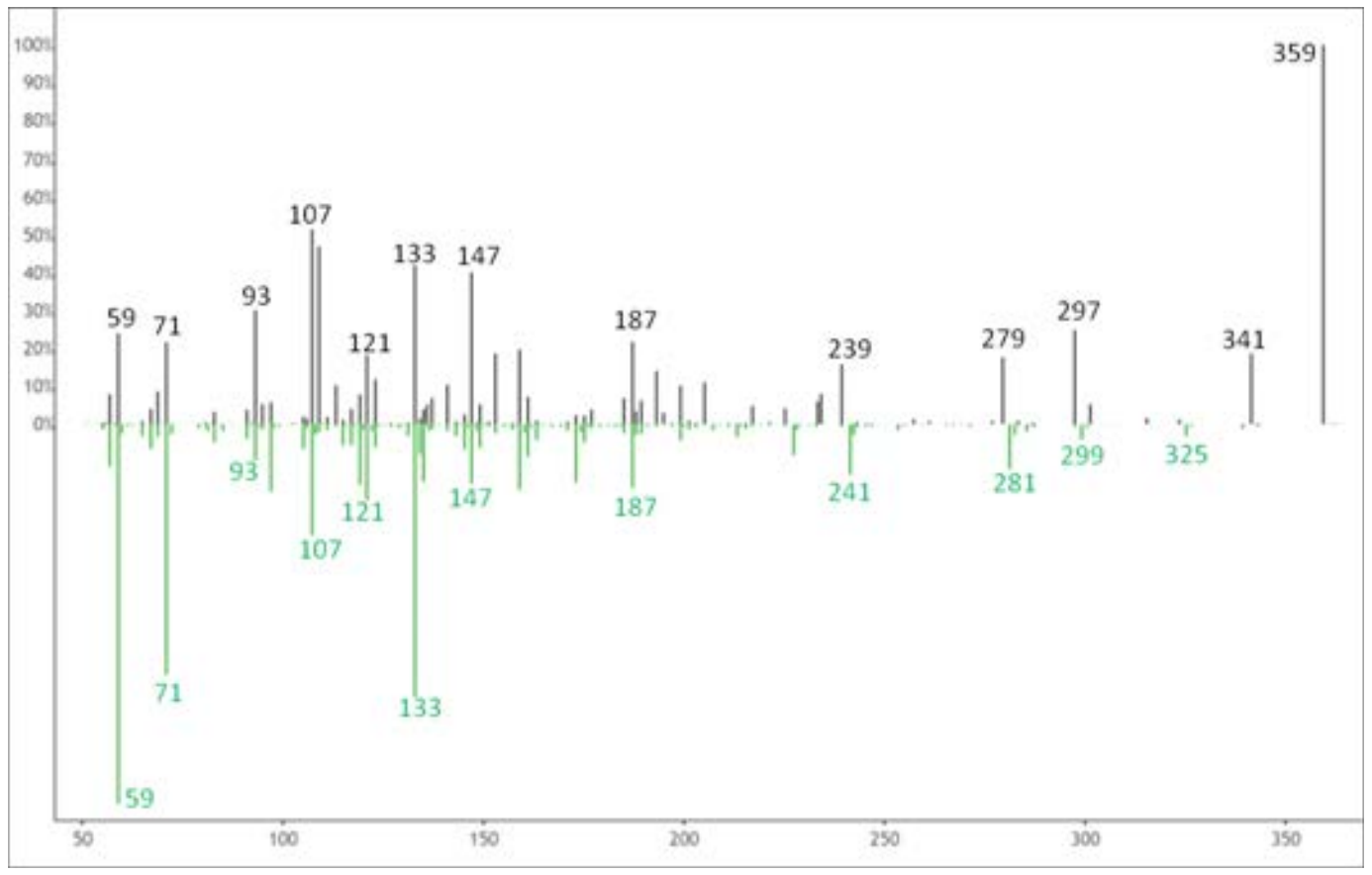

m/z 359.2228
at 2.85 minutes

20-HDoHE
reference spectra

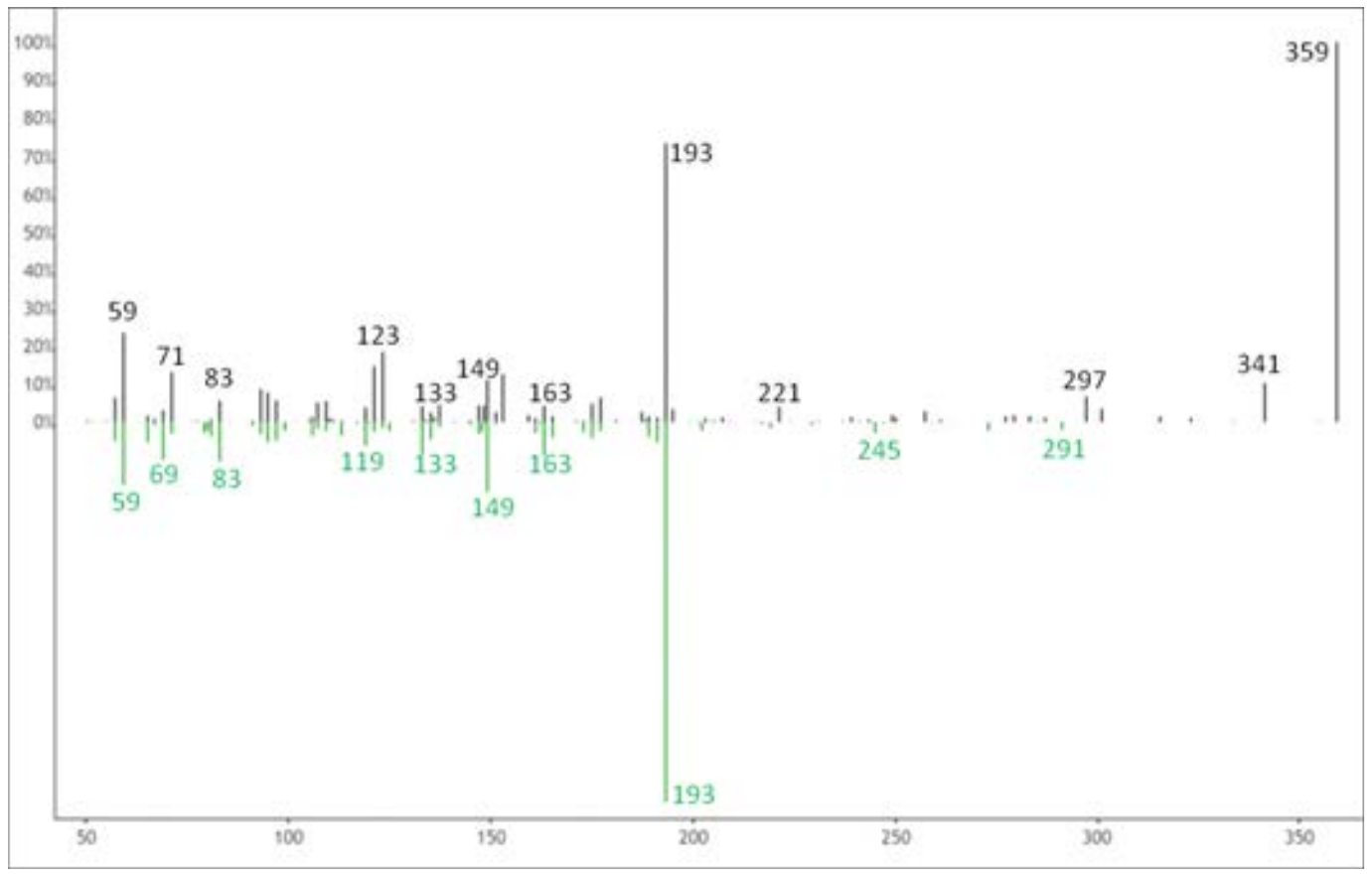

m/z 359.2228
at 2.89 minutes

Prostaglandin F2α
reference spectra

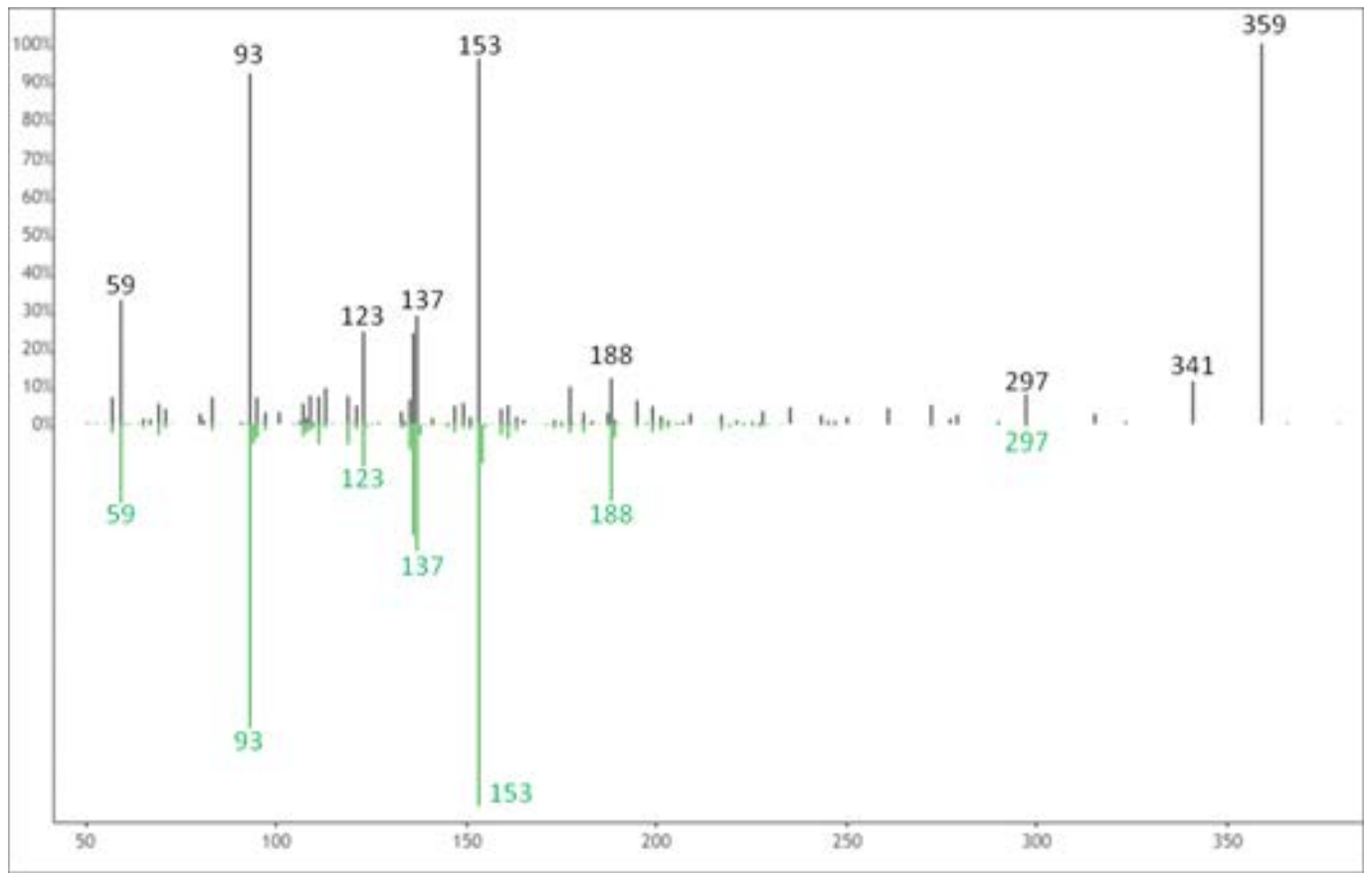

m/z 359.2228
at 3.02 minutes

Protectin D1
reference spectra

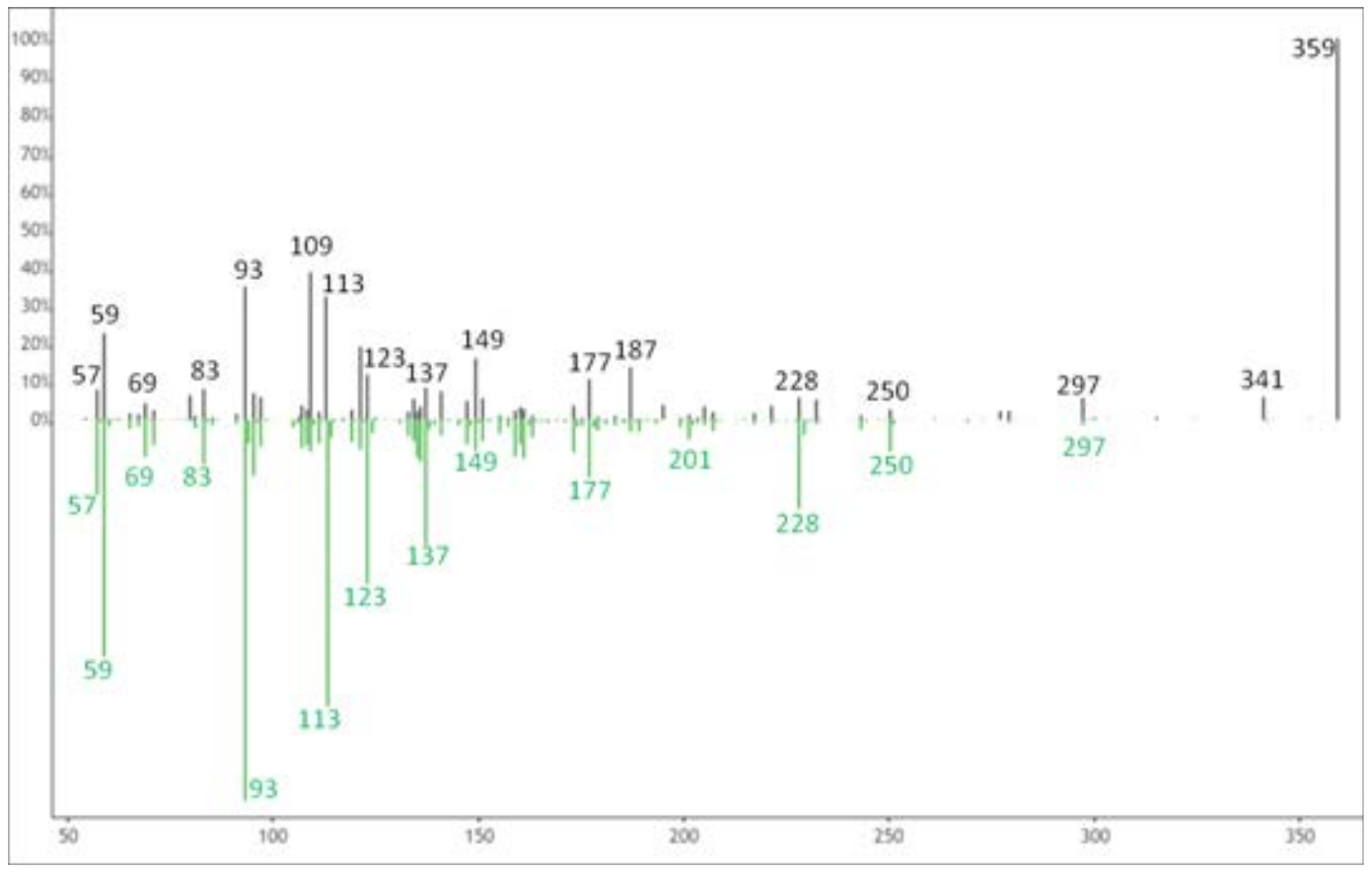

m/z 359.2228
at 3.12 minutes

Maresin 1
reference spectra

## Figure S3 (Continued)

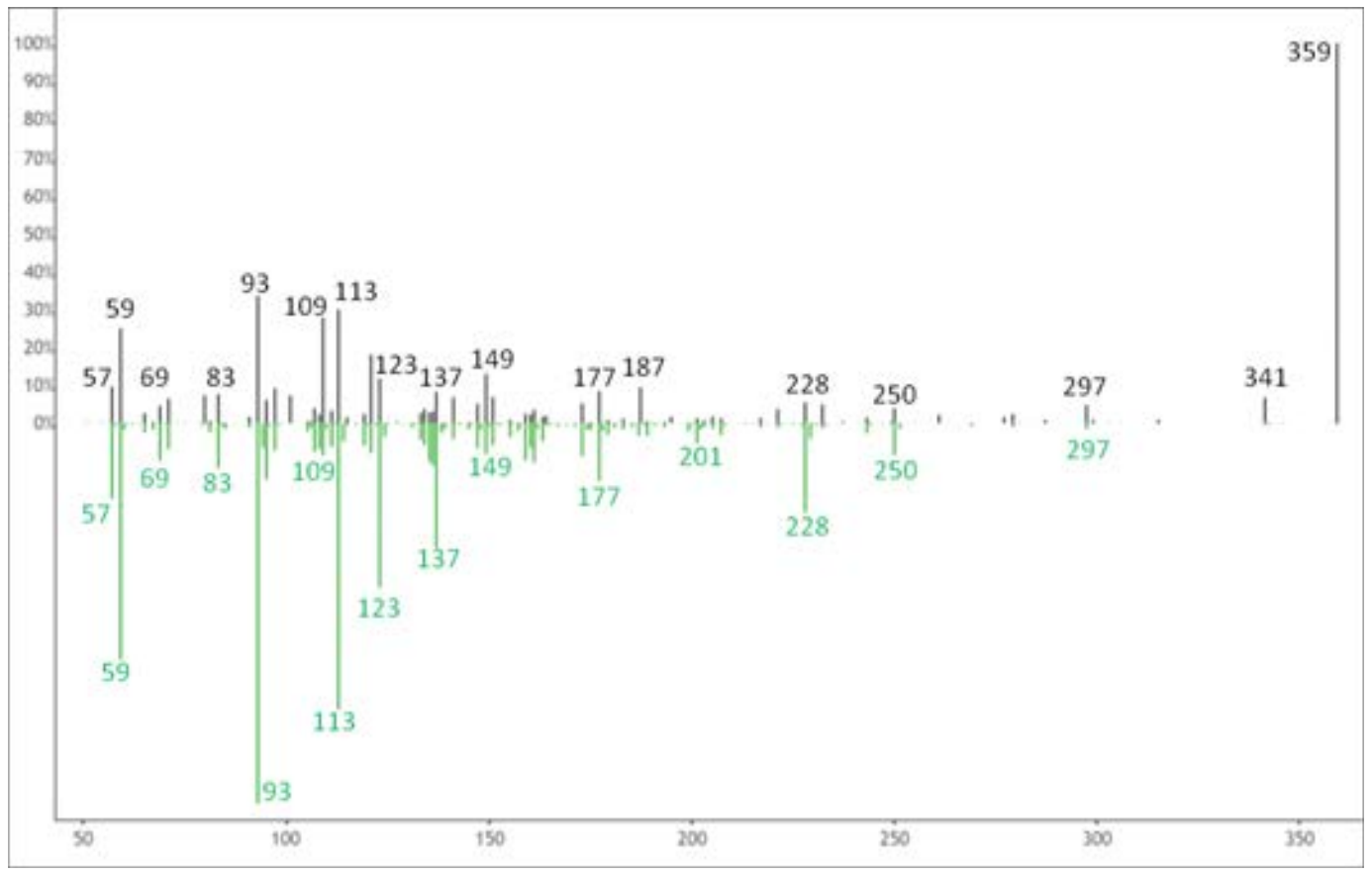

m/z 359.2228
at 3.16 minutes

Maresin 1
reference spectra

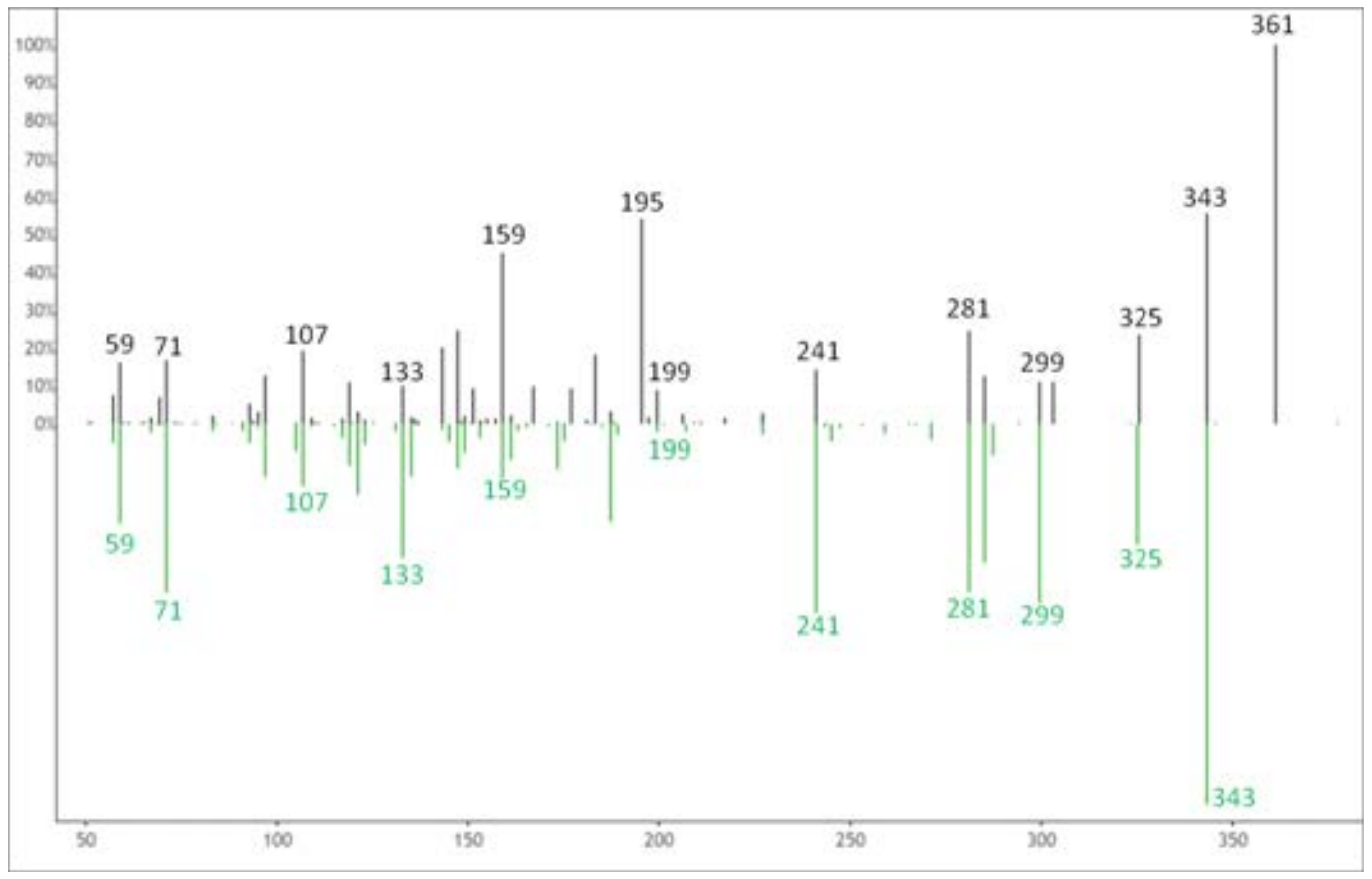

m/z 361.2384
at 2.95 minutes

20-HDoHE
reference spectra

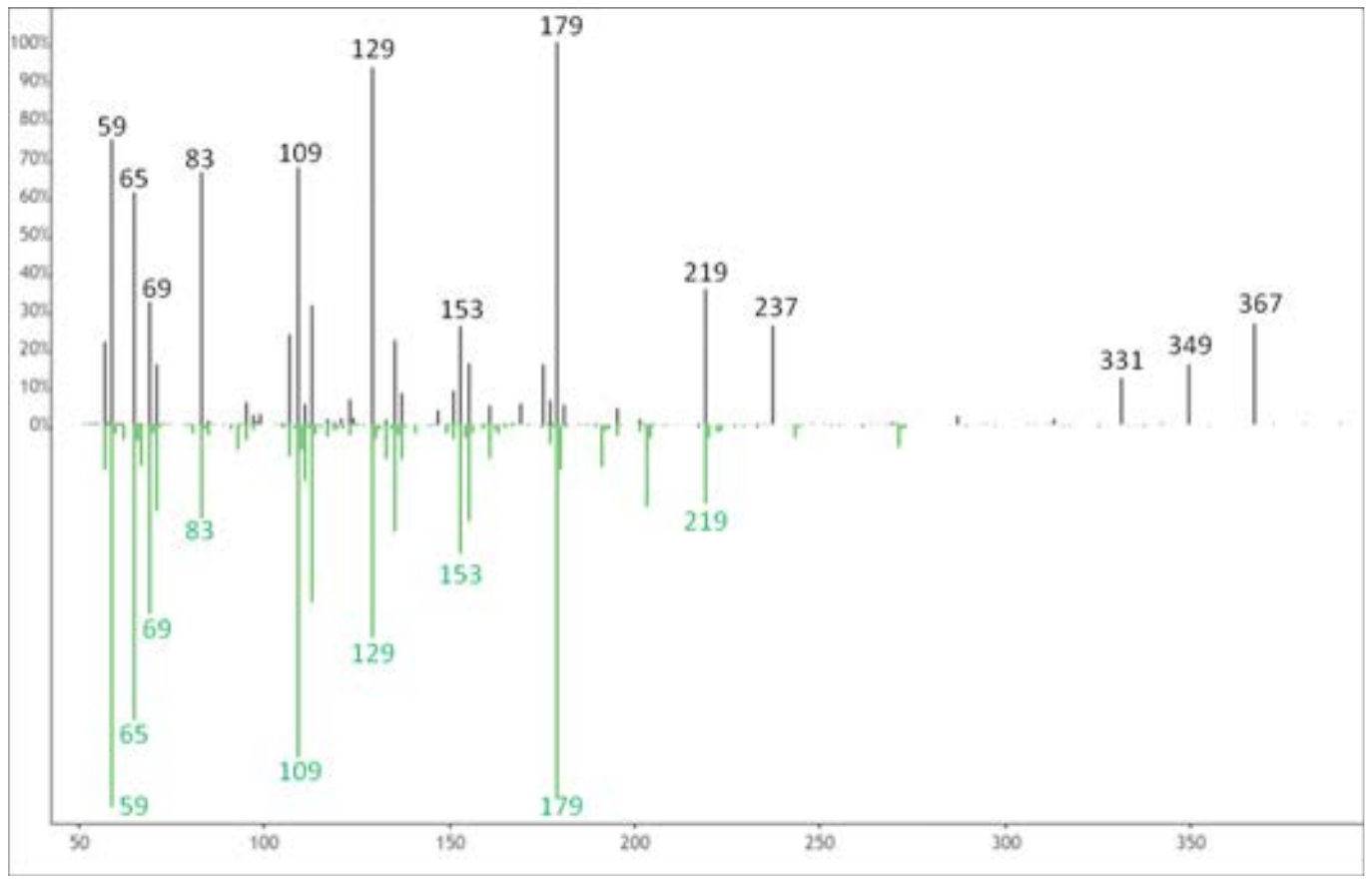

m/z 367.2126
at 1.62 minutes

12-oxo-LTB4
reference spectra

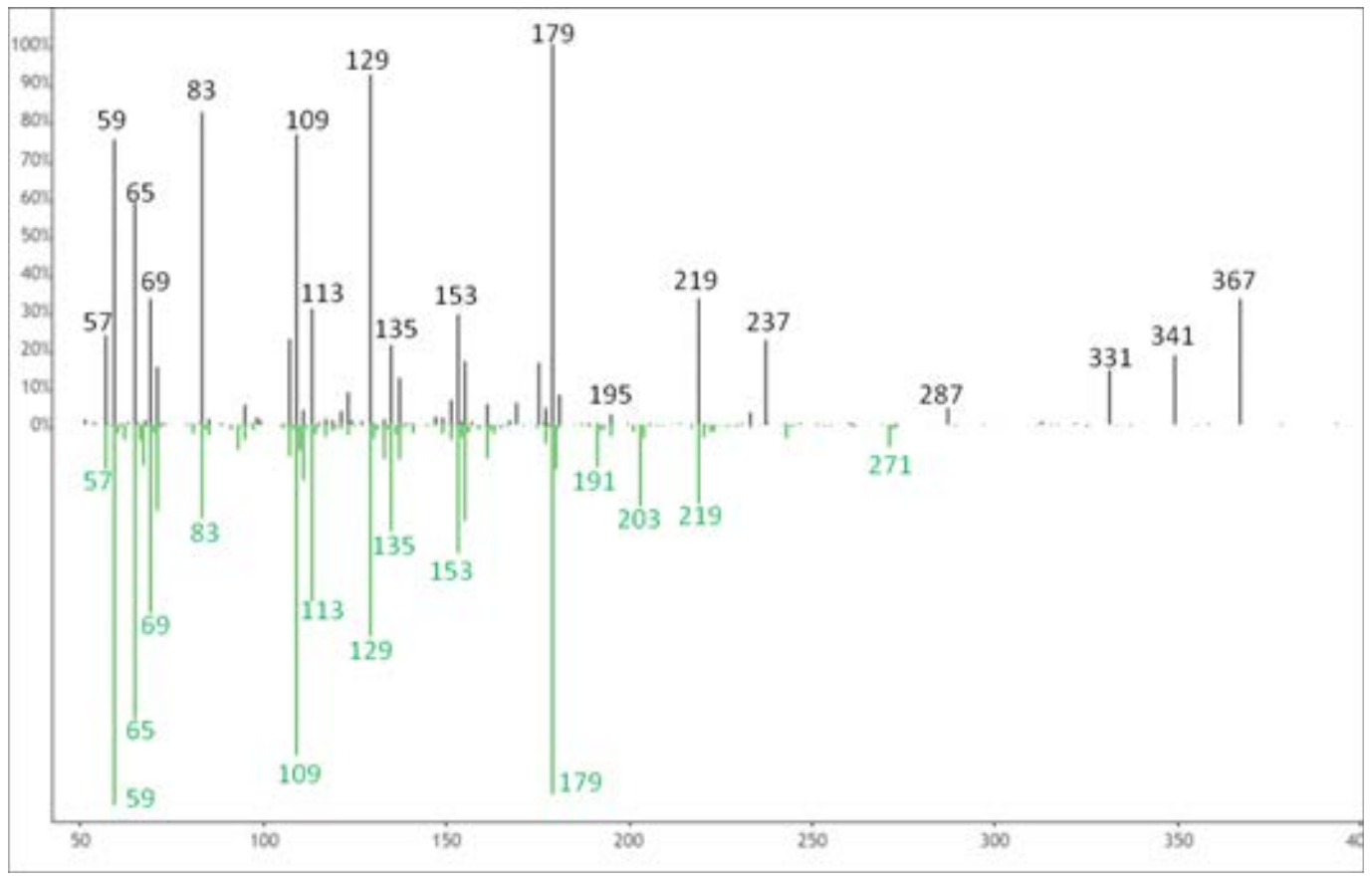

m/z 367.2126
at 1.67 minutes

12-oxo-LTB4
reference spectra

## Figure S3 (Continued)

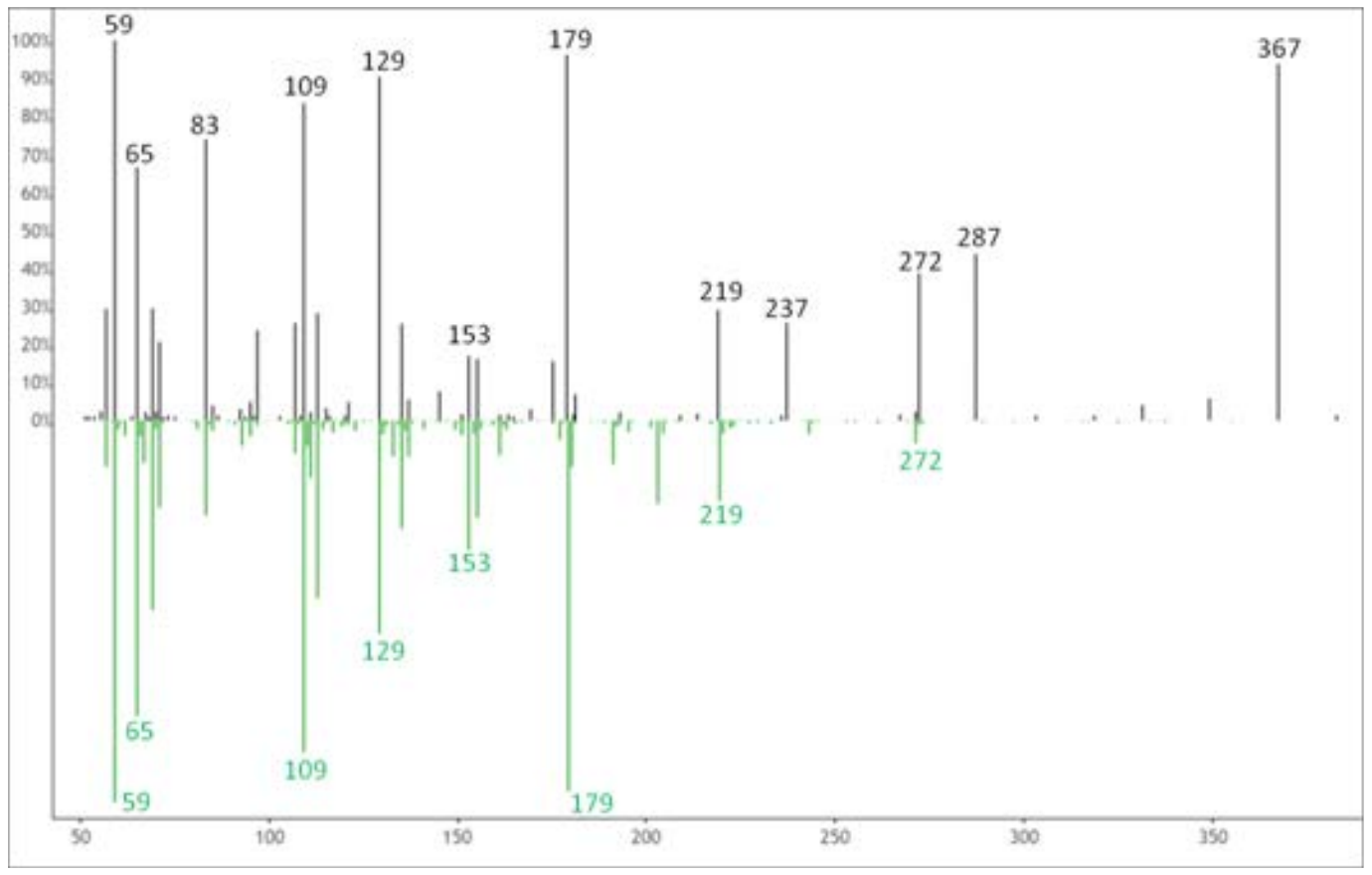

m/z 367.2126
at 1.73 minutes

12-oxo-LTB4
reference spectra

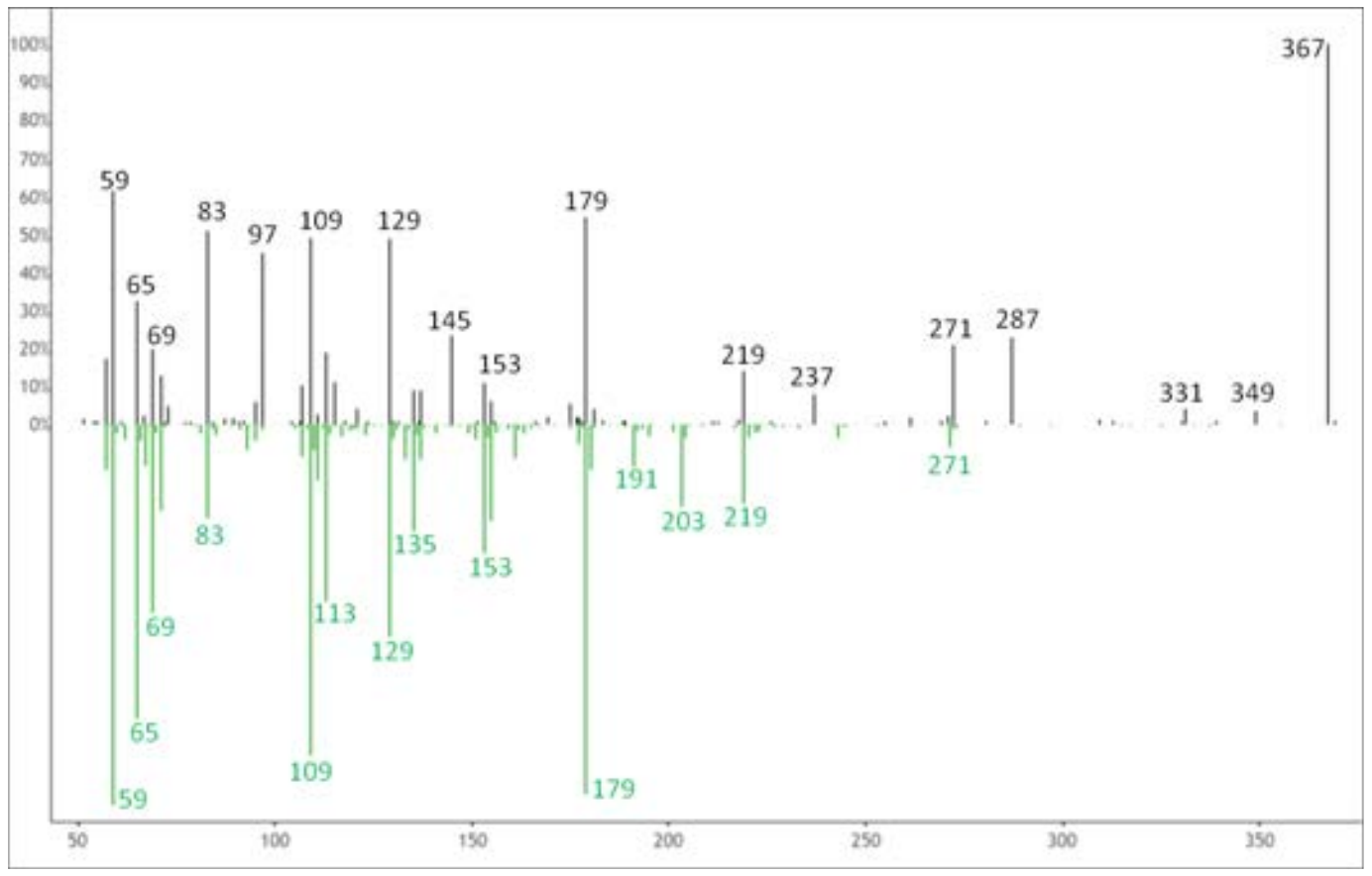

m/z 367.2126
at 1.78 minutes

12-oxo-LTB4
reference spectra

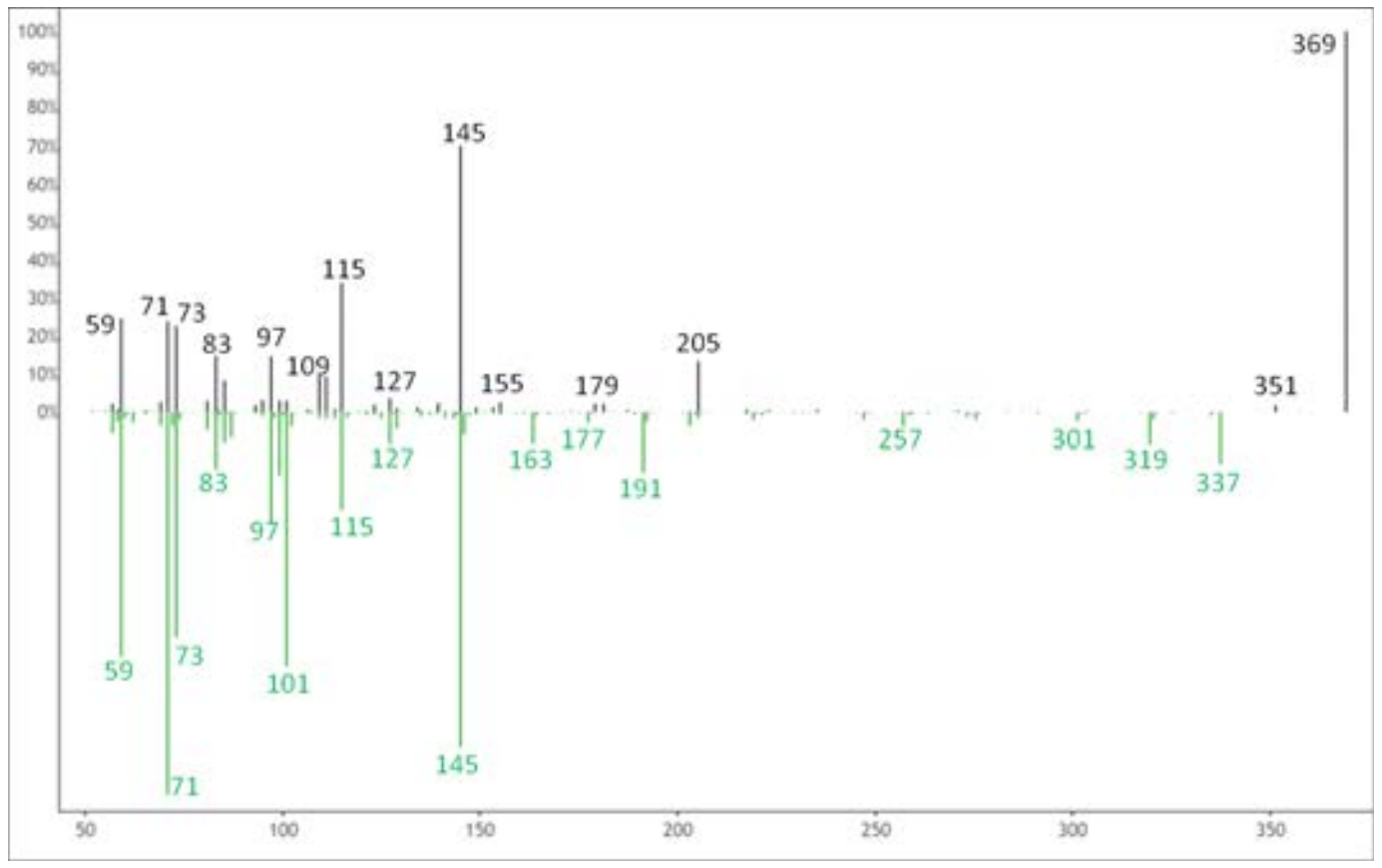

m/z 369.2282
at 2.13 minutes

5,6-DiHETrE
reference spectra

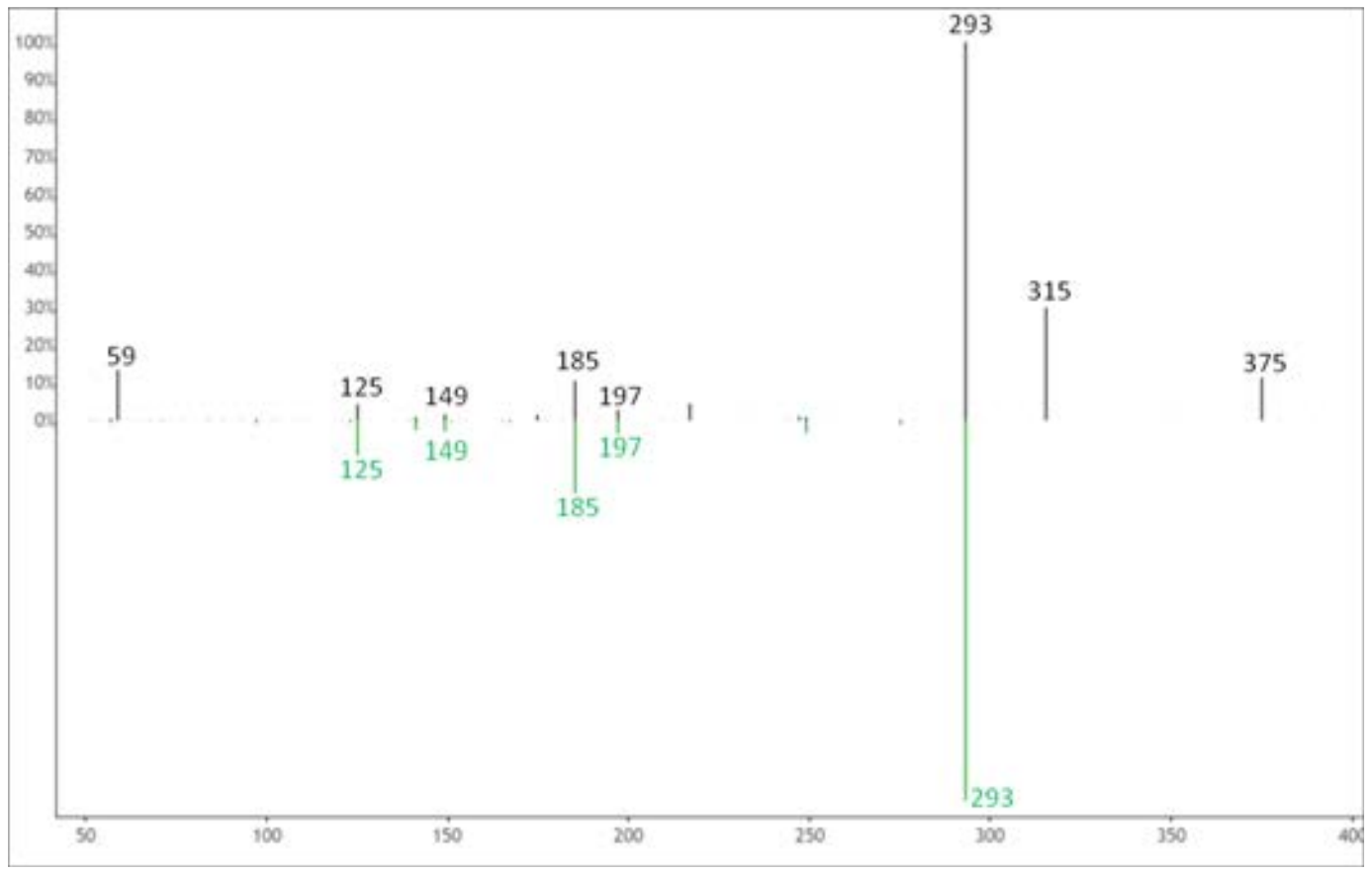

m/z 375.2177
at 4.57 minutes

9-oxo-ODE
reference spectra

# Figure S3 (Continued)

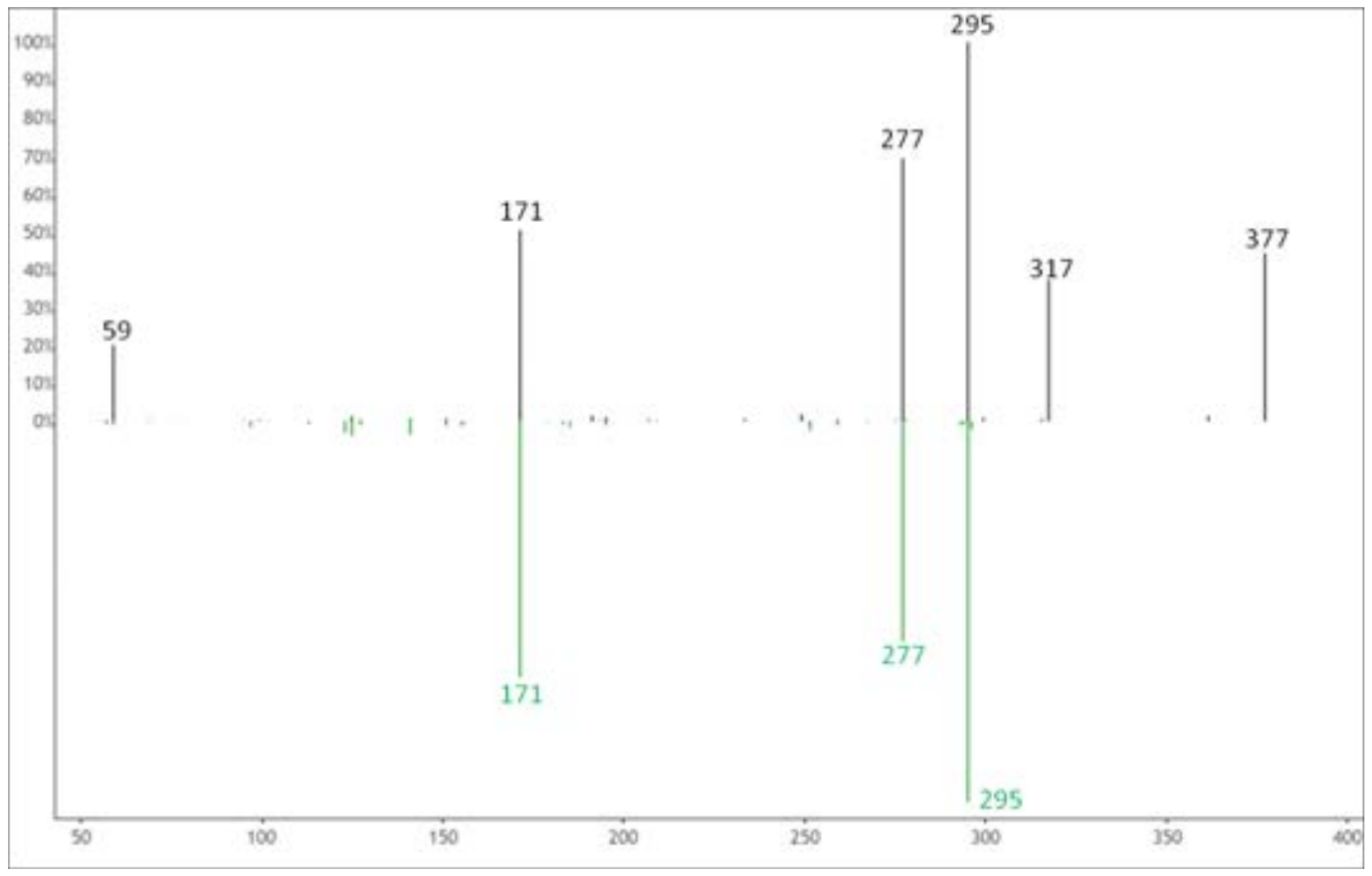

m/z 377.2333
at 4.35 minutes

9,10-DiHOME
reference spectra

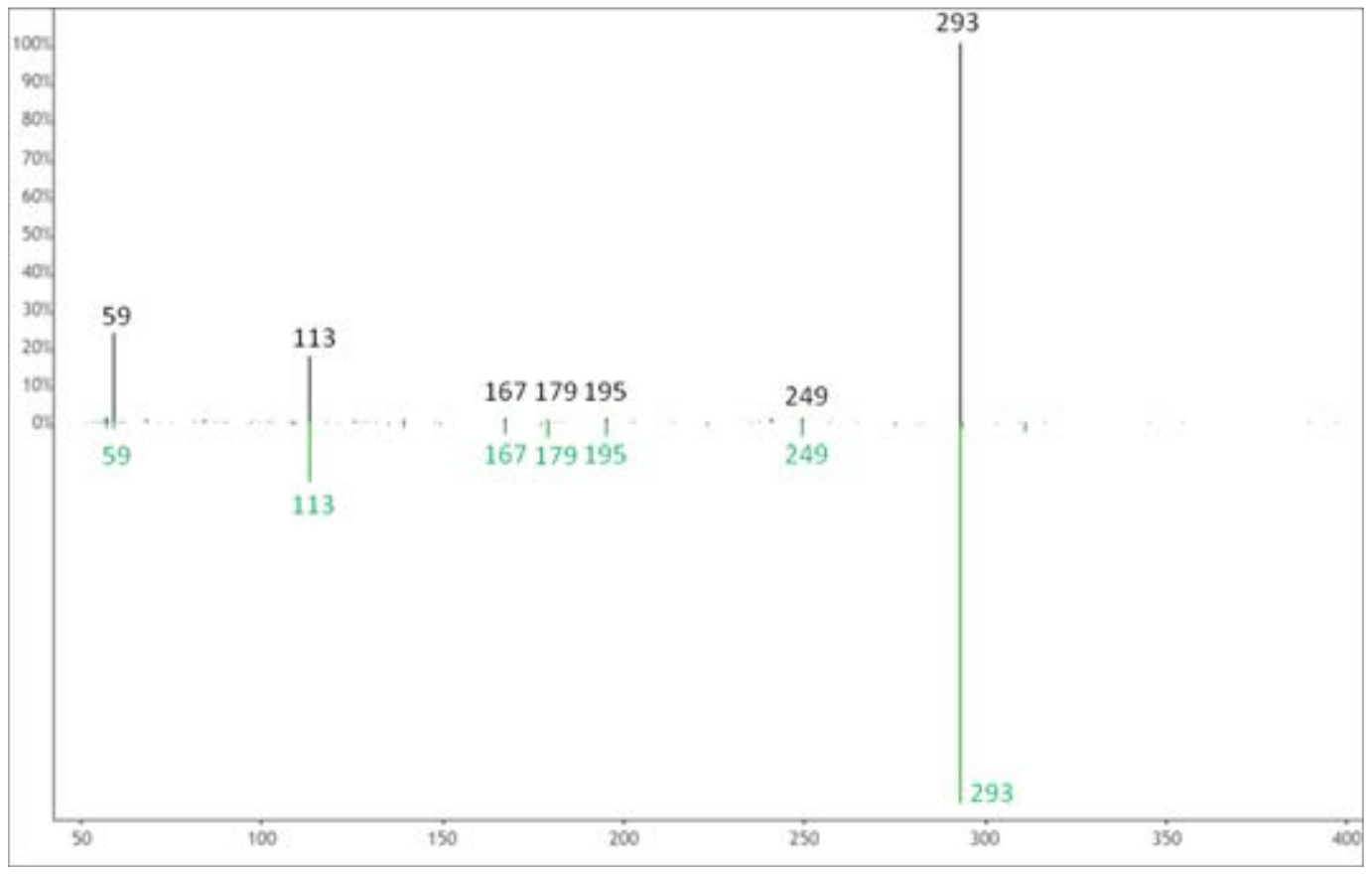

m/z 389.1945
at 4.40 minutes

13-HpODE
reference spectra

**Figure S4.** Mass defect plots used for identification of chemical formulas potentially belonging to novel eicosanoid compounds. (A) For each eicosanoid entry in the LipidMaps library, the nominal mass was plotted against the mass defect to reveal patterns in chemical formulas which compose the 1149 unique compounds. (B) By searching for gaps in the chemical iterations within each chemical family, 'missing' entries were identified and plotted in red. (C) Once data for each of the missing formulas was extracted and searched via chemical spectral networking, all formulas resulting in positive analog matches were left plotted in red relative to all entries from eicosanoids in LipidMaps.

## Figure S4

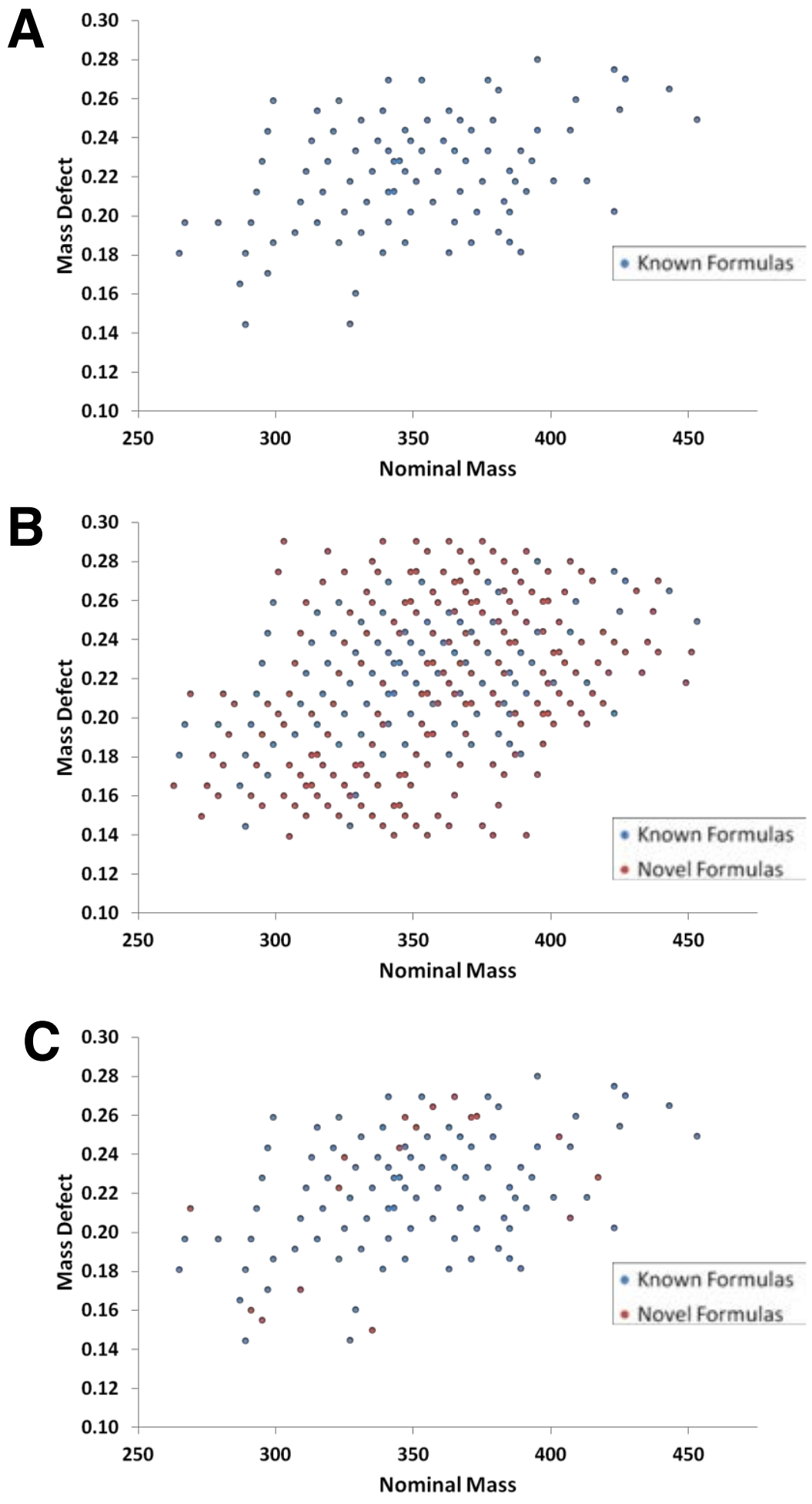

**Figure S5. Related to Figure 4.** Annotation of tandem MS spectra for potentially novel eicosanoids. (A) m/z 347.2591 at 5.58 minutes. (B) m/z 345.2434 at 5.37 minutes. (C) m/z 345.2434 at 5.47 minutes. (D) m/z 345.2434 at 4.94 minutes. (E) m/z 357.2646 at 2.32 minutes.

## Figure S5-A

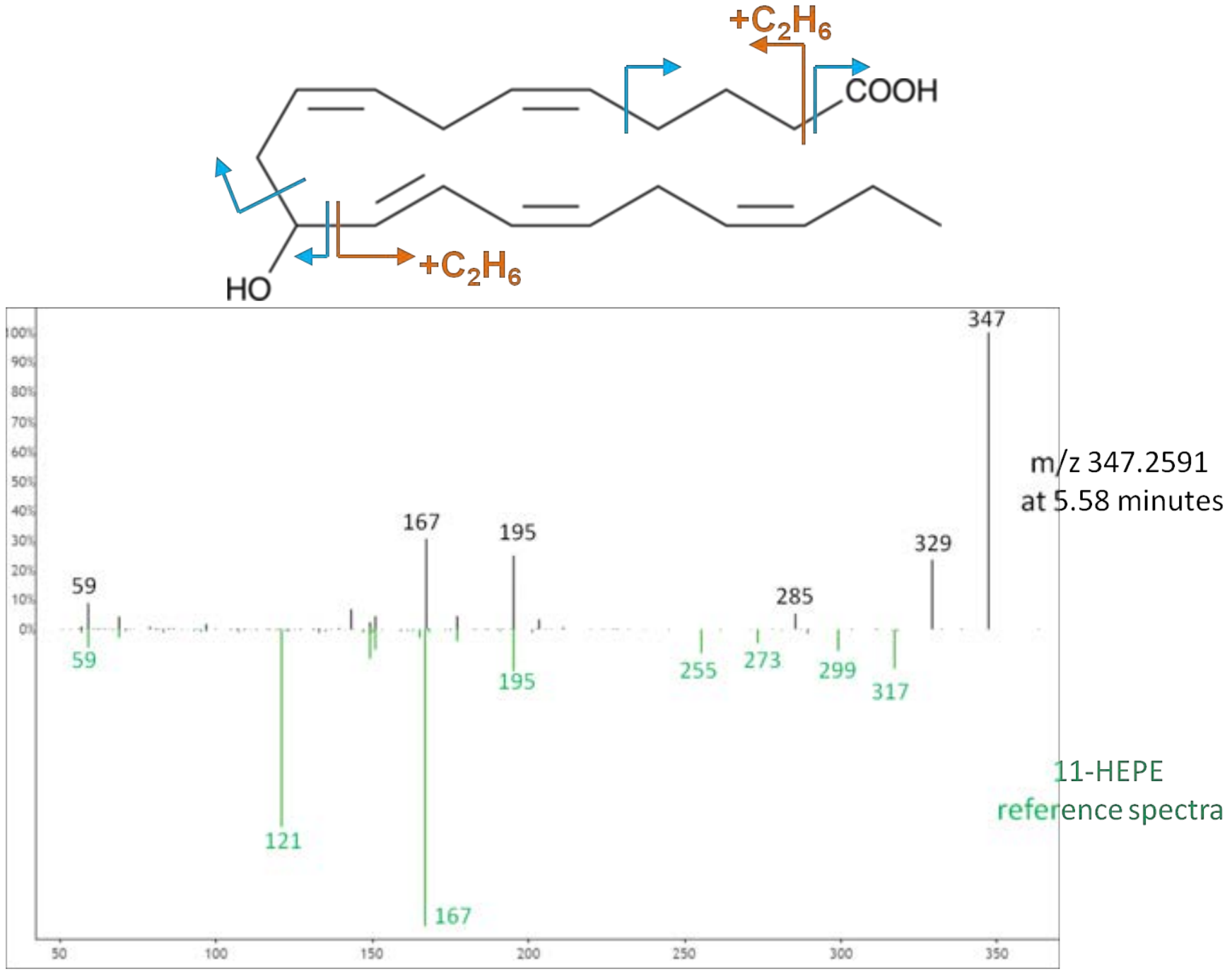

## Figure S5-B

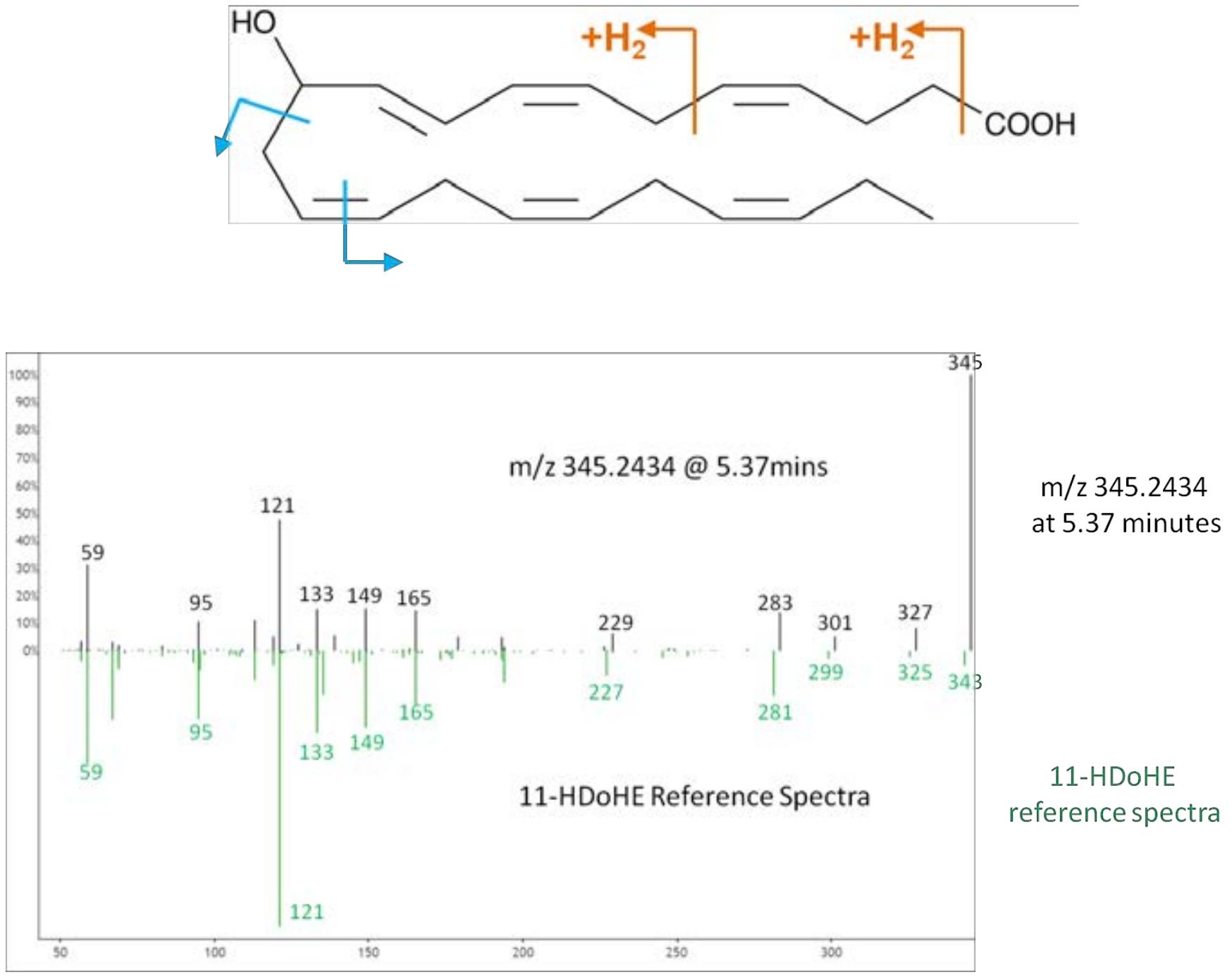

## Figure S5-C

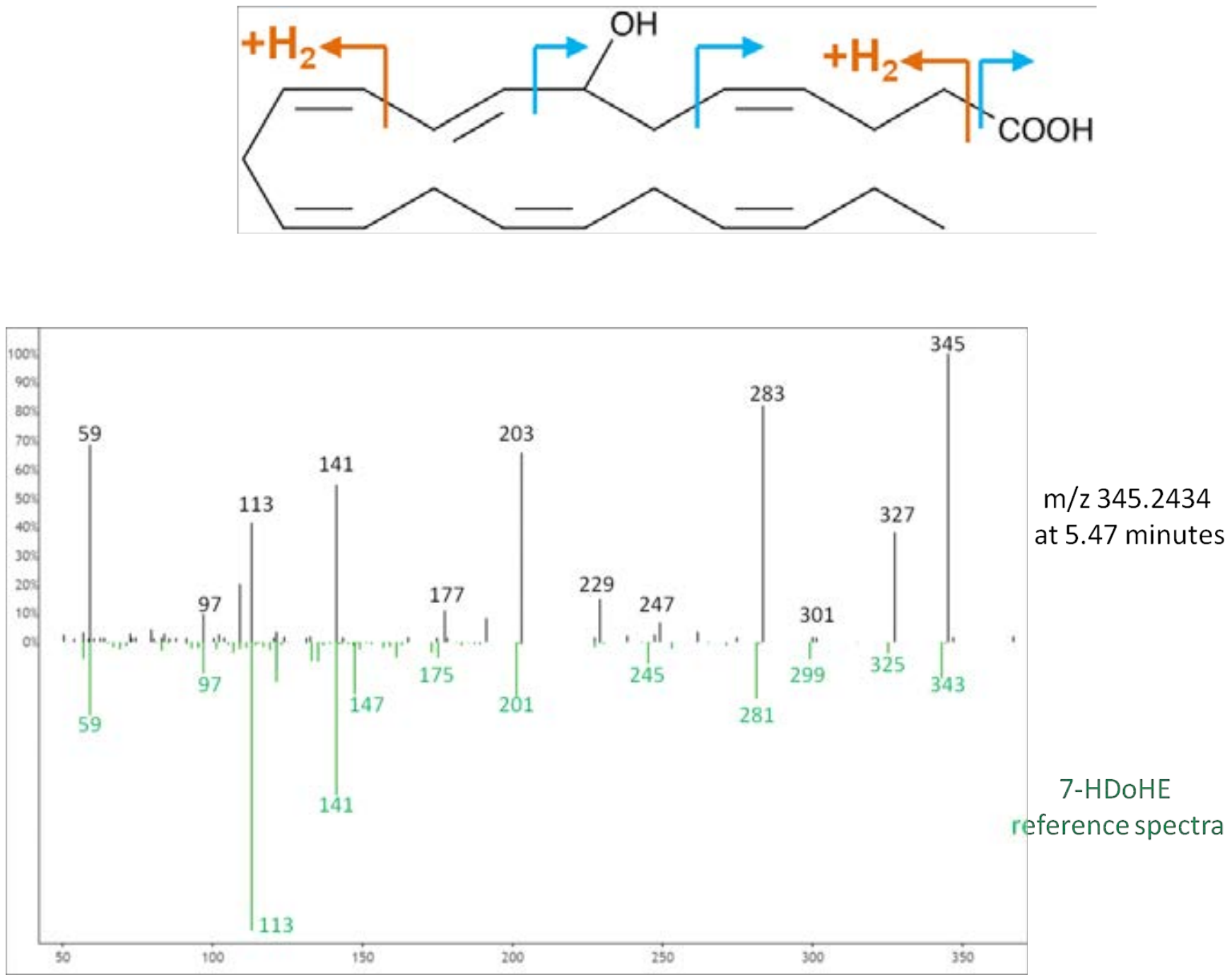

## Figure S5-D

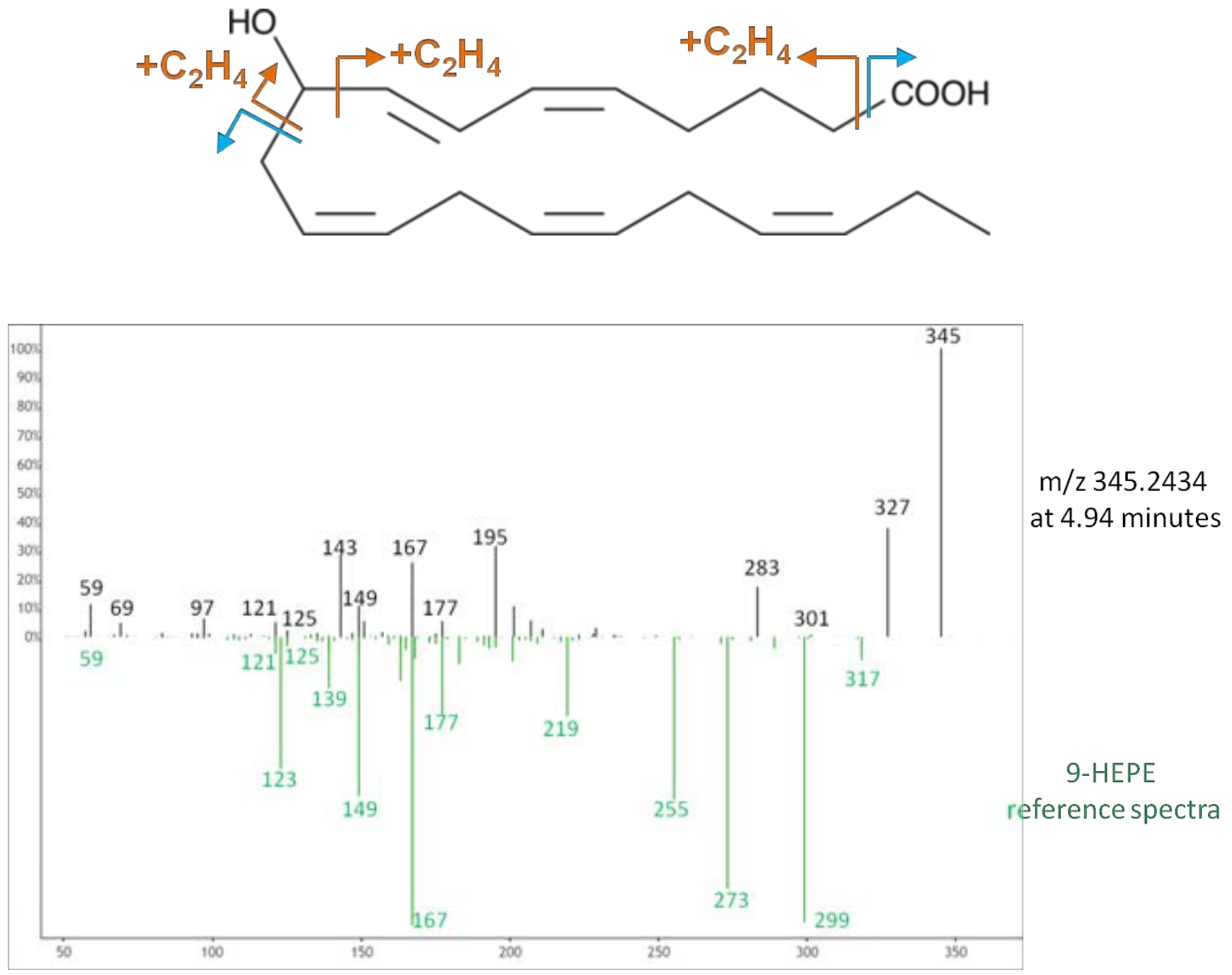

**Figure S5-E**

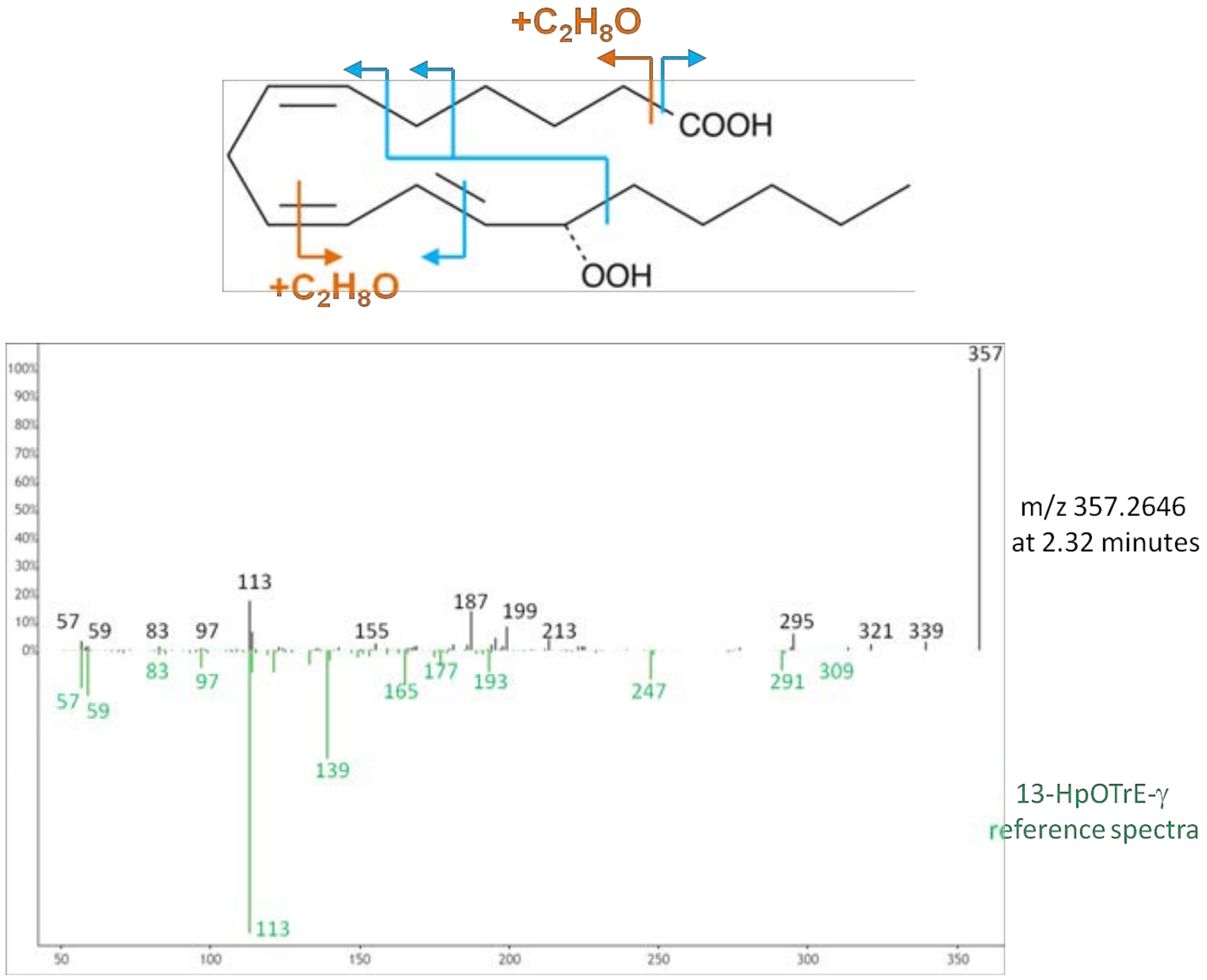